%% file: report_1104.tex
\documentclass{elsart}
\usepackage{graphicx}

\usepackage{amssymb}

\newcommand{\norm}{\frac{2}{(2\pi)^3}}
\newcommand{\qd}[1]{\left[ #1 \right]}
\newcommand{\td}[1]{\left( #1 \right)}

\newcommand{\itg}[1]{\norm \int d^3k f_{#1}(k)}
\newcommand{\ienne}[1]{\itg{n} #1}
\newcommand{\ipi}[1]{\itg{p} #1}
\newcommand{\itau}{\itg{\tau}g(k,\Lambda)}
\newcommand{\rz}{\rho_{_0}}
\newcommand{\ra}{\td{ \frac{\rho}{\rho_{_0}} }}
\newcommand{\umd}[1]{ \td{ \frac{1}{2}+x_{#1} } }

\newcommand{\inew}[1]{\mathcal{I}_#1 }

\begin{document}
\ifx\href\undefined\else\hypersetup{linktocpage=true}\fi 

\begin{frontmatter}
\title
{Reaction Dynamics with Exotic Nuclei} 
\author[LNS]{V. Baran\thanksref{bar}},
\author[LNS]{ M.Colonna},
\author[TEX]{V.Greco},
\author[LNS]{M.Di Toro\corauthref{dit}},

\corauth[dit]{ditoro@lns.infn.it}
\address[LNS]{Laboratori Nazionali del Sud INFN, Via S. Sofia 62,
I-95123 Catania, Italy\\
and Dipartimento di Fisica e Astronomia, Universit\`a di Catania}%
\address[TEX]{Cyclotron Institute, Texas A\&M Univ., College Station, USA}%

\thanks[bar]{On leave from NIPNE-HH and Bucharest University, Romania}

\vskip 0.2cm
{\it Iste magistrorum locus est simul et puerorum,  mittunt quando volunt
 hic res quas perdere nolunt} \cite{Conques}.
\vskip -0.2cm
\begin{abstract}
We review the new possibilities offered by the reaction dynamics of asymmetric
heavy ion collisions, using stable and unstable beams. We show that it 
represents a rather unique tool to probe regions of highly Asymmetric
Nuclear Matter ($ANM$) in compressed as well as dilute phases,
 and to test the in-medium isovector interaction for high momentum nucleons. 
The focus
is on a detailed study of the symmetry term of the nuclear Equation of State
($EOS$) in regions far away from saturation conditions but always under
laboratory controlled conditions.

Thermodynamic properties of $ANM$ are surveyed starting from
nonrelativistic and relativistic effective interactions. In the relativistic
case the role of the isovector scalar $\delta$-meson is stressed. The 
qualitative new features of the liquid-gas phase transition, 
"diffusive" instability and isospin distillation, are discussed.
The results of ab-initio simulations of n-rich, n-poor, heavy ion 
collisions, using
stochastic isospin dependent transport equations, are analysed
 as a function of beam energy and centrality.
The isospin dynamics plays an important role in all steps of the
reaction, from prompt nucleon emissions to the final fragments. 
The isospin diffusion is also of large interest, due
to the interplay of asymmetry and density gradients.
In relativistic collisions, the possibility of a direct study of the
covariant structure of the effective nucleon interaction is shown.
Results are discussed for particle production, collective flows and
iso-transparency.

Perspectives of further developments of the field, in theory as well as 
in experiment, are presented. 
\end{abstract}

\vskip -0.2cm
\end{frontmatter}
\newpage

\tableofcontents\newpage

\renewcommand{\theequation}{\arabic{section}-\arabic{equation}}
\renewcommand{\thefigure}{\arabic{section}-\arabic{figure}}
\setcounter{page}{1}

\include{Chapter-1}

\include{Chapter-2}

\include{Chapter-3}

\include{Chapter-4}

\include{Chapter-5}

\include{Chapter-6}

\include{Chapter-7}

\include{Chapter-8}

\include{Chapter-9}

\include{Bibliography}

\end{document}

%% file: Chapter-1.tex
%\documentclass{elsart}

%\usepackage{graphicx}

%\usepackage{amssymb}

%\begin{document}
\setcounter{page}{1}

\section{Introduction}\label{intro}

\markright{Chapter \arabic{section}: Introduction}

\begin{figure}
\begin{center}
%\epsfysize=6.cm
%\centerline{\epsfbox{iso1.ps}}
\includegraphics*[scale=0.45]{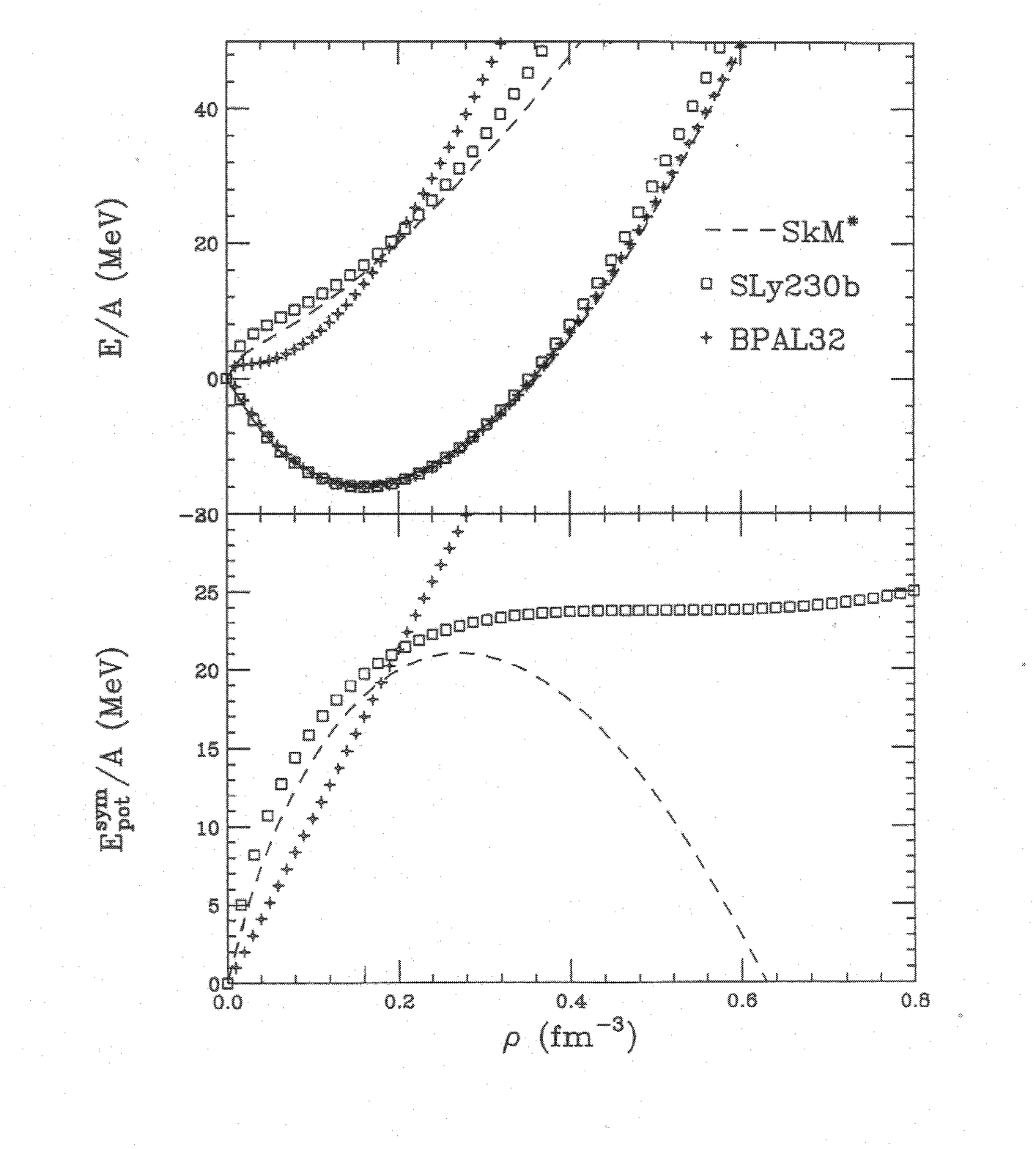}
\caption{$EOS$
for various effective forces.
Top: neutron matter (up), symmetric matter
(down); Bottom: potential symmetry term.}
\label{fig:esym}
\end{center}
%\vskip-1.0cm
\end{figure}

A key question in the physics of
unstable nuclei is the knowledge of the $EOS$ for asymmetric nuclear
matter ($ANM$) away from normal conditions. 
We recall that the symmetry energy at low densities has important effects
on the neutron skin structure, while the knowledge in
high densities region is crucial for supernovae dynamics
and neutron star properties. The paradox is that while we
are planning second and third generation facilities for
radioactive beams our basic knowledge of the symmetry
term of the $EOS$ is still extremely poor.
Effective interactions are obviously tuned to symmetry properties
around normal conditions and any extrapolation can be quite
dangerous. Microscopic approaches based on realistic $NN$
interactions, Brueckner or variational schemes, or on effective
field theories show a rather large variety of predictions.
As an example, in Fig.\ref{fig:esym} we collect the isospin dependence 
of some $EOS$'s
which have {\it the same saturation properties for
symmetric $NM$} (top): $SKM^*$ \cite{VautherinPRC3,KrivineNPA336}, 
 $SLy230b~(SLy4)$ 
 \cite{LyonNPA627,LyonNPA635,LyonNPA665} and $BPAL32$
\cite{BombaciPR242,BombaciPRC55,Rhonote}.

In Fig.\ref{fig:esym} (bottom) we report the density dependence of 
the potential symmetry 
contribution for the three different 
effective interactions.
While all curves obviously cross at normal density $\rho_0$, quite 
large differences are present for values, slopes and curvatures
in low density and particularly in high density
regions.

Moreover even at the relatively well known ``crossing point'' at normal
density the various effective forces are presenting controversial
predictions for the momentum dependence of the fields acting on 
the nucleons and consequently for the splitting of the neutron/proton
effective masses, of large interest for nuclear structure and dynamics.

In the recent years under the stimulating perspectives offered from
nuclear astrophysics and from the new Radioactive Ion Beam ($RIB$) facilities
a relevant activity has started in the field of the isospin degree of freedom
in heavy ion reactions, see the refs.\cite{BaoJMPE7,Isospin01,DitoroEPJA13}.
Here we review the field trying to pin down the most interesting theory
questions and eventually the related key observables.
We will follow non-relativistic and relativistic approaches to construct
effective interactions. In general the physics is not dependent
on the theoretical framework, however we will see some genuine pure
relativistic effects in the dynamics of the isovector part of the $EOS$.

In Section \ref{eos}, we look at the density dependence of the symmetry term
around saturation in order to relate slope and curvature
to physics properties of exotic nuclei, bulk densities, neutron distributions
and monopole frequencies. The momentum dependence of the interactions in the 
isovector channel is thoroughly analysed discussing the expected effects
on the energy-slope of the Lane Potential at normal density and
on the symmetry field seen by high momentum nucleons.

Section \ref{rpa} is devoted to the study of symmetry properties at
low density, i.e. to
the Liquid-Gas Phase
Transition in Asymmetric Matter. We start with a thermodynamical
approach to the mechanical (vs. density variations) and chemical
(vs. concentration variations) instability regions. We show then that
the line of maximum instability for a two component system
goes along a mixing of the two directions with a ``mixing angle''
that depends on microscopic properties of the interaction among
the components.

We pass then to a dynamical approach of great relevance since
it leads to the space-time properties of the unstable normal modes
which give rise to fragments inside the extended spinodal boundary. 
Characteristic coupled dispersion relations must be solved.
We remind that the coupling of the isoscalar and isovector
collective response is well known for stable modes \cite{GR98,HamamotoPRC60}
in neutron rich exotic nuclei, 
with transition densities that show a mixed nature. Here we extend the same
results to the unstable responses of interest in fragmentation reactions. 
The mixed nature of the unstable
density oscillations will naturally lead to the {\it isospin distillation}
effect, i.e. a different {\it concentration} ($N/Z$) between the liquid
and the gas phase 
\cite{MuellerPRC52,BaranPRL86,ColonnaPRL88,MargueronPRC67,ChomazPR389}.
This is indeed the main new feature of the liquid-gas phase transition
which is behind the cluster formation in the expansion stage of the reaction
dynamics. We can expect that quantitatively this effect will be dependent
on properties of the symmetry energy in very dilute matter, well below the
saturation point.

The early emission of particles in heavy ion collisions is
relevant for understanding both the reaction dynamics and the mechanism for
particle production. As discussed before, the density 
behavior of the symmetry potential is strictly related to the value of the
pressure that one obtains in asymmetric nuclear matter.   
The early reaction dynamics is mainly governed by the pressure of 
the excited nuclear matter produced during the initial stage of the collision 
\cite{HandzyPRL75,BaoPRC64,DanielNPA685}. Therefore 
we can expect important symmetry effects
on transport properties ( fast particle emission, collective flows)
in the asymmetric $NM$ that will
be probed by Radioactive Beam collisions at Fermi and intermediate energies.
In Section \ref{fastflows}
we review several observables that are related to the early particle
emission: two-nucleon correlation functions, light cluster formation,
transverse and elliptic collective flows and we will focus on the difference
observed between neutron and proton behavior.  
Results of collision simulations
based on microscopic kinetic equation are discussed, starting from 
realistic effective
interactions widely used for symmetric systems. Different 
parametrizations in the momentum dependence of the isovector channel
are tested,
taking care that the symmetry energy, including its density dependence,
will be not modified. We study in particular the transport effect of
the sign of the $n/p$ effective mass splitting $m^*_n-m^*_p$ in
asymmetric matter at high baryon and isospin density.

In Section \ref{fermi}, we continue the discussion of nuclear 
reactions in the Fermi 
energy domain. We remind that this is the transition
region between a dynamics mainly driven by the mean-field,  
 below $15-20~AMeV$, 
and the one where the nucleon-nucleon collisions play a central role, 
above $100~AMeV$. We can have then a very rich variety of dissipative
reaction mechanisms, with related different isospin dynamics.
One of the new distinctive features is the enhanced production of
Intermediate Mass Fragments $(IMF, 3 \le Z \le 20)$. We analyse then
with great detail the isovector channel effect on the onset of the 
fragmentation
mechanism and on the isotopic content of the produced fragments.
In the reaction simulations it is essential the use of a 
{\it Stochastic Transport Approach} since fluctuations, instabilities and 
dynamical branchings are very important in this energy range.

As we may see, even from Fig.\ref{fig:esym}, the potential symmetry term 
for various effective interactions shows quite different behaviors 
in the region around normal density and at very low densities, where we enter
the spinodal zone and the cluster formation initiates. 
While around $\rho_0$ the density dependence becomes steeper
when we go from asy-soft (Skyrme-like) to asy-superstiff 
(this suggests the names)
at subsaturation densities it manifests an opposite trend.
We will see that with the centrality of the collision we can have different 
scenarios for
fragment production, from the growing instabilities of dilute matter in central
reactions to the cluster formation at the interface between low and normal 
density regions in semicentral collisions. We can then expect a large variety
of isospin effects, probing different regions of the symmetry energy
below saturation. 

The traditional approach to nuclear physics starts from non-relativistic
formalisms in which non-nucleonic degrees of freedom are integrated out
giving nucleon-nucleon potentials. Then nuclear matter is described as a 
collection of quantum nonrelativistic nucleons interacting through an 
istantaneous effective potential. Although this approach has had a great
success, a more appropriate set of degrees of freedom consists of strongly
interacting effective hadron fields, mesons and baryons. These variables
are the most efficient in a wide range of densities and temperatures and
they are the degrees of freedom actually observed in experiments, in
particular in Heavy Ion Collisions at intermediate energies. 
Moreover this framework appears in any case a fundamental ``Doorway Step''
towards a more microscopic understanding of the nuclear matter.
Relativistic contributions to the isospin physics for static properties 
and reaction dynamics will
be discussed in the second part of the report, Sections \ref{qhd}, 
 \ref{relin} and \ref{reldyn}.

The $QHD$ (Quantum-Hadro-Dynamics) effective field model represents a 
very successful attempt to 
describe, in a fully consistent relativistic picture, equilibrium and 
dynamical properties of nuclear systems at the hadronic 
level \cite{WaleckaAP83,SerotANP16,SerotIJMPE6}.
In this report we mostly focus our attention on the dynamical response of 
Asymmetric Nuclear Matter ($ANM$). We present a relativistic kinetic
 theory with the aim
of a transparent connection between the collective and reaction dynamics 
and the 
coupling to various channels of the nucleon-nucleon interaction. 
We show that the same isospin physics described in detail at the
non-relativistic level can be naturally reproduced. Moreover some new
genuine relativistic effects will be revealed, due to the covariant structure
of the effective interactions.

One of the main points of our discussion is the relevance of the coupling
to a scalar isovector channel, the effective $\delta[a_0(980)]$ meson, not
considered in the usual nuclear structure studies
\cite{LiuboPRC65,GrecoPRC67}. 
A related 
feature of interest is the 
dynamical treatment of the Fock terms.
We like to note that recently , see the conclusions of the refs.
\cite{FurnstNPA706,MadlandNPA741}, 
the $\delta$-field coupling has been reconsidered
as an interesting improvement of covariant approaches, in the framework
of an
$Effective~ Field~ Theory$  as a relativistic $Density~ Functional~ Theory$,
since contribution to this channel are mainly
coming from correlation effects. One of the main tasks of
our work is just to try to select the dynamical observables more sensitive
to it. 

In Section \ref{qhd}, this extension of the $QHD$ model is presented.
Equilibrium properties are discussed, like the nuclear $EOS$ and
the corresponding thermodynamical instabilities. Particular attention is
devoted to the expectations for the splitting of the nucleon effective
masses in asymmetric matter, with a detailed analysis of the
relationship between the Dirac masses of the effective field approach and
the Schr\"odinger masses of the non-relativistic models.
A relativistic Dirac-Lane potential is deduced, which shows a structure
very similar to the Lane potential of the non-relativistic optical
model, but now in terms of isovector self-energies and coupling constants.
As a general trend we have a close parallelism between relativistic
and non-relativistic results, as it should be since the main physics is the 
same. We always note this point but we also stress several new features
coming from a consistent use of a field theory approach.

We follow a Relativistic Mean Field ($RMF$) approximation that is allowing
more physics transparent results, often even analytical. We will
always keep a close connection to the more microscopic 
Dirac-Brueckner-Hartree-Fock ($DBHF$) approaches, in their extension to
asymmetric matter, \cite{LeePRC57,JongPRC57,HofmannPRC64,DalenarX0407}
It is
well known that correlations are naturally leading to a density
dependence of the coupling constants, see ref.\cite{JongPRC57} for the
$DBHF$ calculations
and ref.\cite{GrecoPRC64} just for the basic Fock correlations.
Within the $RMF$ model we can get a clear qualitative estimation of the 
contribution
of the various fields to the nuclear dynamics. The price to pay is that
when we try to get quantitative effects we are forced to use different
sets of couplings in different baryon density regions.
In this case we use the $DBHF$ results as guidelines. In particular
for the controversial $\delta$-meson field, expected to be important at
densities above saturation, we fix the corresponding coupling
 from the analysis
of refs.\cite{HofmannPRC64,DalenarX0407}, where it actually appears not 
strongly
density dependent in a wide range of baryon densities.

An important outcome of our work is to show that the two effective couplings,
vector and scalar, in the isovector channel are influencing in a different way
the static (symmetry energy) and dynamic (collective response, reaction 
observables) properties
of asymmetric nuclear matter. All that will open new possibilities for
a phenomenological determination of these fundamental quantities.

In  Section \ref{relin}, we discuss a fully 
relativistic Landau Fermi liquid theory 
based on the 
Quantum Hadro-Dynamics ($QHD$) effective field picture of Nuclear Matter.
 From the linearized kinetic equations we get the dispersion
relations of the propagating collective modes. 
The relation between static properties and the collective response
is analysed stressing the different role of the various channels
present in the effective nuclear interaction.
We focus our attention on the 
dynamical effects of the interplay between scalar and vector field 
contributions. An interesting ``mirror'' structure in the form of the 
dynamical response in 
the isoscalar/isovector degree of freedom is revealed, with a complete 
parallelism in the role respectively played by the compressibility and the
symmetry energy.
In particular we study the influence of a scalar-isovector 
channel (coupling to a $\delta$-meson-like
effective field) on the collective response of asymmetric nuclear 
matter ($ANM$).
Interesting contributions are found on the propagation of isovector-like
modes at normal density and on an expected smooth transition to isoscalar-like
oscillations at high baryon density.

Important ``chemical'' effects on the neutron-proton structure of the 
normal modes are
shown. For dilute $ANM$ we have the isospin distillation mechanism of the 
unstable isoscalar-like oscillations, as already shown in Section \ref{rpa} 
in a 
non-relativistic frame, 
while at high baryon density we predict
an almost pure neutron wave structure of the propagating sounds.

Results for relativistic Heavy Ion Collisions ($HIC$) in the $AGeV$
beam energy region are presented in Section \ref{reldyn}.
We recall that intermediate energy $HIC$'s represent the only way
to probe in terrestrial laboratories the in-medium effective interactions
far from saturation, at high densities as well as at high momenta.
Within a relativistic
transport model it is shown that the
isovector-scalar $\delta$-meson, which affects the high density
behavior of the symmetry term and the nucleon effective mass splitting, 
influences the
isospin dynamics. The effect is largely enhanced by a relativistic
mechanism related to the covariant nature of the fields
contributing to the isovector channel. The possibility is emerging of
a direct measurement of the Lorentz structure of the effective
nuclear interaction in the isovector channel. 

Quantitative calculations are discussed for collective flows, charged
pion production and isospin stopping. Asymmetric systems where some data
are available have been studied. Although the data are mostly of inclusive type
(and the colliding nuclei not very neutron rich), quite clearly a dependence
of some observables on charge asymmetry is emerging.
Very sensitive quantities appear to be the elliptic flow, related to 
the time scale of the particle emissions, and the isospin transparency in 
central collisions.

Finally in Section \ref{out} a general outlook for theory and 
experiment is presented.
Particular attention is paid to a selection of the expected most sensitive
observables to the isospin dynamics, in heavy ion reactions from low
to relativistic energies.

%\include{rep_bib}

%\end{document}

%% file: Chapter-2.tex
%\documentclass{elsart}
%\usepackage{epsfig}

%\usepackage{graphicx}

%\usepackage{amssymb}
%\tightenlines

% nuovi comandi
% 2 su 2pigreco al cubo
%\newcommand{\norm}{\frac{2}{(2\pi)^3}}
% parentesi quadre
%\newcommand{\qd}[1]{\left[ #1 \right]}
% parentesi tonde
%\newcommand{\td}[1]{\left( #1 \right)}

%\newcommand{\itg}[1]{\norm \int d^3k f_{#1}(k)}
% integrale di fn
%\newcommand{\ienne}[1]{\itg{n} #1}
% integrale di fp
%\newcommand{\ipi}[1]{\itg{p} #1}
% integrale di fn g
%\newcommand{\ien}{\ienne{g(k,\Lambda)}}
% integrale di fp g
%\newcommand{\iz}{\ipi{g(k,\Lambda)}}

%\setlength{\unitlength}{1cm}

% integrale di ftau g
%\newcommand{\itau}{\itg{\tau}g(k,\Lambda)}
% integrale di ftau g
%\newcommand{\itaup}{\itg{\tau ^\prime} g(k,\Lambda)}

% rozero
%\newcommand{\rz}{\rho_{_0}}
% rho su rozero
%\newcommand{\ra}{\td{ \frac{\rho}{\rho_{_0}} }}
% densita' di energia per A e B
%\newcommand{\ene}[1]{\qd{
% \td{\frac{1}{2}x_{#1}}\rho^2
%-\td{\frac{1}{2}+x_{#1}} \td{\rho_n^2+\rho_p^2} }}
% 0.5 + x0(x3)
%\newcommand{\umd}[1]{ \td{ \frac{1}{2}+x_{#1} } }

%\newcommand{\inew}[1]{\mathcal{I}_#1 }

%\begin{document}

\setcounter{figure}{0}
\setcounter{equation}{0}

\section{Symmetry term effects on compressibility, saturation density
and nucleon mean field} \label{eos}

\markright{Chapter \arabic{section}: eos}

In asymmetric matter the energy per nucleon, i.e. the equation of state,
will be a functional of the total ($\rho=\rho_n+\rho_p$) and isospin
($\rho_3=\rho_n-\rho_p$) densities.
In the usual parabolic form in terms of the asymmetry parameter
$I \equiv \rho_3/\rho = (N-Z)/A$ we can define 
a symmetry energy $ {E_{sym} \over A} (\rho)$:

\begin{equation}
{E \over A} (\rho,I) = {E \over A} (\rho) + {E_{sym} \over A} (\rho)
 ~ I^2.
\label{sym}
\end{equation}

The symmetry term gets a kinetic contribution directly from the
basic Pauli correlations and a potential contribution from the properties of
the isovector part of the effective nuclear interactions in the medium.
Since the kinetic part can be exactly evaluated we can separate
the two contributions, reducing the discussion just to a function $F(u)$
of the density $u \equiv \rho/ \rho_0$ linked to the interaction:

\begin{equation}
\epsilon_{sym} \equiv {E_{sym} \over A} (\rho) = {\epsilon_F(\rho) \over 3} 
+ {C \over 2} ~ F(u),
\label{kipot}
\end{equation}

with  $F(1)=1$, where $\rho_0$ is the
saturation density and the parameter $C$ is of the order $C \simeq 32MeV$
to reproduce the $a_4$ term of the Bethe-Weisz\"acker mass formula.
The validity of the parabolic form Eq.(\ref{sym}) comes directly form the
isospin structure of the nucleon-nucleon interaction. The effects of
many-body correlations are not much affecting this behaviour, see
ref.\cite{LeePRC57,Lombiso}, well
verified in the nuclear data systematics \cite{MyersADNDT}. Corrections
are neglible up to very high densities, of the order of $5\rho_0$.
The corrections in the kinetic term are also exactly evaluated to be
of $4\%$ for any density.

A traditional expansion to second order around normal density is used 
\cite{LopezNPA483,BaoNPA681,LiuboPRC65}
\begin{equation}
\epsilon_{sym} \equiv {E_{sym} \over A} (\rho) = a_4 + {L \over 3}
~\Big({{\rho - \rho_0} \over \rho_0} \Big) + {K_{sym} \over {18}}
~\Big({{\rho - \rho_0} \over \rho_0} \Big)^2 ,
\label{exp}
\end{equation}
in terms of a slope parameter
\begin{equation}
L \equiv 3\rho_0 \Big({{d\epsilon_{sym}}\over{d\rho}} \Big)_{\rho=\rho_0}
~=~{3 \over \rho_0} P_{sym}(\rho_0)
\label{psym}
\end{equation}
simply related to the {\it symmetry pressure} 
 $P_{sym}= {\rho}^2 d\epsilon_{sym}/d\rho$ at $\rho_0$, and a curvature
parameter
\begin{equation}
K_{sym} \equiv 9 \rho_0^2 \Big({{d^2\epsilon_{sym}}\over{d^2\rho}} 
\Big)_{\rho=\rho_0} ,
\label{csym}
\end{equation}
a kind of symmetry compressibility.
We remark that our present knowledge of these basic properties of
the symmetry term around saturation is still very poor, see the
recent analysis in ref.\cite{FurnstNPA706} and refs. therein. 
In particular we note the uncertainty on the symmetry pressure
at $\rho_0$, of large importance for structure calculations.
These points,
in connection to the possibility of getting a new insight from reaction
data, will be largely discussed in this report.

We first mention some simple considerations
on asymmetry effects on equilibrium density and compressibility, 
 observables related respectively to  bulk densities  and 
monopole resonances in  medium heavy nuclei \cite{YoshidaPRC58}.  
From a linear expansion around the value at symmetry, $I=0$, we get for
the variation of saturation density 
\begin{eqnarray}
\Delta \rho_0(I) = - {{9\rho_0^2} \over K_{NM}(I=0)}~
{d \over {d\rho}} {\epsilon_{sym}} (\rho) \Big\vert
_{\rho=\rho_0}I^2  &&\nonumber\\
 = - {{3\rho_0L} \over {K_{NM}(I=0)}}I^2~<~0,
\label{derho}
\end{eqnarray}
where $K_{NM}(I=0)$ and $\rho_0$ are respectively compressibility
and saturation density of symmetric $NM$.

\begin{figure}
\begin{center}
%\epsfysize=6.cm
%\centerline{\epsfbox{iso1.ps}}
\includegraphics*[scale=0.40]{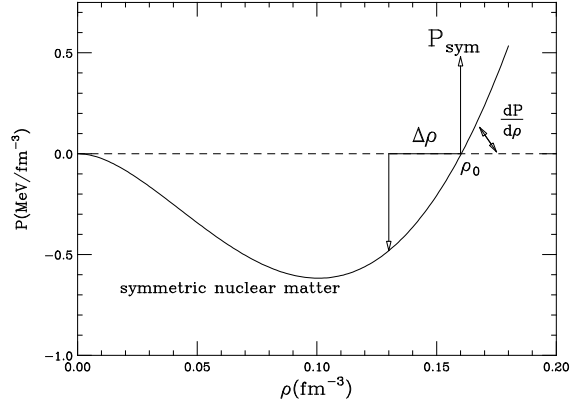}
\caption{Geometric picture of the lowering of the saturation density
in asymmetric matter, Eq.(\ref{derho}).}
\label{fig:rhovar}
\end{center}
%\vskip-1.0cm
\end{figure}

The Eq.(\ref{derho}) has an intuitive {\it geometrical} meaning,
 which is qualitatively
shown in Fig.\ref{fig:rhovar}.
Asymmetry brings an extra pressure $P_{sym}= {\rho}^2 d\epsilon_{sym}/d\rho$
that can be compensated just moving to the left the saturation
point ($P=0$) of the quantity $P_{sym}/(dP/d\rho)$ at $\rho_0$
( we recall that, for symmetric matter at $\rho_0$, 
$9(dP/d\rho) = K_{NM}(I=0)$).
For the compressibility shift we have, after some algebra,
\begin{eqnarray}
\Delta K_{NM}(I) = 9\rho_0 \Big[ \rho_0 {d^2 \over {d\rho^2}}
 - 2 {d \over {d\rho}} \Big] {\epsilon_{sym}} (\rho) \Big\vert
_{\rho=\rho_0}I^2 &&\nonumber\\
 = [K_{sym}-6L]I^2 ~<~0,
\label{decom}
\end{eqnarray}
where we note the interplay between slope and curvature of the symmetry term.

The Eq.(\ref{derho}), and the related geometrical interpretation 
Fig.\ref{fig:rhovar}, is a particular case (for $P=0$) of the general
variation of the density in asymmetric matter corresponding to a fixed
pressure $P$, see the discussion in refs.\cite{PrakashPRC32,PrakashPRC36}.
The corresponding compressibility shift, Eq.(\ref{decom}), is also
evaluated in the isobaric case, where there is the possibility
 to have some experimental information
from Giant Monopole Resonance ($GMR$) data in charge asymmetric nuclei.
Actually Eq.(\ref{decom}) represents an approximate form where
higher order terms in density variations are neglected, see 
refs.\cite{PrakashPRC32,PrakashPRC36}.
The general negative sign of the compressibility shift, Eq.(\ref{decom}),  
is a natural
consequence of Eq.(\ref{derho}), i.e. of the fact that the
saturation density is decreasing with charge asymmetry.

In order to have a quantitative idea, we now show explicitly the influence
on the $L, K_{sym}$ parameters of a different
density dependence in the potential part of the symmetry energy
around saturation, i.e. of the function $F(u)$ of Eq.(\ref{kipot}):
$$
L={2 \over 3}\epsilon_F + {3 \over 2} C {d \over du} F(u) \Big\vert_{u=1}
$$
$$
K_{sym}= -{2 \over 3}\epsilon_F + {9 \over 2} C {{d^2} \over {du^2}} 
F(u) \Big\vert_{u=1}
$$
We obtain the very instructive Table \ref{symslope} for 
various functional forms 
$F(u), u \equiv \rho/\rho_0,$
around $\rho_0$.

\vskip 0.5cm
\begin{table}
\begin{center}
\caption{Symmetry term at saturation}
\begin{tabular}{ c c c c } \hline
$ F(u)$  & $ L $ &~ $ K_{sym} $ &~$ [K_{sym}-6L] $ \\ \hline
$ const=1 $ & $ +25MeV $ & $ -25MeV $ & $ -125MeV $ \\ \hline
$ \sqrt{u} $ & $ +49MeV $ & $ -61MeV $ & $ -355MeV $ \\ \hline
$ u $ & $ +75MeV $ & $ -25MeV $ & $ -475MeV $ \\ \hline
$ u^2/(1+u) $ & $ +100MeV $ & $ +50MeV $ & $ -550MeV $ \\ \hline
\end{tabular}
\label{symslope}
\end{center}
\vskip 0.5cm
\end{table}

The choices in the table of the $F(u)$ behaviors are in fact not arbitrary. 
They 
reflect the wide spectrum of theory predictions for effective forces
in the isovector channel, as discussed in detail in the rest of the report. 
The constant trend of the first row
(around saturation density) is
typical of Skyrme-like forces, the $\sqrt{u}$ behavior is obtained in
variational approaches with realistic $NN$ interactions 
\cite{WiringaPRC38,AkmalPRC58},
the linear dependence in Brueckner-Hartree-Fock ($BHF$)
 non relativistic calculations \cite{BombaciPR242,BombaciPRC55,Lombiso}
 as well as 
in Relativistic Mean Field ($RMF$), \cite{YoshidaPRC58}, and
Dirac-Brueckner-Hartree-Fock ($DBHF$), \cite{LeePRC57}, approaches. Finally
the more repulsive, nearly parabolic, dependence of the fourth row,  
\cite{PrakashPR280,BaoPRL85}, 
can be related to non-relativistic predictions with three-body forces,
either extended $BHF$ or variational \cite{Lombiso,FantoniPRL87} as well as
to other relativistic $DBHF$ estimations, refs. \cite{JongPRC57,HofmannPRC64},
or $RMF$ with scalar isovector meson-like contributions 
\cite{LiuboPRC65,GrecoPRC67}. 

Fixing the same compressibility at saturation for symmetric matter
$K_{NM}(I=0)$, the relativistic effective forces always
predict larger shifts in both equilibrium density and compressibility
for asymmetric matter. This can be seen also in the comparison with 
Skyrme-like 
interactions in the finite nuclei calculations shown 
in ref.\cite{YoshidaPRC58}, even at
relatively small charge asymmetries. 
There are therefore good chances of obtaining some direct experimental 
indications from $GMR$ measurements in $N \not= Z$ nuclei. A recent
systematic study of the isospin dependence of $GMR$'s
in $Sn$ isotopes, \cite{LuiPRC70}, seems to reveal a relatively large 
decrease of the Giant Monopole centroid with increasing asymmetry,
more in agreement with a $stiff$ behavior of the symmetry term 
around saturation, as expected in the relativistic models.
We note that the relation between the ``symmetry compressibility'',
 Eq.(\ref{csym}), 
 and the variation of the compressibility of asymmetric matter is not
trivial: we can have cases where an increase of $K_{sym}$
actually corresponds to a softening of the Equation of State of 
asymmetric matter.

It is finally instructive to evaluate the density gradient of the symmetry
pressure as a function of the slope and curvature of the symmetry
term:
\begin{equation}
{d \over d\rho} P_{sym} = {2 \over 3}L + {1 \over 9}K_{sym},
\label{pressgrad}
\end{equation}
that around normal density gives
$$
{d \over d\rho} P_{sym}   = {10 \over 27}\epsilon_F + C \Big[{d \over du} + 
  {1 \over 2}{{d^2} \over {du^2}}\Big] F(u) \Big\vert_{u=1}
$$
A stiffer symmetry term in general enhances the pressure
gradient of asymmetric matter. We can expect direct effects on the
nucleon emissions in the reaction dynamics, fast particles and collective
flows. Moreover due to the different fields seen by neutrons and protons,
 we shall observe even specific isotopic effects. This 
point will be analysed in detail in Sect.\ref{fastflows}.

\subsection{The symmetry term of Skyrme forces}

Since in this report we will  often show reaction results
from non-relativistic kinetic equations with Skyrme forces 
\cite{VautherinPRC3}, we
will expand a little the discussion on the isospin dependence of these
widely used effective interactions.
In a Skyrme-like parametrization  the symmetry
term has the form:
\begin{equation}
\epsilon_{sym} \equiv {E_{sym} \over A} (\rho) = {\epsilon_F(\rho) \over 3} 
+ {C(\rho) \over 2} ~ {\rho \over \rho_0}
\label{kipotsky}
\end{equation}
with the function $C(\rho)$, in the potential part, given by:
\begin{eqnarray}
{C(\rho) \over {\rho_0}} = - {1 \over 4} \Big[t_0 (1 + 2 x_0) + {t_3 \over 6} 
(1 + 2 x_3)~ \rho^\alpha \Big] \nonumber\\
 + {1 \over {12}}
\Big[t_2 (4 + 5 x_2) - 3 t_1 x_1 \Big] \Big( {{3 \pi^2} \over 2} \Big)
^{2/3}~ \rho^{2/3}
\label{sky}
\end{eqnarray}
with $\alpha > 0$ and the usual Skyrme parameters.
We remark that the second term is related to isospin effects on the momentum 
dependence \cite{LyonNPA627}. 
\begin{figure}
\begin{center}
%\epsfysize=6.cm
%\centerline{\epsfbox{iso1.ps}}
\includegraphics*[scale=0.60]{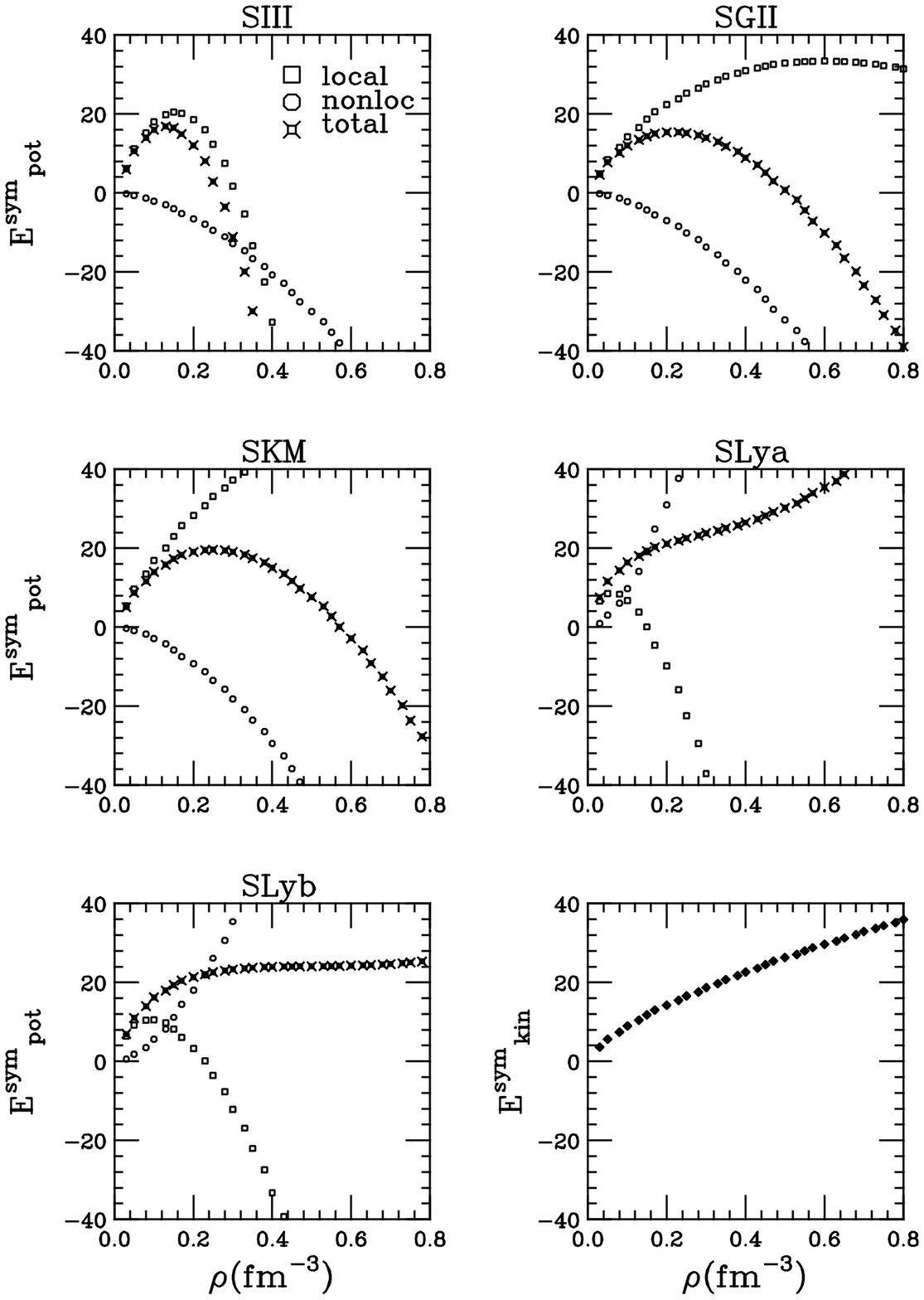}
\caption{Density dependence of the potential symmetry term
for various Skyrme effective forces, see text.
The bottom right panel shows the kinetic contribution}
\label{fig:potsym}
\end{center}
%\vskip-1.0cm
\end{figure}
In the Fig. \ref{fig:potsym} we show the density dependence
of the potential symmetry term of various Skyrme interactions, $SIII$,
 $SGII$, $SKM^*$ (see \cite{KrivineNPA336} and refs. therein) and the 
more recent Skyrme-Lyon forms, $SLya$ and $Slyb$ (or $Sly4$), 
see \cite{LyonNPA627,LyonNPA635}.
We also separately present the local and non-local contributions, first
and second term of the Eq.(\ref{sky}). We clearly see a sharp change
from the earlier Skyrme forces to the Lyon parametrizations, with 
almost an inversion of the signs of the two contributions. The important
repulsive non-local part of the Lyon forces leads to a completely
different behavior of the neutron matter $EOS$, of great relevance 
for the neutron star properties. Actually this substantially modified
parametrization was mainly motivated by a very unpleasant feature
in the spin channel of the earlier Skyrme forces, the collapse of
polarized neutron matter, see discussion in 
\cite{KutscheraPLB325,LyonNPA627,LyonNPA635,LyonNPA665}.

In the Table \ref{skypar} we collect the parameters, of interest for 
Nuclear Matter properties, of the various Skyrme forces used here,
where we have also included the recent $Sly7$ force particularly
tuned for neutron rich systems \cite{LyonNPA665}. Note the transition
to $x_2=-1$ values in the Skyrme-Lyon Forces just to cure the collapse 
of ferromagnetic neutron stars discussed before. In correspondence the
predictions on the isospin effects on the momentum
dependence of the symmetry term are quite different, see Fig.\ref{fig:potsym}.
A very important consequence for the reaction dynamics is the
expected inversion of the sign of the $n/p$ effective mass splitting,
 which will be widely discussed in the next sections.

\vskip 0.5cm
\begin{table}
\begin{center}
\caption{Parameters of Skyrme Forces}
\begin{tabular}{ c c c c c c c} \hline
$ Force$  & $ SIII$ &~ $ SGII$ &~$ SkM^*$ &~$SLya$ &~$SLy4$ &~$SLy7$ 
\\ \hline
$t_0(MeVfm^3) $ & $-1128.75 $ & $-2645.0$ & $-2645.0 $ & $-2490.3$ &
 $-2488.91$ & $-2482.41$ \\ \hline
$t_3(MeVfm^{3+3\alpha}) $ & $14000.0$ & $15595.0$ & $15595.0$ & $13803.0$ &
 $13777.0$ & $13677.0$ \\ \hline
$x_0$ & $ 0.45$ & $ 0.09$ & $ 0.09$ & $1.1318$ & $ 0.8340$ & $ 0.846$ 
 \\ \hline
$x_3$ & $ 1.0$ & $ 0.06044$ & $ 0.0$ & $1.9219$ & $1.3539$ & $ 1.391$ 
 \\ \hline
$t_1(MeVfm^5) $ & $395.0 $ & $340.0$ & $410.0 $ & $489.53$ &
 $486.82$ & $457.97$ \\ \hline
$t_2(MeVfm^5) $ & $-95.0$ & $-41.9$ & $-135.0$ & $-566.58$ &
 $-546.39$ & $-419.85$ \\ \hline
$x_1$ & $ 0.0$ & $-0.0588$ & $ 0.0$ & $-0.8426$ & $-0.3438$ & $-0.511$ 
 \\ \hline
$x_2$ & $ 0.0$ & $1.425$ & $ 0.0$ & $-1.0$ & $-1.0$ & $-1.0$ 
 \\ \hline
$\alpha$ & $ 1 $ & $ 1/6$ & $ 1/6$ & $1/6$ & $1/6$ & $1/6$ 
 \\ \hline
\end{tabular}
\label{skypar}
\end{center}
\vskip 0.5cm
\end{table}

\subsubsection{Mean field and chemical potentials}
%\addtocontents{toc}{\hspace{0.55cm}\thesubsubsection \hspace{0.12cm}
%Mean field and chemical potentials \newline}
We can derive a general Skyrme-like form for 
neutron and proton mean field 
potentials \cite{ColonnaPLB428,BaranNPA632}:
\begin{eqnarray}
U_q ~\equiv~ {{\partial \epsilon_{pot}(\rho_q,\rho_{q'})}\over 
{\partial \rho_q}}~~~~~~~~~~~~~~~~~~~~ \nonumber \\
 =  A\left({\rho \over \rho_0}\right) 
+ B\left({\rho \over \rho_0}\right)^{\alpha +1} 
+ C(\rho)\left({\rho_3 \over \rho_0}\right)\tau_q
+ {1 \over 2} {\partial C \over \partial \rho}
{\rho_3^2 \over \rho_0}~,
\label{field}
\end{eqnarray}
$\epsilon_{pot}$ being the potential energy density.
Here $\rho \equiv \rho_n + \rho_p$ and $\rho_3 \equiv \rho_n - \rho_p$
are respectively isoscalar and isovector densities, 
and $q=n, p$, $\tau_q$ = +1 ($q=n$), -1 ($q=p$).

In the following we will always compare results obtained with
forces that have  {\it the same saturation properties for
symmetric $NM$} \cite{Erice98}.
We will refer to an "$asy-stiff$" $EOS$ (e.g. like $BPAL32$ of Fig.1) 
 when we are 
considering a potential symmetry term linearly increasing with nuclear 
density and to a "$asy-soft$" $EOS$ (e.g. like $SKM^*$ of Fig.1) 
when the symmetry term shows a 
saturation and eventually a decrease above normal density.
In some cases, in order to enhance the dynamical effects, we will
consider also "$asy-superstiff$" behaviours, i.e. with a roughly
parabolic increase of the
 symmetry term above normal density
\cite{Rhonote,PrakashPR280,BaoPRL85}.
We focus our discussion on single particle properties since in
this case the symmetry contribution will be linearly dependent
on the asymmetry of the matter.

\begin{figure}
\begin{minipage}{60mm}
\begin{center}           
%\epsfysize=6.0cm
%\centerline{\epsfbox{snp1.ps}}  
\includegraphics*[scale=0.40]{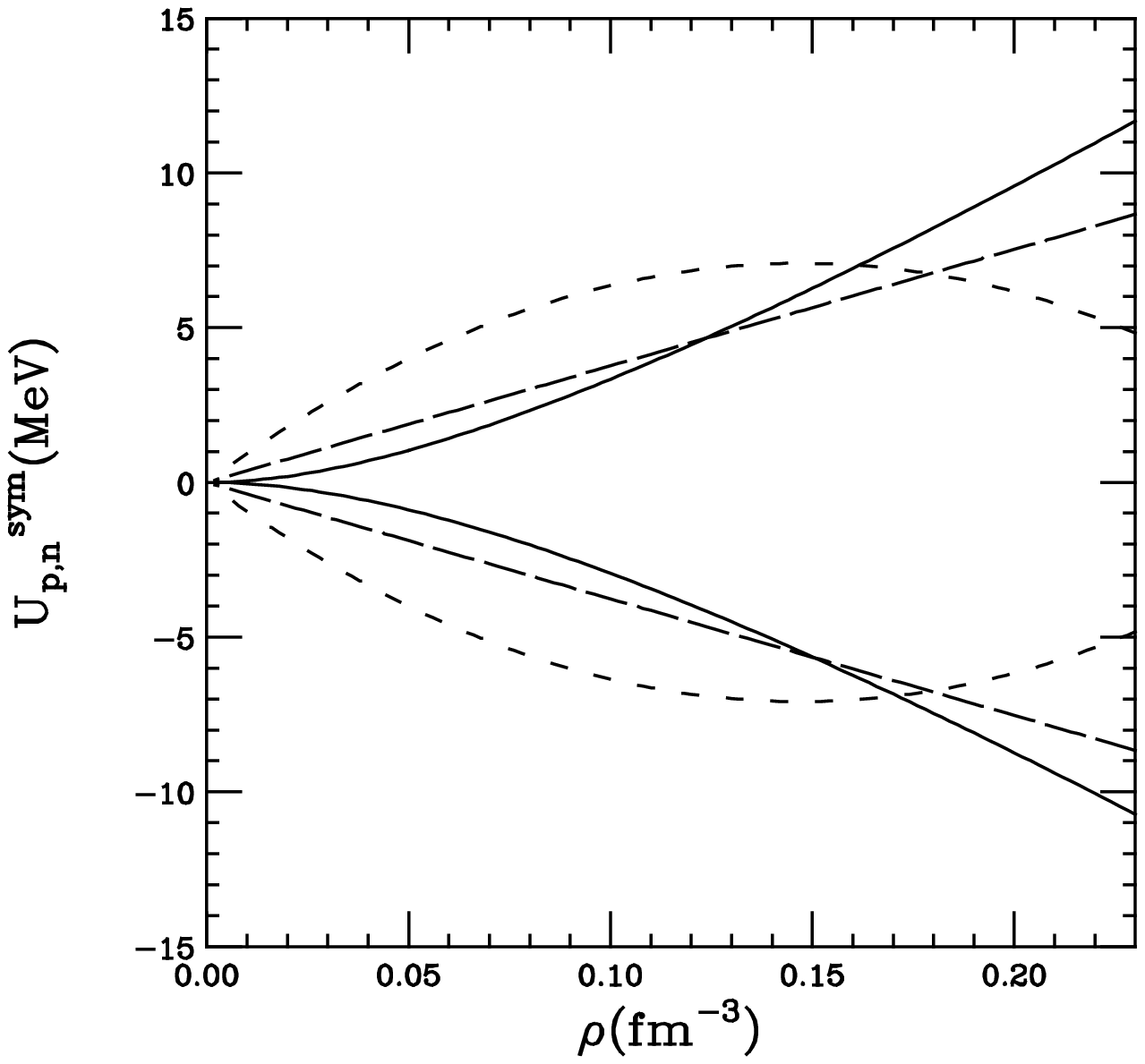}
%\vskip-1.2cm
\caption
{Symmetry contribution to the mean field at $I=0.2$
for neutrons (upper curves) and protons (lower curves):
dashed lines "asy-soft", long dashed 
lines "asy-stiff", solid lines "asy-superstiff"}
\label{fig:mean}
\end{center}
\end{minipage}
\hspace{\fill}
\begin{minipage}{60mm}
\begin{center}
%\epsfysize=5.cm
%\centerline{\epsfbox{snp2.ps}}
\includegraphics*[scale=0.40]{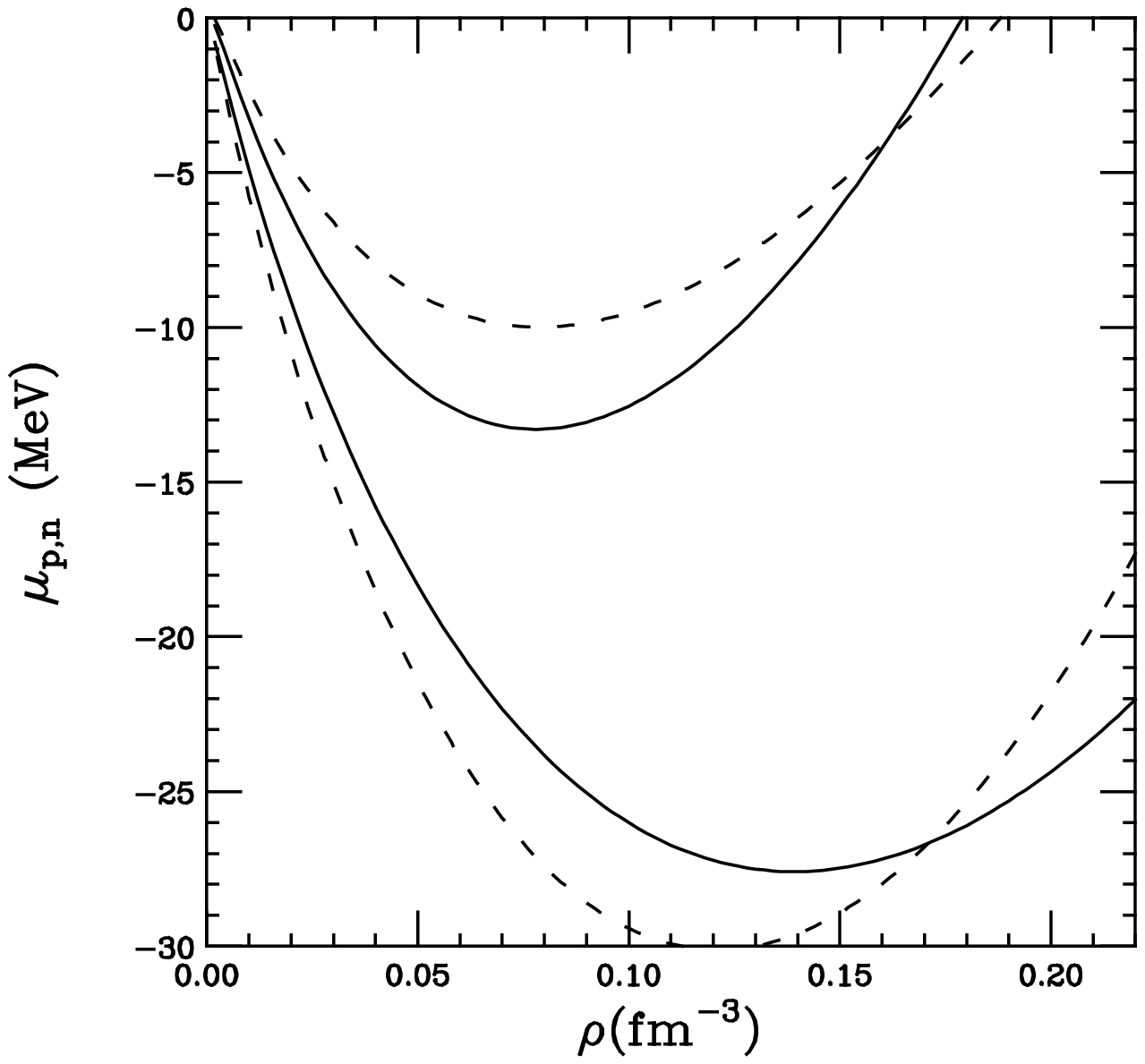}
%\vskip-1.2cm
\caption{Density dependence, for $I=0.2$, 
 of neutron (upper curves) and proton (lower curves) chemical potentials
for asy-superstiff (solid lines) and asy-soft (dashed line) EOS.}
\label{fig:chem}
\end{center}
\end{minipage}
%\vskip-0.6cm
\end{figure}

In Figs.\ref{fig:mean}, \ref{fig:chem} we report, for 
an asymmetry $(N-Z)/A=0.2$ representative of $^{124}Sn$, 
the density dependence of the 
symmetry contribution to the mean-field potential (Fig.\ref{fig:mean}, 
second part of Eq.(\ref{field})) and of the
chemical potentials (Fig.\ref{fig:chem}) for neutrons (top curves)
and protons (bottom curves) , for the different 
effective interactions in the isovector channel.  

From Fig.\ref{fig:mean} we note that in regions just off normal density
the field ``seen'' by neutrons and protons in the three cases is 
very different,
in particular below saturation density. We thus expect important transport
effects during reactions at intermediate energies (prompt particle emissions,
collective flows, $n/p$ interferometry): the interacting asymmetric nuclear 
matter will experience compressed and expanding phases before forming 
fragments around normal density.

A transparent picture of
the isospin dynamics can be obtained from the analysis
of the density dependence of the {\it "local"} values of the
neutron/proton chemical potentials 
$\mu_q \equiv \partial \epsilon(\rho_q,\rho_{q'})/ \partial \rho_q$,
$\epsilon$ being the energy density.
We remind that the chemical
potentials contain all the contributions to the energy per
particle, including the isoscalar part and the kinetic symmetry
term.
In non-equilibrium processes the mass flow
is determined by the differences in the local values of chemical
potential and it is directed from the regions of higher chemical potential to
regions of lower values until equalization. 

From Fig.\ref{fig:chem} we can already predict the {\it Isospin Distillation}
effect and even the differences between the results 
in the isospin dynamics during fragment formation
obtained using different symmetry terms, see Sects.\ref{rpa} and \ref{fermi}. 
The chemical potential for
protons in a system having the  $^{124}Sn$ asymmetry, $I=0.2$, is below the
corresponding value of symmetric nuclear matter while for neutrons is above,
indeed $\mu_n-\mu_p=4\epsilon_{sym}(\rho)I$.
From the density dependence in the low densities region we see that
when the inhomogenities develop both neutrons
and protons have the tendency to move in phase from lower to higher density
regions. This is in qualitative
agreement with the rigorous proof of the fact that the system is unstable
against isoscalar-like fluctuations, see next section.
Since
the variations of the two chemical
potentials are different (larger for protons) we expect a lower asymmetry in
the liquid phase, i.e. in the clusters formed through such bulk instability
mechanism. Moreover from the larger difference in the neutron/proton
slopes in dilute matter, around $1/3\rho_0$, for the $asysoft$ case
 (dashed lines of Fig.\ref{fig:chem}) we
can even expect a larger Isospin Distillation mechanism with such 
symmetry term.

In the case of a clusterization in presence of a contact between 
more dilute and  "normal" density regions in order to understand 
the isospin dynamics we have to look at the density
dependence of proton/neutron chemical potentials in the region
between 0.08 $fm^{-3}$  and 0.16 $fm^{-3}$. We see from Fig.\ref{fig:chem}
that in this range the neutrons have 
the tendency to move towards more dilute regions producing a $n$-enrichment
while the protons will migrate in opposite direction.
Such mechanism is present in the "neck fragmentation",
\cite{ColonnaNPA589,DitNPA681,BaranNPA703,BaranNPA730}:
the neck $IMF's$ will be always more $n-$rich compared to the fragments
produced in
the case of bulk fragmentation. This effect, clearly seen in experiments,
will be discussed in detail in Sect.\ref{fermi} since naturally it appears very
sensitive to the stiffness of the symmetry term around saturation density.

\subsection{Effective masses in neutron-rich matter}

In the Fig.\ref{fig:potsym} we have noticed a dramatic change in
the local vs. non-local contributions to the symmetry energy going
from the {\it old} to the {\it new} Skyrme forces of the Lyon type,
explicitly built for asymmetric matter.
A related interesting effect can be seen on the neutron/proton
effective masses. 

In asymmetric matter we consistently have a splitting
of the neutron/proton effective masses given by:
\begin{equation}
{m^*_q}^{-1} = m^{-1} + g_1 \rho + g_2 \rho_q,
\label{skymass}
\end{equation}
with
$$
\rho_{q=n,p} = {{1 + \tau_q I} \over 2} \rho~~.
$$
%and $I$ the asymmetry parameter $(N-Z)/A$. 
The $g_1,~g_2$ coefficients
are simply related to the momentum dependent part of the Skyrme forces:
\begin{eqnarray}
&& g_1 = {\frac {1}{4 \hbar^2}} [t_1 (2 + x_1) + t_2 (2 + x_2)] \nonumber\\ 
&& \nonumber\\
&& g_2 = {\frac {1}{4 \hbar^2}} [t_2 (1 + 2 x_2) - t_1 (1 + 2 x_1)] 
\label{g1g2}
\end{eqnarray}
This result derives from a general $q-structure$ of the momentum
dependent part of the Skyrme mean field
\begin{equation}
U_{q,MD} = m ( g_1 \rho + g_2 \rho_q) E
\label{momdep}
\end{equation}
where $E$ is the nucleon kinetic energy.

\begin{figure}
\begin{center}
%\epsfysize=6.cm
%\centerline{\epsfbox{iso1.ps}}
\includegraphics*[scale=0.55]{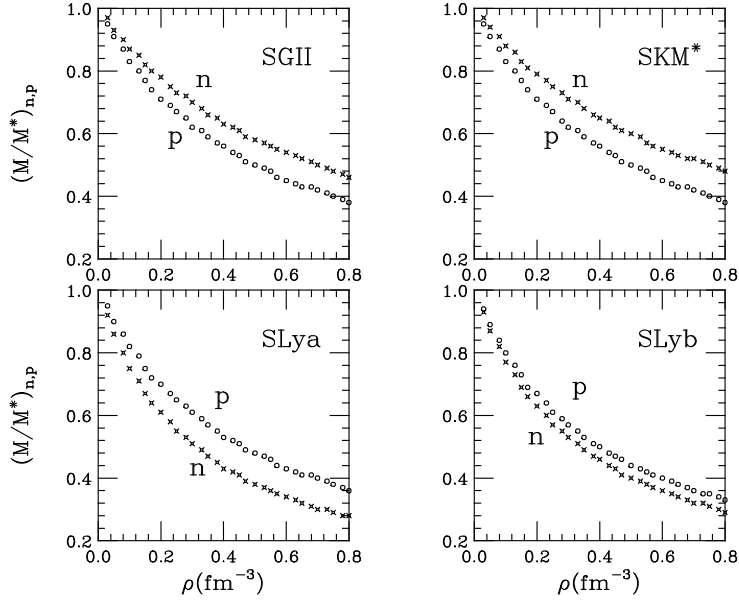}
\caption{Density dependence of the neutron/proton effective
mass splitting
for various Skyrme effective forces, see text.
The asymmetry is fixed at $I=0.2$, not very exotic.}
\label{fig:massplit}
\end{center}
%\vskip-1.0cm
\end{figure}
In the Fig. \ref{fig:massplit} we
show the density behavior of $m^*_{n,p}$ in  neutron rich
matter $I=0.2$ for the same effective interactions.
From the Eqs.(\ref{skymass}, \ref{g1g2}) we see that the sign of the
$g_2$ univocally assigns the sign of the splitting, i.e. $g_2<0$
gives larger neutron masses $m^*_n>m^*_p$ while we have the opposite 
for $g_2>0$.

In the Table \ref{skyg1g2} we report some results obtained with
various Skyrme forces for quantities of interest, around saturation, for the
present discussion. We show also the $E-slope$ of the corresponding
Lane Potential, see later, simply related to the isospin dependent part
of Eq.(\ref{momdep}).
For the effective mass parameters of Eq.(\ref{skymass})
we observe that while the $g_1$ coefficients are always positive, 
corresponding to a decrease of the
nucleon mass in the medium, the isospin dependent part shows different
signs.
In particular we see that in the Lyon forces the $g_2$ values are 
 positive, with neutron effective
masses below the proton ones for n-rich matter as shown in Fig.
\ref{fig:massplit}.

\vskip 0.5cm
\begin{table}
\begin{center}
\caption{Properties at saturation}
\begin{tabular}{ c c c c c c c} \hline
$ Force$  & $ SIII$ &~ $ SGII$ &~$ SkM^*$ &~$SLya$ &~$SLy4$ &~$SLy7$ 
\\ \hline
$g_1~(10^{-3})(MeV^{-1}fm^3) $ & $+3.85$ & $+3.31$ & $+3.53$ & $+10^{-5}$ &
 $+1.67$ & $+1.70$ \\ \hline
$g_2~(10^{-3})(MeV^{-1}fm^3) $ & $-3.14$ & $-2.96$ & $-3.50$ & $+5.78$ &
 $+2.53$ & $+2.76$ \\ \hline
$\rho_0(fm^{-3})$ & $0.150$ & $0.1595$ & $0.1603$ & $0.160$ &
 $0.1595$ & $0.1581$ \\ \hline
$a_4(MeV)$ & $ 28.16$ & $ 26.83$ & $30.03$ & $31.97$ & $32.01$ & $32.01$ 
 \\ \hline
$C(\rho_0)(MeV)$ & $31.72$ & $29.06$ & $35.46$ & $39.40$ & $39.42$ & $39.42$ 
 \\ \hline
$E-slope(Lane Pot.)$ & $-0.22$ & $-0.21$ & $-0.26$ & $+0.43$ &
 $+0.19$ & $+0.20$ \\ \hline
\end{tabular}
\label{skyg1g2}
\end{center}
\vskip 0.5cm
\end{table}

In general we obtain a splitting of the order of $10-15\%$ at normal 
density $\rho_0$,
and increasing with baryon density. Unfortunately from the present nuclear 
data we have a very little knowledge of this effect, due to the low
asymmetries available. This issue will be quite relevant in the
study of drip-line nuclei. 

The sign itself of the splitting is very instructive.
Passing from Skyrme to Skyrme-Lyon we see a dramatic inversion in
the sign of the $n/p$ mass splitting. The Lyon forces predict    
in $n$-rich systems a neutron effective mass always smaller than 
the proton one.
We will come back to this point. Here we just note that the same is predicted
from microscopic relativistic Dirac-Brueckner calculations
\cite{HofmannPRC64} and in general from the introduction of scalar isovector
virtual mesons in $RMF$ approaches \cite{LiuboPRC65,GrecoPRC67}. At variance, 
non-relativistic Brueckner-Hartree-Fock
calculations are leading to opposite conclusions
 \cite{Bombiso,ZuoPRC60}. We remind that a comparison between relativistic
effective ($Dirac$) masses and non-relativistic effective masses requires
some attention. This point will be carefully discussed later in the effective 
field theory approach to the in-medium interactions, Sect.\ref{qhd}.

\subsubsection{Effective masses and Landau Parameters}
%\addtocontents{toc}{\hspace{0.55cm}\thesubsubsection \hspace{0.12cm}
%Effective masses and Landau Parameters \newline}
The quasiparticle nucleon energies in a general Skyrme form are given
by:
\begin{equation}
\epsilon_q (\rho_q,\rho_{q'}, p) = \frac{p^2}{2m} + 
\int {{\frac{d^3p'}{(2\pi\hbar)^3}}(p-p')^2 [g_1 f(p') + g_2 f_q(p')]} 
+ U_q(\rho_q,\rho_{q'})
\label{quasien}
\end{equation}

where $U_q(\rho_q,\rho_{q'})$ is the local part of the mean field, 
 Eq.(\ref{field}), and $f(p),f_q(p)$ are the nucleon momentum distributions.
The corresponding Landau parameters defined by:
\begin{equation}
\delta \epsilon_q \equiv \frac{2}{N_q}
\int {{\frac{d^3p'}{(2\pi\hbar)^3}}[F_{qq}(p,p') \delta f_q(p') + 
 F_{qq'}(p,p') \delta f_{q'}(p')]} 
\label{landpar}
\end{equation}
have the form:
\begin{eqnarray}
&& F_{qq}(p,p') = N_q \Big( \frac{g_1}{2}(p-p')^2 + \frac{g_2}{2}(p-p')^2 +
 \frac {\partial U_q}{\partial \rho_q} \Big) \nonumber \\
%&& \nonumber \\
&& F_{qq'}(p,p') = N_q \Big( \frac{g_1}{2}(p-p')^2 +
 \frac {\partial U_q}{\partial \rho_{q'}} \Big)  
\label{landpar1}
\end{eqnarray}

where $N_q \equiv \frac{m_q^* p_q}{\pi^2 \hbar^3}$ is the energy level density
of the q-nucleons.

From the expansion:
\begin{equation}
F_{qq'} = F_{qq'}^0 +F_{qq'}^1 (\hat p \cdot \hat p')
\label{landpar2}
\end{equation}

we get, at the Fermi momentum, the explicit Skyrme form of the ``local'' and
``non-local'' Landau parameters:

\begin{eqnarray}
&& F_{qq}^0 = N_q(p_F) \Big[ (g_1 + g_2) p_{Fq}^2 +
 \frac {\partial^2 U_q}{\partial \rho_q^2} \Big] \nonumber \\
&& F_{qq'}^0 = N_q(p_F) \Big[ {g_1 p_{Fq}^2} +
 \frac {\partial^2 U_q}{\partial \rho_q \partial \rho_{q'}} \Big]
 \nonumber \\
&& \nonumber \\
&& F_{qq}^1 = - N_q(p_F) (g_1 + g_2) p_{Fq}^2 
 \nonumber \\
&& \nonumber \\
&& F_{qq'}^1 = - N_q(p_F) g_1 p_{Fq} p_{Fq'}  
\label{landparfin}
\end{eqnarray}

The (n,p) effective masses have the compact form:

\begin{equation}
\frac{m_q^*}{m} = 1 + \frac{1}{3} \Big[ F_{qq}^1 + \Big(\frac{p_{Fq'}}
{p_{Fq}} \Big)^2
 F_{qq'}^1 \Big]
\label{landmstar}
\end{equation}

which nicely leads to the $(1+ \frac{1}{3} F^1)$ result for symmetric 
matter \cite{SjoebergNPA265}. For the (n,p) mass splitting in asymmetric
matter we have the expression:

\begin{equation}
\frac{m_n^* -m_p^*}{m} =  \frac{1}{3} \Big[ F_{nn}^1 - F_{pp}^1
 + \Big( \frac{p_{Fp}}{p_{Fn}} \Big)^2 F_{np}^1
 - \Big(\frac{p_{Fn}}{p_{Fp}} \Big)^2 F_{pn}^1 \Big].
\label{landsplit}
\end{equation}
  
Since all the $F_{qq'}^1$ are negative this result has been used to predict
a larger $m_n^*$ mass in n-rich systems due to the larger neutron
Fermi momentum, see the recent ref.\cite{BaoPRC692}. However this is actually 
not generally correct
because in fact all the $F^1$ Landau parameters  also depend on
Fermi momenta and the final balance will be fixed by the microscopic structure
of the effective interaction, i.e. by the interplay between the
$g_1$ and $g_2$ quantities in the case of Skyrme-like forces,  
Eq.(\ref{landparfin}). In particular, we see that when we pass from the
``old'' Skyrme to the ``Lyon'' parametrizations, i.e. from $g_2<0$ to
$g_2>0$, the combination $g_1+g_2$ becomes much larger and the
$F_{nn}^1$ term is dominant in the n-rich case, see Eq.(\ref{landsplit}),
leading to a $m_n^*<m_p^*$ mass splitting, as already discussed.

\subsubsection{Energy dependence of the Lane Potential}
\begin{figure}
\begin{center}
%\epsfysize=6.cm
%\centerline{\epsfbox{iso1.ps}}
\includegraphics*[scale=0.55]{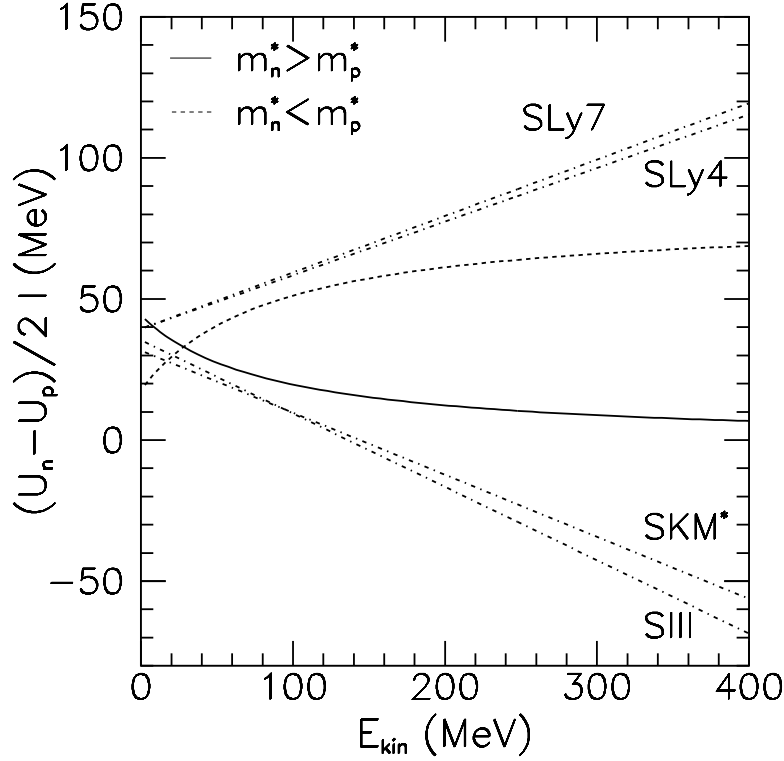}
\caption{
 Energy dependence of the Lane Potential for different Skyrme
forces. The dotted ($m_n^*<m_p^*$) and solid ($m_n^*>m_p^*$) curves correspond
to the more general momentum dependent mean fields $BGBD-1,2$ of the 
ref.\cite{RizzoNPA732}, see text.}
\label{fig:elane}
\end{center}
%\vskip-1.0cm
\end{figure}

%\addtocontents{toc}{\hspace{0.55cm}\thesubsubsection \hspace{0.12cm}
%Energy dependence of the Lane Potential}
We note that the sign of the splitting will directly affect the energy
dependence of the Lane Potential, i.e. the difference between $(n,p)$ optical
potentials on charge asymmetric targets, normalized by the target asymmetry
\cite{LaneNP35}.  
From the Eqs.(\ref{field}, \ref{momdep}) we obtain the explicit Skyrme
form of the Lane Potential:
\begin{equation}
U_{Lane} \equiv {\frac {U_n - U_p}{2I}} = C(\rho_0) + 
 {\frac{m\rho_0}{2}} g_2 E
\label{lane}
\end{equation}
where $C(\rho_0)$ gives the potential
part of the $a_4$ parameter in the mass formula, see Eq.(\ref{kipot}).
We see that the $E-slope$ has just the sign of the $g_2$ parameter, and so
we have opposite predictions from the various Skyrme forces analysed here. 
The change
in the energy slope is reported in the last row of the Table \ref{skyg1g2}.
The difference in the energy dependence of the Lane Potential is quite
dramatic, as we can see from Fig.\ref{fig:elane}.

An important physical consequence of the negative slopes is that the isospin
effects on the optical potentials tend to disappear at energies just above
$100~MeV$ (or even change the sign for ``old'' Skyrme-like forces).
Unfortunately results derived from neutron/proton optical potentials at 
low energies
are not conclusive, \cite{LaneNP35,BecchettiPR182,Hodopt}, since the 
effects appear
of the same order of the uncertainty on the determination of the local
contribution. More neutron data are needed at higher energies, in particular 
a systematics of the energy dependence. 

Moreover we can expect important effects
on transport properties ( fast particle emission, collective flows)
 of the dense and asymmetric $NM$ that will
be reached in Radioactive Beam collisions at intermediate energies.
Indeed at 
supra-saturation density the difference in the predictions 
will be enhanced.

\subsection{Isospin effects on the momentum dependence of the mean field}

In presenting the symmetry energy results of various effective Skyrme forces
we have stressed the conflicting predictions on the isospin dependence of
the effective masses, i.e. on Isospin Momentum Dependent, {\it Iso-MD},
  effects. 
This will be one of the main
questions to address in the reaction dynamics of exotic nuclear systems,
in particular for the close connection to fundamental properties of the
nuclear interaction in the medium.

We will review some results on this direction using non-relativistic and
relativistic (later in Sects. \ref{qhd} and \ref{reldyn}) 
 microscopic kinetic approaches. 
Starting from realistic effective
interactions widely used for symmetric systems we will test very different 
parametrizations in the momentum dependence of the isovector channel,
taking care that the symmetry energy, including its density dependence,
will be not modified. We study in particular the transport effect of
the sign of the $n/p$ effective mass splitting $m^*_n-m^*_p$ in
asymmetric matter at high baryon and isospin density.

In a non-relativistic frame we can extend the general form of  
effective momentum dependent
interactions first introduced by Bombaci et al. \cite{BombaciNPA583,Bombiso}
 for astrophysical
and heavy ion physics applications. The isovector
part can be modified in order to get new parametrizations with different
$n/p$ effective mass splittings while keeping the same symmetry energy
at saturation, including a very similar overall density dependence,
 see ref.\cite{RizzoNPA732}.

\begin{figure}[b]
\begin{picture}(0,0)
\put(1.4,3.9){\mbox{\includegraphics[width=2.8cm]{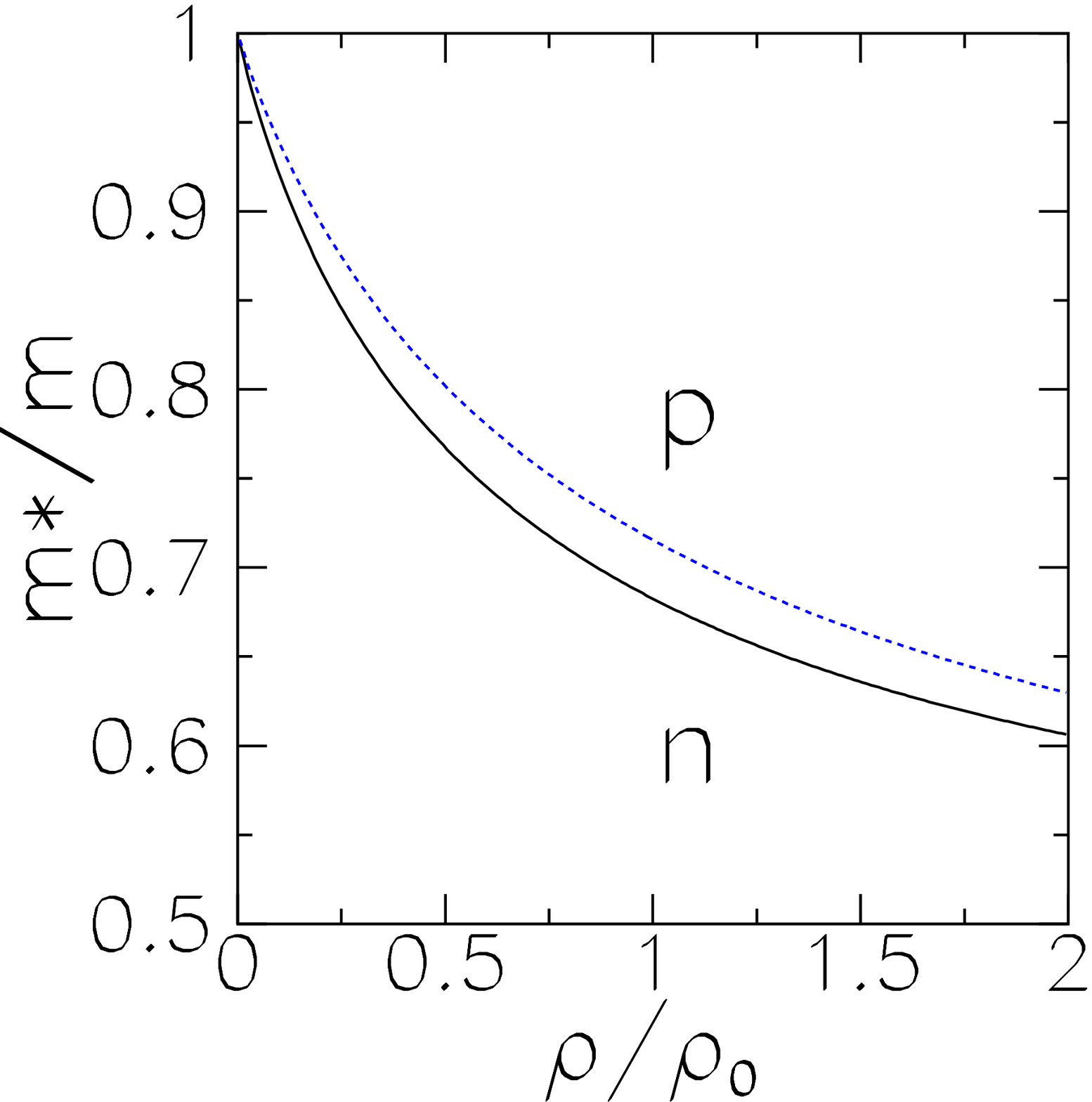}}}
\end{picture}
\includegraphics[width=7.3cm]{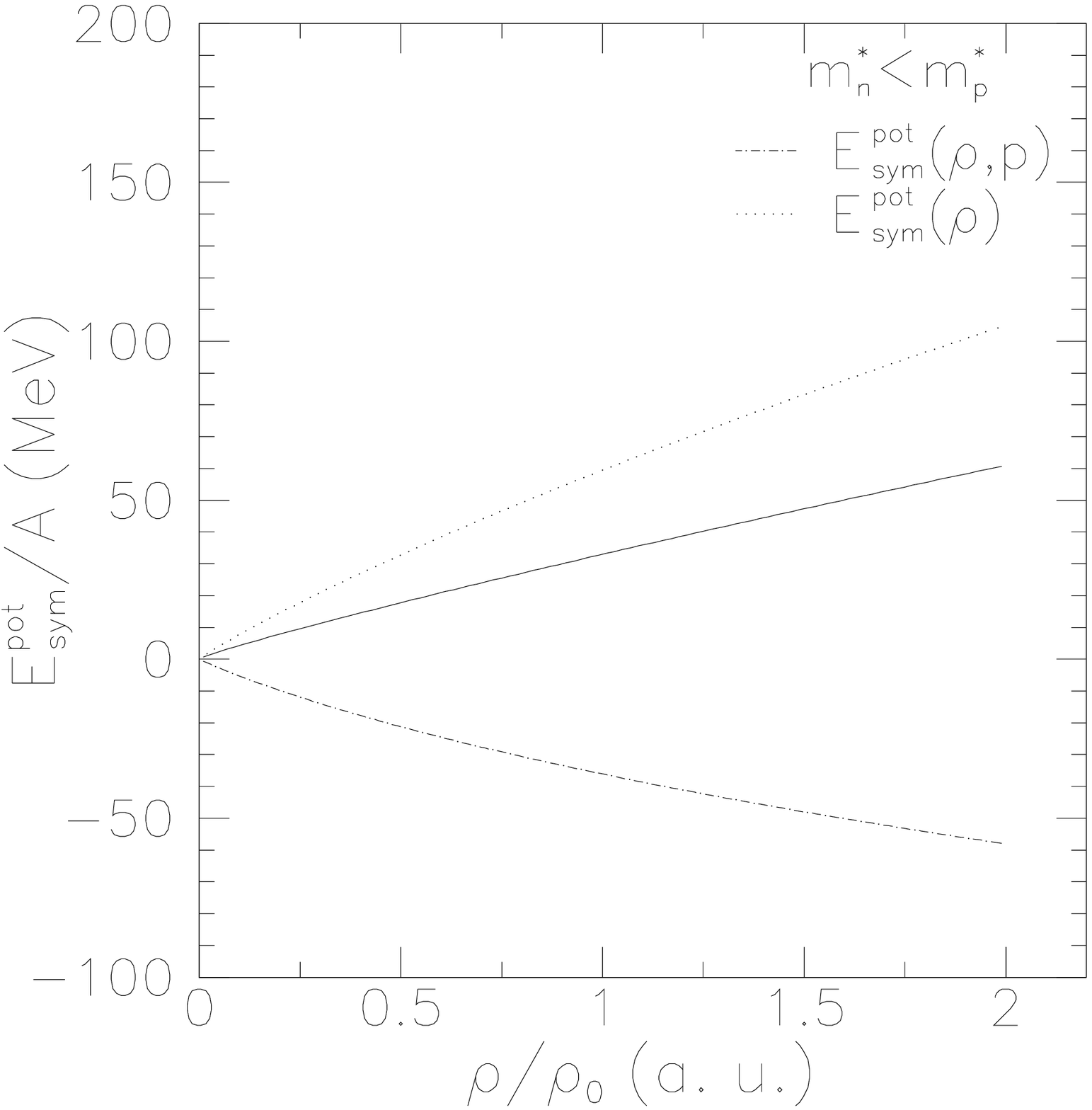}
\includegraphics[width=7.3cm]{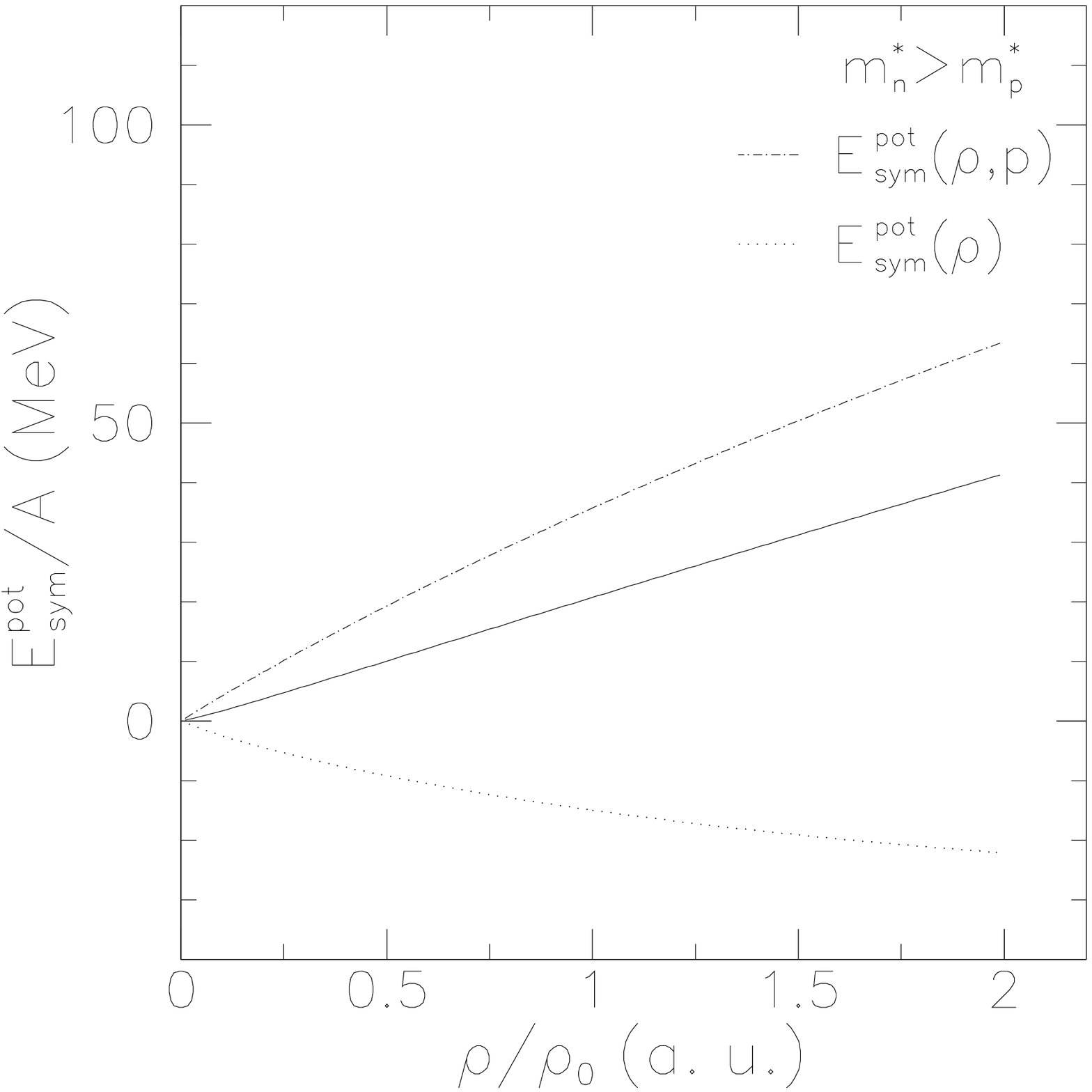}
\begin{picture}(0,0)
\put(8.7,4.3){\mbox{\includegraphics[width=2.8cm]{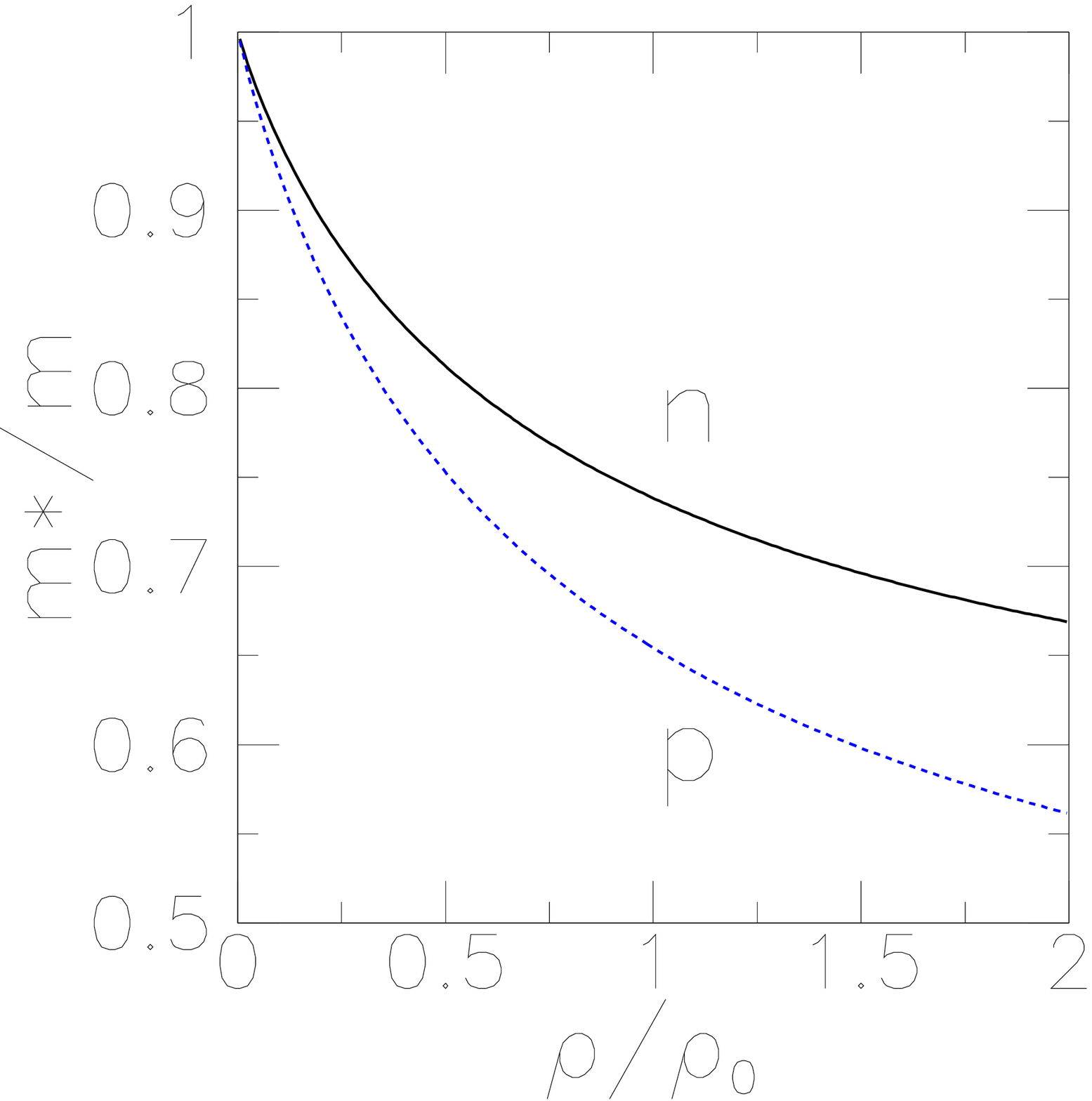}}}
\end{picture}
\caption{Potential symmetry energy as a function of density (solid
line) for $m^*_n<m^*_p$ ($BGBD-1$ set, left) and $m^*_n>m^*_p$ 
($BGBD-2$ set, right). Dotted
lines refer to density-dependent contributions, dashed-dotted to
momentum dependent ones. Small panels indicate the corresponding
behaviour of proton and neutron effective masses as a function of
density, with $I=0.2$.} \label{fig:symMD}
\end{figure}

The energy density as a function of the asymmetry parameter reads:
$I \equiv \frac{N-Z}{A}$:
\begin{eqnarray}\label{densene}
%\begin{Huge}
%$$
&& \varepsilon(\rho_n,\rho_p) = \varepsilon_{kin}+\varepsilon_A
+\varepsilon_B+\varepsilon_{MD}
%&&\nonumber\\
\\
&& \nonumber \\
&& \varepsilon_{kin}(\rho_n,\rho_p) = \ienne{\frac{\hbar^2}{2m}k^2}
+\ipi{\frac{\hbar^2}{2m}k^2}
%\end{eqnarray}
\nonumber\\
&& \nonumber \\
&& \varepsilon_A(\rho, I) = \frac{A}{2}\frac{\rho^2}{\rz}
-\frac{A}{3}\frac{\rho^2}{\rz}\umd{0} I^2\nonumber\\
%&&\nonumber\\
&& \nonumber \\
&& \varepsilon_B(\rho, I) = \frac{B}{\sigma+1}
\frac{\rho^{\sigma+1}}{\rz^\sigma}-\frac{2}{3}\frac{B}{\sigma+1}
\frac{\rho^{\sigma+1}}{\rz^\sigma}\umd{3} I^2\nonumber\\
%&&\nonumber\\
&& \nonumber \\
&& \varepsilon_{MD}(\rho_n,\rho_p, I) = C\frac{\rho}{\rz}
(\inew{n}+\inew{p})+\frac{C-8z_1}{5}\frac{\rho}{\rz}
 I (\inew{n}-\inew{p})\nonumber
%&&\\
%&&\frac{2}{5\rz}(C-8z_1) \qd{ \rho_n \ien + \rho_p \iz
%}\;\;\;\;\;\;\;\;\;\;\nonumber
\end{eqnarray}
where the integrals $\mathcal{I}_\tau(\Lambda)=\itau$
include the momentum dependent part of the mean
field $g(k,\Lambda)= \qd{ 1+\td{\frac{k-<k>}{\Lambda}}^2 }^{-1}$;
the subscript $\tau=n,p$ stands for neutrons
 and protons respectively; $\rz=0.16\,fm^{-3}$ is the normal
 density of nuclear matter.
We refer to this parametrization as the $BGBD-EOS$. For symmetric
nuclear matter ($I=0$),
%equations (\ref{densene})
the energy density Eq.(\ref{densene}) reduces to the parametrization
proposed by Gale, Bertsch and Das Gupta ($GBD$ interaction,
\cite{GalePRC41}). In a sense this interaction represents an extension
of the Gogny non locality since it gives very similar results in the
range $\rho/\rho_0 \leq 1.5$ and $k < 4 fm^{-1}$, see refs.
\cite{GrecoPRC59,SapienzaPRL87}
where nice applications in the Fermi energy region can be found.

The parameters $A,\,B,\,C,\,\sigma$ and $ \Lambda$
take the same values as in \cite{GalePRC41} ($A=-144 \,MeV,\, B=203.3
\,MeV,\, C=-75 \,MeV,\,$ $\sigma=\frac{7}{6},\, \Lambda=1.5
\,p_F^{(0)}$, where $p_F^{(0)}$ is the Fermi momentum at normal
density), and provide a soft $EOS$ for symmetric matter, with a
compressibility $K_{NM} \simeq 210~MeV$. 
The value of  $z_1$ sets the
strength of the momentum dependence ($MD$) in the isospin channel;
%and thus the mass
%splitting of neutrons and protons,
%as we will see in the following.
the remaining parameters $x_0\,$ and $x_3$
can be set to fix the symmetry energy.
 
\begin{figure}[b]
\centering
\includegraphics[width=8cm]{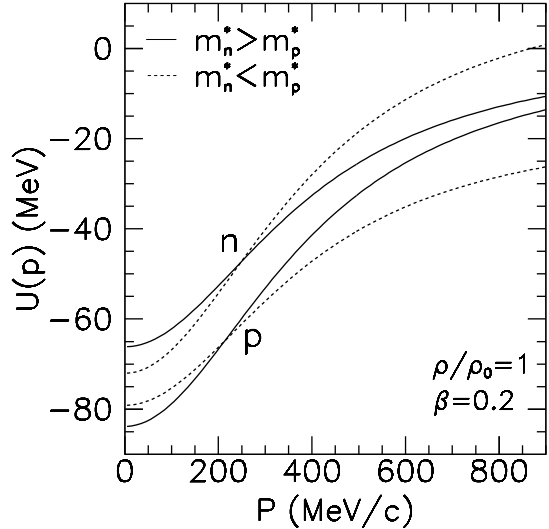}
\caption{Mean field potential as a function of momentum at normal
density, for an asymmetry $I=0.2$. Solid lines refer to the
case $m^*_n>m^*_p$ ($BGBD-1$ set), dashed lines to 
$m^*_n<m^*_p$ ($BGBD-2$ set).} \label{fig:optMD}
\end{figure}

Taking the functional derivative of the energy density
with respect to the distribution function $f_\tau$,
we obtain, apart from the kinetic term $\varepsilon_{kin}$,
the mean field potential for neutrons and protons:

\begin{eqnarray}
U_\tau(k;\rho, I)&=&
%\frac{\hbar^2}{2m}k^2+
A\ra+B\ra^\sigma-\frac{2}{3}(\sigma-1)\frac{B}{\sigma+1}
\umd{3}\ra^{\sigma} I^2
\nonumber\\
&&\pm \qd{-\frac{2}{3}A \umd{0}\ra -
\frac{4}{3}\frac{B}{\sigma+1}\umd{3}\ra^{\sigma}\,} I \\
&& +\frac{4}{5\rz} \left\{ \frac{1}{2} (3C-4z_1) \inew{\tau}
+ (C+2z_1) \inew{{\tau^{\prime}}}\right\}\nonumber\\
&&+ \td{C \pm \frac{C-8z_1}{5} I } \ra g(k)
%\;\;\;\;\;\;\;\;
\label{utau}
\nonumber
\end{eqnarray}

where the subscripts in the integrals are $\tau \neq \tau^\prime$;
the upper signs refer to neutrons, the lower ones to protons.

The last $BGBD$ term
includes, besides the usual $GBD$ momentum dependence, an
isospin-dependent part from which we can get different effective
masses for protons and neutrons. In fact, the effective mass is 
defined as:

\begin{equation}
\frac{m^*_\tau}{m}=\left\{1+
\frac{m}{\hbar^2k}\frac{dU_\tau}{dk}
\right\}
^{-1}_{k=k_F^{\left[ \tau \right]}}
\label{mstar}
\end{equation}

and we see that mass splitting is
determined not only by different
momentum dependence of mean field,
but also by Fermi momenta of neutrons and protons.
In the case considered here we have:
$$
\frac{m^*_\tau}{m}=\left\{1+\frac{-\frac{2m}{\hbar^2} \frac{1}{\Lambda^2}
\td{
 {C \pm \frac{C-8z_1}{5} I}}
 \frac{\rho}{\rho_0}}{
\left[ 1+ \left( \frac{k_{F0}}{\Lambda}
\right)^{^2}   (1 \pm I)^{^{(2/3)}}
(\frac{\rho}{\rho_0})^{^{(2/3)}}\right]^2}  \right\}^{-1}
$$
In order to investigate the effects of mass splitting on
non-relativistic heavy ion collisions
two sets of parameters ($BGBD-1,2$ shown in Table \ref{masses})
which give opposite splitting, but quite similar behaviour of the
symmetry energy, will be considered:

\begin{table}[htb]
\begin{center}
\begin{tabular}{|c|c|c|c|c|}
\hline
 force & mass splitting & $z_1$  & $x_0$  & $x_3$\\
\hline
 BGBD-1  &  $m^*_n<m^*_p$  & 28     & 1.925 & 0.41\\
\hline
 BGBD-2  &  $m^*_n>m^*_p$  & -36.75 & -1.477  & -1.01\\
\hline
\end{tabular}
\caption{Values of the parameters
$z_1,\,x_0$ and $x_3$, for opposite mass splitting but giving the same
$E_{sym}(\rz)=33\, AMeV$, which characterize the two used effective forces.}
\label{masses}
\end{center}
\vskip 0.5cm
\end{table}

Figure \ref{fig:symMD} shows the potential part of the symmetry energy as a
function of density (solid line)
for the two choices of mass splitting. We remark the very similar
overall density dependence.
The dashed-dotted  and dotted lines indicate respectively the
contributions from the momentum dependent
and density-dependent part
of the $EOS$. A change in the relative sign of
mass splitting is related to opposite behaviours of these two
contributions, exactly as already noticed in the Introduction for
the Skyrme-like forces.
The small panels on the top left of each graph illustrate the
corresponding mass splitting as a function
of density for an asymmetry $I=0.2$ (the $^{197}Au$ asymmetry).

We finally discuss the relation between effective mass and
momentum dependence. From the definition Eq.(\ref{mstar}) we see that
the effective mass is inversely proportional to the slope of mean
field at the Fermi momentum:
\begin{itemize}
\item {For n-rich systems the $n/p$ mean field potential difference 
$U_n - U_p$ will increase
with nucleon momentum in the  $m^*_n<m^*_p$ case and decrease in the
opposite $m^*_n>m^*_p$ case.}
  
\item {For nucleons with smaller effective
masses the potential will be more repulsive at momenta higher than the
Fermi one, and more attractive at low momenta.}

\end{itemize}

 Mean field potentials at
normal density as a function of momentum are shown in Fig.\ref{fig:optMD} 
for asymmetry $I=0.2$. As we can see, the parametrizations 
described here give rise to opposite behaviors for low and high
momentum particles. This is a very general feature, due to the
meaning itself of effective mass Eq.(\ref{mstar}). As we will
see in the reaction dynamics this behavior can give rise
to some compensation effects between low and high momenta
contributions. This will lead to an expected larger 
$isospin-MD$ sensitivity of more exclusive measurements,
in particular with a tranverse momentum selection of the
nucleons emitted for a given rapidity.

%\include{rep_bib}

%\end{document}

%% file: Chapter-3.tex
%\documentclass{elsart}
%\usepackage{epsfig}

%\usepackage{graphicx}

%\usepackage{amssymb}
%\tightenlines

% nuovi comandi
% 2 su 2pigreco al cubo
%\newcommand{\norm}{\frac{2}{(2\pi)^3}}
% parentesi quadre
%\newcommand{\qd}[1]{\left[ #1 \right]}
% parentesi tonde
%\newcommand{\td}[1]{\left( #1 \right)}

%\newcommand{\itg}[1]{\norm \int d^3k f_{#1}(k)}
% integrale di fn
%\newcommand{\ienne}[1]{\itg{n} #1}
% integrale di fp
%\newcommand{\ipi}[1]{\itg{p} #1}
% integrale di fn g
%\newcommand{\ien}{\ienne{g(k,\Lambda)}}
% integrale di fp g
%\newcommand{\iz}{\ipi{g(k,\Lambda)}}

%\setlength{\unitlength}{1cm}

% integrale di ftau g
%\newcommand{\itau}{\itg{\tau}g(k,\Lambda)}
% integrale di ftau g
%\newcommand{\itaup}{\itg{\tau ^\prime} g(k,\Lambda)}

% rozero
%\newcommand{\rz}{\rho_{_0}}
% rho su rozero
%\newcommand{\ra}{\td{ \frac{\rho}{\rho_{_0}} }}
% densita' di energia per A e B
%\newcommand{\ene}[1]{\qd{
% \td{\frac{1}{2}x_{#1}}\rho^2
%-\td{\frac{1}{2}+x_{#1}} \td{\rho_n^2+\rho_p^2} }}
% 0.5 + x0(x3)
%\newcommand{\umd}[1]{ \td{ \frac{1}{2}+x_{#1} } }

%\newcommand{\inew}[1]{\mathcal{I}_#1 }

%\begin{document}  

\setcounter{figure}{0}
\setcounter{equation}{0}

\section{Instabilities in Two-component Fluids:
 the Liquid-Gas Phase Transition in Asymmetric Matter}\label{rpa}

\markright{Chapter \arabic{section}: rpa}

Since the fragment production represents a relevant dissipation mechanism
 for intermediate energy reactions and since it is seen as a
consequence of the liquid-gas phase transition in Asymmetric Nuclear
Matter ($ANM$),
we are devoting one section to a detailed
discussion of it. In this way we set some guidelines in order to understand
the physics behind the transport simulation results of Sect.\ref{fermi},
in particular on the possibility of extracting some information on
the nuclear effective forces in the isovector channel.
A very transparent picture of the {\it Isospin Distillation},
 actually {\it Neutron Distillation}, effect is emerging.

One-component systems may become unstable against  
density fluctuations as the result of the strong
attraction between constituents. In symmetric binary systems,
like Symmetric Nuclear Matter ($SNM$), 
one may encounter two kinds of density fluctuations:
i) isoscalar, when the densities
of the two components oscillate in phase with equal amplitude,
ii)  isovector when the two densities fluctuate still with equal
amplitude but out of phase. Then mechanical instability is associated with
instability against isoscalar fluctuations leading to cluster
formation while chemical
instability is related to instability against isovector fluctuations,
of repulsive character, 
leading to species separation.   Turning now to 
asymmetric binary systems, as the $ANM$ of interest here,
this direct correspondence between the nature of fluctuations and the 
occurrence of mechanical or chemical instabilities is lost and we face a more
complicated scenario, where isoscalar and isovector instabilities are 
coupled \cite{MuellerPRC52,BaranPRL86,MargueronPRC67}.

An appropriate framework for the study of instabilities 
is provided by the Fermi liquid theory \cite{LandauJETP5}, which has been 
applied,  
for instance, to 
symmetric binary systems as $SNM$ (the two components being protons and
neutrons) \cite{MigdalBook67} and the liquid $^3He$ 
(spin-up and spin-down components)
\cite{BaymBook78,PethickAP183}.

\subsection{Thermodynamical study}
Let us first discuss the thermodynamical stability of
$ANM$ at $T=0$. We will review here the results obtained in 
\cite{BaranPRL86} with some more discussion on the physics observables.
The introduction of finite temperatures 
is straightforward.

The starting point is an extension to the asymmetric case of the 
formalism introduced in
\cite{BaymBook78}.  
The distribution functions for protons and neutrons are:  
\begin{equation}
f_q (\epsilon_p^q) = 
\Theta(\mu_q - \epsilon_p^q)~, ~~~~~q=n,p         \label{Fd0}
\end{equation}
where $\mu_{q}$ are the corresponding chemical potentials.
The nucleon interaction is characterized by the Landau parameters:
\begin{equation}
F^{q_1 q_2} = N_{q_1} V^2 
\frac{\delta^2 { H}}{\delta f_{q_1} \delta f_{q_2}} = 
 N_{q_1}\frac{\delta^2 { H}}{\delta \rho_{q_1} \delta \rho_{q_2}}
~,~~N_q(T) = \int\,{-2~d{\bf p} \over (2\pi\hbar)^3} 
        {\partial f_q(T) \over \partial\epsilon_p^q}~  
           \label{ld}
\end{equation}
where $ H$ is the energy density, V is the volume and $N_{q}$ is the
single-particle level density at the Fermi energy. 
At $T=0$ this reduces to
$$N_q(0) = mp_{F,q}/(\pi^2\hbar^3) = 3\rho_q / (2\epsilon_{F,q}),$$
where $p_{F,q}$ and $\epsilon_{F,q}$ are Fermi momentum and Fermi energy 
of the $q$-component. Thermodynamical stability for $T=0$ requires the 
energy of the system 
to be an absolute minimum for the undistorted distribution functions, 
so that the relation: 
\begin{equation}
\delta{ H} - \mu_p \delta \rho_p - \mu_n \delta \rho_n > 0~~ \label{varia}
\end{equation}
is satisfied when we deform proton and neutron Fermi seas.

Only monopolar deformations will be taken into account, since we consider 
here
momentum independent interactions, so that
$F^{q_1q_2}_{l=0}$ are the only non-zero Landau parameters. %\cite{foot1}. 
In fact, 
for momentum independent interactions, all the information on 
all possible instabilities of the system is obtained just considering 
density variations.  
However one should keep in mind 
that in the actual dynamical evolution of an unstable system  
in general one observes deformations
of the Fermi sphere, hence the direction taken by the system in the dynamical
evolution is not necessarily the most unstable one defined by the 
thermodynamical analysis.  
 
Then, up to second order in the variations, the condition 
Eq.(\ref{varia}) becomes
\begin{equation}
\delta{ H} - \mu_p \delta \rho_p - \mu_n \delta \rho_n = \frac{1}{2}
(a {\delta\rho_p}^{2} + b {\delta\rho_n}^{2} + 
c \delta\rho_p \delta\rho_n) > 0~~ \label{varia1}
\end{equation}
where 
\begin{eqnarray}
&& a = N_p^{-1}(0)(1 + F_{0}^{pp})~~;~~
b = N_n^{-1}(0)(1 + F_{0}^{nn})~~; \nonumber \\ 
&& c = N_p^{-1}(0) F_{0}^{pn} + N_n^{-1}(0) F_{0}^{np} = 
2 N_p^{-1}(0) F_{0}^{pn}.~~  \label{abc}
\end{eqnarray}
The r.h.s. of Eq.(\ref{varia1}) is diagonalized 
by the following transformation:
\begin{eqnarray}
u &=& cos\beta~ \delta\rho_p + sin\beta~ \delta\rho_n,    \nonumber \\
v &=& - sin\beta~ \delta\rho_p + cos\beta~ \delta\rho_n,         \label{rot}
\end{eqnarray}
where the {\it mixing} angle $0 \le \beta \le \pi/2$ is given by
\begin{equation}
tg~ 2\beta = \frac{c}{a-b} = \frac{N_p^{-1}(0) F_{0}^{pn} + N_n^{-1}(0) 
F_{0}^{np}}
{N_p^{-1}(0)(1 + F_{0}^{pp}) - N_n^{-1}(0)(1 + F_{0}^{nn})}. \label{beta}
\end{equation}
Then Eq.(\ref{varia1}) takes the form
\begin{equation}
\delta{ H} - \mu_p \delta \rho_p - \mu_n \delta \rho_n =
X u^2 + Y v^{2} > 0~~\label{varia2}
\end{equation}
where
\begin{eqnarray}
X &=& \frac{1}{2} (~~ a + b + sign(c) \sqrt{(a-b)^{2} + c^{2}}~~) \nonumber \\ 
&\equiv& \frac{(N_p(0)+N_n(0))^{-1}}{2} (~ 1 + F_{0g}^s~) \label{A}
\end{eqnarray}
and
\begin{eqnarray}
Y &=& \frac{1}{2} (~~ a + b - sign(c) \sqrt{(a-b)^{2} + c^{2}}~~) \nonumber \\
&\equiv& \frac{(N_p(0)+N_n(0))^{-1}}{2} (~ 1 + F_{0g}^a ~), \label{B}
\end{eqnarray}
defining the new generalized Landau parameters $F_{0g}^{s,a}$.

Hence, 
thanks to the rotation Eq.(\ref{rot}), it is possible 
to separate the total variation
Eq.(\ref{varia}) into two independent contributions, the "normal" modes,
characterized by the "mixing angle" $\beta$, which depends on the density of
states and  the details of the interaction.

In the symmetric case, $N_p=N_n \equiv N$, $F_0^{nn}=F_0^{pp}$ 
and $F_0^{np}=F_0^{pn}$,
 Eq.(\ref{varia1}) reduces to
\begin{eqnarray}
\delta{ H} - \mu_p \delta \rho_p - \mu_n \delta \rho_n &=& 
\frac{N(0)^{-1}}{2}
(1 + F_0^s)(\delta\rho_p + \delta\rho_n)^{2}    \nonumber \\
 &+& \frac{N(0)^{-1}}{2}(1 + F_0^a) (\delta\rho_p -
\delta\rho_n)^{2} \label{varia2s}
\end{eqnarray}
where $F_0^s \equiv F_0^{nn} + F_0^{np}$ and 
$F_0^a \equiv F_0^{nn} - F_0^{np}$ are symmetric and antisymmetric
(or isoscalar and isovector) Landau parameters, and the usual 
Pomeranchuk stability conditions for pure isoscalar/isovector
fluctuations are recovered \cite{BaranNPA632}.

In the general case
$u$- and $v$-variations, Eq.(\ref{rot}), can be interpreted as new independent
${\it isoscalar}$-like and ${\it isovector}$-like directions appropriate
for asymmetric systems and 
$F_{0g}^s$ and $F_{0g}^a$, defined by Eqs. (\ref{A},\ref{B}),
can be considered as
generalized symmetric and antisymmetric Landau parameters.
Now, 
the instability  of the system can be studied completely just looking at the
curvatures along the $u$ and $v$ directions.  If the system is stable
against these two directions, it cannot be unstable in any other directions. 
On the contrary, if the system in unstable against $u$ or $v$ 
direction, or both, it can be unstable also in other directions of the
$(\delta\rho_n,\delta\rho_p)$ plane. 
 
Thus the thermodynamical stability requires
$X>0$ {\it and} $Y>0$. Equivalently, the following conditions have 
to be fulfilled:
\begin{equation}
1 + F_{0g}^s > 0~~~~ and~~~~1 + F_{0g}^a > 0,          \label{pomegen}
\end{equation}
They represent Pomeranchuk stability conditions extended
to asymmetric binary systems.

As one can intuitively expect, 
the new stability conditions, Eq.(\ref{pomegen}),
are equivalent to mechanical and chemical stability of a 
thermodynamical state, \cite{LandauSP89}, i.e.
\begin{equation}
\left({\partial P \over \partial \rho}\right)_{T,y} > 0~~~and~~~
\left({\partial\mu_p \over \partial y}\right)_{T,P} > 0
\end{equation} 
where $P$ is the pressure and $y$ the proton fraction. 
In fact, mechanical and chemical stability are very general conditions,
deduced by requiring that the curvatures of thermodynamical potentials, such 
as the free energy (or the entropy)
with respect to the extensive variables are positive (negative).
In the case discussed here, 
it can be proven that \cite{BaranPRL86}: 
\begin{eqnarray}
X Y &=& N_p^{-1}(0) N_n^{-1}(0) [(1 + F_0^{nn})(1 + F_0^{pp}) 
 - F_0^{np}F_0^{pn}]~~~~~~
\nonumber \\
~~~~~~~ \nonumber \\
&=&\frac{ 1 }{(1-y) \rho^{2}} 
\left({\partial P \over \partial \rho}\right)_{T,y}
\left({\partial\mu_p \over \partial y}\right)_{T,P} \label{chimec} 
\end{eqnarray}
and: 
\begin{eqnarray}
\left({\partial P \over \partial \rho}\right)_{T,y} =
\rho y (1-y)(~t~ a + \frac{1}{t}~ b + c~)~~~~~~
\nonumber \\
\propto X(\sqrt{t}cos\beta + \frac{1}{\sqrt{t}}
sin\beta)^2 + Y(\sqrt{t} sin\beta - \frac{1}{\sqrt{t}} cos\beta)^{2}
\nonumber \\~~~~with~~~~
t = \frac{y}{1-y} .~~~~~~~~~~
\label{compres}
\end{eqnarray}

Moreover, from Eq.\ref{compres} (first line), 
it is possible to see that  $(dP/d\rho)_{T,y}$ 
is proportional to the 
variation of the energy along the direction 
$\delta\rho_n/\delta\rho_p = 1/t = (1-y)/y = \rho_n /
\rho_p$.   
In fact, 
from Eq.(\ref{varia1}), along the direction $\delta\rho_p = t~ \delta\rho_n$, 
one has: 
$\delta{ H} - \mu_p \delta \rho_p - \mu_n \delta \rho_n = \frac{1}{2}
t~(a t + b/t  + c ) \delta\rho_n^2$.  
Thus the sign of $(dP/d\rho)_{T,y}$ gives information about the 
stability (or the instability) of the system against variations that 
preserve the initial proton to neutron density ratio.  
However, as we have seen, this 
is not the isoscalar-like direction of the normal mode, that 
is given by $u$.

\subsubsection{Isoscalar-like and isovector-like instabilities}
%\addtocontents{toc}{\hspace{0.55cm}\thesubsubsection \hspace{0.12cm}
%Isoscalar-like and isovector-like instabilities}
From Eq.s (\ref{varia2},\ref{A},\ref{B},\ref{pomegen}), one 
can  define as 
isoscalar-like instability the case when the
state is unstable against $u-$fluctuations, i.e. when 
$1 + F_{0g}^s < 0$
(or $X<0$).
The name isoscalar-like comes from the fact that since the mixing angle
is in the interval  $0 \le \beta \le \pi/2$ the neutron and proton
oscillations are in phase in the normal-mode $u-$direction.
 Analogously we deal with isovector-like instability when the system
is unstable against $v-$fluctuations i.e.
when $1 + F_{0g}^a < 0$ (or $Y<0$).

Let us consider first the case when $X<0$. 
Since, as seen before, the mechanichal instability ($(dP/d\rho)_{T,y}<0$)
is not along the normal-mode $u$-direction,   
when the isoscalar-like instability starts to appear (X = 0),  it 
cannot be a mechanical instability, so it corresponds to
a chemical instability. 
On the contrary, when the isoscalar-like instability becomes stronger, 
we will have mechanical instabilities and the chemical instability
will in turn disappear. However the nature of the unstable mode
has not changed. So the distinction between mechanical and chemical
instabilities is purely semantic. This was shown for the first time
in ref.\cite{BaranPRL86} for nuclear matter and confirmed in
the case of finite nuclear systems in ref.\cite{ColonnaPRL88}.

It is also interesting to observe that when $(dP/d\rho)_{T,y}$ changes 
the sign, passing through zero,   
the quantity
$\left({\partial\mu_p \over \partial y}\right)_{T,P}$ changes
also the sign, but passing through infinity.  
In fact, from Eq.(\ref{chimec}), 
one sees  
that the product 
$\left({\partial\mu_p \over \partial y}\right)_{T,P}
\left({\partial P \over \partial \rho}\right)_{T,y} \propto XY$
is a finite negative number, since $X<0$ and $Y>0$. 

The case of symmetric nuclear matter is easily recovered. 
In fact, as expected, now   
the isoscalar instability, ($X<0,Y>0$), appears as mechanical instability 
and the isovector 
instability, ($X>0,Y<0$), as chemical instability. 
Indeed we have
$t=1$, $a=b$, $\beta=\pi/4$ and so 
$X$ and
$\left({\partial P \over \partial \rho}\right)_{T,y}$  
are proportional (see Eq.(\ref{compres})) in this case. 

A more quantitative analysis can be performed in the case 
of asymmetric nuclear matter considering  
that the quantities $a$ and $b$ remain positive.
In this way one can study the effect 
of the
interaction between the two fluids, given by $c$, on the
instabilities of the mixture.
If $c<0$, i.e. for an attractive interaction between the two components,
from Eq.(\ref{B}) one sees that the system is stable against isovector-like
fluctuations but  it becomes isoscalar
unstable if $c<-2\sqrt{ab}$ (see Eq.(\ref{A})). However
thermodynamically this instability against isoscalar-like fluctuations
will show up as a chemical instability if
$(-t a -b/t ) < c < -2\sqrt{ab}$ or as a
mechanical instability if $c < (-t a -b/t ) < -2\sqrt{ab}$
(see Eq.(\ref{compres})).
So the distinction between the two kinds of instability
(mechanical and chemical) is not really relevant since the nature of density
fluctuations is essentially the same, i.e. isoscalar-like. 
If $c>0$, i.e. when the interaction between the components is repulsive, the
thermodynamical state is always stable against isoscalar-like fluctuation,
but can be isovector unstable if $c > 2\sqrt{ab}$. Since
the system is mechanically stable ($a,b,c>0$, see Eq. (\ref{compres})), the
isovector instability is now always associated with chemical instability.
Such situation will lead to a component separation of the
liquid mixture. 
Following this line a complete analysis of the instabilities of any
binary system can be performed, in connection to signs, strengths and
density dependence of the interactions.

\subsubsection{Nuclear matter instabilities}
%\addtocontents{toc}{\hspace{0.55cm}\thesubsubsection \hspace{0.12cm}
%Nuclear matter instabilities}
We show now quantitative calculations for asymmetric
nuclear matter which illustrate
the previous general discussion on instabilities.  
Let us consider 
a potential energy density of Skyrme type, 
\cite{ColonnaPLB428,BaranNPA632},
\begin{eqnarray}
{ H}_{pot}(\rho_n,\rho_p) &=&
{A \over 2}{(\rho_n+\rho_p)^2 \over \rho_0} +
{B \over \alpha + 2}{(\rho_n+\rho_p)^{\alpha + 2} \over \rho_0^{\alpha + 1}}
\nonumber \\
 &+& (C_1 - C_2(\frac{\rho}{\rho_0})^{\alpha})
 {(\rho_n-\rho_p)^2 \over \rho_0}
                         \label{hpot}
\end{eqnarray}
where $\rho_0 = 0.16~\mbox{fm}^{-3}$ is the 
nuclear saturation density.
The values of the parameters $A=-356.8$ MeV, $B=303.9$ MeV, $\alpha=1/6$,
 $C_1=125$ MeV, $C_2=93.5$ MeV are adjusted to reproduce the saturation 
properties of symmetric nuclear matter and the symmetry energy coefficient.

We focus on the low density region, where phase transitions 
of the liquid-gas type are expected to happen, in agreement with
the experimental evidences of multifragmentation 
\cite{XuPRL85,YennelloHIP97}.
Since $a,b > 0$ and $c<0$,  
we deal only with
instability against isoscalar-like fluctuations, as  
for symmetric nuclear matter.
\begin{figure}[htb]
\centering
\includegraphics[scale=0.65]{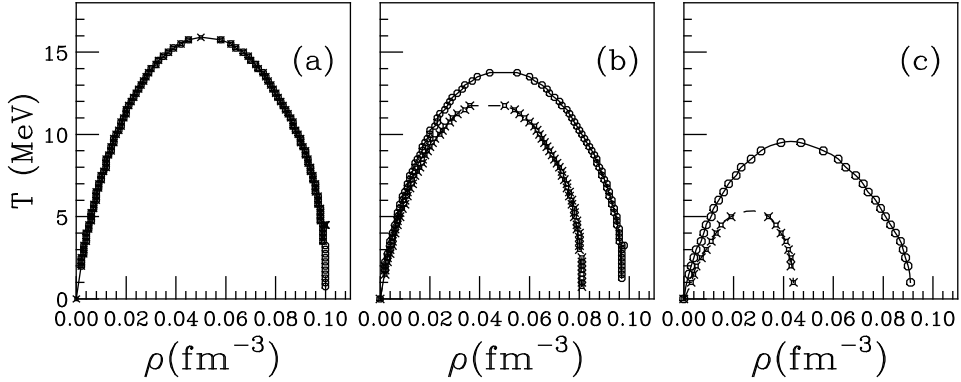} 
\caption{
Spinodal line corresponding to isoscalar-like instability of asymmetric 
nuclear matter (circles) and mechanical instability (crosses) for three
proton fractions: $y=0.5$ (a), $y=0.25$ (b), $y=0.1$ (c).
}
\label{rpa1}
\end{figure}
 In Fig.\ref{rpa1} the circles represent the spinodal line corresponding to
isoscalar-like instability, as defined above, for three values of 
the proton fraction. For asymmetric matter, $y < 0.5$, under this border one
encounters either chemical instability, in the 
region between the two lines, 
or mechanical instability, under the inner line (crosses). The latter is
defined by the set of values ($\rho,T$) for which
$\left({\partial P \over \partial \rho}\right)_{T,y} = 0$.
We observe that the line defining
chemical instability is more robust against the variation of the proton
fraction in comparison to that defining
mechanical instability: reducing the proton fraction it becomes 
energetically less and less favoured to break in clusters with
the same initial asymmetry.

However, we stress again the unique nature of the isoscalar-like instability. 
The change from the chemical to the mechanical character
along this border line is not very meaningful and does not affect 
the properties of the system. Later we will show a quantitative
realistic case of the clusterization of a dilute and heated asymmetric
nuclear matter in a box, Subsect. 3.3.      
\begin{figure}[htb]
%\epsfysize=4.0cm
%\centerline{\epsfbox{baran2.ps}}
\centering
\includegraphics[scale=0.65]{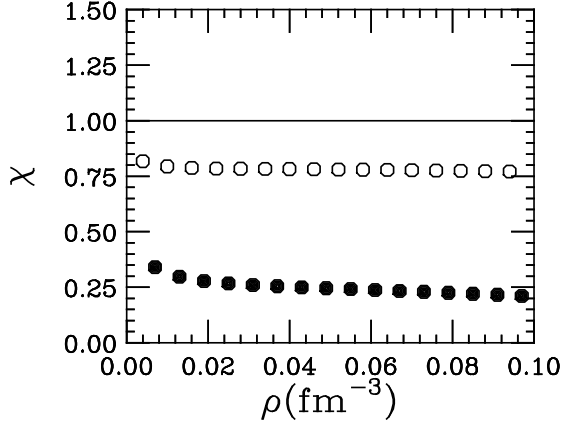} 
\caption{
Density dependence 
of the function $\chi$, see text,  
for three proton fractions, $y=0.5$ (solid), $y=0.4$ (open circles),
and $y=0.1$ (full circles) at $T=1MeV$.
}
\label{rpa2}
\end{figure}
As seen before, in $ANM$ 
isoscalar-like fluctuations are always associated with a 
chemical effect (change of concentration), 
which is responsible for the Isospin Distillation in 
phase transitions \cite{XuPRL85}.
Indeed the variation of the asymmetry $(I=1-2y)= 
(\rho_n-\rho_p)/(\rho_n+\rho_p)$ is
\begin{equation}
\delta I = \delta\rho_p [\frac{(1-I_0)}{(1+I_0)} tg\beta - 1] \equiv
\delta\rho_p [\chi-1]
\end{equation}
where $I_0$ is the initial asymmetry. The defined function $\chi(\rho)$
is reported in Fig.\ref{rpa2} for different asymmetries.
For $y=0.5$, $\delta I=0$,  but for an initial proton fraction $y<0.5$,
one finds $\chi<1$.
Therefore $\delta I < 0$ if $\delta\rho_p>0$, thus when the density 
increases (liquid phase),
the asymmetry decreases.

The effect of the isospin distillation can be connected to the
strength of the symmetry energy and, in particular, to the derivative of 
the symmetry energy coefficient with respect to $\rho$. 
This is easy to demonstrate if we adopt the simplest form for the symmetry
energy: $E_{sym} = {1\over 2} C {\rho\over\rho_0} I^2 = {1\over 2}C_{sym}
(\rho)I^2$.
 According to Eq.(\ref{beta}), the ratio $\delta\rho_p/\delta\rho_n$,
can be expressed as:  $\delta\rho_p/\delta\rho_n = tg\beta' = 1/tg\beta$,
with $tg(2\beta') = -tg(2\beta) = -c/(a-b)$.
The difference $(a-b)$ can be written as: $(a-b) = (N_p^{-1} -  N_n^{-1}) +
 (N_p^{-1}F_0^{pp} - N_n^{-1}F_0^{nn})$.
For the simple interaction considered, the second term of the sum vanishes
and $(a-b)$ is equal to the difference of the inverse of proton and
neutron single-particle level density, which is positive in n-rich matter. 
The quantity $c$ is the derivative of the proton 
potential $U_p = {{\delta H}
\over {\delta\rho_p}}$ (see Eq.(\ref{ld})) with respect to 
the neutron density. 
This is equal to the sum of a negative term coming from the isoscalar
part of the considered effective interaction and a negative term 
($-C/\rho_0$), that comes from the symmetry energy.  
Hence the term $-c$ is positive and increases when $C$ increases. 
In this case $tg2\beta'$
becomes larger and consequently $\delta\rho_p/\delta\rho_n$ increases
leading to a larger distillation effect.  
One can easily notice that $C/\rho_0$ coincides with the derivative of
$C_{sym}$ with respect to $\rho$, hence the distillation effect is related
to the derivative of the symmetry energy coefficient. 

Within this simple model even the effect 
of the Coulomb repulsion on the distillation effect can be easily understood.  
Taking into account the Coulomb interaction, the term 
$(N_p^{-1}F_0^{pp} - N_n^{-1}F_0^{nn})$ does not vanish anymore, but it is
equal to the derivative of the Coulomb potential with respect to the proton 
density, that is a positive term.  Hence the difference $(a-b)$ increases
and consequently  the distillation effect 
decreases when including the Coulomb interaction, as intuitively expected.    
   
\begin{figure}[htb]
%\epsfysize=4.0cm
%\centerline{\epsfbox{baran3.ps}}
\centering
\includegraphics[scale=0.65]{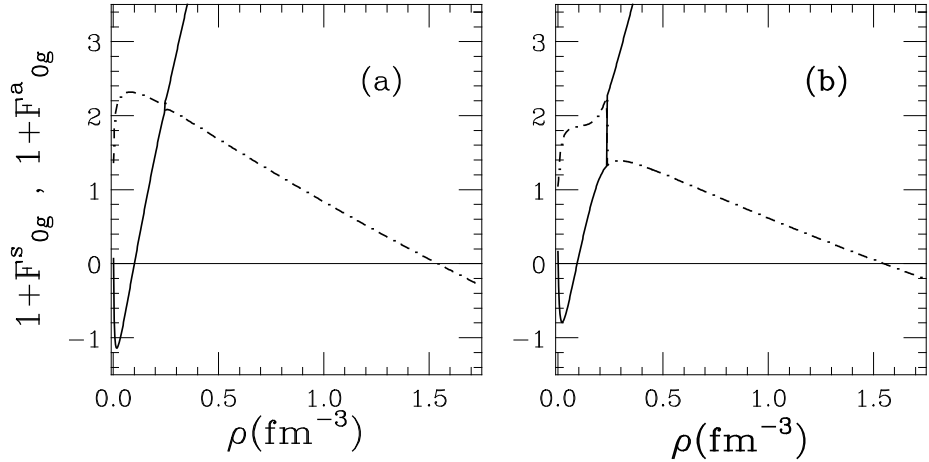} 
\caption{
Density dependence of the generalized Landau parameters for two proton
fractions, $y=0.4$ (a) and $y=0.1$ (b) (symmetric, solid, and 
antisymmetric, dashed)
at $T=1MeV$.
}
\label{rpa3}
\end{figure}

Investigations on instabilities in nuclear matter can be 
extended also to the high density region. Fig.\ref{rpa3} shows 
the density dependence of the generalized Landau parameters Eqs.(\ref{A}),
(\ref{B}).  
For the interaction considered here the system exhibits another
instability at high density, around $1.5 fm^{-3}$
 (far away from the validity of this $NM$ model), where the quantity
$c$ is positive and 
$1 + F_{0g}^a < 0$ (or $Y<0$).
From Eq.(\ref{chimec}) one finds that this is again a chemical
instability. However now it
results from isovector-like fluctuations, in contrast to the low density
instability. The
reason is the change in the character of the interaction between
the two components. Since the interaction becomes repulsive the nuclear phase
can become unstable against proton-neutron separation.
We also notice in Fig.\ref{rpa3} that the generalized 
Landau parameters display 
a discontinuity where the quantity $c$ changes the sign.
Other effective forces, with more repulsive symmetry terms, will not show
this high density chemical instability \cite{ColonnaPLB428}, that 
actually could
be of interest for other many body systems.

\subsection{Dynamical analysis}

The dynamical behaviour of a two-fluid system can be described, at the
semi-classical level, by considering 
two Vlasov equations, for neutrons
and protons in the nuclear matter 
case \cite{HaenselNPA301,MateraPRC49,ColonnaPLB428,BaranNPA632},
coupled through the self-consistent nuclear field :
\begin{equation}
{\partial f_q({\bf r},{\bf p},t) \over \partial t} 
+ {{\bf p} \over m}{\partial f_q \over \partial {\bf r}}  
- {\partial U_q({\bf r},t) \over \partial {\bf r}}
{\partial f_q \over \partial {\bf p} } = 0~,~~~~~~q=n,p.  
\label{ve}
\end{equation}
 
For simplicity effective mass corrections are neglected.  
In fact in the low density region, of interest for our analysis of spinodal 
instabilities,  effective mass corrections should not be large.  

$U_q({\bf r},t)$ is the self-consistent mean field potential 
in a Skyrme-like form \cite{ColonnaPLB428,BaranNPA632}~:
\begin{eqnarray}
U_q = {\delta{ H}_{pot} \over \delta\rho_q} =
  A\left({\rho \over \rho_0}\right) 
+ B\left({\rho \over \rho_0}\right)^{\alpha +1} 
+ C\left({\rho_3 \over \rho_0}\right)\tau_q
\nonumber \\
+ {1 \over 2} {d C(\rho) \over d \rho}
{\rho_3^2 \over \rho_0} - D\triangle\rho
                       + D_3\tau_q\triangle\rho_3~,         \label{Uq}
\end{eqnarray}
where
\begin{equation}
{ H}_{pot}(\rho_n,\rho_p)=
{A \over 2}{\rho^2 \over \rho_0} +
{B \over \alpha + 2}{\rho^{\alpha + 2} \over \rho_0^{\alpha + 1}} +
               {C(\rho) \over 2} {\rho_3^2 \over \rho_0}
+ {D \over 2}(\nabla\rho)^2
- {D_3\over 2}(\nabla\rho_3)^2~                           \label{hpotsurf}
\end{equation}
is the potential energy density (see Eq.\ref{hpot}), 
where also surface terms are included;
 $\rho = \rho_n + \rho_p$ and 
$\rho_3 = \rho_n - \rho_p$ are respectively the total (isoscalar) and the 
relative (isovector) density; 
$\tau_q$ = +1 ($q=n$), -1 ($q=p$).

The value 
of the parameter  
$D=130$ MeV$\cdot$fm$^5$ is adjusted to reproduce 
the surface
energy coefficient in the Bethe-Weizs\"acker mass formula $a_{surf}=18.6$ MeV.
The value 
$D_3= 40$ MeV$\cdot$fm$^5 \sim D/3$ 
is chosen according to Ref. \cite{BaymNPA175}, and 
is also close to the value $D_3 = 34$ MeV$\cdot$fm$^5$ given by 
the SKM$^*$ interaction \cite{KrivineNPA336}. 

Let us now discuss the linear response analysis 
to the Vlasov Eqs. (\ref{ve}),
 corresponding to a semiclassical $RPA$ approach.
For a small amplitude perturbation of 
the distribution functions $f_q({\bf r},{\bf p},t)$  ,
periodic in time,  $\delta f_q({\bf r},{\bf p},t) \sim \exp(-i\omega t)$, 
Eqs. (\ref{ve})~ can be linearized leading to the following form:
\begin{equation}
-i\omega\delta f_q 
+ {{\bf p} \over m}{\partial\delta f_q \over \partial {\bf r}}
- {\partial U_q^{(0)} \over \partial {\bf r}}
  {\partial\delta f_q \over \partial {\bf p}}
- {\partial\delta U_q \over \partial {\bf r}}
  {\partial f_q^{(0)} \over \partial {\bf p}} = 0~,   \label{lve}
\end{equation}
where the superscript $(0)$ labels stationary values and 
$\delta U_q$ is the dynamical component of the mean field potential.
The unperturbed distribution function 
$f_q^{(0)}$ is a Fermi distribution at finite temperature~:
\begin{equation}
f_q^{(0)}(\epsilon_p^q) = 
{1 \over \exp{(\epsilon_p^q-\mu_q)/T} + 1}~.          \label{Fd}
\end{equation}

Since we are dealing with nuclear matter, 
$\nabla_r U_q^{(0)} = 0$ in Eq. (\ref{lve}) and 
$\delta f_q \propto \exp(-i\omega t + i{\bf k r})$. 
Following the standard Landau procedure \cite{PethickAP183,ColonnaPLB428}, 
 one can derive 
from Eqs. (\ref{lve}) the following system of two equations for 
neutron and proton density perturbations~:
\begin{eqnarray}
& & [1 + F_0^{nn}\chi_n]\delta\rho_n 
  + [F_0^{np}\chi_n]\delta\rho_p      =   0~,       \label{eq1} \\
& & [F_0^{pn}\chi_p]\delta\rho_n
  + [1 + F_0^{pp}\chi_p]\delta\rho_p  =   0~,       \label{eq2}
\end{eqnarray}
where:
\begin{equation}
\chi_q(\omega,{\bf k}) = {1 \over N_q(T)} 
\int\,{2~d{\bf p} \over (2\pi\hbar)^3} 
      {{\bf kv} \over \omega + i0 - {\bf kv}}
      {\partial f_q^{(0)} \over \partial\epsilon_p^q}~,  \label{Lfun}
\end{equation}
is the long-wavelength limit of the Lindhard function \cite{PethickAP183}, 
 ${\bf v}={\bf p}/m$ and
\begin{equation}
F_0^{q_1q_2}(k) = N_{q_1}(T){\delta U_{q_1} \over \delta \rho_{q_2}}
~,~~~~~q_1=n,p,~~~~q_2=n,p                                \label{Lpar}
\end{equation}
are the usual zero-order Landau parameters, as already introduced 
in Eq.(\ref{ld}), 
where now the $k$-dependence is 
due to the presence of space derivatives in the potentials (see 
Eq.(\ref{Uq})).
For the particular choice of potentials given by Eq.(\ref{Uq}), 
the Landau parameters are 
expressed as:
\begin{eqnarray}
F_0^{q_1q_2}(k) = N_{q_1}(T)\left[ {A \over \rho_0} 
+ (\alpha + 1)B{\rho^\alpha \over \rho_0^{\alpha + 1}}
+          Dk^2  
+ ( {C \over \rho_0} - D'k^2 )\tau_{q_1}\tau_{q_2}\right. \nonumber \\
&&\nonumber\\
 + \left. {d C \over d \rho}{\rho' \over \rho_0}
           (\tau_{q_1} + \tau_{q_2})
+          {d^2 C \over d \rho^2}{\rho'^2 \over 2\rho_0}
      \right]~.                                             \label{Lpar1}
\end{eqnarray}
Multiplying the first equation by $N_n^{-1}\chi_p$ and the second one 
by $N_p^{-1}\chi_n$, we are led to define the following functions: 
\begin{eqnarray}
&& a(k,\omega) = N_p^{-1}(1 + F_{0}^{pp}\chi_p)\chi_n~~;~~
b(k,\omega) = N_n^{-1}(1 + F_{0}^{nn}\chi_n)\chi_p~~; \nonumber \\ 
&& c(k,\omega) = (N_p^{-1}F_{0}^{pn} + N_n^{-1}F_{0}^{np})\chi_n \chi_p = 
2 N_p^{-1}F_{0}^{pn}\chi_n \chi_p,~~ 
\end{eqnarray}
in some analogy with Eqs.(\ref{abc}) 
and we obtain the following system of equations:
\begin{eqnarray}
&& a \delta \rho_p + c/2~ \delta \rho_n = 0; \nonumber \\
&& c/2~ \delta \rho_p + b \delta \rho_n = 0 
\end{eqnarray}
The system can be diagonalized with eigenvalues $\lambda_s$ and $\lambda_i$
solutions of  the equation:
$$ (a-\lambda_{s,i})(b-\lambda_{s,i}) - c^2/4 = 0~.$$ 
Formally we obtain for $\lambda_{s,i}$ the same expressions as given in
Eqs.(\ref{A},\ref{B}) for $X$ and $Y$, 
but now $a,b$ and $c$
depend on $\omega$.   
The unstable solutions for $\omega$ are  obtained by
solving the equations:  
$\lambda_s = 0$ (for isoscalar-like fluctuations),  $\lambda_i = 0$
(for isovector-like fluctuations). 
This problem is completely equivalent to solve the equation: $c^2(\omega,k) = 
4a(\omega,k)b(\omega,k)$, i.e. the dispersion relation 
\begin{equation}
(1 + F_0^{nn}\chi_n)(1 + F_0^{pp}\chi_p) 
- F_0^{np}F_0^{pn}\chi_n\chi_p = 0~,                     \label{drel}
\end{equation} 
that is also obtained  directly by imposing the determinant 
of the system of Eqs.(\ref{eq1}), (\ref{eq2}) equal to zero. 

The dispersion relation is quadratic in $\omega$ and one 
finds two independent solutions (isoscalar-like  and isovector-like 
solutions): $\omega_s^2$ and $\omega_i^2$.   
Then the structure of the eigenmodes can be determined and one finds:
$$\delta \rho_p/\delta \rho_n = -2b(\omega_s,k)/c(\omega_s,k),$$ 
 for the isoscalar-like modes and
$$\delta \rho_p/\delta \rho_n = -2b(\omega_i,k)/c(\omega_i,k),$$ 
for isovector-like oscillations.  
However, it is important to notice that the corresponding angles 
$\beta_{s,i}$ 
are not equal to the angle $\beta$ determined in the thermodynamical 
analysis, Eq.(\ref{beta}),  because of the $\omega$ dependence 
in $a,b$ and $c$. 
They only coincide with $\beta$ when $\omega = 0$ (and thus $\chi_{n,p} = 1$), 
i.e. at the border of the unstable region.   

\subsubsection{Illustrative results for liquid-gas phase transitions}
\begin{figure}  
\centering
\includegraphics[scale=0.65]{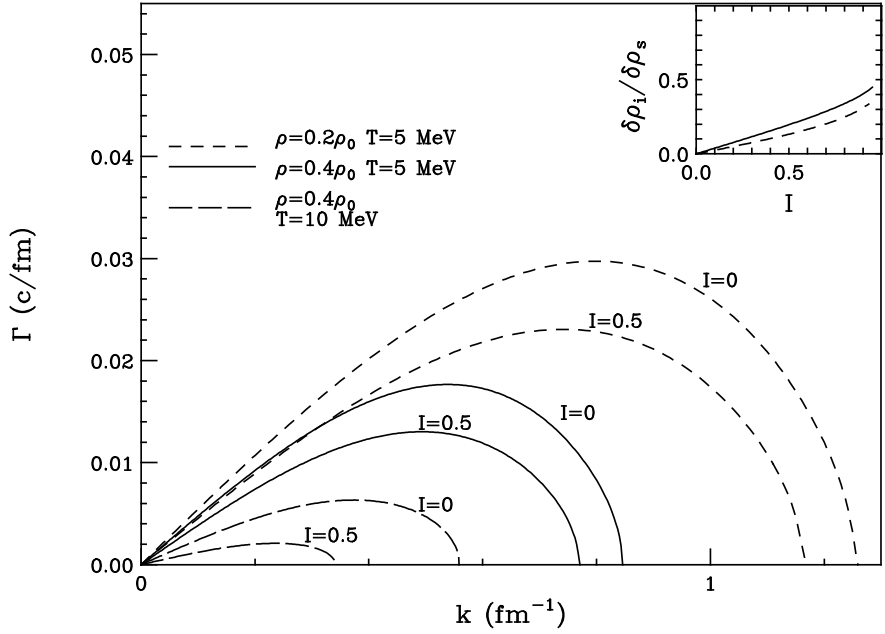} 
\caption
{Growth rates of instabilities as a function of the wave vector, as calculated
from the dispersion relation Eq.(\ref{drel}), for three situations inside the 
spinodal region. Lines are labelled with the asymmetry value $I$. 
The insert shows the asymmetry of the perturbation 
$\delta \rho_I / \delta\rho_S$, 
as a function of the asymmetry $I$ of the initially uniform system,
 for the most unstable mode, in the case $\rho = 0.4 \rho_0$, $T=5~MeV$.}
\label{rpa4}
\end{figure}

%\addtocontents{toc}{\hspace{0.55cm}\thesubsubsection \hspace{0.12cm}
%Illustrative results for liquid-gas phase transitions}
The dispersion relation, Eq.(\ref{drel}), have been solved 
for various choices of the initial density,
temperature and asymmetry of nuclear matter.
Fig.\ref{rpa4} reports the 
 growth rate $\Gamma~=~{\rm Im}~\omega(k)$
 as a function of the wave vector $k$,  for three situations inside the
spinodal region.  Results are shown for symmetric ($I=0$) and asymmetric 
 ($I=0.5$)
nuclear matter.
The growth rate has a maximum $\Gamma_0=0.01\div0.03$ c/fm corresponding 
to a wave-vector value around $k_0=0.5\div1~\mbox{fm}^{-1}$ and 
becomes equal to zero
at $k\simeq1.5k_0$, due to $k$-dependence of the Landau parameters, as 
discussed above. 
One can see also that instabilities are reduced when increasing the
temperature, an effect also present in the symmetric N = Z case 
\cite{ColonnaPRC49,ColonnaNPA567,ChomazPR389}.

At larger initial asymmetry the development 
of the spinodal instabilities is slower, decrease of the maximum of the
growth rate. One should  expect also 
an increase of the size of the produced 
fragments, decrease of the wave number corresponding to the maximum growth 
rate. 
From the long dashed curves of Fig.\ref{rpa4} we can predict the
asymmetry 
%dependence of both variables
%$\Gamma_0$ and $\lambda_0$ is 
effects to be 
more pronounced at higher temperature,
when in fact the system is closer to the boundary of the spinodal region.

\subsubsection{Coulomb effects on instabilities}
\begin{figure}  
\centering
\includegraphics[scale=0.50]{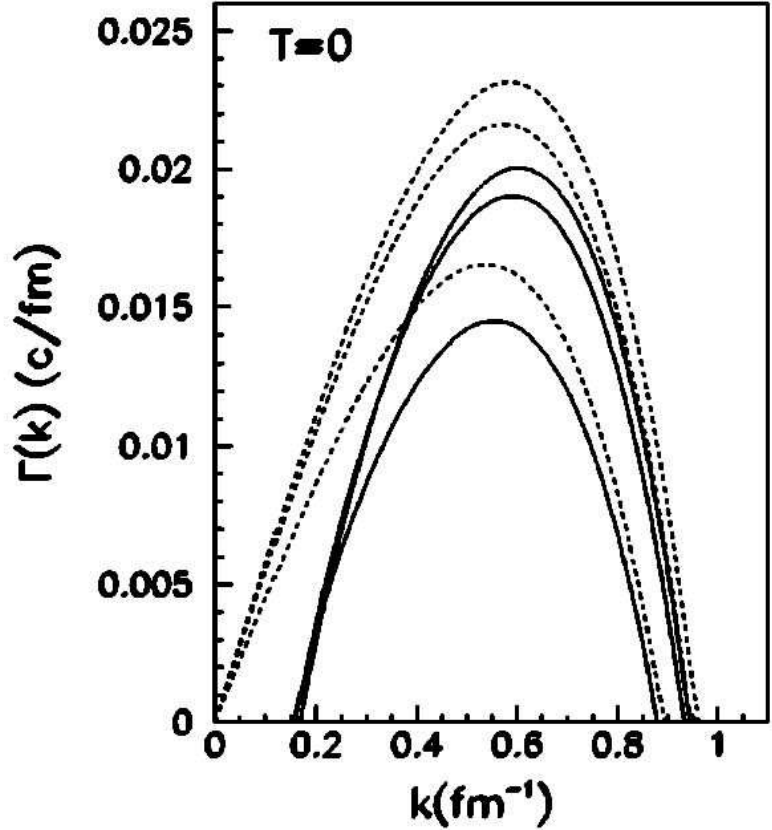} 
\caption
{Growth rates of the unstable modes for T=0, $\rho = 0.4~\rho_0$ and three
different values of the asymmetry parameter $I$ (from top to bottom 
$I = 0, 0.3, 0.6$). Results including the Coulomb interaction (full line) 
and without the Coulomb interaction (dashed) are shown. Taken from 
\cite{FabbriPRC58}} 
\label{rpa5}
\end{figure}

%\addtocontents{toc}{\hspace{0.55cm}\thesubsubsection \hspace{0.12cm}
%Coulomb effects on instabilities}
%%%%%  Coulomb effects 
The influence of Coulomb effects on the growth rates 
%in asymmetric nuclear matter
can be easily investigated within the formalism outlined above, as done in
\cite{FabbriPRC58}.
It suffices to add to the energy density, Eq.(\ref{hpotsurf}), the 
Coulomb energy density, that can be calculated in the 
Hartree-Fock approximation,
with the Fock term evaluated in the local density approximation:
\begin{equation}  
{ H}^{(C)}({\bf r})=\frac{e^2}{2}\rho_p({\bf r})\,
\int d{\bf r}^\prime\frac{\rho_p({\bf r}^\prime)}
{\vert {\bf r}-{\bf r}^\prime\vert}-\frac{3}{4}\Bigl(\frac{3}{\pi}\Bigr)
^{\frac{1}{3}}e^2\rho_p^{\frac{4}{3}}~.\label{coulen}
\end{equation}
This only modifies the $F_0^{pp}$ Landau parameter, implemented
by adding the term:
\begin{equation}
\frac{4\pi e^2}{k^2} - \frac{1}{3}\Big({\frac{3}{\pi}}\Big)^{\frac{1}{3}}e^2
{\rho_p}^{-\frac{2}{3}}
\end{equation}
Fig.\ref{rpa5} 
shows a comparison between the results obtained with and without 
the Coulomb interaction, for density $\rho = 0.4 \rho_0$ and
$T=0$.
The Coulomb force causes an overall decrease of growth rates. This decrease is
almost independent of the asymmetry. Moreover, as discussed in 
\cite{FabbriPRC58}, it
depends only slightly on the temperature. 
It is also observed that, when the Coulomb force is included, the wave 
vector $k$ must exceed a certain value $k_{min}$ in order to observe 
instabilities. 
The two effects (the decrease of the growth rate and the appearance of 
$k_{min}$) are due to the competition between the Coulomb and the nuclear 
forces. In fact, the Coulomb forces push the protons towards 
regions of lower density, the nuclear forces instead push the neutrons 
in this direction (neutron distillation).  
 
\subsubsection{Isospin distillation}
\begin{figure}[htb]
%\epsfysize=8.0cm
%\centerline{\epsfbox{freccia.ps}}
\centering
\includegraphics[scale=0.60]{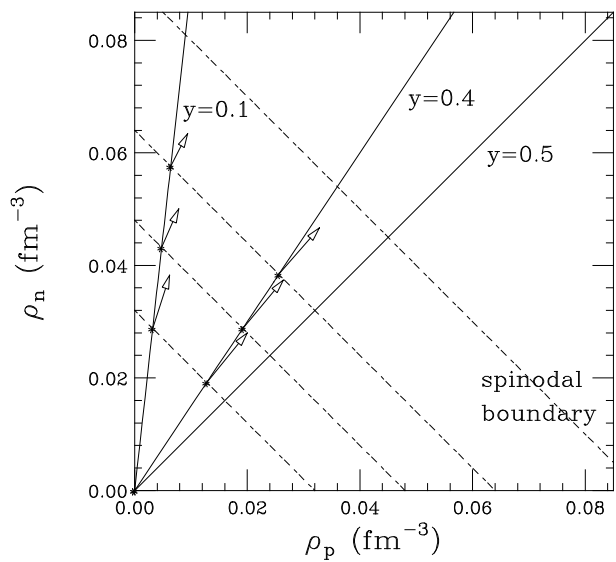} 
\caption{
Direction of $\delta\rho_n/\delta\rho_p$ in several points of the 
 $(\rho_p,\rho_n)$ plane.
}
\label{rpa6}
\end{figure}

%\addtocontents{toc}{\hspace{0.55cm}\thesubsubsection \hspace{0.12cm}
%Isospin distillation}
A better understanding of liquid-gas phase transitions  
in a two-component system can be achieved
by studying the chemical composition of the growing mode. 
This is shown in the insert of Fig.\ref{rpa4}, where
%Fig. 4 shows  
the asymmetry of the perturbation 
$I_{pt}=(\delta\rho_n - \delta\rho_p)/(\delta\rho_n + \delta\rho_p)$
= $\delta \rho_I / \delta\rho_S$, as obtained by solving the 
system of Eqs.(\ref{eq1}), (\ref{eq2}),  
is diplayed
as a function of the asymmetry of the initially uniform system
$I=(\rho_n^{(0)} - \rho_p^{(0)})/(\rho_n^{(0)} + \rho_p^{(0)})$, for the
dominant (most unstable) mode (for the case $\rho = 0.4~\rho_0$, $T=5~MeV$). 
Without any chemical processes it should be $I_{pt}=I$.
However one obtains $I_{pt} \le 0.5~I$. This means that a growing mode
produces more symmetric high-density regions (liquid phase) and
less symmetric low-density regions (gas phase). Hence, during the 
fragmentation,
a collective diffusion of protons from low-density regions to high-density
regions takes place.   This is the isospin fractionation (distillation) 
effect already
discussed in the context of liquid-gas phase transitions in two-component
systems. 
%It is evident from Fig. 4 that the chemical effect becomes stronger 
%with increasing
%density. 
This is essentially due to the increasing behaviour of the symmetry 
energy per nucleon with density, in the density region considered here.
The distillation effect is represented in Fig.\ref{rpa6} , where the ratio
$\delta\rho_n/\delta\rho_p$, as obtained by solving the dispersion 
relation, is displayed (direction and magnitude of the arrows) for some 
points of the $(\rho_p,\rho_n)$ plane inside the spinodal region. 
One observes that the distillation effect is more
pronounced at large asymmetry (i.e. smaller proton fraction $y$).

We recall that the {\it Isospin Distillation} has been revealed even
in n-rich finite nuclei, performing quantal $RPA$ calculations
\cite{ColonnaPRL88} (see also \cite{ChomazPR389} for a thorough discussion
on instabilities in finite systems).

As discussed before, the size of the effect is related to the 
stiffness of the symmetry 
energy, as illustrated in Table \ref{dist}, where the ratio 
$\delta\rho_n/\delta\rho_p$, as obtained using a stiff 
 or a soft (in brackets) parameterization of the
symmetry energy,  is reported. As expected, the neutron distillation
effect (reduction of the asymmetry in the formed clusters) is
systematically larger at lower densities in the asy-soft case
while the opposite is seen already at $\rho = 0.4 \rho_0$.
This is nice evidence of the dependence on the {\it slope} of
the symmetry term in the low density region (see Fig.1, Sect.\ref{intro}).

\begin{table}[t]
\begin{center}
\begin{tabular}{|c|c|c|} \hline 
  $\rho/\rho_0$   & $I=0.2~(N/Z=1.5)$   & $I=0.8~(N/Z=9.0)$   \\ 
\hline\hline
   0.2 &  1.23~(1.2)   &    3.15~(2.61)      \\ \hline
   0.3  &  1.19~(1.186)   &    2.44~(2.4)      \\ \hline
   0.4  &  1.15~(1.18)   &    1.99~(2.32)          \\ \hline
\end{tabular}
\end{center}
\vskip 0.2cm
\caption{\label{dist} 
Isospin content of the clusters, density 
variations $\delta\rho_n/\delta\rho_p$, formed 
in dilute matter at different initial densities for two initial asymmetries,
$N/Z=1.5~and~9.0$. The values in brackets are obtained with a softer
symmetry term (asy-soft $EOS$)}
\end{table}

\subsection{Simulation results: heated nuclear matter in a box}

The previous analytical study is restricted to the onset of 
Fragmentation, and related Isospin Distillation, in Nuclear Matter,
 in a linearized approach.
Numerical calculations have been performed in order to study all
stages of the fragment formation process \cite{BaranNPA632,BaoNPA699}.
We report on the results of ref.\cite{BaranNPA632} where the same
effective Skyrme interactions have been used.

In the numerical approach 
the dynamical response of nuclear matter is studied in a cubic box of 
size $L$ imposing periodic boundary conditions. 
The Landau-Vlasov 
dynamics is simulated following a phase-space test particle method,
 using gaussian wave packets 
\cite{GregoireNPA465,BonaseraPR243,BaranPPNP38}. 
 The dynamics of nucleon-nucleon collisions is included by solving 
the Boltzmann-Nordheim collision integral using a Monte-Carlo method
\cite{BonaseraPR243}.
The width 
of the gaussians is chosen in order
to correctly reproduce the surface energy value in finite systems. 
In this way a cut-off appears in the short wavelength unstable modes,
preventing the formation of too small, unphysical,  clusters 
\cite{ColonnaPRC49}. 
  The calculations are performed using 80 gaussians per nucleon and the
number of nucleons inside the box is fixed in order to reach the initial 
uniform density value. An initial temperature is introduced by distributing the
test particle momenta according to a
Fermi distribution.

\begin{figure}[htb]
\centering
\includegraphics[scale=0.55]{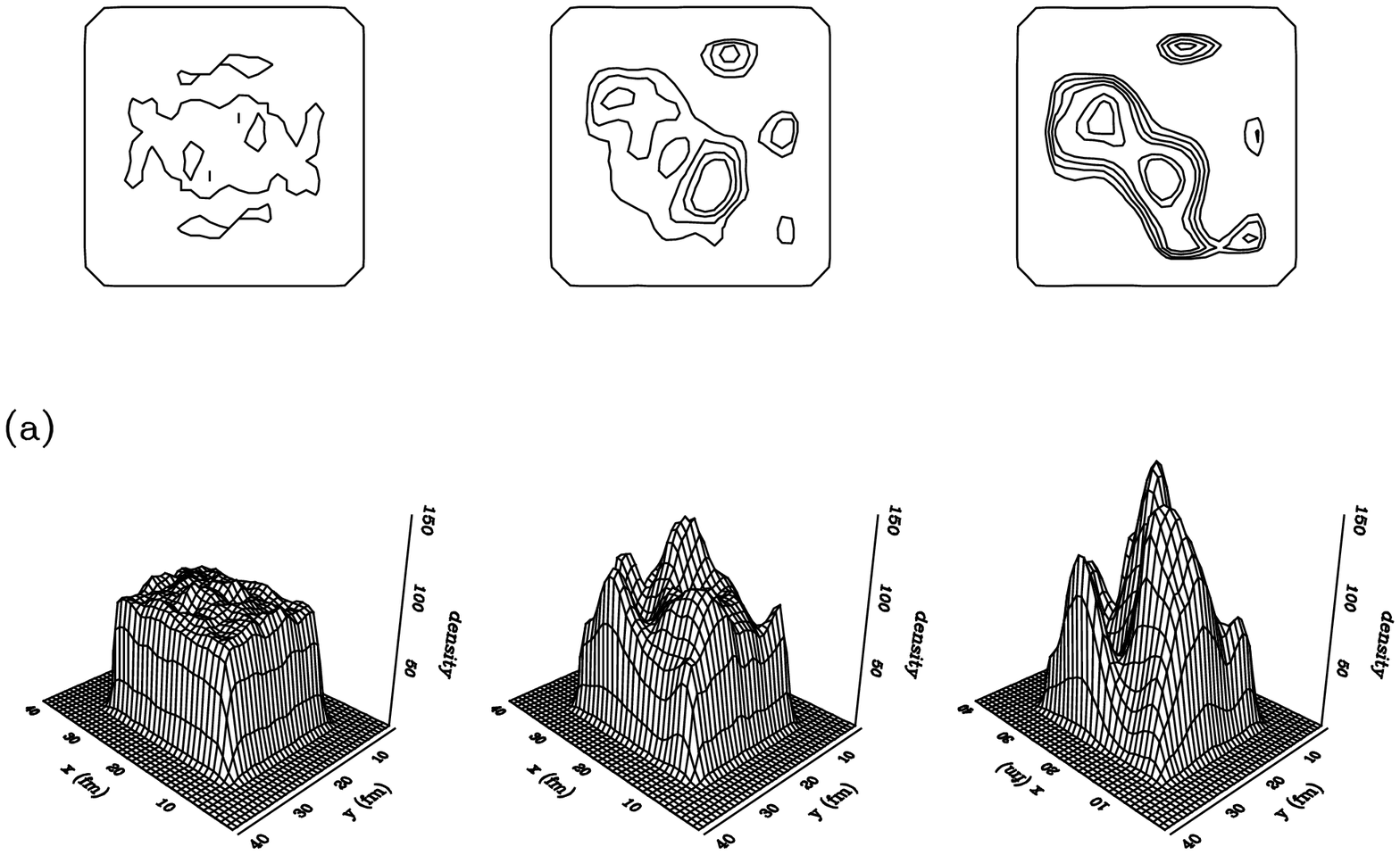}
\vskip 0.2cm 
\includegraphics[scale=0.55]{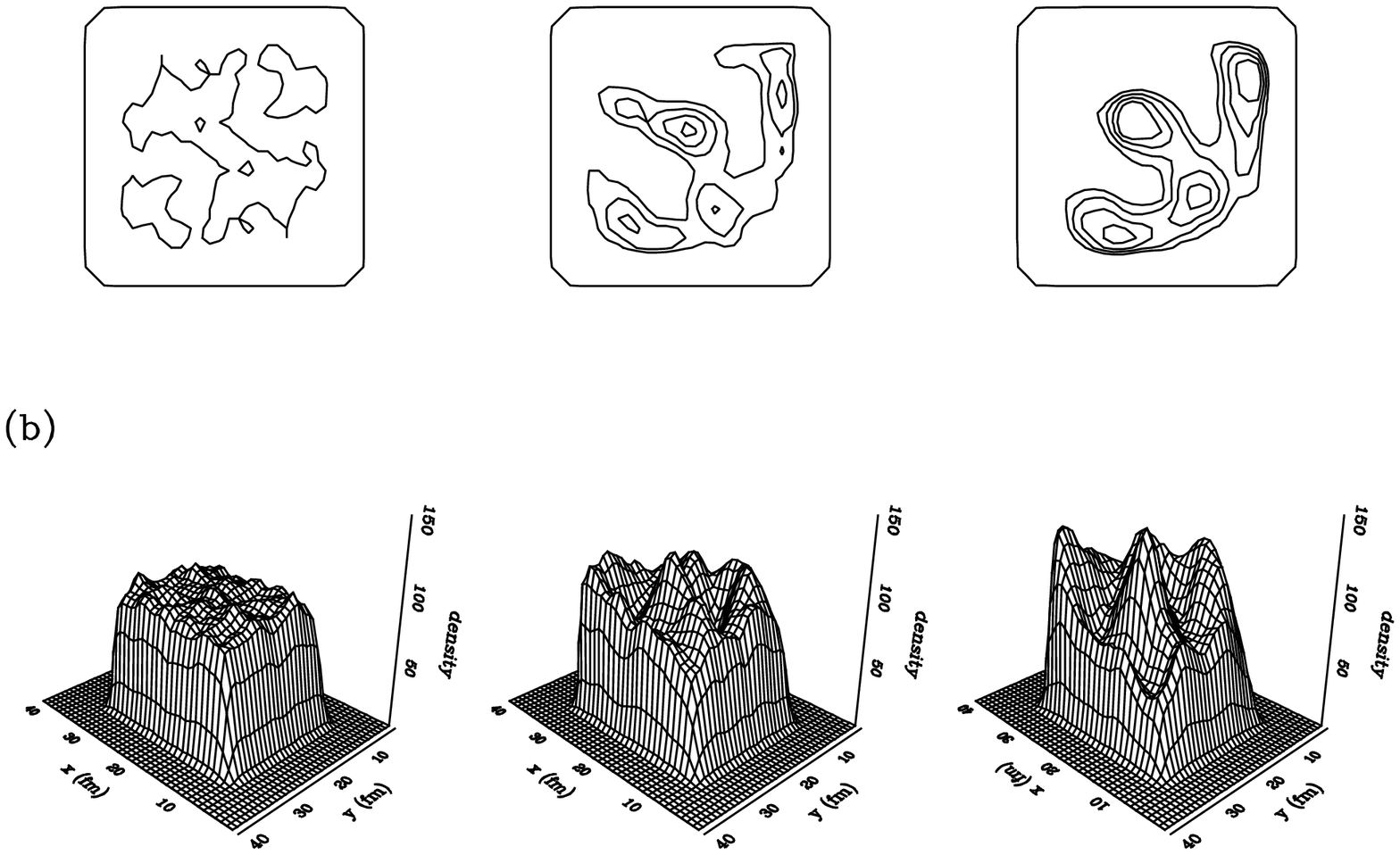}
\caption{Time evolution ($t=0.0,~100.0~and~200.0~fm/c$) of the density 
$\rho(x,y)$ in the plane $z=0$. Initial conditions: $\rho=0.06fm^{-3}$,
$T=5MeV$ and asymmetries $I=0.0~(a);I=0.5~(b)$.
Upper panels: contour plots; Lower panels: corresponding two-dimensional 
surfaces. Densities in $fm^{-3}$.
}
\label{rpa7}
\end{figure}

 We have followed the space-time
evolution of test-particles in a cubic box with side $L=~24fm$ 
for three values of 
the initial
asymmetry $I=0,~0.25$ and $0.5$, at initial density 
$\rho^{(0)}=0.06fm^{-3} \simeq 0.4\rho_0$ and temperature $T=5$ MeV. 
The initial density 
perturbation is created automatically due to the random choice of 
test-particle positions. 
Results for the initial asymmetries $I=0$ and
$I=0.5$, are reported in Fig. \ref{rpa7}, (a) and (b) respectively.  
The figure  shows density distributions in the 
plane $z=0$, which contains the center of the box, at three time steps
$t=0,~100$ and $200$ fm/c, corresponding respectively to initial 
conditions, intermediate and final stages of the cluster formation. 
Clearly, a 
growth of the small initial density perturbations takes place,
this time non-linear and hard particle collision effects are
included. 
%It is seen, 
%also that the "$SD$" is delayed 
%and results in a larger size of clusters for a larger asymmetry.

\begin{figure}[htb]
\centering
\includegraphics[scale=0.55]{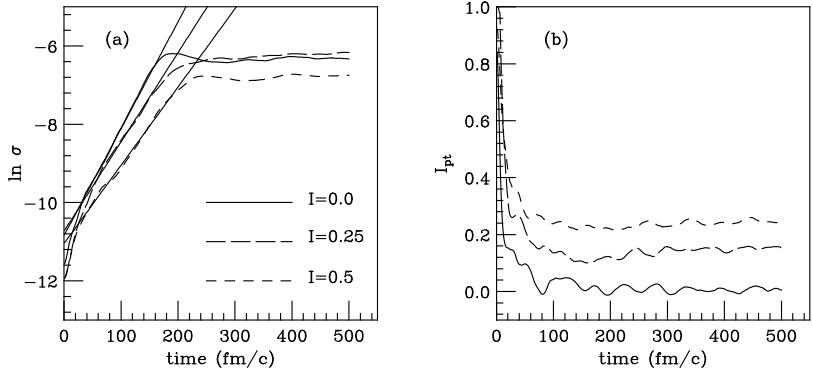}
\caption{Time evolution  of the density variance (a) and of the
perturbation asymmetry (b) (see text) for the box simulations
at initial density $\rho=0.06fm^{-3}$, $T=5MeV$ and asymmetries
 $I=0.0,~0.25,~0.5$. The straight lines in (a) show linear
logarithmic fits in the initial stage of the cluster formation. 
}
\label{rpa8}
\end{figure}

We can compare the dynamical evolution with the analytical predictions of 
the previous sections. 
To do this, two variables are 
constructed: the total density variance (see \cite{ColonnaPL307})
\begin{equation}
         \sigma = < (\rho - \rho^{(0)})^2 >_{all}        \label{totvar}
\end{equation}
and the correlation function between proton and neutron density perturbations,
normalized to the neutron density variance,
\begin{equation}
         R_{pn} = {< (\rho_p - \rho_p^{(0)}) (\rho_n - \rho_n^{(0)})
         >_{all}   \over < (\rho_n - \rho_n^{(0)})^2 >_n }~.
                                                         \label{Rpn}
\end{equation}
In Eqs. (\ref{totvar}),(\ref{Rpn}) $<...>_{all}$ denotes the average
over all test particles, while $<...>_n$ denotes the average over
neutrons only.  The densities $\rho$, $\rho_n$ and $\rho_p$ 
are calculated in the position of the test particle considered by taking 
contributions from gaussians of all test particles. For a 
dominant plane-wave perturbation we have the limit:
\begin{equation}~~~~~~~~~
\sigma \propto \exp(2\Gamma t),~~~~~ 
R_{pn} = {\delta\rho_p \over \delta\rho_n}~.      
\end{equation}
Fig.\ref{rpa8} shows the evolution of 
%$\mbox{ln}\,\sigma$ (a) and of the
$\sigma$ (a) and of the
(test-particle) perturbation asymmetry $I_{pt}=(1-R_{pn})/(1+R_{pn})$
 (b)
for the same initial conditions discussed above, i.e. $T=5~MeV$,
$\rho^{(0)}=0.4\rho_0$ and asymmetries $I=0.0,~0.25,~0.5$.
A general feature is the
 clear linear increase of $ln(\sigma)$ in the time
interval
$50 < t < 150$ fm/c. During the first $50~fm/c$ the system is
quickly "self-organizing", selecting the most unstable normal
mode. Afterwards the variance (Eq. (\ref{totvar})) increases exponentially
with a time scale given by $\Gamma={\rm Im}~\omega(k)$, finally it saturates.
In correspondence (see Fig.\ref{rpa8}b), the perturbation asymmetry $I_{pt}$
reveals also a quick saturation at 
$t \sim 50$ fm/c. At times before $50fm/c$ the proton and
neutron density perturbations are not correlated 
($I_{pt}(t=0) \simeq 1.0$ since nicely $R_{pn}(t=0)\simeq 0.0$) , but at 
$t > 50$ fm/c the correlation of plane-wave type 
$(\delta\rho_p/\delta\rho_n=\mbox{const} > 0)$ develops.

We notice that the time scales necessary to reach the asymmetry value
characteristic of the most important growing modes, which are quite 
short in our calculations, generally depend on the structure of the 
initial noise put in the neutron and proton densities. In our calculations
all modes are nearly equally excited. This causes the quick appearance of 
the features associated with the dominant mode. In agreement with analytical 
calculations, the instability grows slower in the case of larger asymmetry.

For an initial asymmetry $I=0.5$,
the extracted values of growth time $\Gamma \simeq 0.01$ c/fm and perturbation
asymmetry $I_{pt} \simeq 0.24$ (see Fig.\ref{rpa8}), and of wavelength 
$\lambda \simeq 12$ fm (from the distance between the density distribution 
maxima in Fig.\ref{rpa7}b),
are in good agreement with the analytical 
results presented before.

The Spinodal Decomposition Mechanism, $SD$, leads to a fast 
formation of the liquid 
(high density) and gaseous (low density) phases in the matter. 
Indeed this dynamical mechanism of clustering will roughly end
when the variance (Eq. (\ref{totvar})) saturates \cite{ColonnaNPA580}, i.e.
around $250~fm/c$ in the asymmetric cases (see Fig.\ref{rpa8}a).
\begin{figure}[htb]
\centering
\includegraphics[scale=0.55]{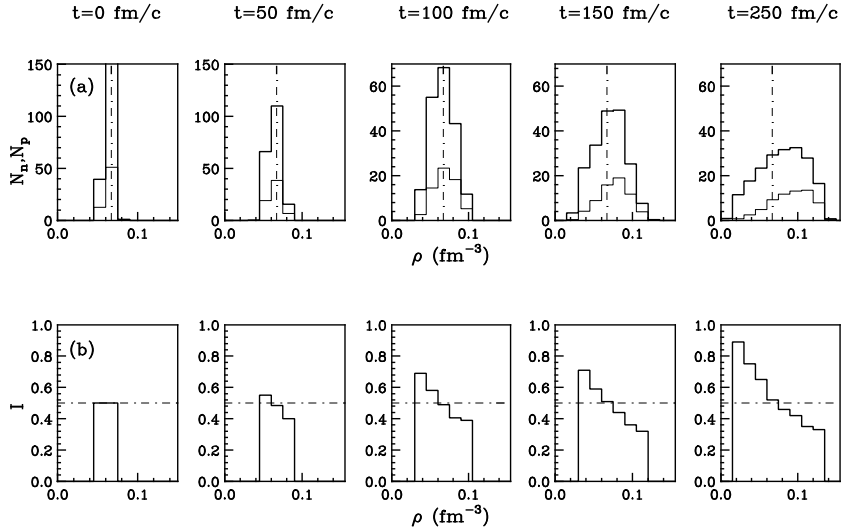}
\caption{Time evolution  of neutron ({\it thick lines}) and proton
({\it thin lines}) abundances (a) and of asymmetry (b) in different density 
bins. The calculation refers to the case of $T=5MeV$, with
 initial average density $\rho=0.06fm^{-3}$ and asymmetry
 $I=0.5$ (see the first panel of the (b) plots).
}
\label{rpa9}
\end{figure}
We also discuss the "chemistry" of the liquid
phase formation. In Fig.\ref{rpa9} we report the time evolution of
neutron (thick histogram in Fig. \ref{rpa9}a) 
and proton (thin histogram in Fig.\ref{rpa9}a)
abundances and of asymmetry (Fig.\ref{rpa9}b) in various density bins.
The dashed lines respectively shows the initial uniform density
value $\rho \simeq 0.4\rho_0$ (Fig.\ref{rpa9}a) and the initial asymmetry
$I=0.5$ (Fig.\ref{rpa9}b). The drive to higher density regions is clearly 
different for neutrons and protons: at the end of the dynamical 
clustering mechanism we have very different asymmetries in the 
liquid and gas phases (see the panel at $250fm/c$ in Fig.\ref{rpa9}b).

It was shown in Refs. \cite{BaymNPA175,MuellerPRC52,BaoNPA618}, on 
the basis of 
thermodynamics, that the two phases should have different asymmetries,
namely, $I_{gas} > I_{liquid}$, and actually a pure neutron gas
was predicted at zero temperature if the initial global asymmetry 
is large enough ($I>0.4$) \cite{BaymNPA175}. Here we are studying this 
chemical effect in a non-equilibrium clustering process, on very short 
time scales, and we confirm the predictions of a linear response approach
discussed before.

\begin{figure}[htb]
\centering
\includegraphics[scale=0.55]{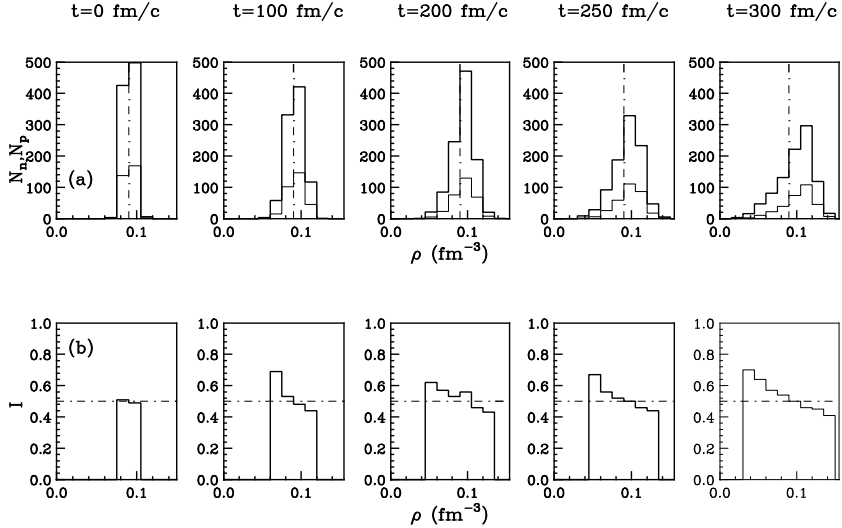}
\caption{Same calculation as in Fig.\ref{rpa9} but with
 initial average density $\rho=0.09fm^{-3}$, inside the {\it chemical}
instability region.
}
\label{rpa10}
\end{figure}

We can directly check the important result on the unique
nature of the most unstable mode, independent of whether we start 
from a {\it mechanical}
or from a {\it chemical} instability region.
The isospin distillation dynamics presented in the 
Fig.\ref{rpa9}  refers to the initial conditions of $T=5MeV$,
 average density $\rho=0.06fm^{-3}$ and asymmetry
 $I=0.5$, i.e. we start from a point well inside the {\it mechanical}
instability region of the used $EOS$, see Fig.\ref{rpa1}(b).
We can repeat the calculation at the same temperature and initial asymmetry,
but starting from an initial average density $\rho=0.09fm^{-3}$,
i.e. inside the {\it chemical} instability region of  Fig.\ref{rpa1}(b).
The results for the {\it Isospin Distillation Dynamics} are shown
in the Fig.\ref{rpa10}.
The trend is the same as in the previous Fig.\ref{rpa9}. 
This nicely shows the uniqueness of the unstable modes in the
spinodal instability region, as discussed in detail in the previous 
subsection. 
Such result is due to gross properties of the {\it n/p} interaction, 
thus it should be not dependent on the use of a particular 
effective force. This has been clearly shown recently in the linear 
response frame, \cite{MargueronPRC67}, and in full transport simulations,
\cite{BaoNPA699}.

As intuitively expected, and as confirmed by the $RPA$ analysis (see 
\cite{BaranNPA632}), 
the isospin distillation effect becomes more important when increasing
the initial asymmetry of the system. 
At the same time, the instability growth rates become smaller for the 
more asymmetric systems, see Fig.\ref{rpa4}. 

Moreover it is possible to observe
a rather smooth 
and continuous transition from 
the trend observed at $\rho=0.06fm^{-3}$ (mechanical unstable region)  
to the trend observed at $\rho=0.09fm^{-3}$ (chemical unstable region),  
 thus indicating that  there is no qualitative change between the two kinds of 
instabilities. In fact they actually correspond to the same mechanism, 
the amplification of isoscalar-like fluctuations, with a significant 
chemical component (change of the concentration).   

The conclusion is that the fast spinodal decomposition mechanism in  
neutron-rich matter will dynamically form more symmetric fragments 
surrounded by a less symmetric gas. 
Some recent experimental observations from fragmentation reactions
with neutron rich nuclei at the Fermi energies seem to be in 
agreement with this result on
the fragment isotopic content : nearly symmetric Intermediate Mass
Fragments ($IMF$) have been detected in connection to very neutron-rich
light ions \cite{XuPRL85}, \cite{YennelloHIP97}.  
This will be extensively discussed in Sect.\ref{fermi}.

%\include{rep_bib}

%\end{document}

%% file: Chapter-4.tex
%\documentclass{elsart}
%\usepackage{epsfig}

%\usepackage{graphicx}

%\usepackage{amssymb}
%\tightenlines
% nuovi comandi
% 2 su 2pigreco al cubo
%\newcommand{\norm}{\frac{2}{(2\pi)^3}}
% parentesi quadre
%\newcommand{\qd}[1]{\left[ #1 \right]}
% parentesi tonde
%\newcommand{\td}[1]{\left( #1 \right)}

%\newcommand{\itg}[1]{\norm \int d^3k f_{#1}(k)}
% integrale di fn
%\newcommand{\ienne}[1]{\itg{n} #1}
% integrale di fp
%\newcommand{\ipi}[1]{\itg{p} #1}
% integrale di fn g
%\newcommand{\ien}{\ienne{g(k,\Lambda)}}
% integrale di fp g
%\newcommand{\iz}{\ipi{g(k,\Lambda)}}

%\setlength{\unitlength}{1cm}

% integrale di ftau g
%\newcommand{\itau}{\itg{\tau}g(k,\Lambda)}
% integrale di ftau g
%\newcommand{\itaup}{\itg{\tau ^\prime} g(k,\Lambda)}

% rozero
%\newcommand{\rz}{\rho_{_0}}
% rho su rozero
%\newcommand{\ra}{\td{ \frac{\rho}{\rho_{_0}} }}
% densita' di energia per A e B
%\newcommand{\ene}[1]{\qd{
% \td{\frac{1}{2}x_{#1}}\rho^2
%-\td{\frac{1}{2}+x_{#1}} \td{\rho_n^2+\rho_p^2} }}
% 0.5 + x0(x3)
%\newcommand{\umd}[1]{ \td{ \frac{1}{2}+x_{#1} } }

%\newcommand{\inew}[1]{\mathcal{I}_#1 }

%\begin{document}
\setcounter{figure}{0}
\setcounter{equation}{0}

\section{Symmetry term effects on fast nucleon emission and collective flows}
\label{fastflows}

\markright{Chapter \arabic{section}: fastflows}

\subsection{Pre-equilibrium dynamics}

The early reaction dynamics is mainly governed by the pressure of 
the excited nuclear matter formed in the initial stage of the collision,
 \cite{HandzyPRL75,BaoPRC64,DanielNPA685}. 
We recall that  
a $stiff$ symmetry energy gives a larger gradient of pressure than
the $asy-soft$ case, as shown in Sect.\ref{eos}, Eq.(8). This
should result in an overall faster emission of particles. 
Moreover, the soft behaviour of the symmetry energy leads (at low density) 
to a larger repulsion of neutrons, with respect to the stiff case
(see Fig.\ref{fig:mean} of Sect.\ref{eos}). 
Thus neutrons are expected to be emitted at earlier times, with respect
to protons, in the $asy-soft$ case. 
The isotopic content of the pre-equilibrium emission is also sensitive
to the symmetry part of the $EOS$. For the same reasons explained above
more neutrons are emitted in the soft case.

These ideas have lead
to investigate the effect of the symmetry energy on the nucleon 
emission. Of course we do not have direct access to the particle 
emission time experimentally, 
but two-particle correlation functions, through final-state
interactions and quantum statistics effects, have been shown to be a
sensitive probe to the temporal and spatial distribution of emission
sources during the reaction
dynamics of heavy ion collisions at incident energies ranging from intermediate
\cite{BoalRMP62,BauerARNPS42,ArdouinIJPE6} 
to RHIC energy \cite{WiedemannPR319}.

To discuss isospin effects on pre-equilibrium emission, we will review 
the results obtained by Chen et al., \cite{ChenPRL90,ChenPRC68}, using
an isospin dependent transport code, $IBUU$. This is a one-body dynamics
approach without fluctuating terms \cite{BertschPR160} but the results
discussed here can be reliable since mainly due to the average nuclear 
dynamics.

We will consider a typical reaction with radioactive beams:
$^{52}$Ca + $^{48}$Ca at $E=80$ MeV/nucleon \cite{ria}, central collisions. 
This study has been performed for two kinds of density dependence, 
labelled as ``soft'' and ``stiff'',
that correspond to $\gamma=0.5$ and
$\gamma=2.0$ respectively, in the parametrization of the symmetry energy
 (total, kinetic $+$ potential) 
$$E_{\mathrm{sym}}(\rho )=E_{\mathrm{sym}}(\rho _{0})\cdot u^{\gamma }. $$

 It is important to note that, in this case, ``soft'' refers to a slope 
 around normal density $L=52$ MeV, which
is close to the common value,  
while in the
stiff case the slope is $L=210$ MeV, which is much larger than what has been
labelled as ``super-stiff'' ($L \simeq 100$ MeV) in the previous 
Sect.\ref{eos}. 

\subsubsection{Average emission times}
\begin{figure}[htb]
\centering
%\centerline{\epsfbox{emTimeP.eps}}
%\includegraphics[height=3.0in,width=3.0in]{emTimeP.eps}
\includegraphics[scale=0.8]{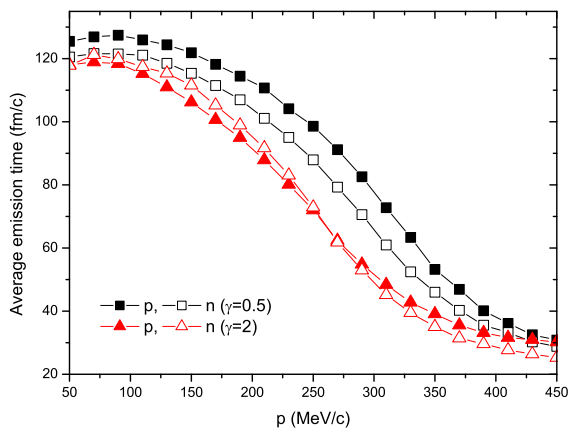} \vspace{-0.5cm}
\caption{Average emission times of protons and neutrons as
functions of their momenta for different symmetry energies, for 
 $^{52}$Ca+$^{48}$Ca at 80 MeV/nucleon b=0 fm, see text
 (from \cite{ChenPRC68}).}
\label{emTimeP}
\end{figure}

%\addtocontents{toc}{\hspace{0.55cm}\thesubsubsection \hspace{0.12cm}
%Average emission times}
For the two symmetry energy parametrizations we report
in Fig. \ref{emTimeP} ($soft$: squares, $stiff$: triangles) the 
average emission 
time of protons and neutrons
 as a function of their momenta. It is seen that the average emission
time of nucleons with a given momentum is earlier for the stiff symmetry
energy than for the soft one. This is expected because of the 
overall larger pressure gradient in the stiff case. 
We also notice that the relative 
emission sequence of neutrons and protons is, however, determined by the 
difference in their symmetry potentials. In other words if we average over
neutron and proton the emission time is faster for the stiff case, but,
on top of this there is a difference in the neutron-proton emission that
is driven by the symmetry potential which is more attractive for protons
causing an emission delay with respect to neutrons.
 
The neutron-proton difference is larger for the soft case because
the isovector mean field at low densities, where most nucleons 
are emitted, is larger for the soft symmetry energy than for the stiff 
symmetry potential. In such a reasoning we have disregarded the effect of
Coulomb interaction which reduces the delay of protons with respect to 
neutrons.
For $^{52}$Ca+$^{48}$Ca the effect is quite small, but it grows with 
the size of the colliding system 
modifying
the $p-n$ difference in the emission time, $|t_p-t_n|$. 
Therefore for heavier systems one can expect
a reduction of the $|t_p-t_n|$ for the soft case and an increase of the 
difference for the stiff case, because the stronger Coulomb effect will shift 
down the proton average emission time.
We will discuss the consequences on the correlation functions and 
the cluster formation in the corresponding subsections. 

We notice the relation between larger momenta and earlier emission.
For the reaction considered here, it is shown that most
of the particles with momemtum $P \geq 250$ MeV/c are emitted within  
$t\leq 70$ fm/c. This allows to connect the momenta of the emitted particles
with the time evolution of the reaction, giving hints also about
the average density at which particles are emitted.
The crossing point
between the neutron-proton emission time is connected to the 
crossing in the symmetry potential, including Coulomb, at density 
around $0.8\rho_0$.
We can say, for example, that particles with momenta 
of the order of $300$ MeV show features of the interaction already
at density around and below the saturation one.

Using a coalescence model also the emission time of light clusters, such
as $d$, $t$, $^3He$ has been studied \cite{ChenNPA729}. The authors  
have found that 
light cluster production occurs mainly from the pre-equilibrium stage
and has a shorter duration than nucleon emission. We notice that this 
is different from the cluster formation at lower energies, where the 
multifragmentation process is dominant, see the previous Sect.\ref{rpa}.
A stiffer symmetry energy causes an earlier emission of nucleons that 
reflects the earlier formation of clusters. 
The average emission time
decreases with increasing kinetic energy as for the nucleons, moreover
 the heavier is the cluster and earlier it is formed. 
This has a simple explanation: the probability to coalesce decreases with
the average density, moreover we can have a cluster break-up in
the expansion phase. The dependence is stronger for clusters with
more nucleons. 

Therefore the study of cluster isobars 
has two advantages: they are  
easier to detect with respect to 
neutrons and they carry information from higher 
density.
However the influence of the symmetry energy on the average emission time
is reduced with respect to the nucleon one. This leads to a reduced sensitivity
of observables like the  correlation function of light clusters, see next
 subsections.
  
In summary a simple analysis of the average emission time nicely shows how
 the particle emission is regulated
by the symmetry potential. However, as mentioned above, the particle
emission time cannot be directly accessed by experiments, but 
the nucleon emission function can be extracted from
two-particle correlation functions, 
see, e.g., Refs. \cite{BoalRMP62,BauerARNPS42,ArdouinIJPE6,WiedemannPR319} 
for reviews.
In most studies, only the
two-proton correlation function has been measured 
\cite{GongPRL65,GongPRC43,GongPRC47,KundePRL70,HandzyPRL75}. 
In particular, the 
correlation function of two nonidentical particles
has been found to depend on their relative space-time distributions at 
freeze out and thus provides a useful tool for measuring the emission 
sequence, time delay, and separation between the emission sources for 
different particles 
\cite{GelderloosPRL75,LednickyPLB373,VoloshinPRL79,ArdouinPLB446,PrattPRL83,GourioEPJ7}.
Recently, data on two-neutron and
neutron-proton correlation functions have also become available. The
neutron-proton correlation function is especially useful as it is free of
correlations due to wave-function anti-symmetrization and Coulomb
interactions. Indeed, Ghetti \textit{et al.} have deduced from measured
neutron-proton correlation function the time sequence of neutron and proton
emissions \cite{GhettiNPA674,GhettiPRL87}.

\subsection{Nucleon-nucleon correlation functions}
In the standard Koonin-Pratt formalism 
\cite{KooninPLB70,PrattPRL53,PrattPRC36}, the
two-particle correlation function is obtained by convoluting the emission
function $g(\mathbf{p},x)$, i.e., the probability for emitting a particle
with momentum $\mathbf{p}$ from the space-time point $x=(\mathbf{r},t)$,
with the relative wave function of the two particles, i.e., 
\begin{equation}
C(\mathbf{P},\mathbf{q})=\frac{\int d^{4}x_{1}d^{4}x_{2}g(\mathbf{P}%
/2,x_{1})g(\mathbf{P}/2,x_{2})\left| \phi (\mathbf{q},\mathbf{r})\right|^{2}%
}{\int d^{4}x_{1}g(\mathbf{P}/2,x_{1})\int d^{4}x_{2}g(\mathbf{P}/2,x_{2})}.
\label{Eq1}
\end{equation}%
In the above, $\mathbf{P(=\mathbf{p}_{1}+\mathbf{p}_{2})}$ and $\mathbf{q(=}%
\frac{1}{2}(\mathbf{\mathbf{p}_{1}-\mathbf{p}_{2}))}$ are, respectively, the
total and relative momenta of the particle pair; and $\phi (\mathbf{q},%
\mathbf{r})$ is the relative two-particle wave function with $\mathbf{r}$
being the relative position, i.e., $\mathbf{r=(r}_{2}\mathbf{-r}_{1}%
\mathbf{)-}$ $\frac{1}{2}(\mathbf{\mathbf{v}_{1}+\mathbf{v}_{2})(}t_{2}-t_{1}%
\mathbf{)}$. This approach has been very useful in studying 
effects of nuclear equation of state and nucleon-nucleon cross sections on
the reaction dynamics of intermediate energy heavy-ion collisions 
\cite{BauerARNPS42}. The extension to the study of symmetry energy 
from a theoretical point of view
is instead quite recent \cite{ChenPRL90,ChenPRC68}. 

\begin{figure}[htb]
\centering
%\centerline{\epsfbox{CFsym.eps}}
\includegraphics[scale=1.1]{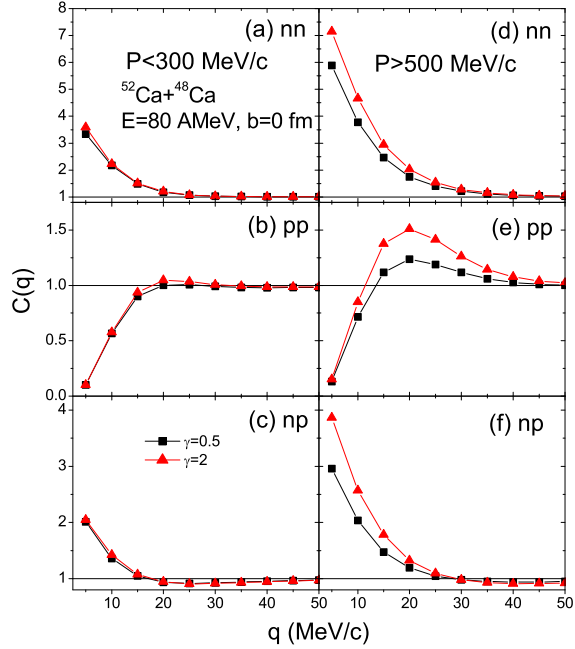} \vspace{-1cm}
\caption{Two-nucleon correlation functions gated on total
momentum of nucleon pairs using the soft (filled squares) or stiff (filled
triangles) symmetry energy. Left panels are for $P<300$ MeV/c while right
panels are for $P>500$ MeV/c (from \cite{ChenPRL90}).}
\label{CFsym}
\end{figure}

Shown in Fig. \ref{CFsym} are two-nucleon correlation functions
gated on total momentum $P$ of nucleon pairs from central collisions of $%
^{52}$Ca + $^{48}$Ca at $E=80$ \textrm{MeV/nucleon}. The left and right
panels are for $P<300$ \textrm{MeV/c} and $P>500$ \textrm{MeV/c},
respectively. Both neutron-neutron (upper panels) and neutron-proton
(lower panels) correlation functions peak at $q\approx 0$ \textrm{MeV/c%
}. The proton-proton correlation function (middle panel) is, however, peaked
at about $q=20$ \textrm{MeV/c} due to the strong final-state s-wave attraction,
but is suppressed at $q=0$ as a result of Coulomb repulsion and
anti-symmetrization of two-proton wave function. These general features
are consistent with those observed in experimental data from heavy ion
collisions \cite{GhettiNPA674}.

Since emission times of low-momentum nucleons do not change much with the
different $E_{\mathrm{sym}}(\rho )$ used in IBUU model as shown in Fig. %
\ref{emTimeP}, two-nucleon correlation functions are not much affected by
the stiffness of the symmetry energy. On the other hand, the emission times of 
high-momentum nucleons, which are dominated by those with momenta near $250$ 
\textrm{MeV/c}, differ appreciably for the two symmetry energies considered
here. Correlation functions of high-momentum nucleon pairs thus show an
important dependence on nuclear symmetry energy. Gating on nucleon pairs with
high total momentum allows one to select those nucleons that have short
average spatial separations at emission and thus exhibit enhanced
correlations. 
The strength of
the correlation function is stronger for the stiff symmetry energy than
for the soft symmetry energy: about $30\%$ and $20\%$ for neutron-proton
pairs and neutron-neutron pairs at low relative momentum $q=5$ MeV/c,
respectively, and $20\%$ for proton-proton pairs at $q=20$ \textrm{MeV/c}.

The neutron-proton correlation function thus exhibits the highest
sensitivity to variations of $E_{\mathrm{sym}}(\rho )$, however one should
heed that proton-proton correlation function is much easier to measure
thanks to the proton charge and to the larger relative momentum at which
the effect is expected.
As shown in Fig. 
\ref{emTimeP} and discussed earlier, the emission sequence of neutrons and
protons is sensitive to $E_{\mathrm{sym}}(\rho )$. With a 
stiff $E_{\mathrm{sym}}
(\rho )$, high momentum neutrons and protons are emitted
almost simultaneously, and they are thus temporally strongly correlated,
leading to a larger neutron-proton correlation function. 
At variance in the case of soft $E_{\mathrm{sym}}(\rho )$ we have a
 delay in the    
proton emission, larger attractive field, which reduces the temporal 
correlation with neutrons.

Furthermore, both neutrons and protons are emitted earlier with 
stiff $E_{\mathrm{sym}}(\rho)$, so they have smaller spatial separation. 
This
demonstrates that correlation functions of nucleon pairs with high
total momentum can indeed reveal sensitively the effect of nuclear symmetry
energy on the temporal and spatial distributions of emitted nucleons.
However, one must be aware that such an effect is delicate and can vanish 
as a function of impact parameter, beam energy, colliding system. 

The impact
parameter dependence has been studied again for the $^{52}$Ca + $^{48}$Ca 
at $E=80$ \textrm{MeV/nucleon}. 
\begin{figure}[htb]
%\centerline{\epsfbox{Edep.eps}}
\centering
\includegraphics[scale=1.1]{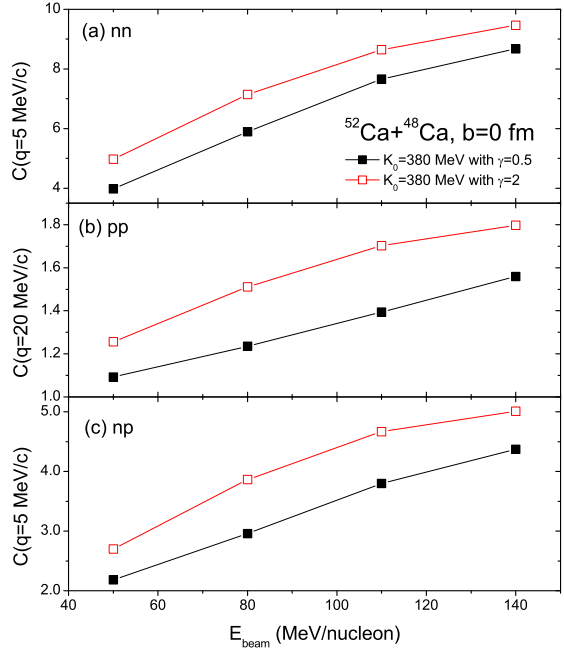} \vspace{-0.5cm}
\caption{{\protect\small Dependence of nucleon-nucleon correlation functions
on the incident energy for hight total momentum ($P>500$ MeV)
neutron-neutron (a) and neutron-proton (c) pairs with relative momentum }$%
q=5 ${\protect\small \ MeV/c and proton-proton (b) pairs with relative
momentum }$q=20${\protect\small \ MeV/c by using }$K_{0}=380${\protect\small %
\ MeV and free \textsl{N}-\textsl{N} cross sections with the soft (filled 
squares) or stiff (open circles) symmetry energy.} 
 (from \cite{ChenPRC68}).}
\label{Edep}
\end{figure}
It is seen that the symmetry energy effect is stronger in central and 
semi-central 
collisions and becomes weaker in semi-peripheral and peripheral collisions. 
This reflects the decrease in nuclear compression with the consequent
change in the reaction mechanism.

The dependence of nucleon-nucleon correlation functions on the incident energy
of heavy ion collisions is shown in Fig.\ref{Edep}. It is seen that the values 
of nucleon-nucleon correlation functions increase with increasing incident 
energy. This is understandable since nuclear compression increases with 
increasing incident energy, and nucleons are also emitted earlier, leading 
to a smaller source size. The difference due to the symmetry energy 
does not change
significantly with the incident energy at least in the range 
$50-150$ MeV/nucleon.

A dependence on the mass of the colliding system has been
 already studied in Ref.\cite{ChenPRC68} for central collisions. 
The authors have looked at 
 $^{132}$Sn + $^{124}$Sn at $E=80$ \textrm{MeV/nucleon}. It is seen that, even
if the isospin  asymmetry is similar to $^{52}$Ca + $^{48}$Ca system, the 
effect is reduced
by about a factor two for {\it pp} and {\it np}, while {\it nn} 
is less affected.
This can be understood with the increased Coulomb potential which is 
expected to reduce the difference in $C_{pn}$.
Considering the emission times in  $^{52}$Ca + $^{48}$Ca as a reference, 
for heavier systems the Coulomb potential  
increases the correlation for the soft case but decreases it for the 
stiff case, where for $^{52}$Ca + $^{48}$Ca neutrons and protons 
have almost equal emission times.

In conclusion the symmetry term effect on nucleon nucleon correlation
finctions shows up only if particular conditions of impact
parameter and colliding systems are matched. An experimental plan to pursue
the investigation of such an observable needs a careful choice of the system
and the possibility to have a good centrality selection.
Futhermore it is important to choose the various colliding systems
in such a way that the isospin asymmetry is as much different as possibile
while the total mass is unchanged \cite{ColonnaPRC57,ScalonePLB461}.
Moreover this kind of comparative analysis is in general 
more reliable than the study of absolute values of correlation functions,
 which 
themselves
are not perfectly reproduced by $BUU$ simulations already for isospin
symmetric systems, see ref.\cite{HandzyPRL75}.

\subsubsection{Isospin Momentum Dependence}
%\addtocontents{toc}{\hspace{0.55cm}\thesubsubsection \hspace{0.12cm}
%Isospin Momentum Dependence}
During the preparation of this report,
phenomenological potentials extracted from the Gogny effective interaction 
\cite{BaoPRC69} or extensions of the Bertsch-Gale-Das Gupta ($BGBD$)
\cite{RizzoNPA732,BaoNPA735}  have been considered
to improve $IBUU$ and $BNV$ transport codes, to include the effect of a 
momentum dependence in
the isovector channel, namely a splitting of the effective masses of 
neutrons and protons, see the discussion in Sect.\ref{eos}.
A detailed study of particle emission with a momentum
dependent mean field was firstly done 
in \cite{GrecoPRC59,SapienzaPRL87}, but without including isospin effects.
Generally the momentum dependence causes a faster and richer pre-equilibrium
 emission. This reflects in an increasing of the two nucleon correlation 
function and of the yield of the clusters formed.

A first analysis of the effect of the
momentum dependence on the particle emission has shown a weakening of the 
role of the symmetry energy leading to a weaker dependence of the correlation
function on the symmetry energy \cite{ChenPRC69}. 
In fact such a study has employed an effective interaction with a splitting 
of the effective masses as $m^*_n(\rho) > m^*_p(\rho)$, which leads
to a reduction of the effective action of the 
symmetry energy at higher density. 
Indeed the difference of the neutron-proton potentials decreases 
with momentum \cite{RizzoNPA732,BaoPRC69}, see Fig.\ref{fig:optMD} of 
Sect.\ref{eos}.
However,
if a momentum dependence in the isovector channel corresponding to a splitting
in the opposite direction is considered, $m^*_n(\rho)<m^*_p(\rho)$, a strong
effect on the correlation function can be envisaged, as we will 
see in the results concerning collective flows discussed later in the Section.

Since this is a stimulating open problem, directly related to the
$E-$slope of the Lane Potential of Sect.\ref{eos}, a systematic study of
the beam energy dependence of the $(n,p)$ correlation functions
appears to be very promising. A related interesting observable would
also be just the isotopic content of the free nucleon emissions
and/or of light isobars, see following. 

\subsubsection{Light Clusters}
%\addtocontents{toc}{\hspace{0.55cm}\thesubsubsection \hspace{0.12cm}
%Light Clusters}
Another observable of interest for the $E_{sym}(\rho)$ search
is the yield and the energy spectrum of light clusters that differ
in the isospin content. As mentioned above, the light clusters (isobars)
offer 
the possibility of having particles with different isospin that are 
charged and therefore can 
be detected in a much simpler way with respect to neutrons. Moreover, as  
already discussed, 
their production time is shorter and they represent good
probes of the higher density phase.

Isospin effects on cluster production and isotopic ratios in heavy ion
collisions have been previously studied using either the lattice gas model
\cite{ChomazPLB447} or a hybrid of $IBUU$ and statistical 
multifragmentation
model \cite{TanPRC64}. These studies are, however, at lower energies, 
 where multifragmentation processes, as a result of
possible liquid-gas  phase transition, play an important role, 
see Sect.\ref{rpa}.

Except deuterons, both tritons and $^{3}$He are only a small fraction
of the total multiplicity in heavy ion collisions at beam energy
around $100$ \textrm{AMeV}.
In this energy range the reaction dynamics can be described in terms
of a  relatively large 
compression followed by a faster expansion and vaporization. 
In such a case it is quite suitable to 
apply a coalescence model for determining the cluster production.
However 
a simple coalescence model can only be applied if the role of correlations play
a negligible role and the binding energy of the formed clusters
can be discarded. The last
issue is expected to affect more the absolute value of the yields and can 
be overtaken looking at the ratios of particle yields or spectra.
In fact the $t/^3He$ ratio which is relevant for the isospin physics
is expected to be independent on the binding energy effect, that 
is quite similar for the two clusters. 

Within a coalescence model, using the freeze-out distributions 
obtained in $IBUU$, the behavior of $d$, $t$, $^3He$ production 
has been studied
again for the  $^{52}$Ca + $^{48}$Ca system in central collisions,
 ref.\cite{ChenPRC69}. 
The feed-down from heavier fragments is not taken into account, 
however especially at large kinetic energies, its contribution 
is less relevant.
It is found that the yield
of light clusters is quite sensitive to the symmetry potential.
As it can be inferred 
from the result on the particle emission the stiff symmetry
energy leads to a larger yield of clusters,
in particular the yield of  $^3He$ can be a factor two larger than in the soft
case. The stiff symmetry energy induces a stronger pressure gradient causing 
an earlier
emission of neutrons {\it and} protons. Furthermore in the stiff case neutrons
and protons are more correlated in time, see Fig.\ref{emTimeP} and \ref{CFsym},
enhancing the probability of  $^3He$ formation in phase space.
If one looks at 
the $t/^3He$ ratio, the
effect of the symmetry energy is reduced because both yields are modified
so that the increase in the ratio is just about
$15\%$.

In Ref.\cite{BaoPRL78} the ratio of neutron to proton as a function
of their kinetic energies was found to be quite sensitive to the symmetry 
energy. This appears also in the light cluster production 
as recently found by Chen et al. \cite{ChenPRC68b,ChenPRC69}.  
In particular the $t/^3He$ ratio shows a quite different qualitative 
behaviour:
 while it increases with the kinetic
energy for the soft symmetry energy, it decreases for the stiff one.
This is due to the fact that, using a soft symmetry energy, a larger number
of neutrons is emitted during the early stage of the collision, thus enhancing
the production of tritons with high kinetic energy. 

Since clusters are expected to be good probes of the high density
region reached during the reaction dynamics all the previous
features will be very sensitive to the Momentum Dependence ($MD$) of
the effective forces in the isovector channel.
For the same  $^{52}$Ca + $^{48}$Ca system
in ref.\cite{ChenPRC68b} it has been shown that symmetry energy effects
on the $t,^3He$ yields are largely reduced when a Gogny effective
interaction is used. This is not surprising since the isospin-$MD$
is in the $m_n^*>m_p^*$ direction, which implies a quenching of 
symmetry effects for high momentum nucleons. Of course an enhancement
is expected in the case of an opposite effective nucleon mass splitting.
Similar information can be obtained from the beam energy dependence 
of the yields.

\subsection{Collective Flows}

In the search for the EOS of the nuclear matter in various conditions
of density and temperature a chief role is played by the study of the
collective flow, which is a motion characterized by space-momentum correlation
of dynamical origin. 

The build up of sideward and elliptic flow is realized
around the higher density stage of the reaction and thus it is a powerful
tool for the search of the high density behaviour of the symmetry energy.
It represents a mean of investigation very general, giving information on the 
dynamical response of excited nuclear matter in heavy-ion collisions,
from the Fermi energies up to the ultrarelativistic ones, in the search for 
a phase transition to QGP, 
\cite{StoeckerPR137,DanielPRC46,DasguptaPT46,DanielNPA685,DanielSCI298,RHIC-v2}.
The collective motion can be characterized in several ways that pin down
different kind of space-momentum correlation that can be generated 
by the dynamics. The kind of collective flows that have been suggested 
and employed to get information on the equation of state
can be divided into three categories: radial, sideward and elliptic. 

The sideward and elliptic flow have been and are currently useful tools
for the study of the compressibility of symmetric nuclear matter.
In the search for the density behaviour of the symmetry energy similar 
concepts can be exploited but highlightening the difference between
neutrons and protons or light clusters with different isospin.
We will define now the different types of collective flow and we will discuss
the current status of the effects expected due to different $E_{sym}(\rho)$,
and related momentum dependence.   
We will see that first experimental results with stable beam already
show hint of the effect of the symmetry energy.
Thus future, more exclusive, experiments with radioactive
beams should be able to set stringent
constraints on the density dependence of the symmetry energy far
from ground state nuclear matter.

\subsubsection{Definitions}
%\addtocontents{toc}{\hspace{0.55cm}\thesubsubsection \hspace{0.12cm}
%Definitions}
Sideward flow is a deflection of forwards and backwards moving particles,
within the reaction plane \cite{DanielPLB157}. It is formed because 
for the compressed and excited
matter it is easier to get out on one side of the beam axis than on the other.
The sideward flow is often represented in terms of the average in-plane 
component of the transverse momentum at a given rapidity $<p_x(y)>$:
\begin{equation}\label{dirfl}
F(y)\equiv \frac{1}{N(y)} \sum_{i=1}^{N(y)} p_{x_{i}} \equiv 
 \langle \frac{p_x}{A} \rangle
\end{equation}
The particular case in which the slope of the transverse flow is 
vanishining in a region
around midrapidity is referred as balance energy. It comes out from a 
balance between the attraction of the mean field and the repulsion of 
the two-body collisions. 

For the isospin effect the sum over the particles in Eq.\ref{dirfl} is 
separated
into protons and neutrons. In Ref.\cite{BaoPRL82} also
the neutron-proton differential flow
$F^{pn}(y)$ has been suggested as very useful probe of the isovector part of 
the $EOS$ since it appears rather insensitive to the isoscalar potential
and to the in medium nuclear cross section and, as we will discuss, it combines
the isospin distillation effects with the direct dynamical flow effect.  
The definition of the differential flow $F_{pn}(y)$ is
\begin{equation}\label{dif_dir}
F_{pn}(y)\equiv \frac{1}{N(y)} \sum_{i=1}^{N(y)} p_{x_{i}} \tau_i
\end{equation}
where $N(y)$ is the total number of free nucleons at the
rapidity $y$, $p_{x_{i}}$ is the transverse momentum of
particle $i$ in the reaction plane, and $\tau_i$ is +1 and -1
for protons and neutrons, respectively.

The flow observables can be seen
respectively as the
first and second coefficients from the Fourier expansion of the
azimuthal distribution \cite{OlliPRD46}:
$$\frac{dN}{d\phi}(y,p_t)=1+V_1cos(\phi)+2V_2cos(2\phi)$$
where $p_t=\sqrt{p_x^2+p_y^2}$ is the transverse momentum and $y$
the rapidity along beam direction. 

The transverse flow can be also
expressed as: 
$$V_1(y,p_t)=\langle \frac{p_x}{p_t} \rangle$$ 
It provides information on the azimuthal anisotropy of the transverse 
nucleon emission
and has been used to study the $EOS$ and cross section
sensitivity of the balance energy \cite{BaoPRL82}.

The second  coefficient of the expansion defines the elliptic
flow $v_2$ that can be expressed as
$$V_2(y,p_t)=\langle \frac{p_x^2-p_y^2}{p_t^2} \rangle$$
It measures the competition between in-plane and out-of-plane emissions. 
 The sign of $V_2$ indicates the azimuthal anisotropy of emission:
particles can be preferentially emitted either in the reaction
plane ($V_2>0$) or out-of-plane ($squeeze-out,~V_2<0$)
\cite{OlliPRD46,DanielNPA673}.

The $p_t$-dependence of $V_2$,
 which has been recently investigated by various groups
\cite{DanielSCI298,DanielNPA673,LarionovPRC62,GaitanosEPJA12}, is
very sensitive to the high density behavior of the $EOS$ since highly
energetic
particles ($p_t \ge 0.5$) originate from the initial compressed and
out-of-equilibrium phase of the collision, see e.g. ref.\cite{GaitanosEPJA12}.
Also at high energy it is allowing to get insight of the partonic stage
and hadronization mechanism in ultrarelativistic heavy-ion collisions 
\cite{RHIC-v2}.

\subsubsection{Collective Flows at the Fermi Energies}
%\addtocontents{toc}{\hspace{0.55cm}\thesubsubsection \hspace{0.12cm}
%Collective Flows at the Fermi Energies}
The isospin dependence of the transverse collective flow near the balance
energy was first pointed out in Ref.\cite{BaoPRL76}, where it is shown
that the reactions involving neutron-rich nuclei have a significant stronger
attractive flow and consequently a higher balance energy.
A discussion of the effect of
a different density dependent symmetry energy started only some year later
\cite{ScalonePLB461,BaoPRL85,BaoPRC64}.
It was also stressed that the effects are expected to be more evident
for semi-peripheral collisions (see also Fig. \ref{fig:flowbred}) 
around the balance energy.
Clear signals of the mean field momentum dependence ($MD$) were
also revealed 
(\cite{ScalonePLB461,ZhangPRC50,DanielNPA673}). In the range of the 
Fermi energies it is indeed
essential an appropriate treatment of the momentum dependence of the mean
field \cite{DasguptaPT46}.  

Further studies that look at more n-rich systems investigating the effect 
of different density 
behaviour of the symmetry energy and based on a more exclusive analysis 
(like $p_t$ dependence), 
have shown that isospin effects on collective flows can 
be found  also at higher energies up to beam energies of $1\, AGeV$ 
\cite{BaoNPA708,GrecoPLB562,RizzoNPA732}.

The search for the density dependence of $E_{sym}(\rho)$ has also to 
 disentangle effects coming from the isospin dependence of the
nucleon-nucleon 
cross sections \cite{LiPRC48,LiPRC69}.
In fact  little as been
done till now to address quantitatevly this problem, the best strategy is
of course to look for observables that mainly depend on the collision rate.
A possibility is offered by the pre-equilibrium gamma radiation, as
mentioned in Ref.\cite{GrecoPRC59}. Another possibility is the measurement
of the radial flow for system with the same mass but different isospin 
asymmetry. 
Some observables have been found to be weakly sensitive to an overall change 
of the absolute value of the $pp$, $nn$, $pp$ cross sections. We will
pay particular attention at these features. 

\subsubsection{Isospin effects around the Balance Energy}
%\addtocontents{toc}{\hspace{0.55cm}\thesubsubsection \hspace{0.12cm}
%Isospin effects around the Balance Energy}
In Ref.\cite{ScalonePLB461} the investigation of the isospin effect 
on transverse flows at Fermi energies has been 
carried out, including also a momentum dependence  of the mean field.
The main structure of this 
interaction has been already introduced in Sect.\ref{eos}.
It is a $GBD-MD$ form in a Skyrme-like effective mean field, 
\cite{GalePRC41,PrakashPRC37}. For a large dynamical range, in particular 
at the Fermi
energies, it gives results quite similar to non-local forces of Gogny type
\cite{DechargePRC21}, that well describe collective flows in symmetric systems
with realistic compressibilities around $K \simeq 220 MeV$,
 \cite{PrakashPRC37,VirginiaPRC46}. 
For the local part a general $Skyrme$ form is used, with
two choices for the symmetry term: i) {\it asysoft} with
the $SIII$-$C({\rho})$ parametrization and ii) {\it asystiff} with the 
choice $C(\rho)=const=32 MeV$ (see Sect.\ref{eos}). In this calculation 
isospin effects on the momentum dependence are not included.

The microscopic transport $BNV$ equations are solved following a test
particle evolution on a lattice \cite{ScalonePLB461,GrecoPRC59,twingo}.
Isospin effects on nucleon cross sections and Pauli blocking are consistently
evaluated. Medium effects on the $NN$ cross sections are included
following the prescriptions of ref.\cite{LiPRC48}. A dependence on 
$\sigma_{NN}$ variations is also analysed.

We present $(p,n)$ transverse flow results for the collisions
$^{58}Fe+^{58}Fe$ (n-rich) and $^{58}Ni+^{58}Ni$ (n-poor) in the energy range
$55-105~AMeV$ where nice data are available for $Z=1,2,3$ flows
\cite{WestfallNPA630,Pak1PRL78,Pak2PRL78} and some other calculations 
are existing 
for the proton
flows \cite{BaoJMPE7,Pak2PRL78}.
Certainly the asymmetry $(N-Z)/A$ of $^{58}Fe$ is quite small, 
nonetheless
isospin effects have been found both experimentally 
\cite{WestfallNPA630,Pak1PRL78,Pak2PRL78} and 
theoretically to be sizeable. In particular the comparison of
{\it iso-transport} simulations
 with
 experimental results seem to show that a stiff behaviour
of $E_{sym}(\rho)$ can give a good description of the difference in the proton
transverse flows between $^{58}Fe+^{58}Fe$ (n-rich) and $^{58}Ni+^{58}Ni$.

\begin{figure}
\begin{center}
\includegraphics[scale=0.60]{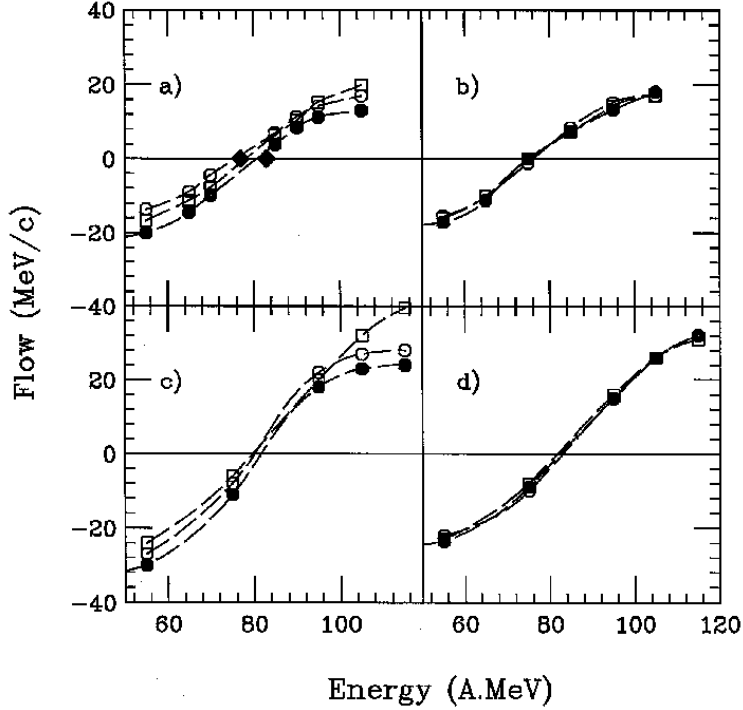}
\caption{Energy dependence of flows at $b_{red}=0.45$ \cite{bred}:
(full circles) $Fe-Fe$ protons; (open circles) $Ni-Ni$ protons; (squares)
$Fe-Fe$ neutrons - (a) asystiff; (b) asysoft; (c),(d) same for 
$\sigma_{NN} = 2fm^2$ no isospin dep. - The full diamonds in (a)
represent the proton balance energy data of ref.\cite{WestfallNPA630}
for the $Fe-Fe$ (right) and $Ni-Ni$ systems}
\label{fig:flowbal}
\end{center}
\end{figure}

\begin{figure}
\begin{center}
\includegraphics[scale=0.60]{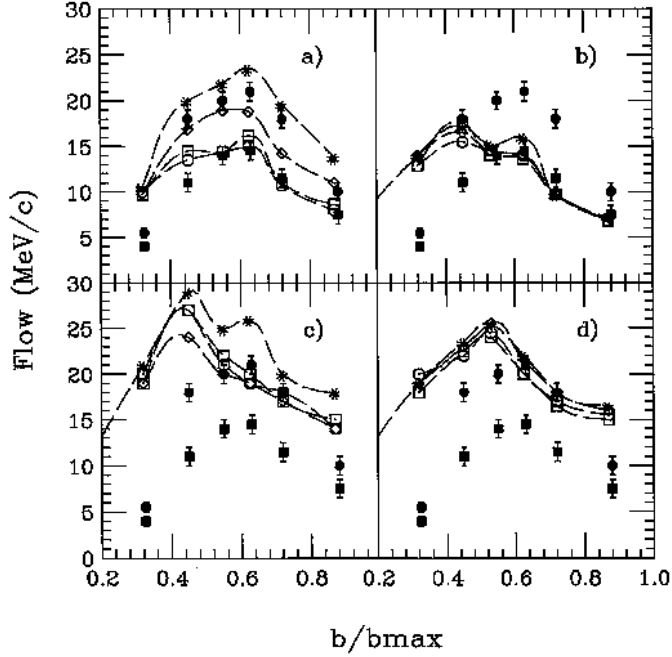}
\caption{Centrality dependence of flows at beam energy $55~AMeV$:
(stars) $Fe-Fe$ protons; (open circles) $Ni-Ni$ protons; (open diamonds)
$Fe-Fe$ neutrons; (open squares) $Ni-Ni$ neutrons - (a) asystiff; (b) asysoft; 
(c),(d) same for 
$\sigma_{NN} = 2fm^2$ no isospin dep. - Experimental points: (full circles)
 $Fe-Fe$ protons; (full squares) $Ni-Ni$ protons.}
\label{fig:flowbred}
\end{center}
\end{figure}

In Figs.\ref{fig:flowbal}, \ref{fig:flowbred} we clearly see the 
isospin effects 
on the tranverse flows, in particular the sensitivity to the density
dependence of the symmetry term of the nuclear $EOS$: an {\it asystiff}
behavior, more attractive for protons above normal density for the Fe
asymmetric case, gives a clear shift in the balance energy as well
as a larger (negative) flow at $55~AMeV$. i.e. below the balance.
Both effects are in agreement with the data and are disappearing in the
{\it asysoft} choice. Of course also the isospin and density dependence of the
$NN$ cross sections is important (see the (c) plots) but we note
that a noticeable sensitivity to the isovector part of the $EOS$
is still present. In particular we see that the isospin dependence of 
the mean field is able to keep
the transverse flow difference between protons in $Fe-Fe$ and $Ni-Ni$.
However a systematic
study over different systems with more ``exotic'' isospin content is necessary
to confirm this result.

\begin{figure}
\begin{center}
\includegraphics[scale=0.60]{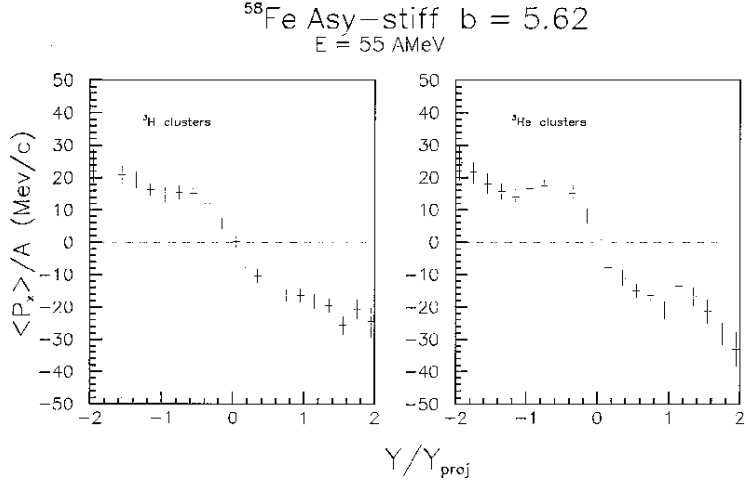}
\caption{Mean transverse momentum in the reaction plane vs. reduced rapidity
for light $3He-3H$ isobars in the $Fe-Fe$ collisons at $55~AMeV$ beam energy
(i.e. below the balance) for semicentral impact parameter, $b_{red}=0.6$.
 Asystiff parametrization.}
\label{fig:flowisobar}
\end{center}
\end{figure}

An important effect predicted by the simulations is the clear difference
between neutron and proton flows. Due to the difficulties in measuring
neutrons this should be seen in a detailed study of light isobar flows.
Moreover we like to recall that clusters are better probing the
higher density regions.
This point is quantitatively shown in the Fig.(\ref{fig:flowisobar}) where we 
present the transverse momentum vs. rapidity distributions for 
$^3He-Triton$ clusters in semicentral $Fe-Fe$ collisions at $55~AMeV$,
i.e. below the balance energy \cite{ScalonePLB461}.
We can estimate a $20\%$ larger (negative) flow for the $^3He$ ions,
just opposite to what expected from Coulomb effects.
This appears to be a clear indication of the contribution of a much
reduced (negative) neutron flow in the case of an {\it asystiff} force, i.e. a
more repulsive symmetry term just above $\rho_0$. The effect 
would disappear in an {\it asysoft} choice. 
Of course even for this observable the $Isospin-MD$ will modify
the symmetry energy dependence, as already noted before for the cluster 
yields.

\subsubsection{Differential Flows}
%\addtocontents{toc}{\hspace{0.55cm}\thesubsubsection \hspace{0.12cm}
%Differential Flows}
It has been suggested that in the study of symmetry energy by mean of the
neutron and proton collective flow one can even exploit the difference 
in the so 
called isospin fractionation, i.e. the difference in the neutron-proton 
pre-equilibrium emission. 
In fact it has been pointed out that a soft/stiff symmetry energy 
leads to a different $N/Z$ emission \cite{ColonnaPRC57,BaoPRL85,BaranNPA703}, 
see also previous section. 
This can be attributed to the
density pressure gradient, which is quite different in the two cases.

We note that the gradient of the pressure should not be confused with
the so called $K_{sym}$, see also the discussion in Sect.\ref{eos}.
Therefore 
the interpretation given in \cite{BaoPRL85} which relies entirely of the 
different $K_{sym}$, around saturation, for the two used parametrization 
(from $-69$ to $+61$ MeV) should be taken with caution. In fact between 
the soft and
stiff symmetry energy
there is, at the same time, a change in the slope $L$, from $54$ to $95$ MeV, 
that indeed gives a larger contribution to the pressure gradient, according
to :  
\begin{equation}
\frac{dP_{asy}}{d\rho}=\left(2\rho \frac{d(E/A)_{sym}}{d\rho}+
\rho^2  \frac{d^2(E/A)_{sym}}{d\rho^2}\right) I^2= 
\left(\frac{2}{3}L + \frac{1}{9} K_{sym}\right) I^2
\end{equation}
where $I \equiv (N-Z)/A$ is the asymmetry parameter.

Despite the possibile interpretation, in order to make the analisis of 
collective flow more sensitive to the symmetry 
potential, the $neutron-proton$ differential flow,
defined in Eq.\ref{dif_dir}, has been introduced
\cite{BaoPRL82}. In such a way one combines
constructively the difference in the neutron-proton collective flow
and the difference in the number of protons and neutrons emitted. At 
the same time the influences of the isoscalar potential and the
in-medium nucleon-nucleon cross sections are also reduced. 
However the measurement of such a differential flow demands not only for the
measurement of neutron collective flow but also for a precise assessment
of their number, which most likely is impossible.
On the other hand the idea to combine more than one isospin contribution
 in one observable
is certainly important for elusive effects as those coming from the
symmetry energy. Moreover the usual problems caused by the neutrons can be
overtaken looking at clusters.
For example one can use the definition of differential collective flow
and apply it to the $^3H-^3He$ isospin doublet, see previous subsection.

\subsection{Effective Mass Splitting and Collective Flows}

The problem of momentum dependence in the isospin
channel is still open. 
Intermediate energies are
important in order to have high momentum particles and to test regions
of high baryon (isoscalar) and isospin (isovector) density during the
reactions dynamics.
Now we
present some qualitative features of the dynamics in heavy ion
collisions related to the splitting of nucleon effective masses.

We report here on expected effects of the isospin $MD$, studied by means of 
the Boltzmann-Nordheim-Vlasov $BNV$ 
transport code, Ref. \cite{GrecoPRC59}, implemented with 
a $BGBD-like$ mean field 
with a different $(n,p)$ momentum dependence, as already introduced in 
Sect.\ref{eos}, see ref.\cite{RizzoNPA732}.
Free (isospin dependent) isotropic cross sections are used in the collision 
term. 
In order to enhance the role of the energy dependence of the mean field 
we focus on
semi-peripheral $Au+Au$ collisions at the energy of $250\, AMeV$ and 
on transverse and elliptic flows of neutrons and protons in high
rapidity regions, see \cite{GiessenRPP56}.

\subsubsection{Tranverse flows}
%\addtocontents{toc}{\hspace{0.55cm}\thesubsubsection \hspace{0.12cm}
%Tranverse flows}

\begin{figure}[b]
\centering
\includegraphics[width=8cm]{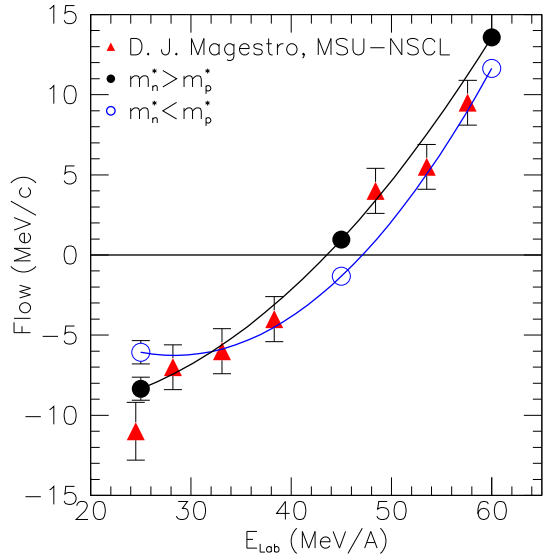}
\caption{Balance energy in $Au+Au$ collisions for intermediate
impact parameters. Solid circles refer to the
case $m^*_n>m^*_p$, empty circles to $m^*_n<m^*_p$.
Triangles represent the data from ref.\cite{MagesPRC61}} \label{fig:balau}
\end{figure}

The evaluation of the balance energy for the proton more inclusive
{\it Directed Flow}, integrated over all transverse momenta, 
in $Au+Au$ collisions represents a very good test
for the used effective interactions at lower energies.
The results are shown in Fig.\ref{fig:balau} and compared with NSCL-MSU
data. The proton directed flows are calculated for semicentral collisions,
i.e. $b_{red} = 0.5$ \cite{bred}.

Both parametrizations are well reproducing the experimental balance 
energy region. As expected we see a slightly earlier balance of the flow
in the $m^*_n>m^*_p$ case due to some more repulsion for fast protons
but the effect is hardly appreciable at these energies, see also
the previous subsection. We will now look
more carefully at the mass-splitting effects with increasing beam energy 
and for more exclusive observables.

\begin{figure}[t]
%\centering
\includegraphics[width=7cm]{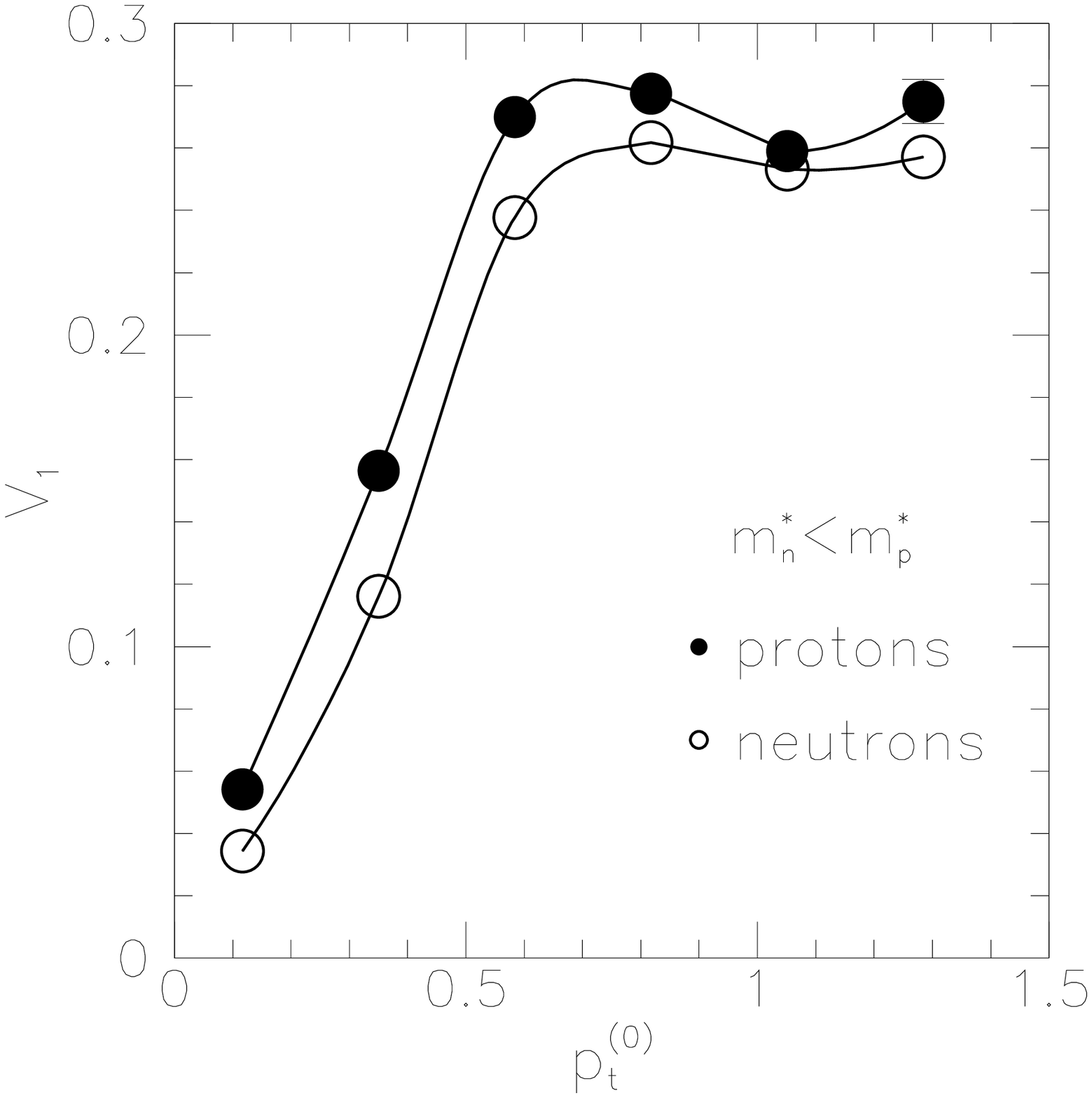}
\includegraphics[width=7cm]{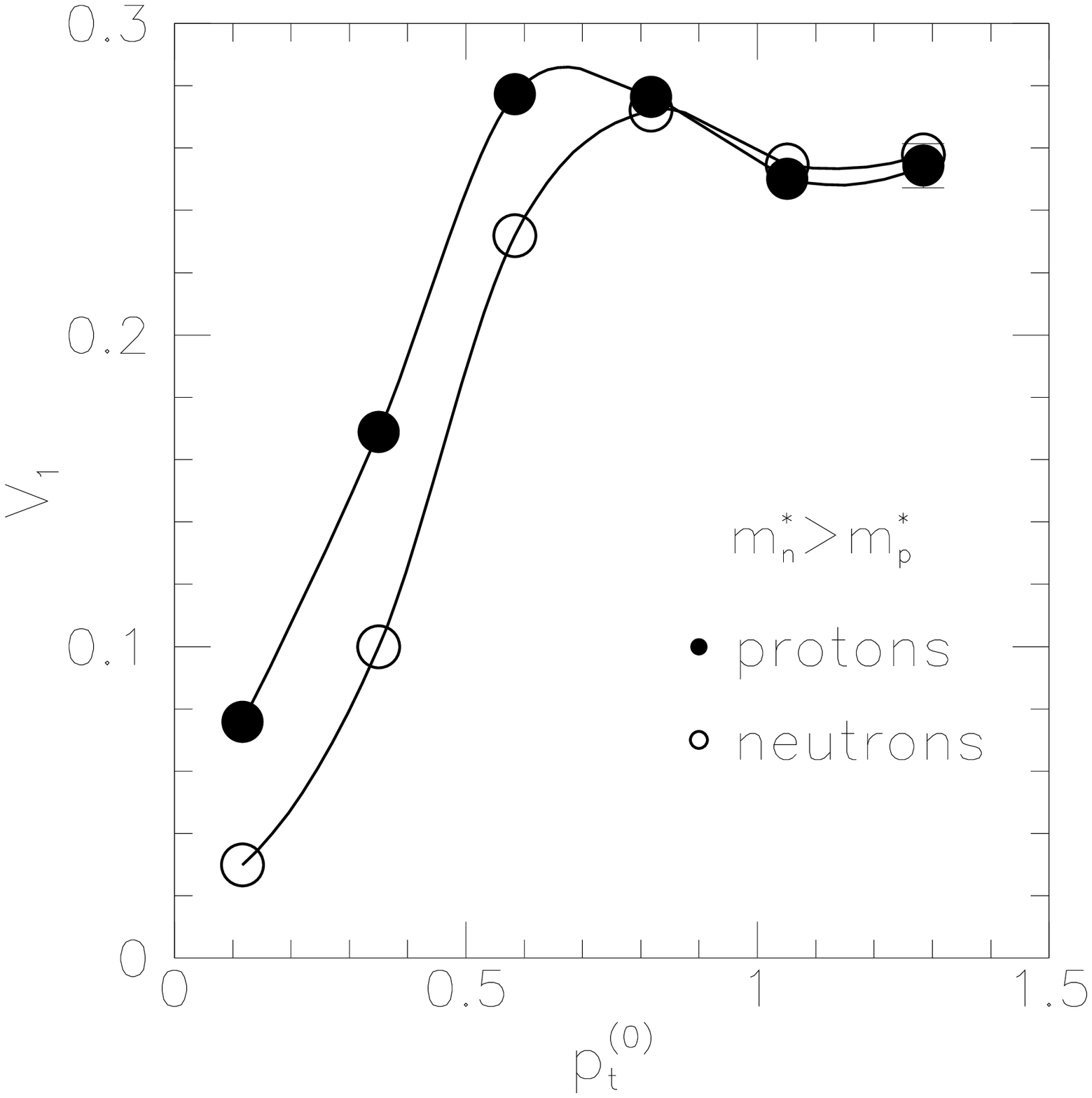}
\caption{Transverse flow of protons and neutrons
for Au+Au reactions at $250\, AMeV$,
$b/b_{max}=0.5$, in the rapidity interval
$0.7\leq |y^{(0)}| \leq 0.9$.}
\label{fig:v1np}
\end{figure}

$n/p$ transverse flow results for semicentral
$^{197}Au+^{197}Au$ collisions at $250~AMeV$ beam energy, for
particles in regions of relatively high rapidities
are shown in Fig. \ref{fig:v1np} where the left and
right panels illustrate the two
 $m^*_n<m^*_p$ and $m^*_n>m^*_p$ cases.
It is useful to work with momenta and rapidities normalized to
projectile ones in the center of mass system ($cm$), defined as
$y^{(0)} \equiv (y/y_{proj})_{cm}$ and
$p^{(0)}_t \equiv (p_t/p_{proj})_{cm}$.

\begin{figure}[b]
\centering
\includegraphics[width=8cm]{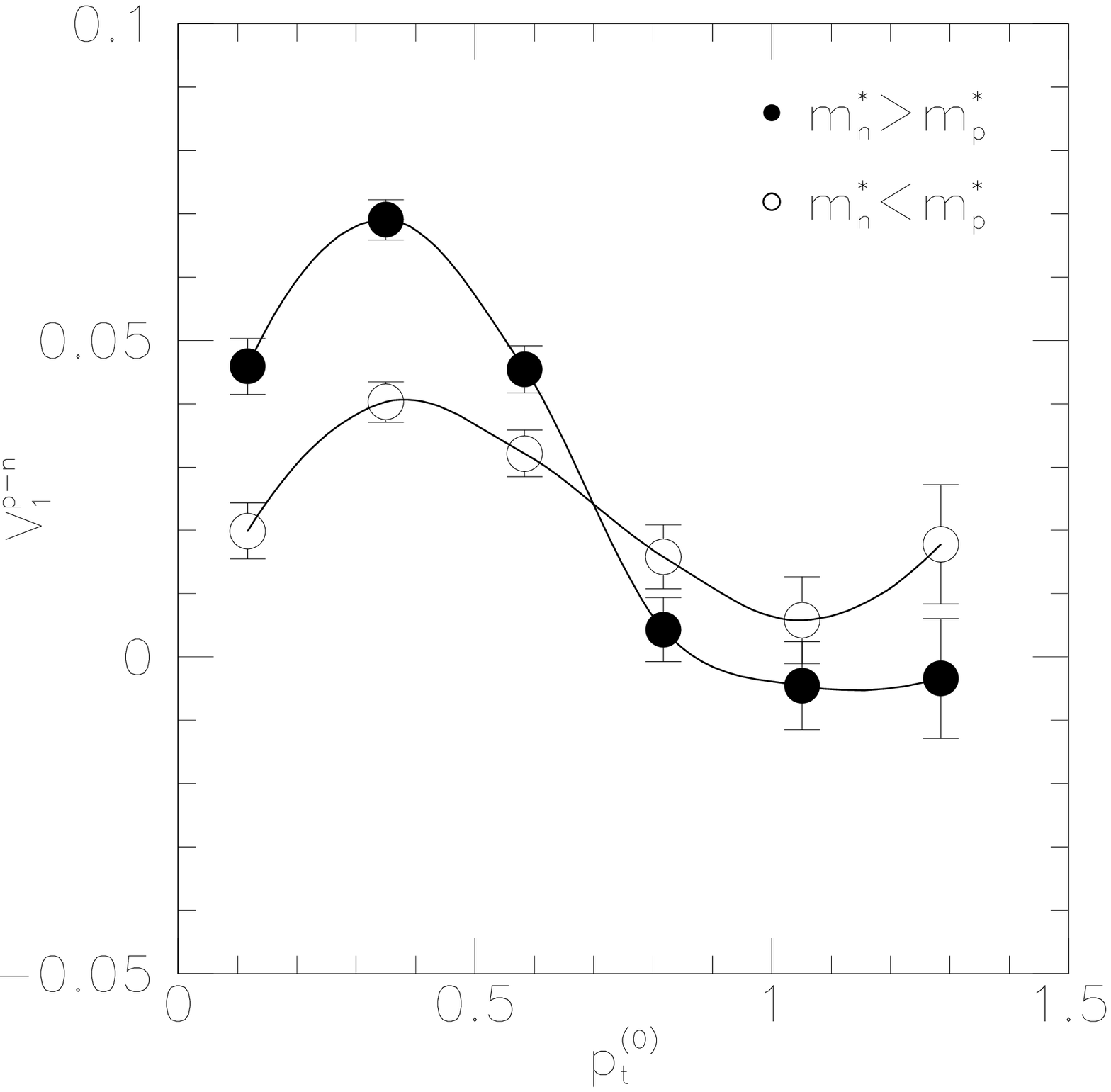}
\caption{$p_t$ dependence of the difference between neutron and proton 
transverse flows in $Au+Au$ 
collisions, same energy, centrality and rapidity selection as in 
Fig.\ref{fig:v1np}. Solid circles refer to the
case $m^*_n>m^*_p$, empty circles to $m^*_n<m^*_p$.} 
\label{fig:difv1}
\end{figure}

High rapidity selections allow us to find effects of the mean field
$MD$, which become more important with increasing momenta. Indeed
the different behaviour of mean field depending on mass splitting
is evident all over the range of transverse momentum shown here.
We emphasize that the average momentum of these particles is
generally beyond the Fermi momentum. Indeed a rough estimate in the center
of mass system
gives already for the beam parallel component
$$p_z\approx m_0 \cdot 0.8(y_{proj})_{cm} \simeq 260-280\,
MeV/c$$ 
In this momentum region a smaller effective mass
determines greater repulsive interaction. When $m^*_n<m^*_p$
neutrons feel greater repulsion than protons and their deflection
in the reaction plane for $p_t^{(0)}\leq 0.6$ is close to
protons, where the Coulomb field is also acting. 
At variance in the case $m^*_n>m^*_p$ the proton flow is greatly
enhanced if compared with neutrons.

We like to note that in both plots of Fig. \ref{fig:v1np} the proton flow
is above the neutron one, particularly at low transverse momenta.
This is a clear effect of the Coulomb repulsion. This is also responsible
for the non-symmetric behavior of neutron and proton flows in the two cases 
of opposite mass splitting. 

At high transverse momentum we must take into account the increase
of the out-of-plane component, so in this case increased repulsion
determines a greater reduction of transverse flow. It causes a
crossing between proton and neutron flow in the case
$m^*_n>m^*_p$. This behaviour is clearly seen if we look at the
different slope of the curves in  Fig. \ref{fig:v1np} (right panel) 
around $p_t^{(0)}\geq
0.6$. 

A quite suitable way to observe such effective mass splitting effect
on the tranverse flows is to look directly at the difference between
neutron and proton flows 
$$V_1^{p-n}(y,p_t) \equiv V_1^{p}(y,p_t) -  V_1^{n}(y,p_t)$$   
shown in Fig.\ref{fig:difv1}.

The variation due to the mass splitting choice is quite evident in 
the whole $p_t$ range of emitted 
nucleons. Since the statistics is much smaller at high $p_t$'s (large
error bars in the evaluation) the effect should be mostly observed in the
$p_t^{(0)} \leq 0.5$ range.

\subsubsection{Elliptic flows}
%\addtocontents{toc}{\hspace{0.55cm}\thesubsubsection \hspace{0.12cm}
%Elliptic flows}
\begin{figure}[b]
\centering
\includegraphics[width=8cm]{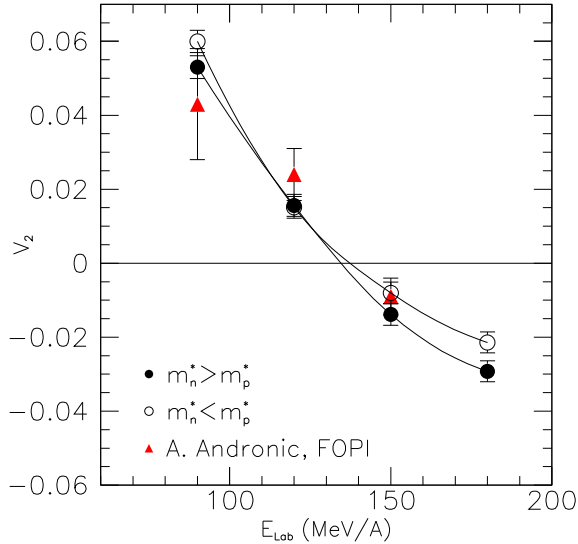}
\caption{Energy dependence of the elliptic flow in $Au+Au$ 
semicentral collisions at mid-rapidity, $|y^{(0)}| \leq 0.1$,
 integrated over the $p_t^{(0)} \geq 0.8$ range. $FOPI$ data with
similar selections, see ref. \cite{AndronNPA661},
are given by the full triangles. 
Solid circles refer to the
case $m^*_n>m^*_p$, empty circles to $m^*_n<m^*_p$.} 
\label{fig:v2au}
\end{figure}
Also in this case a good check of our effective interaction choices
is provided by some more inclusive data in the medium energy range.
In Fig.\ref{fig:v2au} we report a comparison with the $FOPI-GSI$ data in the
interesting beam energy region of the change of sign of the elliptic flow for
protons, integrated over large transverse momentum contributions.
The agreement is reasonable for both parametrizations. Also in this
case we see a slightly larger overall repulsion (earlier zero-crossing) 
for the $m^*_n>m^*_p$ case.
\begin{figure}[t]
\includegraphics[width=7cm]{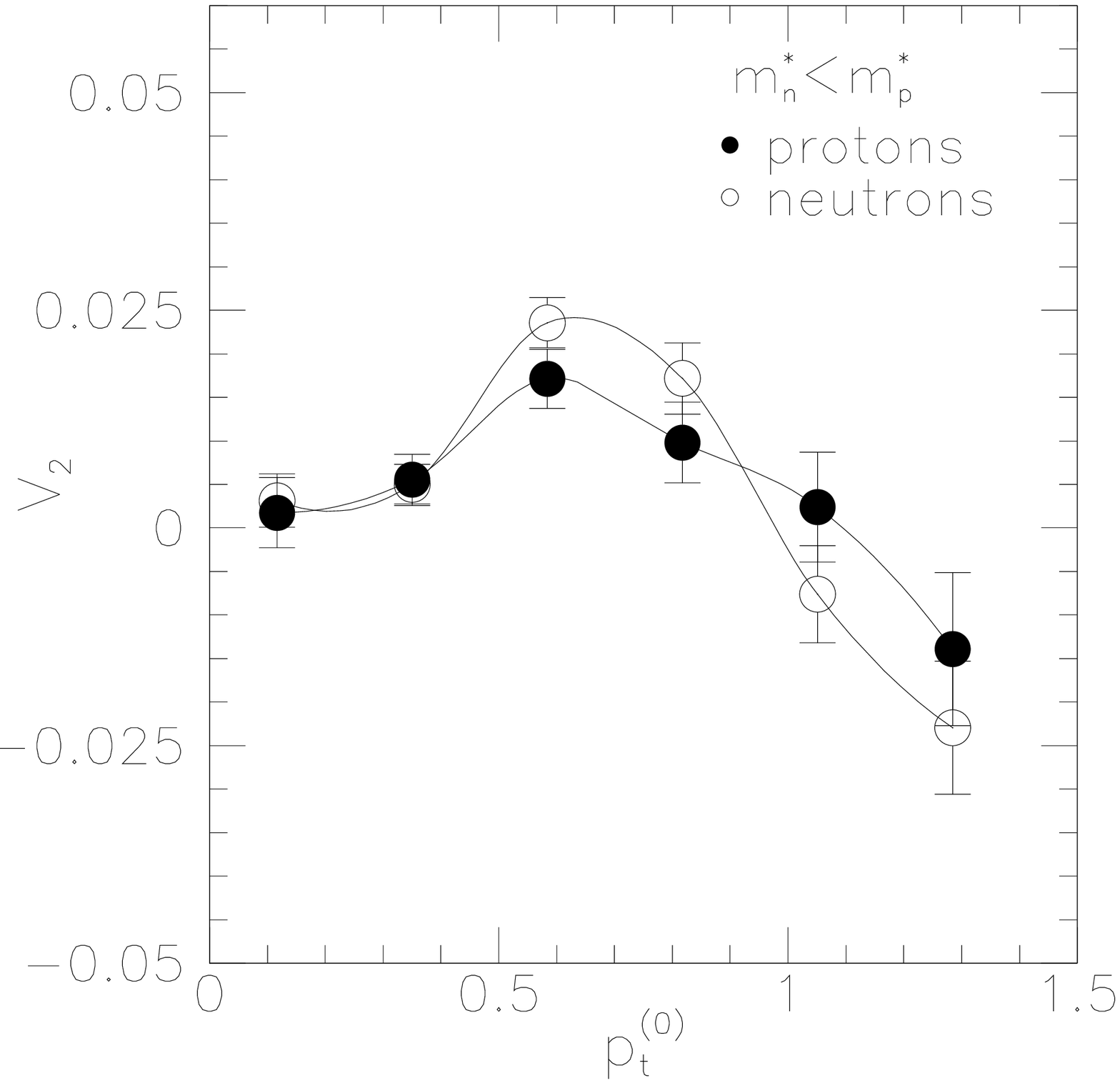}
\includegraphics[width=7cm]{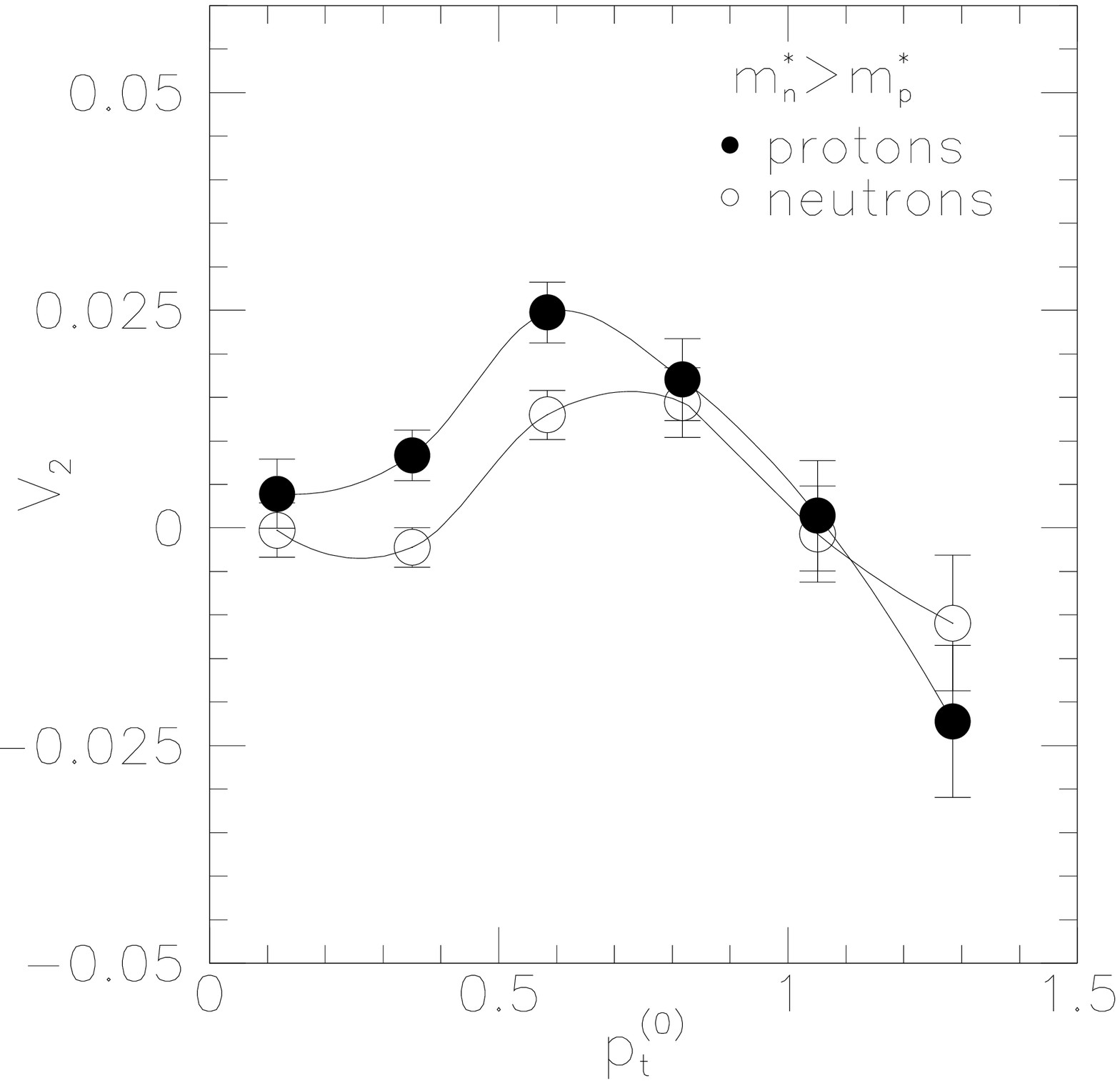}
\put(-12.8,2.85){\mbox{\line(1,0){5}}}
\put(-5.75,2.85){\mbox{\line(1,0){5}}} \caption{Elliptic flow of
protons and neutrons in Au+Au reaction at $250\, AMeV$,
$b/b_{max}=0.5$, in the rapidity interval $0.7\leq |y^{(0)}| \leq
0.9$.} \label{fig:v2np}
\end{figure}

\begin{figure}[b]
\centering
\includegraphics[width=8cm]{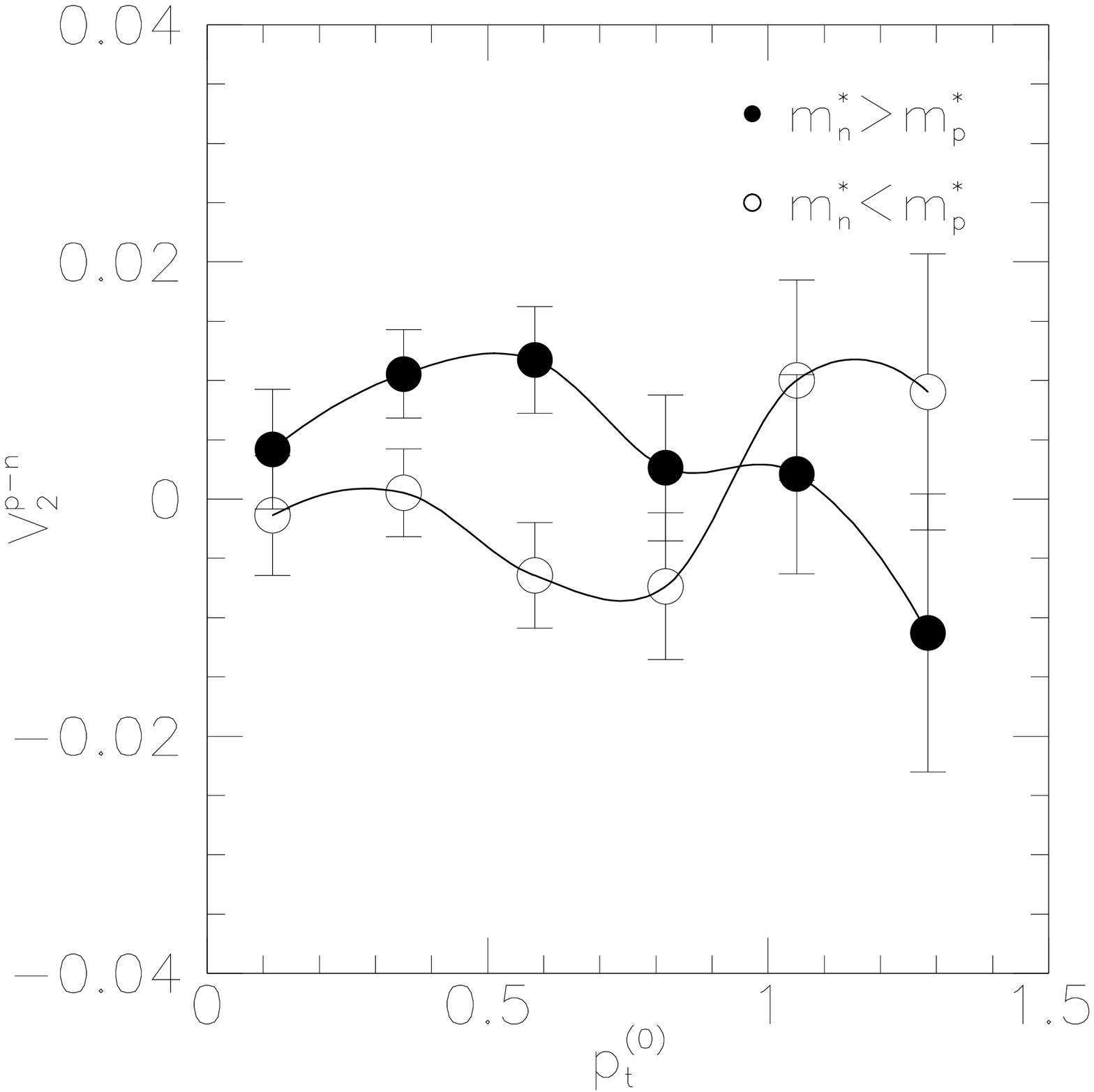}
\caption{$p_t$ dependence of the difference between neutron and proton 
transverse flows in $Au+Au$ 
collisions, same energy, centrality and rapidity selection as in 
Fig.\ref{fig:v2np}. Solid circles refer to the
case $m^*_n>m^*_p$, empty circles to $m^*_n<m^*_p$.} 
\label{fig:difv2}
\end{figure}

Results on the $p_t$ dependence of the neutron/proton
elliptic flows in $Au+Au$ reactions at $250~AMeV$ are shown in Fig.
\ref{fig:v2np}. We can see that the $p_y$ component rapidly grows at
momenta $p_t^{(0)}>0.6$, causing the flows to fall down towards
negative values. At low momenta, emitted particles undergo a
deflection in the reaction plane, more pronounced for nucleons
with smaller effective mass.
 In fact, comparing the two panels of Fig.\ref{fig:v2np} we see that
 around the maximum in-plane deflection 
$0.4\lesssim p_t^{(0)}\lesssim 0.8$ neutron and proton flows are inverted. We
also find different slopes for protons and neutrons around $p_t^{(0)}\geq
0.7$, depending on
the relative sign of mass splitting, in agreement with the
previous results in the transverse flows.
 Around $p_t^{(0)} \simeq 1$,
i.e. projectile momentum in the $cm$,  the elliptic flow finally takes 
negative
values, with slight differences for the two parametrizations.
Again the proton and neutron behaviors are not perfectly inverted
in the two cases because of the proton Coulomb repulsion, in
particular at low $p_t$.
Also in this case the best observable to look at is the $p_t$
dependence of the difference between neutron and proton elliptic flows
$$V_2^{p-n}(y,p_t) \equiv V_2^{p}(y,p_t) -  V_2^{n}(y,p_t)$$ 
shown in the Fig.\ref{fig:difv2}.

Again the difference is quite evident in the whole $p_t$ range of emitted 
nucleons. We note that also here the statistics is much smaller 
at high $p_t$'s 
(large error bars in the evaluation). Anyway the mass splitting effect 
appears very clearly  in the low
$p_t^{(0)} \leq 0.5$ range where the results are more reliable.

As a general comment we stress the importance of more exclusive data,
with a good $p_t$ selection. Indeed from Figs. \ref{fig:v1np}, \ref{fig:difv1}
(for $V_1$) and Figs. \ref{fig:v2np}, \ref{fig:difv2} (for $V_2$) 
we clearly see
that more inclusive $V_1(y)$, $V_2(y)$ data, integrated over
all $p_t$'s of the emitted nucleons at a given rapidity,
will appear to be less sensitive to $isospin-MD$ effects. As already discussed
from general features of the mean field $MD$ we expect a kind of compensation
between low and high $p_t$ contributions.
As already remarked the Coulomb repulsion acting on protons in this heavy 
system tends to reduce the dynamical effects of nucleon mass splitting,
in particular for low momentum particles. In spite of that the signals
appear quite clear in the simulations. In any case this represents
a further indication of the importance of momentum selections.

\subsubsection{Changing the stiffness of the symmetry term}
%\addtocontents{toc}{\hspace{0.55cm}\thesubsubsection \hspace{0.12cm}
%Changing the stiffness of the symmetry term}
It is well known that the collective flows for asymmetric systems are 
sensitive to the density dependence of the symmetry term of
the nuclear EOS, in particular at high transverse momentum 
\cite{ScalonePLB461,BaoPRL85,BaoPRC64}.
 For this reason it is important to be sure that signals previously
discussed are mainly due to the $n/p$ effective mass splitting and 
not strictly 
depending on the choice of the stiffness of the symmetry energy. 
In order to check this point
we show some calculations using a symmetry energy with a 
density behaviour very different from the one used before, 
{\it but keeping the same $n/p$-effective mass splittings}. 
We refer to it as 
{\it asy-soft}, compared with the {\it asy-stiff} 
previously used. 
For details see ref.\cite{RizzoNPA732}.
\begin{figure}[b]
%\centering
\includegraphics[width=7cm]{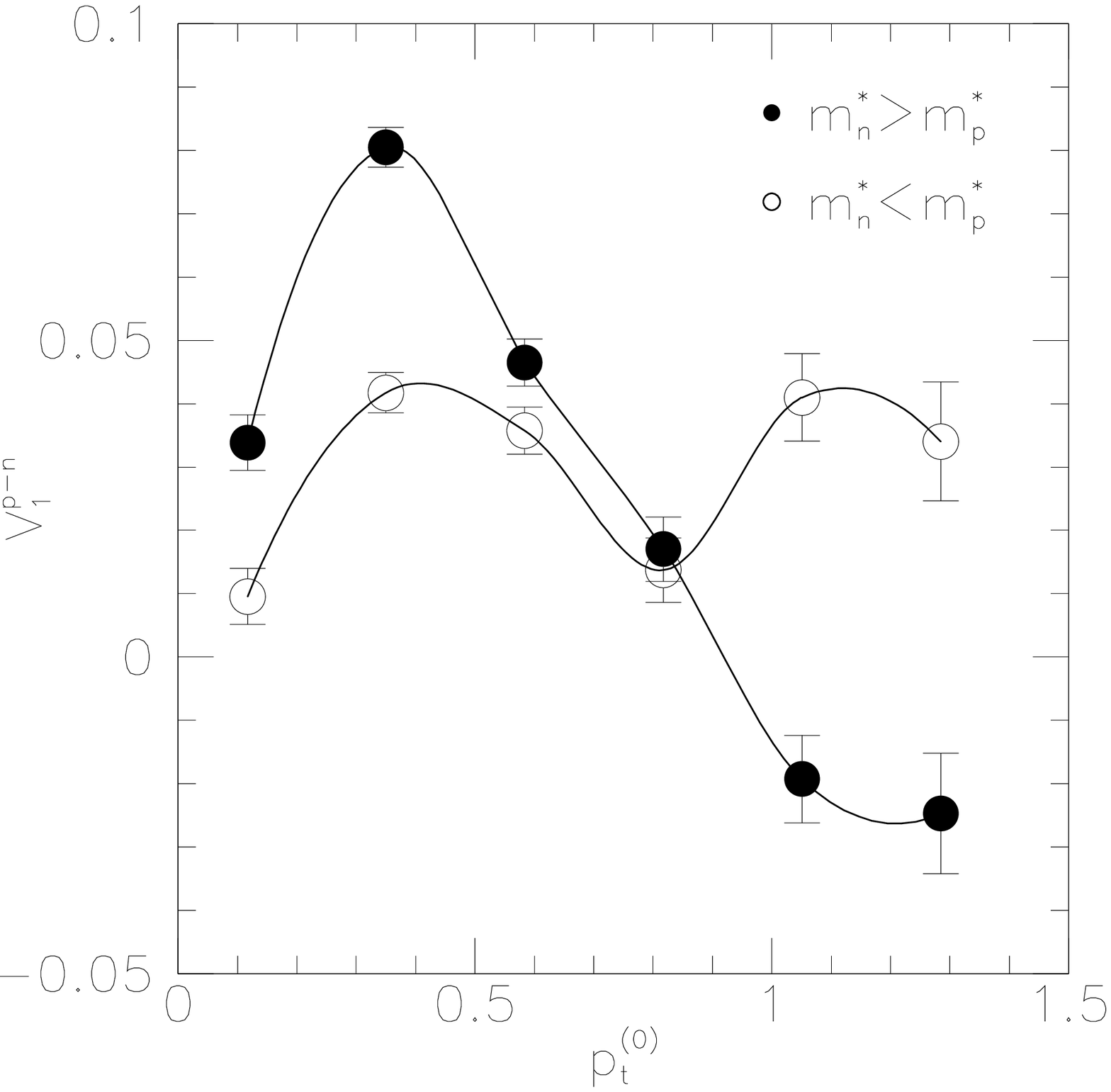}
\includegraphics[width=7cm]{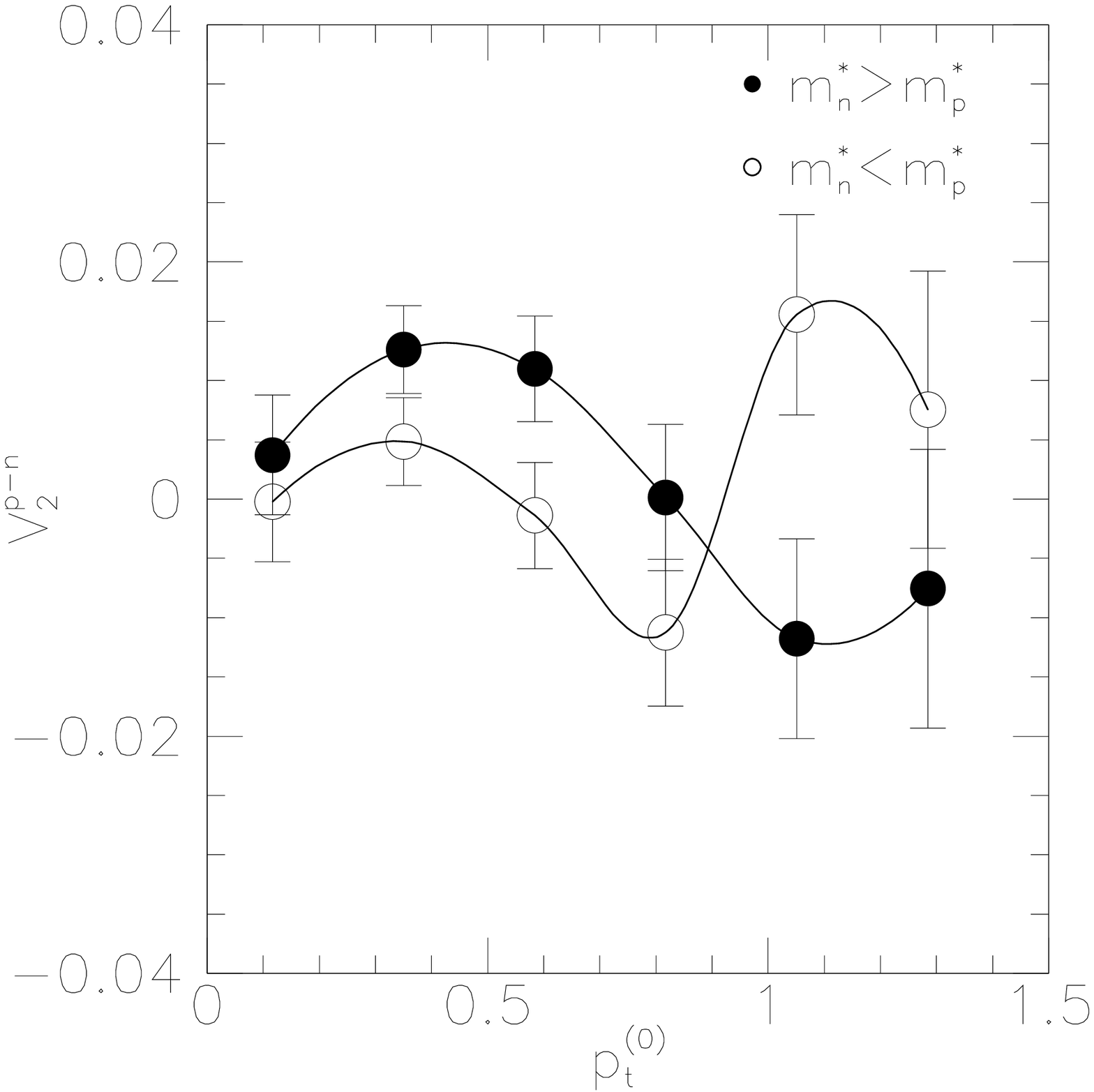}
\caption{$p_t$ dependence of the difference between neutron and proton
transverse (left panel) and elliptic flow (right panel)
in $Au+Au$ collisions at $250\, AMeV$,
$b/b_{max}=0.5$, in the rapidity interval
$0.7\leq |y^{(0)}| \leq 0.9$. Calculations are performed using the $asy-soft$
symmetry energy. Solid circles refer to the
case $m^*_n>m^*_p$, empty circles to $m^*_n<m^*_p$.}
\label{fig:difsoft}
\end{figure}
Our results show that while the nucleon flows are depending on
the stiffness of the symmetry term, like in other calculations
\cite{ScalonePLB461,BaoPRL85,BaoPRC64}, the transverse momentum behavior
of the differences keeps the same sensitivity to the effective mass
splittings. 
Fig. \ref{fig:difsoft} illustrates the $n/p$-flow differences as a function of
transverse momentum for the same energy, impact parameter and high rapidity
selections used before. Both transverse $V_1^{p-n}(y,p_t)$ and 
elliptic $V_2^{p-n}(y,p_t)$ flow 
differences show very small
variations from the $asy-stiff$ case (i.e. Figs. 
\ref{fig:difv1} and \ref{fig:difv2}) over the whole $p_t^{(0)}$ range, in
particular
in the region  
$p_t^{(0)}\leq0.5$, of interest for the better statistics. 
This result represents an important indication that we are 
selecting experimental observables that
directly probe the isospin effects on the momentum dependent part
of the nuclear $EOS$.

%\include{rep_bib}

%\end{document}

%% file: Chapter-5.tex
%\documentclass{elsart}

%\usepackage{graphicx}

%\usepackage{amssymb}

%\begin{document}

\setcounter{figure}{0}
\setcounter{equation}{0}

\section{Isospin Dynamics at the Fermi energies}\label{fermi}

\markright{Chapter \arabic{section}: fermi}

\subsection{Stochastic BNV transport theory}

We will show that Heavy Ion Collisions in the Fermi energy  
domain are rather sensitive to the
different density/momentum dependences of the symmetry term in a 
quite transparent way.
We present studies with stable and unstable ions in order
to increase the asymmetry range of the nuclear systems.

Since dynamical instabilities are playing an essential role in the reaction
dynamics at Fermi energies it is essential to employ a stochastic transport 
theory.
An approach has been adopted based on microscopic transport equations
of Boltzmann-Nordheim-Vlasov ($BNV$) type
\cite{GregoireNPA465,BertschPR160,BonaseraPR244,BonaseraPLB221,twingo}
where asymmetry effects are
suitably accounted for \cite{ColonnaPRC57,ScalonePLB461}
and the dynamics of fluctuations is included \cite{Erice98,ColonnaNPA642}. 

The transport equations,
with Pauli blocking consistently evaluated,
are integrated following a
test particle evolution on a lattice \cite{twingo,GuarneraPLB373,GrecoPRC59}.
A parametrization of free $NN$ cross sections is used, with
isospin, energy and angular dependence. 
The same symmetry term is
utilized even in the initialization, i.e. in the ground state construction of 
two colliding nuclei.

We consistently include isospin effects in the treatment
of the stochastic term of the transport equation.
Indeed the evolution under the influence of 
fluctuations is described by a transport equation with a 
stochastic fluctuating term, the so-called 
Boltzmann-Langevin equation ($BLE$)
\cite{AyikPLB212,AyikNPA513,RandrupNPA514,ColonnaPRC49,ChomazPRL73}. 
We  follow two methods to 
include the fluctuations effects. The Stochastic Mean Field ($SMF$) approach, 
\cite{Erice98,ColonnaNPA642,ColonnaNPA742},
and a simplified procedure, computationally much easier 
\cite{ColonnaNPA580,GuarneraPLB373,ColonnaPRC47,ColonnaNPA630}, 
based on the introduction of density fluctuations by a random sampling
of the phase space, (Phase Space Sampling, $PSS$). The amplitude of the 
noise is
gauged to reproduce the dynamics of the most 
unstable modes \cite{ColonnaNPA580}.

The $SMF$ method is based on
a fully self-consistent dynamical treatment of the fluctuations 
during the time 
evolution.
We notice that $\sigma^2 = <(f-\bar f)^2>$, the variance around the 
mean trajectory of the system,
 given by the Boltzmann-Vlasov evolution in phase space 
$\bar f({\bf r}, {\bf p}, t)$, 
 in each phase space cell
obeys the equation of motion \cite{ColonnaNPA567,ColonnaNPA642}:
\begin{equation}
\frac{d}{dt}\sigma^2 = -\frac{2}{\tau(t)}\sigma^2 + 2~D(t).
%{\bf p}, t) .
\label{sigm}
\end{equation}
with $2~D(t)$  the correlation function of the fluctuating term
 and the relaxation time $\tau(t) = 1 / (w^++w^-)$, 
where $w^+$ and $w^-$ are the transition probabilities 
into and out of the cell. 
The statistical 
value $\sigma_0^2 = f_0(t_{eq})(1-f_0(t_{eq}))$ at equilibrium 
suggests an ansatz for the correlation 
function of the fluctuation term of the form:
\begin{equation}
2D(t) = (1-\bar f) w^+ + \bar f w^-
\label{corr}
\end{equation}
i.e. the magnitude of the  fluctuations is given by the 
total number of collisions  
(fluctuation-dissipation theorem) \cite{RandrupNPA514}.  
Then we obtain for the time evolution of the 
difference relative to local statistical fluctuations
$\Delta \equiv \sigma^2 - \bar f (1- \bar f)$:
\begin{equation}
\frac{d}{dt}\Delta = -\frac{2}{\tau(t)}\Delta .
\label{diff}
\end{equation}
Since $\Delta = 0$ is a solution of Eq.(\ref{diff}),
if the variance is initially locally set equal to its statistical value, it  
will always follow the evolution of the average distribution function 
according to the relation $\sigma^2 =  \bar f (1- \bar f)$.
Thus from our knowledge of the mean ``trajectory'' 
$\bar f({\bf r}, {\bf p}, t)$, at any time step  we can extract local 
statistical
fluctuations in each phase space cell.

From a projection over coordinate space we derive local density
variances wich are implemented with a Montecarlo method at
suitably chosen time intervals, see ref.\cite{ColonnaNPA742}. 
In this way we can have also the  
trajectory branchings in regions of instability. The procedure is correct 
if we assume a {\it local} 
thermal 
equilibrium in each phase space cell. This is  appropriate for the 
problems discussed 
here, namely fragment production in the expansion/separation phase, and 
it can be tested numerically.

For each system we have checked the equivalence of the $SMF$ approach
with the sampling $PSS$ method
in the description of the collision dynamics, from
fast particle emissions to the fragment production.
The analysis of the results, prsented in the following, is
based on events collected with both numerical procedures.
We note that in this way the transport simulation will represent a
kind of event generator. In general we have accumulated $400-500$
events for each physical case.

In order to simplify the analysis of the observables most sensitive
to isospin effects we keep fixed the isoscalar part of the $EOS$,
corresponding to a  $soft$ behavior of symmetric nuclear matter($SNM$) with 
compressibility $K=201MeV$. We will usually consider three choices
for the density dependence of the symmetry term.
From Fig.\ref{fig:mean} of Sect.\ref{eos} these are 
 {\it asy-soft} (dashed line), 
{\it asy-stiff} 
(solid line) 
and {\it asy-superstiff} (long-dashed line).
In this way we try to disentangle dynamical symmetry term effects.

\subsection{Main mechanisms of dissipative collisions: Fusion vs. 
Deep-Inelastic Competition}

We start the analysis at relatively low energies. Moreover we will
also consider medium-light unstable beams eventually available
in this energy range fron $SPIRAL/I-II$ and other new 
Radioactive Ion Beam $(RIB)$ facilities. Here we investigate
isospin effects on the
competition between binary events and heavy residue formation 
which occurs at intermediate impact parameter $b \simeq 0.4b_{max}$.

\begin{figure}
\begin{center}
\includegraphics[scale=0.90]{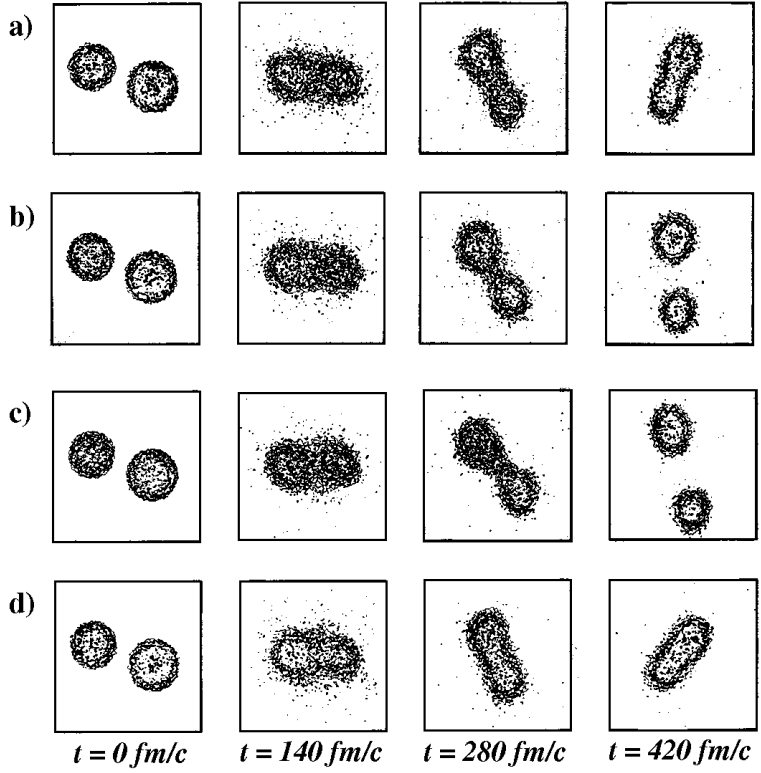}
\caption{Density plots at different times in a reaction between
 neutron-rich ions ($a,b: ^{46}Ar+^{64}Ni$) and neutron-poor ions
($c,d:^{46}V+^{64}Ge$). ($a,c$): $asysoft$ symmetry term;
($b,d$): $asystiff$. See text}
\label{fig:rib}
\end{center}       
\end{figure}

In Fig.\ref{fig:rib} we show plots, for two choices of the symmetry term,
 of the time-evolution of the 
density projection on 
the reaction plane for the mean trajectory (averaged over several events)
in semicentral collisions ($b=4fm$) of neutron rich 
($^{46}Ar(N/Z=1.56)+^{64}Ni(N/Z=1.29)$)
and $N=Z$, neutron poor ($^{46}V+^{64}Ge$) ions at $30AMeV$
 \cite{ColonnaPRC57}.

The effect on the main reaction mechanism of the different
density dependences of the symmetry term is quite evident
in Figs.\ref{fig:rib}$(a,b)$. It can be understood
in terms of the amount of repulsion present during the interaction 
of the two surfaces, i.e. in a region below normal density.
For neutron rich colliding ions the asy-soft choice in general leads 
to a less repulsive dynamics since the symmetry field for protons
is more attractive, see Fig.3 of Sect.\ref{eos}.
We see a stronger interaction between the two partners, a larger
dissipation of the relative energy, and thus the system will more likely
enter a fusion path, Fig.\ref{fig:rib}(a). Experimentally this means a 
larger incomplete fusion
cross section at medium energies. 

For neutron poor systems, Figs.\ref{fig:rib}$(c,d)$
we see just the opposite. Naively we would expect more repulsion
since we have larger Coulomb and direct $n-p$ collision contributions.
Actually in the asy-soft case we have now a dominant
binary deep-inelastic mechanism, Fig.\ref{fig:rib}$(c)$.
At variance, in the asy-stiff choice Fig.\ref{fig:rib}$(d)$ we get
a clear fusion dominance. This result is amazing for two reasons:
i) The dynamics of a symmetric $N=Z$ system appears to be isospin-dependent;
ii) The stiffer symmetry term leads to a larger attraction.
A way to understand this is that the two exotic neutron poor systems 
develop
a ``proton skin'' which is overlapping in the interaction zone. The protons 
in the asy-stiff case see a stronger repulsive field and are more easily 
promptly emitted. As a consequence the two partners feel a smaller Coulomb
repulsion and more likely fusion can be reached. Consistently a larger yield
of fast proton emission is predicted with the asy-stiff parametrization
\cite{ColonnaPRC57}. We like to recall that some sensitivity to isospin
effects in deep inelastic collisions was already shown in ref.
\cite{FarineZPA339} from Landau-Vlasov simulations of $Ca$-isotope
collisons at $15AMeV$.  

These results suggest that in low energy dissipative collisions an observable
sensitive to the stiffness of the symmetry term can be just the relative yield
of incomplete fusion vs. deep-inelastic events.
Moreover for n-rich systems in the asy-soft case we expect more interaction 
time available for charge equilibration. This means that even the binary 
events will show a sensitivity through a larger isospin diffusion. At variance,
in the asy-stiff case the two final fragments will keep more memory of the
initial conditions. This point will be futher discussed later in the Section.

The dependence of the interaction time on the stiffness of the symmetry term  
will also influence the competition between binary
and $neck-fragmentation$ events \cite{BertschPLB141,ColonnaNPA589}, 
 where Intermediate Mass Fragments ($IMF$, in the range $3 \le Z \le 10$) are
formed in the overlapping region, roughly at mid-rapidity in semicentral
reactions. This represents the new leading dissipative mechanism at the
Fermi energies, as we will discuss in detail later on. Therefore the
expected sensitivity to the isovector $EOS$ appears very stimulating.

\subsection{Isospin in Fragmentation Dynamics: Survey of experimental results}

In the last decade the experimental interest on the
isospin dynamics in heavy ions collisions at Fermi energies and
$EOS$ symmetry term effects on various reaction mechanisms
has been steadly rising, largely stimulated by the perspective of
the second generation
radioactive beam facilities. 

A large variety of systems and phenomena
have been investigated. This clearly shows the richness of the field
but also reveals some uncertainty in focussing on the most relevant physics
questions. One of the aims of this report is just to contribute
on that direction, even trying to select the most sensitive
observables. 
In this respect we will first organize a discussion of the
performed experiments, in a precise correspondence to 
theoretical analyses and suggestions presented in the
following subsections.

\subsubsection{Isospin equilibration and fragment production}
%\addtocontents{toc}{\hspace{0.55cm}\thesubsubsection \hspace{0.12cm}
%Isospin equilibration and fragment production}
The systems $^{40}Ca, ^{40}Ar$ ,$^{40}Cl +$  $^{58}Fe, ^{58}Ni$
have been among the first devoted to study the
role of the isospin degree of freedom at Fermi energies \cite{YennelloPLB321}.
Let us note that the $Cl$-projectile is a radioactive 
secondary beam. The total mass of composite
system $(CS)$ is the same in all cases allowing to
explore the effects of the $N/Z$ content on the $IMF$ production.
At $53AMeV$, the isotopic ratios
of 
 $^{11}B/ ^{10}B$, $^{8}Be/^{7}Be$, $^{7}Li/ ^{6}Li$,
exhibit an increasing trend with the $(N/Z)_{CS}$. 
Differences in the yield 
in central selections, 
 formed with the two beam-target combinations,
$^{40}Ar +  ^{58}Ni$ and $^{40}Ca +  ^{58}Fe$, show that 
the $IMF$ emission
occurs before the isospin degree of freedom is 
fully equilibrated.
Similar studies were performed at lower energies,
at $33$ and $45 AMeV$, in ref. \cite{JohnstonPLB371},
for n-rich vs. n-poor isobaric $IMF$ ratios.
Differences in the isotopic ratios were seen
at $45 AMeV$ but not at $33 AMeV$. 
The authors claim that with increasing beam energy we see a transition 
from a regime
in which the isospin equilibration is reached before fragment production
to one when $IMF$ emission happens before
charge equilibration. This could be a nice evidence of isospin transparency 
for these medium-light systems,
but a more accurate analysis of the reaction mechanism would be
required (centrality cuts).

An example of the importance of the centrality cuts can be seen
from the results obtained at low energy, $E_{beam}= 30 AMeV$, 
for the heavier $A=58$ systems $^{58}Fe(n-rich)/^{58}Ni(n-poor)$,
 in various projectile/target combinations. 
The isobaric and isotopic ratios $^{11}B/ ^{11}C$, $^{10}Be/^{10}B$,
$^{7}Li/$$^{7}Be$, $^{12}C/$ $^{11}C$, $^{11}B/$$^{10}B$,
$^{7}Li/$$^{6}Li$, measured at various laboratory angles
scale aproximately with the $(N/Z)_{CS}$, \cite{RamakrishnanPRC57},
 clearly showing that both  projectile and target nucleons
contribute to the source composition.
However, for the $^{58}Ni+ ^{58}Fe$ and  $^{58}Fe+ ^{58}Ni$ mixed cases,
different values of $^{12}C/ ^{11}C$ or
$^{11}B/ ^{11}C$ at $40^0$ seems to suggest an incomplete
charge equilibration even at this energy 
for the source situated at midrapidity. We will see from the transport
simulations that at these forward angles in semicentral collisions we can 
have fragments produced in the neck region but still driven from the
projectile remnants, so the discussion on isospin equilibration
is more involved.
 
For the same $Fe,Ni$ systems
the correlation between the mean number of $IMF$s 
and charged particle multiplicity, a rough indication of centrality, 
has been investigated as a function of
incident energy in \cite{MillerPRL82}.
Neutron rich combinations systematically produce more $IMF$s for 
the same value of charged particle multiplicy.
The effect is robust at the three lower energies, $45, 55, 75 AMeV$,
as clearly evidenced when an appropriate normalization to the system
total charge, accounting for the larger number of protons
in the $Ni$ case, is introduced. However at higher energy, $105AMeV$,
the isospin dependence of multifragment production appears much weakened.
This could be an interesting indication of the onset of different
mechanisms for fragment production at higher energies, well above the Fermi
region discussed in this section.

\subsubsection{Isospin Distillation and Isoscaling}
%\addtocontents{toc}{\hspace{0.55cm}\thesubsubsection \hspace{0.12cm}
%Isospin Distillation and Isoscaling} 
In other collisions the formed composite
systems differ not only in $N/Z$ but also in mass.
In these cases, for a proper interpretation of
isospin effects, attention is payed even to the mass dependence of observables.
The n-rich/n-poor $^{124}Sn/^{112}Sn$ systems have been extensively
investigated with outstanding results.

The first evidence of the Isospin Distillation (i.e. different 
asymmetry in the liquid and gas phases) in central collisions
was in fact observed in
$^{124}Sn+^{124}Sn$ and $^{112}Sn+^{112}Sn$ reactions at $40AMeV$,
\cite{KundePRL77}, where 
the correlations between the average number $<N_{IMF}>$ of $IMF$s
and the multiplicities of charged
particles $N_{C}$, of light-charged particles $N_{LC}$ $(Z \le 2)$,
and  neutrons $N_{n}$ were measured.
 It was shown,
\cite{KundePRL77,KortemeyerPRC55}, that in the n-rich 
$^{124}Sn+^{124}Sn$ case
the multiplicity of the emitted 
neutrons $N_{n}$ for the maximum value of $<N_{IMF}>$ was
much larger than expected from the neutron excess of the $124+124$
system. A clear {\it Neutron Distillation} effect was seen for the first
time, associated to the largest yield of fragment formation: very n-rich
gas phase coexisting with a more symmetric cluster phase, the 
 so-called  {\it healthier fragments} of ref.\cite{KortemeyerPRC55}.

The $Sn$ reactions in all possible n-rich/n-poor target/projectile
combinations have been extensively studied at the Cyclotron Lab. of MSU
at the beam energy $E_{beam}= 50 AMeV$, \cite{XuPRL85,TsangPRL86}.
The Isospin Distillation in central collisions was observed and analysed
using an approach aimed to reduce as much as possible the
effects of sequential decays
by normalizing the isotopic yield
from different $Sn+Sn$ systems to that of the more charge-symmetric case. 
It was then realised that the yield ratio of a given isotope in two different
reactions $R_{21}(N,Z)=Y_2(N,Z)/Y_1(N,Z)$  obey an {\it Isoscaling} law:
its logarithm 
is a linear function on the number of neutrons and protons,
with the slope given by the
the isoscaling parameter $\alpha$ and $\beta$ respectively, 
 \cite{XuPRL85,TsangPRL86}.

The double ratio of the measured
isotope and isotone yields can be used, in a grand-canonical statistical 
approach, to extract the isospin
composition of the free nucleons, i.e. the proton/neutron density ratios..
Observing its dependence
on the asymmetry of the composite system, $(N/Z)_{CS}$, it was clearly 
established that a neutron enrichement
of the free nucleonic gas takes place.
Within the same thermodynamical model the absolute
 value of $\rho_n/\rho_p$ in the nucleonic gas can be evaluated.
The output is in fact
significantly larger than the initial neutron to proton ratio:
between $1.7$ to $3.4$ for $^{112}Sn+^{112}Sn$ (initial $1.24$) and 
$2.8$ to $8.2$ for $^{124}Sn+^{124}Sn$ (initial $1.48$). 
This effect is certainly present but it appears overestimated within the
grand canonical picture. In the same scheme the isoscaling parameters,
being related to the difference of neutron and proton chemical potentials,
can provide an evaluation of the symmetry term of the $EOS$ 
in dilute matter \cite{TsangPRL86,TanPRC64}. Very low values are deduced,
with a large uncertainty due to the determination of the temperature
of the equilibrated source. Alternative interpretations have been 
recently proposed,
e.g. see ref. \cite{LiuPRC69}, were the isospin dynamical aspects play a 
greater role and no global equilibration is required. Information even on 
the density gradient of the symmetry energy can be reached, see the following
discussions based on transport simulations.

Reactions for similar systems, 
$^{124}Sn,^{124}Xe+$ $^{124}Sn,^{112}Sn$, have been recently studied
at lower energies, $28AMeV$
\cite{ShettyPRC68}.
Isotopic
yield ratios  measured at $40^0$ laboratory
angle show a reduced isospin distillation effect. This could be an 
indication of the quenching at lower energies of the liquid-gas transition 
for the fragment production. However the centrality selection of the events
should be accurately checked since in semiperipheral collisions other
mechanisms, with different isospin dynamics, can produce fragments in the same
angular range, as discussed below.
In any case we expect a rise and fall of the liquid-gas mechanism with
increasing beam energy for central collisions, and consequently of the 
related isoscaling signal. This trend has been recently observed
for the $^{58}Fe,^{58}Ni$ reactions systematically studied at 
$30, 40$ and $47 AMeV$ beam energies, 
 \cite{YennelloHIP97,ShettyPRC68R,ShettyPRC70}. 

Isoscaling behaviour and neutron enrichement
of the free nucleon gas have been revealed
also for $^{112,124}Sn+$ $^{58,64}Ni$ central collisions at $35AMeV$
\cite{GeraciNPA732}. 
Finally we have to note that the isospin distillation was also noticed
in multifragmentation reactions induced by high energy protons
\cite{MartinPRC62}.

\subsubsection{Isospin Dynamics for Neck Fragmentation in 
 Semiperipheral Collisions}
%\addtocontents{toc}{\hspace{0.55cm}\thesubsubsection \hspace{0.12cm}
%Isospin Dynamics for Neck Fragmentation in 
% Semiperipheral Collisions}
It is now quite well established that the largest part of the reaction
cross section for dissipative collisions at Fermi energies goes
through the {\it Neck Fragmentation} channel, with $IMF$s directly
produced in the interacting zone in semiperipheral collisions on very short
time scales. We can expect different isospin effects for this new
fragment formation mechanism since cluster are formed still in a dilute
asymmetric matter but always in contact with the regions of the
Projectile-Like and Target-Like remnants almost at normal densities.

A first evidence of this new dissipative mechanism was suggested
at quite low energies, around $19AMeV$, in semicentral $^{100}Mo+^{100}Mo$,
$^{120}Sn+^{120}Sn$ reactions \cite{CasiniPRL71,StefaniniZFA351}.
A transition from binary, deep-inelastic, to ternary events was observed,
with a fragment formed dynamically 
that influences the fission-like decay of the primary 
projectile-like ($PLF$) and target-like ($TLF$) partners. 
From in-plane fragment angular distribution a decrease of 
scission-to-scission lifetimes with the mass asymmetry
of the $PLF/TLF$ ``fission-fragments'' from $3000fm/c$ 
down to $200fm/c$ was deduced.
Similar conclusions were reached in \cite{TokePRL75}.
Consistent with
the dynamical scenario was the anisotropic azimuthal distribution of IMF's.  
In fact the $IMF$ {\it alignement} with respect to the ($PLF^*$) velocity
direction has been one clear property of the ``neck-fragments'' first
noticed by Montoya et al.\cite{MontoyaPRL73} for 
$^{129}Xe + ^{63,65}Cu$ at $50AMeV$.

A rise and fall of the neck  mechanism for mid-rapidity fragments
with the centrality,  with a maximum for intermediate
impact parameters $b \simeq \frac{1}{2} b_{max}$, 
as observed in \cite{LukasikPRC55},
suggests the special
physical conditions required.
The size of the participant zone is of course important but it also 
appears that a good time matching between the reaction and the 
neck instabilities time-scales is required, as suggested in
refs. \cite{BertschPLB141,ColonnaNPA589}. 
In fact a simultaneous presence, in noncentral collisions,
of different $IMF$ production mechanisms
at midrapidity was inferred in several experiments 
\cite{LefortNPA662,GingrasPRC65,DavinPRC65,BocageNPA676,ColinPRC67,PaganoNPA734}. 

An accurate analysis of charge, parallel velocity,
and angular distributions has been extended to high fragment multiplicities
in \cite{ColinPRC67} by Colin et al. 
They have noticed a ''hierarchy effect'':
the ranking in charge induces on average a ranking in the velocity
component along the beam, $v_{par}$, and in the 
angular distribution. This means that the
heaviest $IMF$ formed in the mid-rapidity region is the fastest 
and the most forward peaked, consistently with 
the formation and breakup of a neck structure or a strongly deformed
quasiprojectile.
A very precise and stimulating study of the time scales in neck fragmentation
can be carried out using the new $4\pi$ detectors with improved performances
on mass resolution and thresholds for fragment measurements. Such kind of data
are now appearing from the $CHIMERA$ collaboration 
\cite{PaganoNPA681,PaganoNPA734}.
 
We have reported here the experimental evidences
for ``neck fragmentation'' since we can immediately expect
large isospin dynamical effects from the presence of large density
gradients and from the possibility of selecting various time-scales
for the fragment formation. This will be widely discussed in the
theory part of this Section.
The first evidences of isospin effects in neck fragmentation
were suggested by Dempsey et al. in \cite{DempseyPRC54} from
semiperipheral collisions of the systems  $^{124,136}Xe +$ $^{112,124}Sn$ 
at $55AMeV$, where
 correlations between the average number of $IMF$'s,
$N_{IMF}$, and  neutron and charged particle multiplicities 
were measured. 
The variation of the relative yields of
$^{6}He/$$^{3,4}He$, $^{6}He/Li$
with  $v_{par}$ for several $Z_{PLF}$ gates
shows that the fragments produced in the midvelocity
region are more neutron rich than are the fragments
emitted by the $PLF$. 
Enhanced $triton$ production an midrapidity was considered
in  ref.\cite{LukasikPRC55}, and more recently in \cite{PoggiNPA685},
 as an indication of a neutron neck enrichement.

P.M. Milazzo et al. \cite{MilazzoPLB509,MilazzoNPA703} analyzed the
the $IMF$ parallel velocity distribution for $^{58}Ni$$+ ^{58}Ni$
semiperipheral collision at $30AMeV$. The two-bumps structure 
for $IMF$s with $5 \le Z \le 12$,
located around the center of mass velocity and close
to the quasiprojectile ($PLF^*$) source respectively,
was explained assuming the simultaneous presence of two 
production mechanisms:
the statistical disassembly of an equilibrated $PLF^*$ and 
the dynamical fragmentation of the participant region.
The separation of the two contributions allows for several
interesting conclusions. The average elemental event multiplicity $N(Z)$
exhibits a different trend for the two processes: in particular, the
fragments with $5 \le Z \le 11$ are more copiously produced
at the midvelocity region.
This experiment has a particular importance since 
isospin effects were clearly observed, {\it in spite of the very low
inital asymmetry}.
The measured isotopic content of the fragments is clearly different in the
two mechanisms. The experimental heavy isotope/light isostope yield ratios, 
$^{14}C/ ^{12}C$, $^{12}B/^{10}B$, $^{10}Be/ ^{7}Be$, $^{8}Li/^{6}Li$,
show a systematic decreasing trend as a function of parallel velocity
from c.m. to $PLF$ values.  
All these results indicate a neutron enrichement of the 
neck region, {\it even when initially the system $N/Z$ is close to unity}.

Plagnol et al. \cite{PlagnolPRC61}
have examined, for the system 
$Xe+Sn$ between $25$ and $50AMeV$, the competition between
midvelocity dynamical emission and equilibrium
evaporation as well as its evolution with incident energy.
The onset of the neck emission takes place around
$25AMeV$ and rises with the energy while the evaporative part
remains quite invariant for a selected centrality. Neck matter
is found to be more charge asymmetric: more neutron rich isotopes are
favored at midvelocity in comparison to evaporation.  
Evidence of a neck-like structure and its neutron enrichement
has been seen even in collisions with a rather light symmetric target,
 $^{58}Ni +$ $^{12}C$, $^{24}Mg$,
at $34.5 AMeV$ \cite{LarochellePRC62}.
The average $N/Z$ ratio for isotopes with $Z=3,4$
exhibits a clear increase from the $PLF$
to the midrapidity zone.

We close 
with a couple of comments that indicate the need of new, possibly
more exclusive, data.
The reasons for a preponderance of neutron-rich isotopes emitted from
the neck region are a matter of debate. Possible explanations being, 
apart the density dependence of
symmetry energy, also a fast light 
cluster production, especially of $\alpha$-particles, 
which promptly leads to an amplification of neutron excess
in the participant matter \cite{SobotkaPRC55}.
For completeness we have to mention that different analyses even
give conflicting results on the n-enrichement of the clusters
produced at mid-rapidity \cite{XuPRC65,SobotkaPRC62}. 
This shows that the reaction dynamics is in general very complicated,
and we can have different isospin effects even in competition. 
The point will be clearified in the next subsections.

\subsubsection{Isospin Diffusion}
%\addtocontents{toc}{\hspace{0.55cm}\thesubsubsection \hspace{0.12cm}
%Isospin Diffusion}
The isospin equilibration appears of large interest
also for more peripheral collisions, where we have 
shorter interaction times, less overlap and a competition
between binary and neck-fragmentation processes. The 
specific feature at Fermi energies
is that the interaction times are close to the
specific time scales for isospin transport  
allowing a more detailed investigation of isospin diffusion
and equilibration in reaction between nuclei with different $N/Z$
asymmetries. The low density
neck formation and the  preequilibrium emission are adding 
essential differences with respect to what 
is happening in the lower energies regime.
Tsang et al. \cite{TsangPRL92} probed the isospin
diffusion mechanism for the systems $^{124}Sn +$ $^{112}Sn$ at
$E=50AMeV$ in a peripheral impact parameter range $b/b_{max}>0.8$,
observing the isoscaling features of the light
isotopes $Z=3-8$ emitted around the projectile rapidity. 
An incomplete equilibration was deduced. The
value of the isoscaling parameter $\alpha=0.42 \pm 0.02$
for $^{124}Sn +$ $^{112}Sn$ differs substantialy from
 $\alpha=0.16 \pm 0.02$ for $^{112}Sn +$ $^{124}Sn$.
The isospin imbalance ratio \cite{RamiPRL84}, defined as

\begin{equation}\label{imb}
R_{i}(x) = \frac{ 2 x - {x}^{124+124} - {x}^{112+112}}
{{x}^{124+124} - {x}^{112+112}}
\end{equation}

($i=P,T$ refers to the projectile/target rapidity measurement,
and $x$ is an isospin dependent observable, here
the isoscaling $\alpha$ parameter) was estimated to be
around $R_P(\alpha)=0.5$ 
(vs.$R_P(\alpha)=0.0$ in full equilibration). 
The authors
pointed out that this quantity can be sensitive to the
density dependence of symmetry energy term since the
isospin transfer takes place through the lower density
neck region.

The isoscaling analysis of heavy residues identified
in peripheral reactions $^{86}Kr +$ $^{124,112}Sn$ 
at $E=25AMeV$ was employed to trace the degree
of $N/Z$ equilibration as a function of the excitation
energy of the primary residue, \cite{SouliotisPLB588}.
The evolution of isospin exchange with the
centrality or interaction time, can be established.
It was concluded that for the most damped binary events,
corresponding to the largest excitation energies, charge
equilibration could be attained.

\subsection{Theoretical interpretation: overview.}

Using the above described Stochastic Mean Field ($SMF$) approach 
 dissipative collisions of
$Sn,Sn$ and $Sn,Ni$ isotopes at Fermi energies, in various 
asymmetry combinations, have investigated in great detail . 
In the transport simulations
the isovector part of the effective forces is only modified,
following the general philosophy of the research project which is
behind this report. We have seen before the amazing
richness of the data on isospin effects in reactions, still
remaining in the stable beam sector. Here we will try to select
the key observables in order to extract precise information
on the isovector part of the nuclear $EOS$ in regions far from
the normal conditions.
In particular we report on a study of the $50AMeV$ collisions of the systems 
$^{124}Sn+^{124}Sn$ $^{112}Sn+^{112}Sn$ and $^{124}Sn+^{112}Sn$, 
 \cite{BaranNPA703}, 
where data are available from $NSCL-MSU$ exps. for fragment production
and isospin diffusion, \cite{XuPRL85,TsangPRL86,TsangPRL92}. 

We  also analyse the neutron rich, mass asymmetric
reaction $^{124}Sn+^{64}Ni$, at $35 AMeV$ bombarding energy,
and the corresponding neutron poor reaction 
$^{112}Sn+^{58}Ni$, to look in detail at the neck fragmentation
dynamics and its isoscaling properties, \cite{BaranNPA730}.
These reactions have been experimentally 
studied with the $CHIMERA$ detector at $LNS$, Catania 
\cite{PaganoNPA681,PaganoNPA734}. As already remarked, the great advantage 
of the $CHIMERA$
detector for semicentral collision studies is the possibility of
measuring target-rapidity data, due to the low threshold of the 
detection telescopes, togheter with a large precision in $IMF$ mass separation.
As we will see this will allow new correlation
measurements of fundamental importance for understanding the reaction 
mechanism and the related isospin dynamics.

\subsection{From central to semi-central collisions: bulk fragmentation.}

\subsubsection{Fragmentation dynamics}
\begin{figure}
\centering
\includegraphics*[scale=0.45]{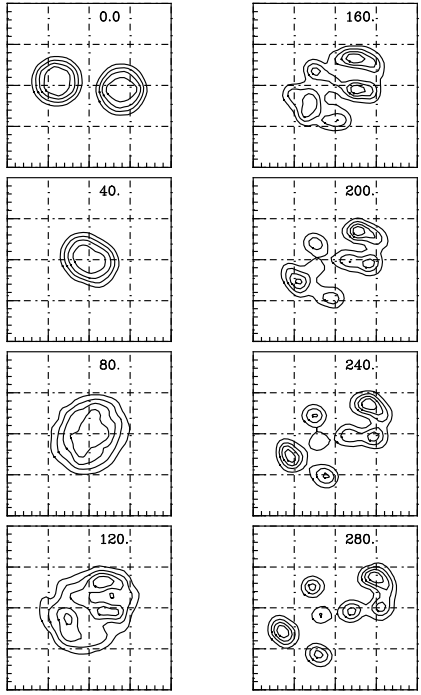}
\caption{$^{124}Sn+^{124}Sn$ collision at $50AMeV$: time evolution of the
nucleon density projected on the reaction plane.
Central $b=2fm$ collision: approaching, compression and
separation phases. The times are written on each figure.
The iso-density lines are plotted every $0.02fm^{-3}$
starting from $0.02fm^{-3}$.}
\label{iso5}
\end{figure}

%\addtocontents{toc}{\hspace{0.55cm}\thesubsubsection \hspace{0.12cm}
%Fragmentation dynamics}
At central impact parameters in this energy range 
the reaction mechanism manifests the features of bulk fragmentation, 
 as obseved in various dynamical simulations,
\cite{ZhangPRC60,DitNPA681,LiuNPA726}.
One can identify quite generally three main stages of the collision,
as observed also from the density contour plot of a typical
event at $b=2fm$ displayed in Fig. \ref{iso5}:
(1.) In the early compression stage, during the first $40-50fm/c$, the
density in the central region can reach values
around $1.2-1.3$ normal density;
 (2.) The expansion phase, up to $110-120fm/c$,
brings the system to a low density state. The physical
conditions of density and temperature reached during this
stage correspond to an unstable nuclear matter phase;
 (3.) In the further expansion fragmentation is observed.

According to stochastic mean field simulations, the fragmentation mechanism
can be understood in terms of the growth of  
density fluctuations in the presence of instabilities. 
The volume instabilities have time to develop through
spinodal decomposition leading to the formation
of a liquid phase in the fragments and a gas of nucleons and light 
clusters. As seen in the Figure, the fragment
formation process typically takes place up to a {\it freeze-out} time 
(around $260-280 fm/c$).
This time is well defined in the simulations since as the time of 
saturation of the average
number of excited primary fragments.  The clusters are rather far
apart with a negligible nuclear interaction left among them.
 
An intermediate behaviour, between the fragmentation mechanism 
illustrated above and the formation of an elongated ``neck'' region
between the two collisional partners, extensively discussed later, 
is observed at $b=4fm$
(see also Fig. \ref{iso6}). The {\it "Freeze-Out time"}, when the nuclear
interaction among clusters disappears, is decreasing with impact parameter.  
This gradual transition suggests that it may be
unappropriate to think in terms of a
unique fragmentation mechanism and more so to
assign a fixed size or shape for a multifragmenting source,
even passing from $b=1-2fm$ to $b=4-5fm$.

Guided by the density contour plots we can investigate the behaviour of
some characteristic quantities which give information on the isospin
dynamics in fragment formation. 

In many of the following figures (i.e. see Fig.\ref{iso8}) we report as a 
function of time: 
\vskip-0.2cm
(a) {\it Mass} $A$ in the liquid phase (solid line and dots)
and gas phase (solid line and dots).
\vskip-0.4cm
(b) {\it The asymmetry parameter} $I=(N-Z)/(N+Z)$ in
the gas "central" (solid line and squares), gas total (dashed+squares), 
liquid "central" (solid+circles) and $IMF$'s 
(clusters with $3<Z<15$ , stars). 
The horizontal line indicates the initial
average asymmetry. ``Central'' means a box of dimension $20$ fm
 around the $C.M.$ of the total system. 
\vskip-0.4cm
(c) {\it Mean fragment multiplicity} $Z \geq 3$ whose saturation
defines the freeze-out time and configuration.
\vskip-0.2cm
Moreover we show properties of the "primary" fragments in the
{\it Freeze-Out Configuration}:
\vskip-0.2cm
(d) {\it Charge distribution probability P(Z)}, 
\vskip-0.4cm
(e) {\it Average asymmetry distribution $I_{av}$(Z)} and
\vskip-0.4cm
(f) {\it Fragment multiplicity distribution P(N)} (normalized to $1$).

\subsubsection{Isospin distillation}
\begin{figure}
\begin{minipage}{65mm}
\begin{center}            
\includegraphics*[scale=0.45]{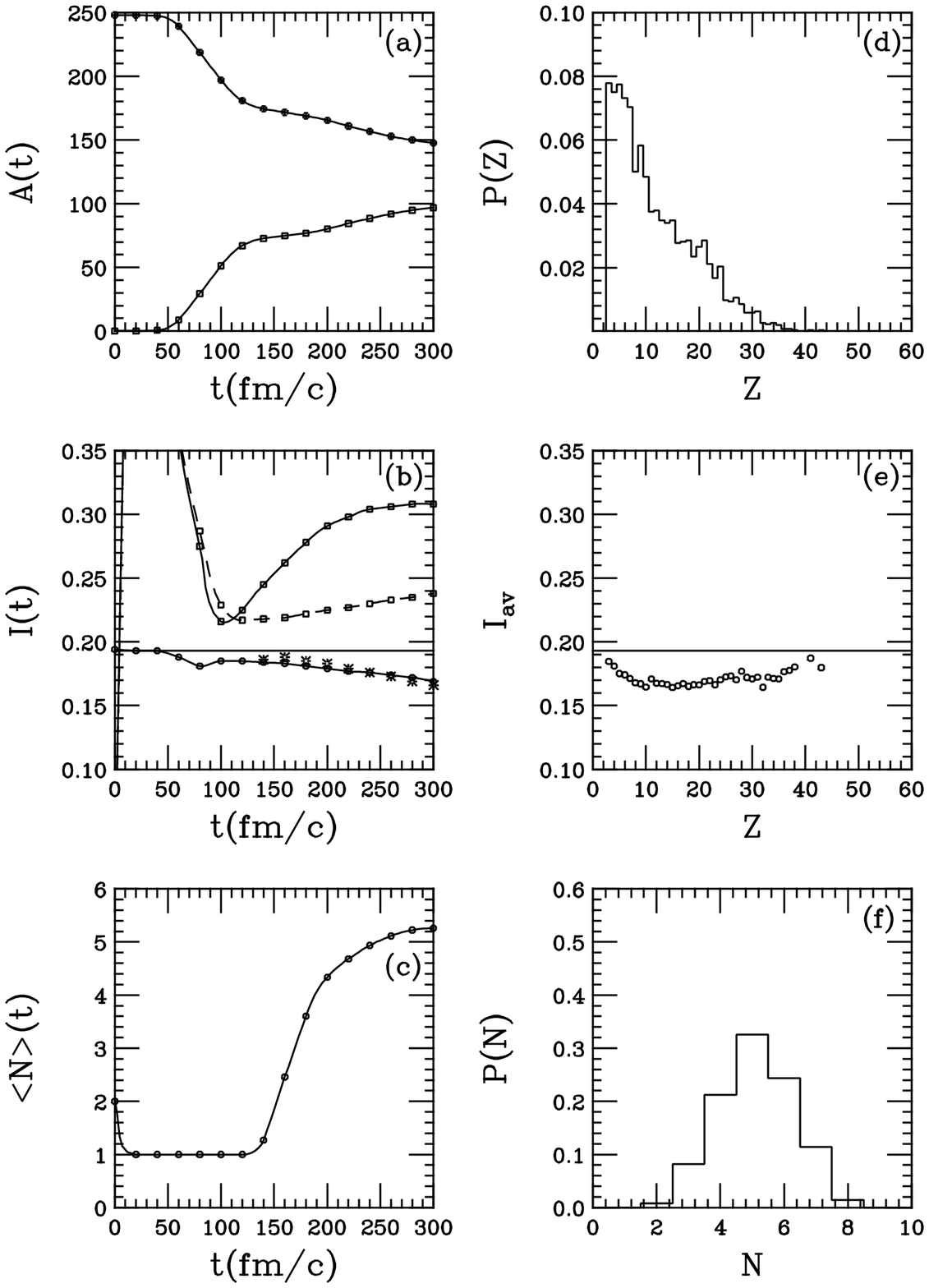}
\caption
{$^{124}Sn+$$^{124}Sn$ $b=2fm$ collision:time evolution
(left) and freeze-out properties (right),ASY-STIFF EOS.}
\label{iso8}
\end{center}
\end{minipage}
\hspace{\fill}
\begin{minipage}{65mm}
\begin{center}
\includegraphics*[scale=0.45]{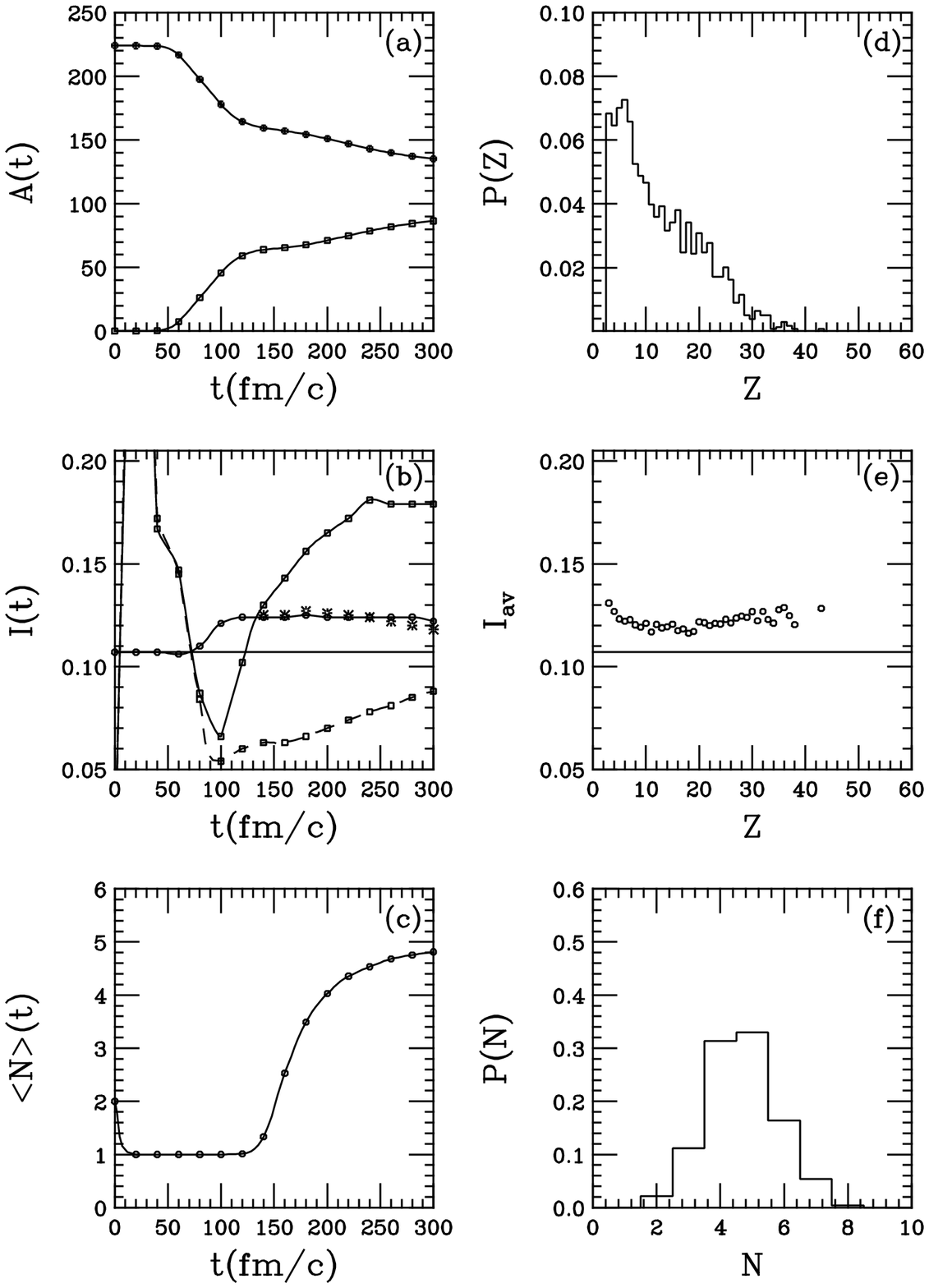}
\caption{$^{112}Sn+$$^{112}Sn$ $b=2fm$ collision:time
evolution (left) and freeze-out properties (right), ASY-STIFF EOS.}
\label{iso14}
\end{center}
\end{minipage}
\end{figure}

%\addtocontents{toc}{\hspace{0.55cm}\thesubsubsection \hspace{0.12cm}
%Isospin distillation}
We look first at the influence of the mass-to-charge
ratio of the colliding ions on fragment production comparing,
for a fixed asy-EOS (asy-stiff),
the neutron-rich and neutron-poor $Sn+Sn$ systems.
The results are plotted in the
Figures \ref{iso8} and \ref{iso14}.
For $^{124}Sn +$ $^{124}Sn$ we notice a neutron
dominated pre-equilibrium particle 
emission during the first 50fm/c. The liquid
phase becomes more symmetric during the compression
and expansion. At variance, the system $^{112}Sn+^{112}Sn$ is
on the p-rich side of the valley of stability
and along the fast nucleon emission the asymmetry of the liquid
increases.
Hence the pre-equilibrium stage appears very important
to assess the effective asymmetry of the system that will
eventually fragment. 

From the beginning of the fragment formation  phase of the evolution, 
 between $110$ and $280fm/c$,
the trends of the liquid and gas phase asymmetry
of the two systems  become similar, Fig. \ref{iso8}, \ref{iso14}(b).
In the ''central region'' the liquid asymmetry decreases while 
an {\it isospin burst} of the gas phase is observed. This behaviour
is consistent with the kinetic spinodal mechanism
in dilute asymmetric nuclear matter leading to the
{\it Isospin Distillation} between the liquid and the gas phase.

The effects of this process are clearly seen in the $IMF$
isospin content, in both
cases lower than at the beginning of the spinodal
decomposition, Fig. \ref{iso8},\ref{iso14}(e).  
Opposite
trends for fragments with charge above and below $Z\approx 15$
can be observed.
For heavier products the average asymmetry increases
with the charge, a Coulomb related effect.
However, the asymmetry rises again for lighter fragments.
This can be a result of the differences in density and
isospin between the regions in which the fragments grow, 
 due to the fact that not all of them form simultaneously,
as shown in the density contour plot. 

The charge distribution of primary fragments has a
rapidly decreasing trend, typical of a multifragmentation 
process, Fig. \ref{iso8}, \ref{iso14}(d). The neutron rich system favors
events with larger $IMF$'s multiplicities,
see Fig. \ref{iso8}, \ref{iso14}(c, f).
This  feature was evidenced also in experiments,
as discussed before.

\subsubsection{Isoscaling analysis}\label{isoscal}
%\addtocontents{toc}{\hspace{0.55cm}\thesubsubsection \hspace{0.12cm}
%Isoscaling analysis}
In recent years isospin effects on fragment production
have been discussed in terms of an isoscaling behaviour, revealed
first in multifragmentation data \cite{XuPRL85,TsangPRL86}. 
It was experimentally observed that
comparing two different reactions, 
one neutron rich (label $2$) and one neutron poor (label $1$), 
the ratio between the
yields of a given $N,Z$ isotope $R_{21}=Y_2(N,Z)/Y_1(N,Z)$
follows the relation:

\begin{equation}
\ln R_{21}~=~C + N \alpha + Z \beta
\end{equation}
with $\alpha$, $\beta$ isoscaling parameters and $C$ a constant.

Much work has been done on the interpretation of the isoscaling
parameters in the framework of statistical approaches and the 
relation to the value of the symmetry energy coefficient. 
For instance, 
in a grand-canonical statistical approach, these parameters are
related to the neutron (proton) chemical potential differences
in the nuclear environments, assumed to be at the same temperature, 
where the fragments are  
created in the two reactions. It can be also related 
to the ratio of the free neutron (proton) density in the two sources:
\begin{equation}
~~~~~\alpha~\equiv~\frac{\Delta \mu_n}{T}~~~~,
~~~~\beta~\equiv~\frac{\Delta \mu_p}{T}~~~~~,
\end{equation}
In turn, these are related to the symmetry energy properties:
\begin{eqnarray}\label{eqsym}
\Delta \mu_n~=~\rho \frac{\partial \epsilon_{sym}}{\partial\rho} 
(I_2^2 - I_1^2)
 + 2 \epsilon_{sym} [(I_2-I_1) - \frac{(I_2^2 - I_1^2)}{2}] \nonumber \\
\Delta \mu_p~=~\rho \frac{\partial \epsilon_{sym}}{\partial\rho} 
(I_2^2 - I_1^2)
 - 2 \epsilon_{sym} [(I_2-I_1) + \frac{(I_2^2 - I_1^2)}{2}] \nonumber 
\end{eqnarray}
where $\epsilon_{sym}$ is the symmetry energy per nucleon (see Sect.\ref{eos}).
We note that only in the limit
$~~\frac{\partial \epsilon_{sym}}{\partial\rho}~\simeq~0~~$
we get the ``liquid-drop'' results used in refs.\cite{TsangPRC64,OnoPRC68}:
\begin{eqnarray}\label{isodrop}
\Delta \mu_n~=~4 \epsilon_{sym}[(\frac{Z_1}{A_1})^2 - 
(\frac{Z_2}{A_2})^2]~~~and~~~ \Delta \mu_p~=~4 \epsilon_{sym}
[(\frac{N_1}{A_1})^2 - (\frac{N_2}{A_2})^2] \nonumber \\
~~~with~~~\vert {\Delta \mu_p} \vert > \vert {\Delta \mu_n} \vert~.
\end{eqnarray}
Since we have always the relation:
\begin{equation}
\ln \Big(\frac{N_2/Z_2}{N_1/Z_1}\Big)~\equiv~\alpha-\beta~=
~\frac{4}{T} \epsilon_{sym}(\rho) (I_2 - I_1)
\label{equil}
\end{equation}
a large interest is rising on the possibility of a direct measurement
of the symmetry energy in the fragment source from the isoscaling
$\alpha,\beta$ parameters \cite{TsangPRC64,OnoPRC68} .

\begin{figure}
\centering
\includegraphics*[scale=0.45]{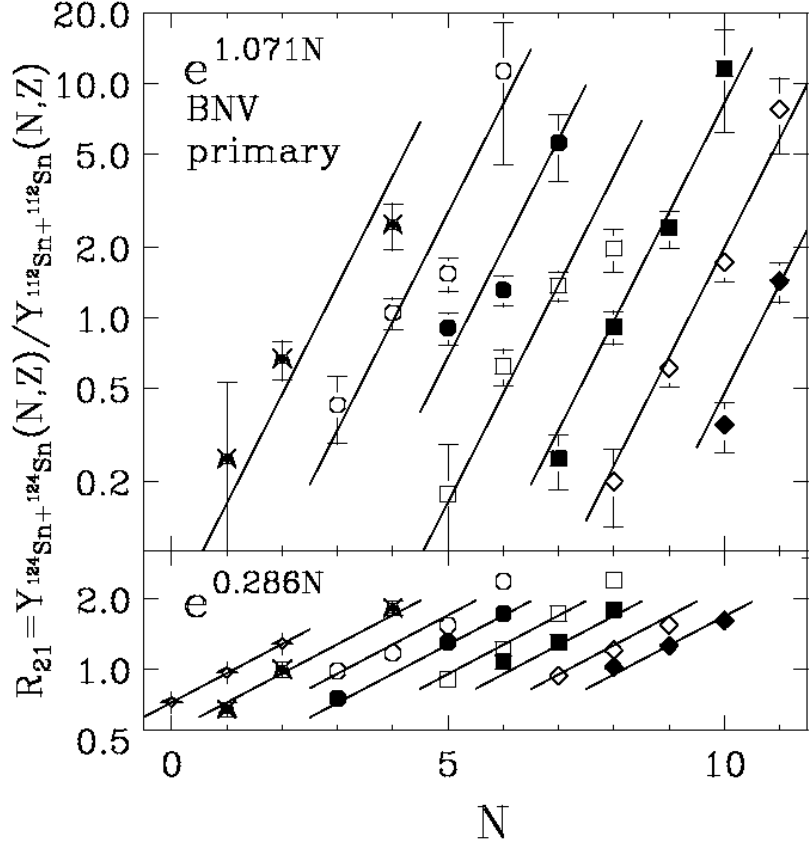}
\caption{Stochastic transport simulations of $Sn+Sn$ central collisions 
at $50AMeV$. Isoscaling behaviors: primary fragments (upper curves);
 after sequential decays (lower curves).}
\label{isodyn}
\end{figure}

However , it appears that an isoscaling behavior does not necessarily
require the existence of a fully equibrium, i.e. a grand canonical 
ensemble formula.
An isoscaling law is in fact observed also in dynamical simulations of
the stochastic mean-field type \cite{LiuPRC69}.
In Fig.\ref{isodyn} we report the isoscaling behavior of
primary fragments 
observed in stochastic transport simulations of central
$Sn+Sn$ collisions at $50~AMeV$. In the lower part
the isoscaling plots are shown after a secondary evaporation calculation.
It is seen that the sequential
neutron emission reduces the $\alpha$ slope considerably.

Actually one can easily show that isoscaling is expected also in the case
where the fragment isotopic distributions in the two reactions have a 
gaussian shape 
around a given average asymmetry. Then the isoscaling parameters are
determined by the difference of those averages divide by the widhts of
the distributions.
The isoscaling parameters
are expected to be larger in the dynamical relatively to the statistical 
approach, since the 
widths of the isotopic distributions are smaller than the ones obtained at
equilibrium.   

\subsubsection{Symmetry Energy effects}
\begin{figure}
\begin{minipage}{65mm}
\begin{center}           
\includegraphics*[scale=0.45]{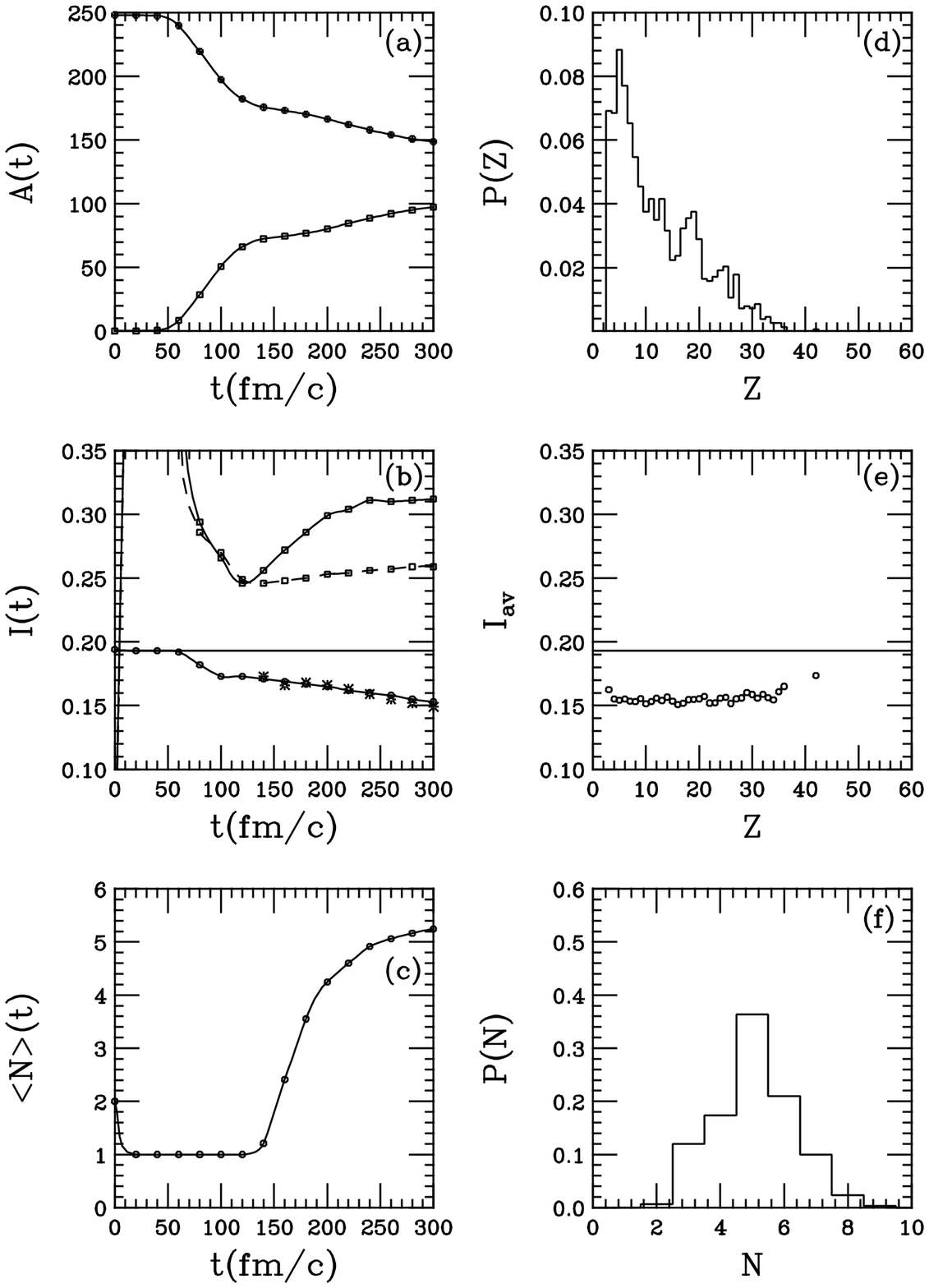}
\caption
{$^{124}Sn+$$^{124}Sn$ $b=2fm$ collision:time evolution
(left) and freeze-out properties (right),ASY-SOFT EOS.}
\label{iso10}
\end{center}
\end{minipage}
\hspace{\fill}
\begin{minipage}{65mm}
\begin{center}
\includegraphics*[scale=0.45]{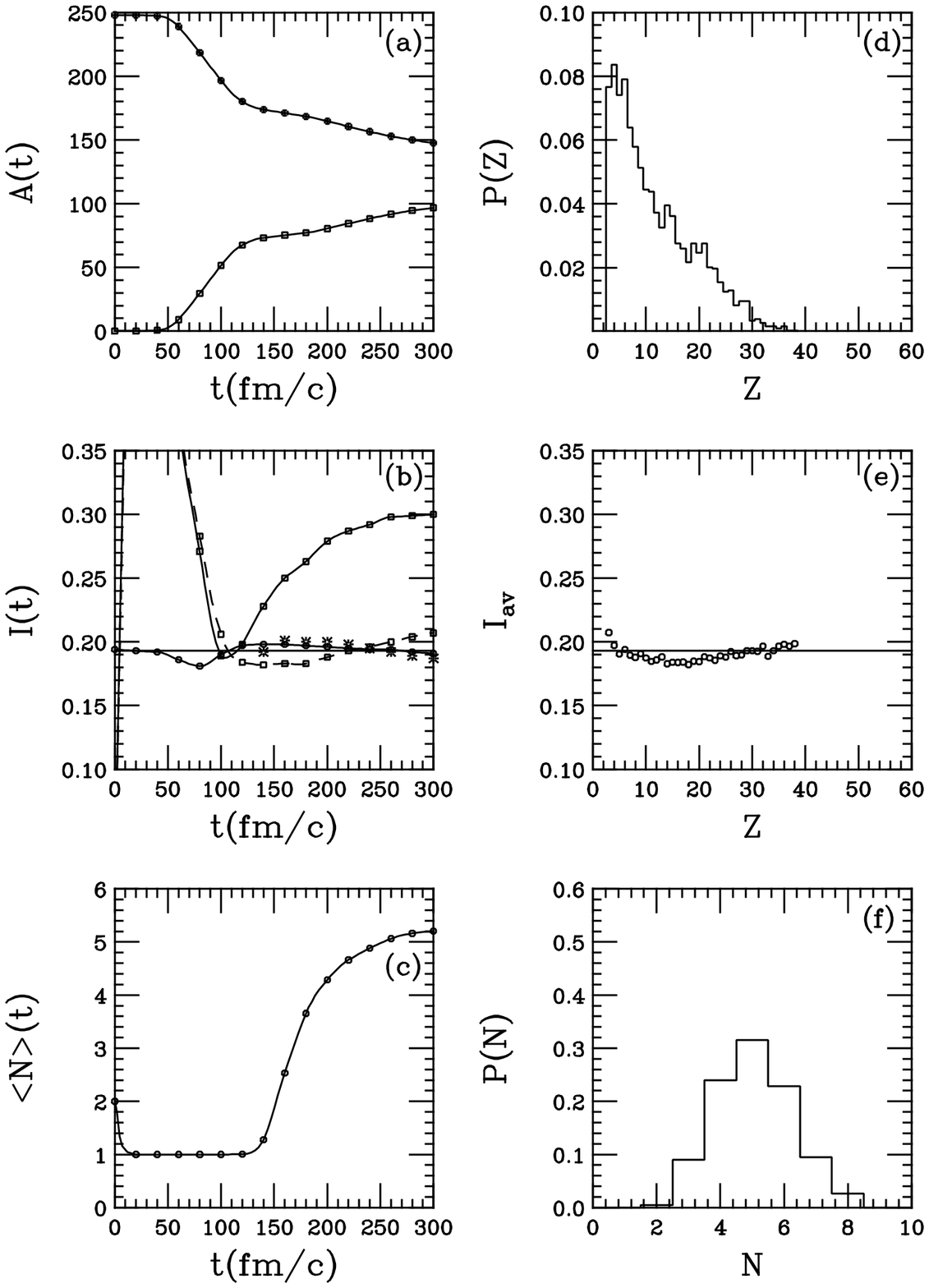}
\caption{$^{124}Sn+$$^{124}Sn$ $b=2fm$ collision:time
evolution (left) and freeze-out properties (right), ASY-SUPERSTIFF EOS.}
\label{iso12}
\end{center}
\end{minipage}
\end{figure}

%\addtocontents{toc}{\hspace{0.55cm}\thesubsubsection \hspace{0.12cm}
%Symmetry Energy effects}
Let us now investigate how different assumptions
on the density dependence of the symmetry energy may influence the
observables discussed above. In Figs. \ref{iso10}, \ref{iso12}
we report the results 
obtained respectively with the asy-soft and the
asy-superstiff symmetry term. 
We see that 
the average value of $N/Z$ of the fast particles
emitted during the expansion phase is affected by the
symmetry term.
In the asy-soft case, below $\rho_0$,
neutrons are less bound than in the asy-superstiff case 
(opposite for protons).
A more neutron-rich prompt particle emission 
with the soft asy-$EOS$ leads to
a lower asymmetry of the initial dilute matter 
undergoing spinodal decomposition. 
This is seen from Figs.\ref{iso10}(b) (asy-soft) and \ref{iso12}(b) 
(asy-superstiff).
As expected the results of Fig.\ref{iso8} (b) (asy-stiff) 
are somewhat in between.

The influence of the symmetry term is clearly seen in the isospin 
content of the primary 
intermediate mass fragments. The asy-soft choice is more effective for
the isospin distillation effect, producing the
most symmetric $IMF$'s, as the comparison of
Figs. \ref{iso10} ,\ref{iso12}(b),(e) shows.
This can be explained energetically observing that this
choice has a larger symmetry energy at very low densities. 
The differences between the results observed using
the two parametrizations can also be directly interpreted in terms of
the corresponding density behaviours of the $n,p$-chemical
potentials, see the comments on Fig.\ref{fig:chem} of Sect.\ref{eos}. 

\subsection{From semicentral to peripheral collisions: neck fragmentation}

Summarizing the main experimental observations, we would like to 
stress the following  
peculiarities of a ''dynamical'' $IMF$ production mechanism in semi-peripheral
collisions:  

1. An enhanced emission is localized in the mid-rapidity region,
intermediate between $PLF$ and $TLF$ sources,
especially for $IMF$'s with charge $Z$ from $3$ 
to $15$ units.
\vskip-0.3cm
2. The $IMF$'s relative velocity distributions with respect to $PLF$ 
(or $TLF$) cannot be explained in terms of a pure
Coulomb repulsion following a statistical decay.
A high degree of decoupling from the $PLF$ ($TLF$) is also invoked.
\vskip-0.3cm
3. Anisotropic $IMF$'s angular distributions are indicating preferential
emission directions and an alignment tendency.
\vskip-0.3cm
4. For charge asymmetric systems the light particles and $IMF$ emissions
 keep track of a neutron enrichment 
process that takes place in the neck region.

\begin{figure}
\centering
\includegraphics*[scale=0.45]{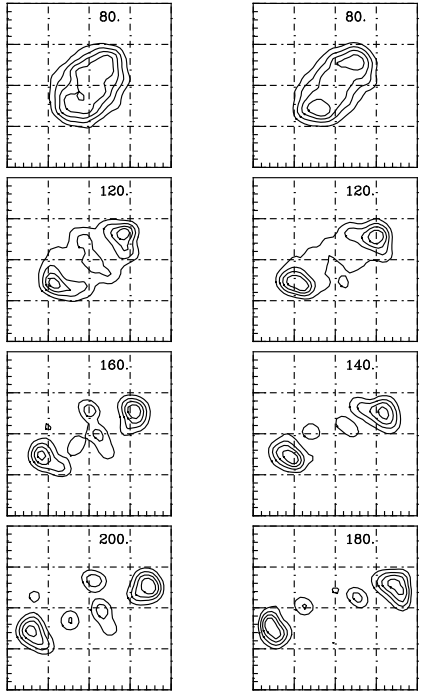}
\caption{$^{124}Sn+^{124}Sn$ collision at $50AMeV$: time evolution of the
nucleon density projected on the reaction plane.
Left column: $b=4fm$. Right column: $b=6fm$.}
\label{iso6}
\end{figure}

A fully consistent physical picture of the processes 
that can reproduce observed characteristics is still
a matter of debate and several physical phenomena 
can be envisaged, ranging from the formation of a transient
neck-like structure that would break-up due to Rayleigh instabilities
or through a fission-like process, to the statistical decay of 
a hot source, triggered by the proximity with PLF 
and TLF \cite{BrosaPR197,LukasikPLB566,BotvinaPRC59}.
Dynamical transport models suggest since long time
the possibility of observing neck emission 
\cite{BertschPLB141,ColonnaNPA541,ColonnaNPA589,SobotkaPRC50},  
\cite{DempseyPRC54,LukasikPRC55,SobotkaPRC55}.
We show in Fig. \ref{iso6} the density contour plots
of a neck fragmetation event at $b=6fm$, for the
reaction $^{124}Sn+$$^{124}Sn$ at $50AMeV$, obtained
from the numerical calculations based on the stochastic transport
approach described before, \cite{BaranNPA730}.

\subsubsection{Neck dynamics and IMF properties: the iso-migration}
%\addtocontents{toc}{\hspace{0.55cm}\thesubsubsection \hspace{0.12cm}
%Neck dynamics and IMF properties: the iso-migration}
In the following we will review results regarding the
fragmentation mechanism in semiperipheral collisions,
as well as the related isospin dynamics
and effects of the symmetry term density dependence.
First we examine again the $Sn+Sn$ collisions
at $50AMeV$, but at impact parameter $b=6fm$. We discuss the
same observables as in central collisions.

The development of a neck stucture in the overlap
region of the two colliding nuclei is evidenced
in Fig. \ref{iso6}. During the interaction time this zone
heats and expands but remains in contact with the denser and colder
regions of $PLF$ and/or $TLF$. The surface/volume
instabilities of a cylindrically shaped neck
region and the fast leading motion of the $PLF$
and $TLF$ will play an important role in the fragmentation dynamics.
At the freeze-out time, with the neck rupture at about
$140 fm/c$, intermediate mass fragments are produced
in the mid rapidity zone. In some events fragments form
very early while, in others, they can remain for a longer
time attached to the leading $PLF$'s or $TLF$'s.
In the same figure we note an intermediate
behavior between multifragmentation
and neck fragmentation at $b=4fm$.  
 
\begin{figure}
\begin{minipage}{65mm}
\begin{center}             
\includegraphics*[scale=0.45]{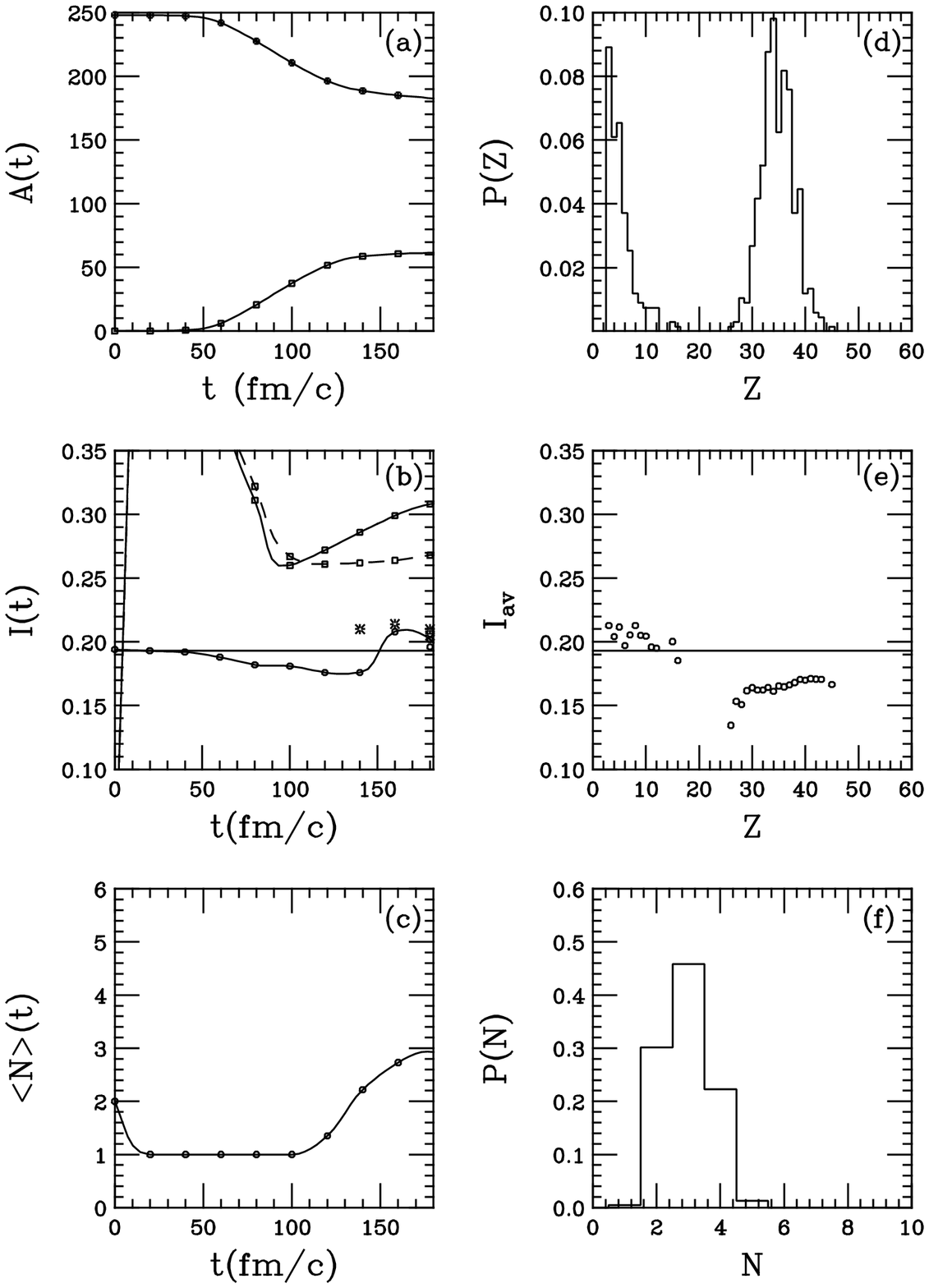}
\caption
{$^{124}Sn+$$^{124}Sn$ $b=6fm$ collision:time evolution
(left) and freeze-out properties (right), ASY-STIFF EOS.}
\label{iso9}
\end{center}
\end{minipage}
\hspace{\fill}
\begin{minipage}{65mm}
\begin{center}
\includegraphics*[scale=0.45]{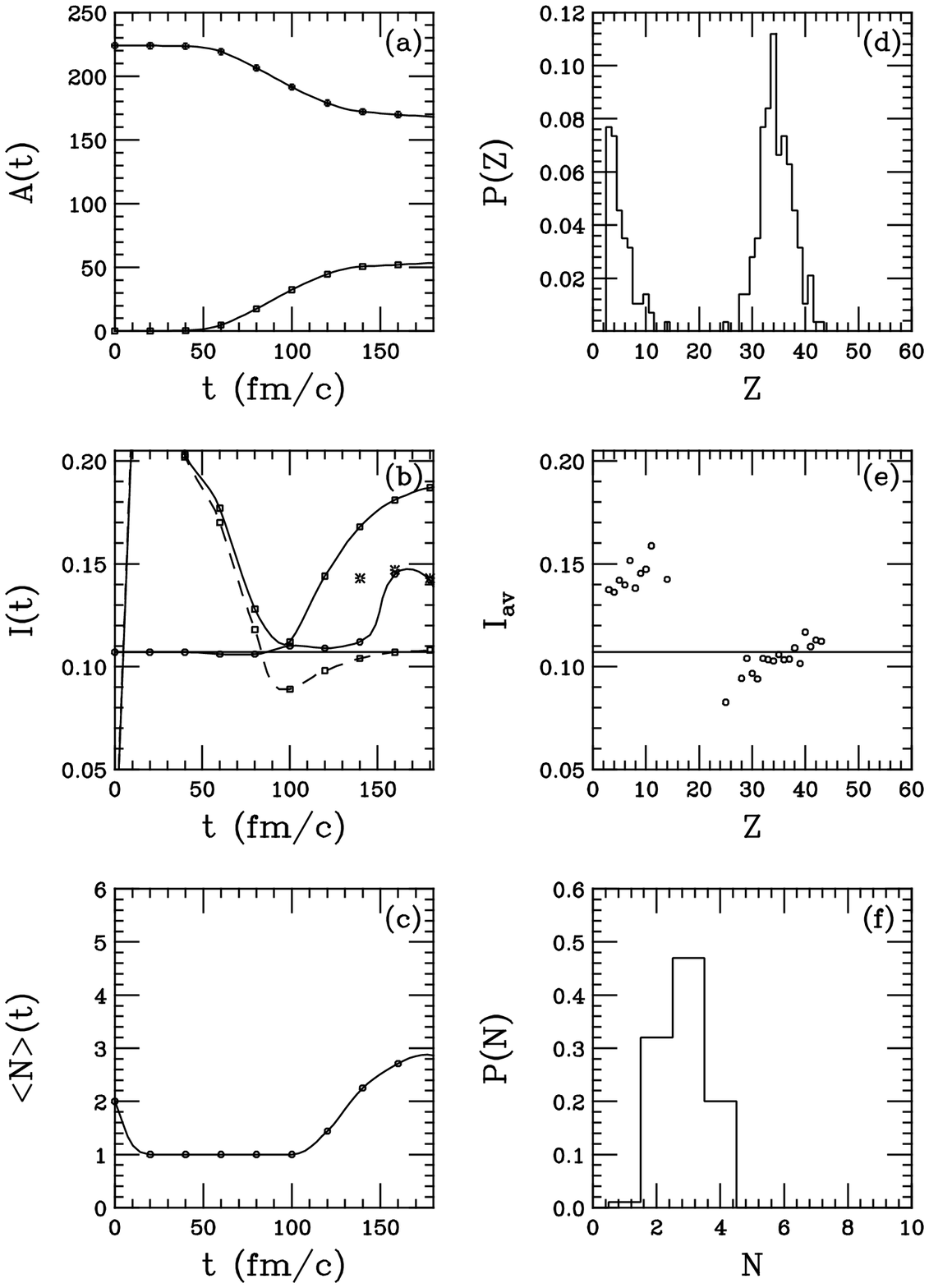}
\caption{$^{112}Sn+$$^{112}Sn$ $b=6fm$ collision:time
evolution (left) and freeze-out properties (right), ASY-STIFF EOS.}
\label{iso15}
\end{center}
\end{minipage}
%\vskip-0.6cm
\end{figure}

In the Figs.\ref{iso9}, \ref{iso15} we plot the same quantities as before
 (Sect.5.5.2) for n-rich and n-poor $Sn+Sn$ collisions for impact
parameter $b=6fm$ (asy-stiff $EOS$).
In the charge distribution probability, Figs. (\ref{iso9}, \ref{iso15})(d),
 one can clearly separate the $PLF/TLF$ residues and the $IMF$'s
with upper limit charge $Z\approx 15$.
We call the $IMF$'s produced by such a mechanism Neck Intermediate 
Mass Fragments ($NIMF$'s).
For neutron-rich systems events with two or even three $NIMF$'s
appear, are slightly more likely, Figs. (\ref{iso9}, \ref{iso15})(f). 
The isospin content reveals new distinctive features as seen
in Figs. (\ref{iso9}, \ref{iso15})(e).
The $IMF$'s formed in the neck region are much more neutron rich
than the corresponding fragments produced in semicentral collisions.
Moreover the $PLF$ and $TLF$  residues have definitely a
lower asymmetry than $NIMF$'s.
To interpret this behaviour we have to keep in mind
that the neck region is always in contact with a normal-density phase 
(the spectators)
during the fragment formation and this determines a different isospin dynamics.
Moreover the clusters are formed in a nuclear matter not very dilute
relative to saturation density. 
We also notice that 
the neck breaking leads, in some events, to
fragments closer to one of the spectators with increased interaction
between the two. Thus there should be a smooth transition
to $PLF/TLF$ fast-fission type of events.

In neck fragmentation we are testing the symmetry energy
in different regions of nucleon density. 
It is energetically more favorable to have 
protons (neutrons) migration from (to)
the neck region to (from) the more dense spectators,
leaving the nuclear matter in the neck neutron-rich at the time of breaking.
Clearly the isospin dynamics is ruled by the same energetic arguments as in the
case of isospin distillation in central collisions.
The main difference relative to bulk
fragmentation is related to the density gradient between 
the lower density neck and the spectators, which can trigger 
proton and neutron flows in opposite directions (isovector mode).
Therefore, we propose to call this phenomenon isospin
migration in contradistinction to the distillation phenomenon
driven by isoscalar-like unstable fluctuations.

\subsubsection{Neck dynamics at lower energy}
\begin{figure}
\centering
\includegraphics*[scale=0.50]{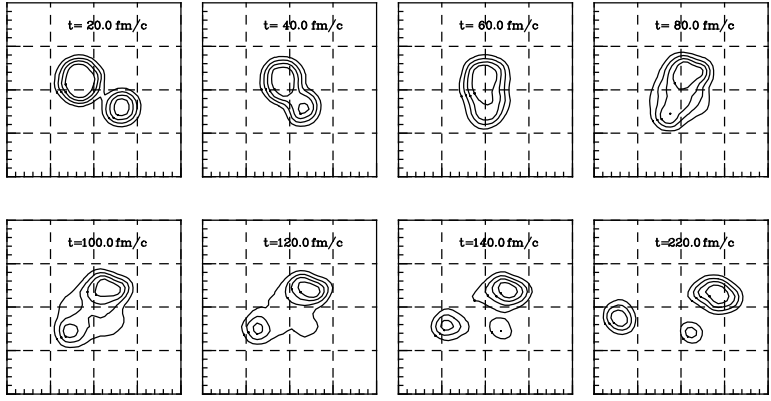}
\caption{$^{124}Sn$+$^{64}Ni$ at $35 AMeV$. Typical evolution of the 
density contour
plot for a neck fragmentation event at $b=6fm$.}
\label{neck2}
\end{figure}

%\addtocontents{toc}{\hspace{0.55cm}\thesubsubsection \hspace{0.12cm}
%Neck dynamics at lower energy}
It is also interesting to study the features of the neck dynamics 
at lower energy. Here we consider the 
$^{124}Sn +$ $^{64}Ni$ semiperipheral collisions at
incident energy around $35 AMeV$, where nice data are appearing from
the $CHIMERA$ collaboration at the $LNS$ \cite{PaganoNPA681,PaganoNPA734}.
The reaction evolution at impact parameter
$b=6fm$ is shown in Fig.\ref{neck2}.
Calculations are performed using the asy-stiff parameterization.
The dynamics corresponds to a mainly
two-center system. We notice the superimposed 
motion of the $PL$ and $TL$ pre-fragments linked to
the formation of a neck-like structure
with a fast changing geometry between $40fm/c$ and $140-160fm/c$.

\begin{figure}
\centering
\includegraphics*[scale=0.50]{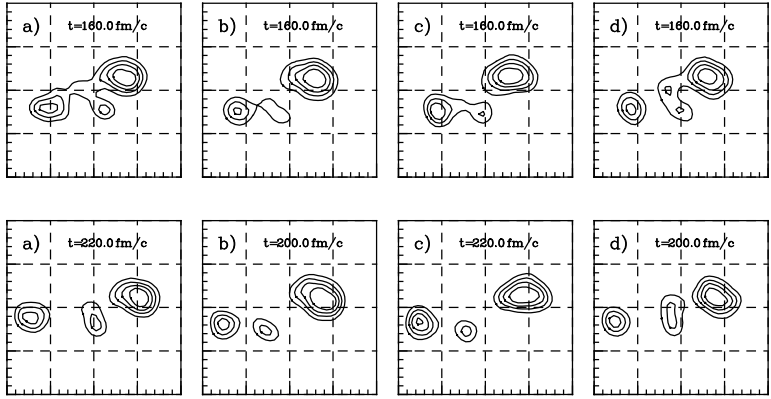}
\caption{Density contour plots as in Fig.\ref{neck2}. Early stage of fragment 
formation (top)
and same events close to the freeze-out (bottom) for four ternary cases a), b), c), d)
in neck fragmentation.}
\label{neck3}
\end{figure}

Neck instabilities favour the appearance of $NIMF$'s,
after $150fm/c$, in a variety of places and ways as can be seen
by looking at Fig.\ref{neck3}. For four events, we illustrate 
two characteristic stages, the early phase of the fragment formation process
and the configuration close to freeze-out.

\begin{figure}
\centering
\includegraphics*[scale=0.50]{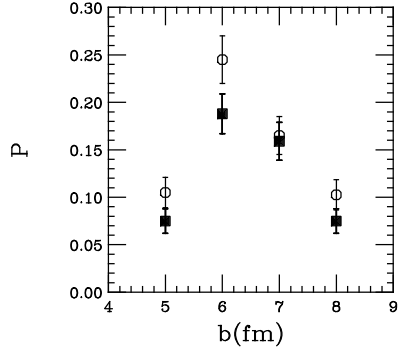}
\caption{ $Sn+Ni$ reactions: Impact parameter dependence of
the neck fragmentation probability (ternary events).
 White circles: neutron rich reaction.
Black circles: neutron poor reaction. Asystiff $EOS$.}
\label{neck4}
\end{figure}

Apart from the events corresponding to the fast $NIMF$ production 
we have a large fraction of binary events, with a
presence in the exit channel of only
primary excited $PLF$'s and $TLF$'s. Induced deformations
suggest that they can split, also asymmetrically, on 
time scales long with respect to neck fragmentation, but
much shorter that in an equilibrium fission process.
In  Fig. \ref{neck4} we plot
the dependence on the impact parameter of the
ternary event probability.
The maximum value, about  $25 \%$, is attained 
around mid-centrality, between $b=6-7fm$,
decreasing on both sides 
to about $10\%$ for $b=5fm$ and $b=8fm$. 
It is even smaller at $b=4fm$, in spite of
a stronger dissipation. The longer interaction time 
favors a reabsortion of the neck matter
at this energy. This is
a different behaviour in comparison to 
the dynamics at $50AMeV$, where, as we have already emphasized,
the higher available energy makes possible the 
transition to multifragmentation
with increasing centrality.
 At greater impact parameters, $b=9fm$,
a smaller overlap and a faster separation are also
suppressing this mechanism.

\begin{figure}
\centering
\includegraphics*[scale=0.50]{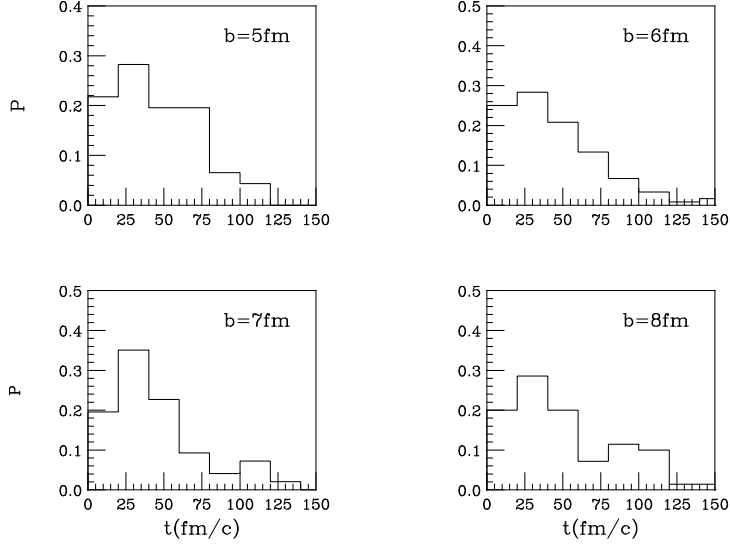}
\caption{The probability distribution of scission-to-scission time
in the neck fragmentation for impact parameters from $5$ to $8 fm$.
Asystiff $EOS$.} 
\label{probsciss}
\end{figure}

From the simulations we can extract an interesting information on
the time scale of the Neck-$IMF$ production.
In Fig.\ref{probsciss} we show, for different impact parameters,
the probability distribution of the
time interval between the instant of the first separation of
the dinuclear system and the moment when a Neck-$IMF$ is
identified (scission-to-scission time). A large part of the
$NIMF$s are formed in short time intervals, within $50fm/c$. 

Finally we would like to remark that 
the neck fragmentation shows a dependence on 
the nucleon-nucleon cross sections and the $EOS$-compressibility.
The latter point is particularly interesting since it  seems to indicate 
the relevance
of volume instabilities even for the dynamics of neck. This appears consistent
with the short time scales shown before, see also the
discussion in ref. \cite{BaranNPA730}.

\subsubsection{Analysis of kinematical observables}
%\addtocontents{toc}{\hspace{0.55cm}\thesubsubsection \hspace{0.12cm}
%Analysis of kinematical observables}
The nonstatistical features of the $NIMF$ production
are revealed in various kinematic correlations.
The corresponding observables can be measured in 
exclusive experiments.

An interesting kinematical observable is the asymptotic relative 
velocitity of 
the neck-produced $IMFs$ with respect to the  $PLF$ ($TLF$),
$v_{rel}(PLF,TLF) \equiv {\vert {\bf v_{PLF,TLF}} - {\bf v_{IMF}} \vert}$.
 This is compared  
with the relative velocity reached in a pure Coulomb-driven separation, 
signature of a statistical fission process of a compound $PLF^*$ or $TLF^*$
system, as provided by the Viola systematics \cite{ViolaPRC31,HindeNPA472}:
\begin{equation}
 v_{viola}(1,2)= \sqrt{\frac{2}{M_{red}}(0.755 \frac{Z_{1}Z_{2}}
{A_{1}^{1/3}+A_{2}^{1/3}}+7.3)}
\end{equation}
where $A_{1},A_{2},Z_{1},Z_{2}$ are the mass and charge numbers of 
the fission products and $M_{red}$ is the corresponding reduced mass.

\begin{figure}
\centering
\includegraphics*[scale=0.45]{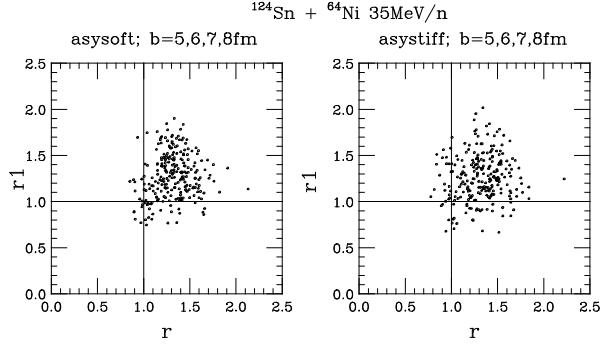}
\caption{Wilczynski-2 Plot: correlation between deviations from Viola 
systematics, see text. Results are shown for two $asy-EOS$.}
\label{wilcz2}
\end{figure}

For each Neck-$IMF$ we can evaluate the ratios
$r=v_{rel}(PLF)/v_{viola}(PLF)$, ($r1=v_{rel}(TLF)/v_{viola}(TLF)$).
In  Figure \ref{wilcz2}
we plot $r1$ against $r$ for each $NIMF$. We call such a representation
a $Wilczynski-2$ plot \cite{wilcz}. 
The solid lines represent the loci of the $PL$-($r=1$) and $TL$-($r1=1$)
 fission events respectively.
The values $(r,r1)$  appear simultaneously 
larger than $1$ suggesting a weak $NIMF$ correlation with {\it both} 
$PLF$ and $TLF$, in contrast to a statistical fission mechanism.
The process has some similarities
with the participant-spectator scenario. However the dynamics appear much 
richer than in the simple sudden abrasion model, where the locus of
the $r-r1$ correlation should be on the bisectrix, apart the $Goldhaber$
widths, see ref.\cite{LukasikPLB566}. Here the wide distributions
of Fig.\ref{wilcz2} reveal 
a broad range of fragment velocities, typical of the
instability evolution in the neck region
that will lead to large dynamical fluctuations on $NIMF$ properties. 

Another very selective correlation is provided by the anisotropy
in the fragment emission, measured by the in-plane azimuthal angle, 
$\Phi_{plane}$, defined as the angle between
the projection of the $PL-IMF$ scission axis onto the reaction plane and 
the separation axis between $PLF$ and $TLF$, \cite{StefaniniZFA351}. 
A $\vert \Phi_{plane} \vert \simeq 0$ collects events
corresponding to asymmetric ``fissions'' of the $PL$-system very aligned
along the outgoing $PL-TL$ separation axis. At variance a statistical 
fission dominance 
would correspond to a flat $\Phi_{plane}$ behavior.
The $\Phi_{plane}$ distributions, corresponding to the same $NIMF$
events analysed before, are covering a quite
limited angular window, close to the full alignment configuration,
$\Phi_{plane}=0^{\circ}$. The distribution becomes wider when
we approach the two $r,r1~=~1$ lines of the Fig.\ref{wilcz2}.

We can  transform in $time-scales$ the correlations
discussed before.
In fact an induced asymmetric, fast fission
can also manifest some deviations from Viola systematics.
From our simulations, this mechanism takes place 
on longer time scales
compared to neck fragmentation. If a light fragment escapes
later from its $PL/TL$ partner, friction certainly will attenuate 
the dynamical effects and relative velocities will
deviate less from Coulomb values.
In the plane $r-r1$ such events are located closer to the line
$r=1$ or $r1=1$. A prolonged contact with the $PLF$ or $TLF$  will 
induce also different angular distributions in 
comparison to the neck fragments (less alignement, wider $\Phi_{plane}$ 
distribution).
As a limiting case we recover the statistical fission mechanism for a
$IMF$ production from an equilibrated $PL/TL$ excited residue,
with no Viola-deviations and no alignement.
We conclude that an analysis based on the Wilczynski-2 plot, 
Fig.\ref{wilcz2}, of the $IMF$ produced in semicentral collisions
will be able to select and study fragments formed on a large variety of 
interaction time-scales, of particular interest for the mass and
isospin dynamics.

\subsubsection{Symmetry term effects}
%\addtocontents{toc}{\hspace{0.55cm}\thesubsubsection \hspace{0.12cm}
%Symmetry term effects}
From the simulations of ref.\cite{BaranNPA730} for
the $^{142}Sn+^{64}Ni$ system it appears that
the symmetry term
of the $EOS$ does not seem to influence sensitively the
main features of the neck mechanism, i.e.  
probability of ternary events, deviations from Viola 
systematics and $NIMF$ angular distributions.
However, one has to consider that with the studied system we cannot reach
high charge asymmetries, the asymmetry parameter $I \equiv (N-Z)/A$
ranging from $0.193$ for the projectile to $0.125$ for the target, with
an average $I=0.17$. 

\begin{figure}
\centering
\includegraphics*[scale=0.55]{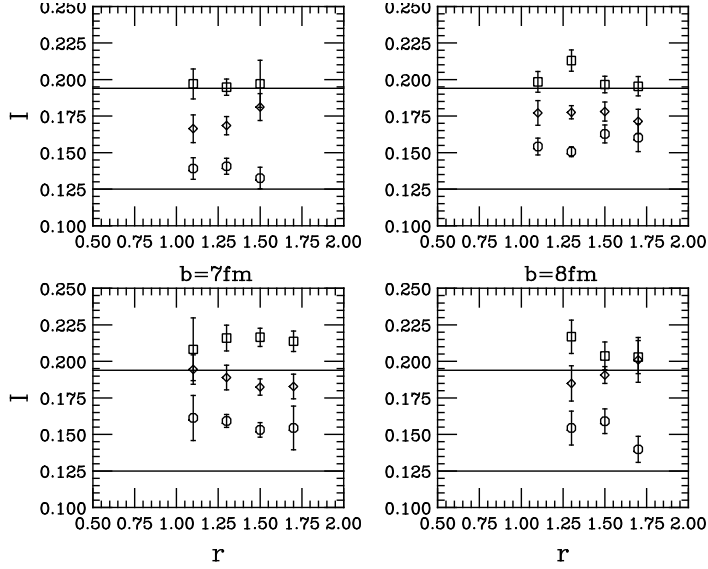}
\caption{$NIMF$ isospin content for $asysoft$ (circles), $asystiff$ (rombs) 
and $superasystiff$ (squares)
$EOS$ as a function of $r$-deviation from Viola systematics, 
at impact parameters from $b=5fm$ to $8fm$. The two solid lines
represent the mean asymmetries of the projectile (top) and target (bottom). }
\label{papisorr}
\end{figure}

A more promising observable seems to be the isotopic content
of the Neck-$IMF$.
For the three $asy-EOSs$ introduced in Sect.\ref{eos} we plot in 
Fig.\ref{papisorr} the average isotopic composition $I$ of
the $NIMF$'s as a function of the $PLF$ $r$-deviation from Viola
systematics. At all impact parameters clear differences 
are evident. The average asymmetry does not depend
strongly on $r$. We notice however that it increases 
with the stiffness of the symmetry potential around and below saturation. 
The {\it superasystiff}
parametrization, i.e. with an almost parabolic increasing behavior
around $\rho_0$, produces systematically more neutron rich
$NIMF$s.
This effect is clearly due to a different neutron/proton migration
at the interface between $PL/TL$ ''spectator'' zone of normal
density and the dilute neck region where the $NIMF$s are 
formed, as it was already observed for the reactions at $50~AMeV$.

From the Fig.\ref{papisorr} we see that the isospin content of $NIMF$s
carries important information on the isovector part of the
effective interaction and the related isospin dynamics in the early stages of
the reaction. We expect a different isospin pattern for
the $IMF$'s produced at later times from induced and/or statistical 
fission of the
more charge symmetric $PLF^*$'s and $TLF^*$'s. The Wilczynski-2 plot will
help to make the selections.  
We have to remind that all the results presented here refer to properties of
primary (excited) fragments. The neutron excess signal appears likely to be
washed out from later evaporation decays. A reconstruction of the
primary fragments with neutron coincidence measurements would be very
important.

\subsubsection{Isoscaling analysis}
%\addtocontents{toc}{\hspace{0.55cm}\thesubsubsection \hspace{0.12cm}
%Isoscaling analysis}
Relative to the earlier discussion
it is of interest to test whether the isoscaling behavior can manifest
itself even in the neck-fragmentation, which is a clear dynamical process,
 related to short characteristic time scales.  

\begin{figure}
\begin{minipage}{65mm}
\begin{center}           
%\epsfysize=6.0cm
%\centerline{\epsfbox{snp1.ps}}  
\includegraphics*[scale=0.45]{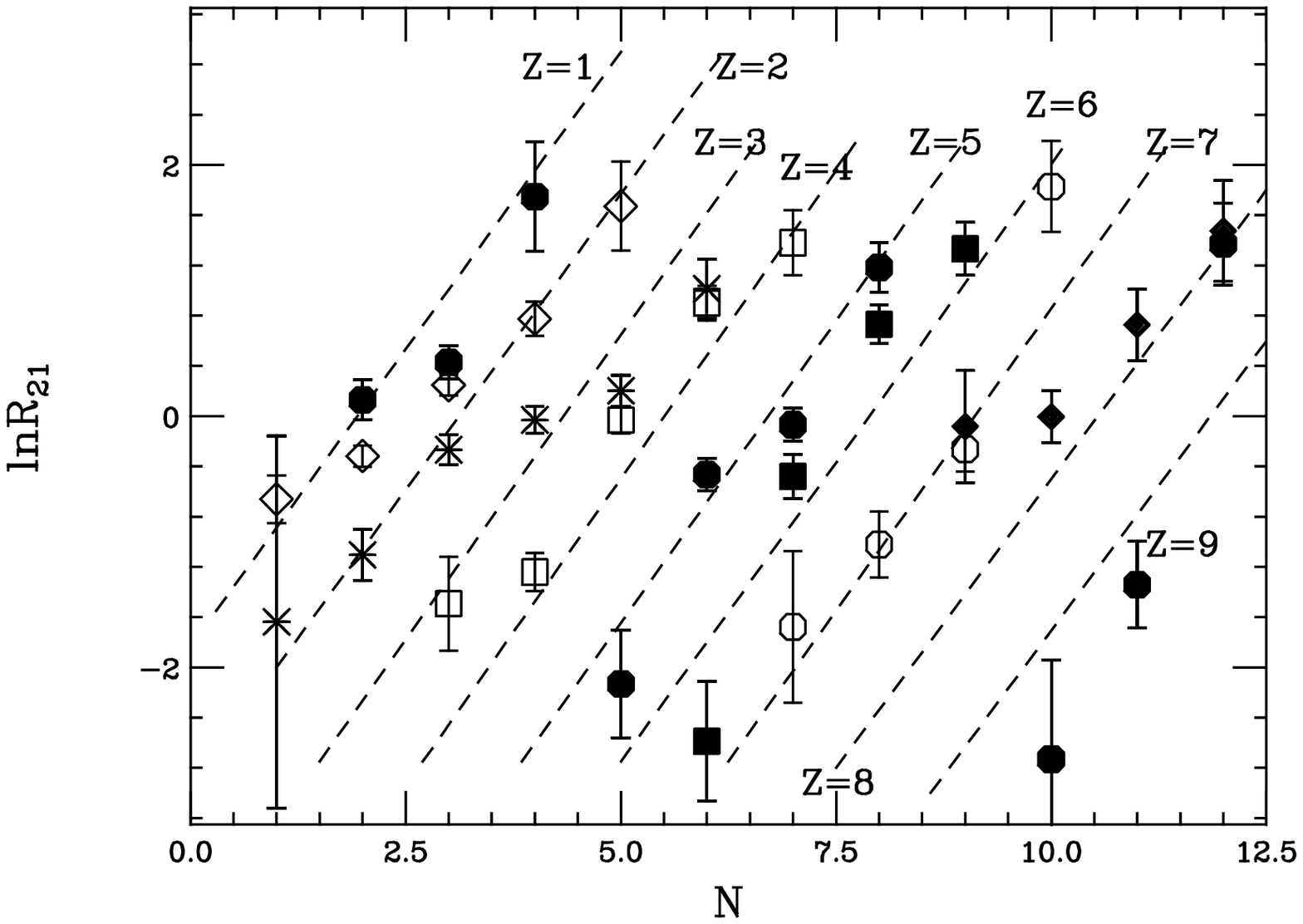}
%\vskip-1.2cm
\caption
{Isoscaling in neck fragmentation: $lnR_{21}$ dependence on N.}
\label{fig:isoalpha}
\end{center}
\end{minipage}
\hspace{\fill}
\begin{minipage}{65mm}
\begin{center}
%\epsfysize=5.cm
%\centerline{\epsfbox{snp2.ps}}
\includegraphics*[scale=0.45]{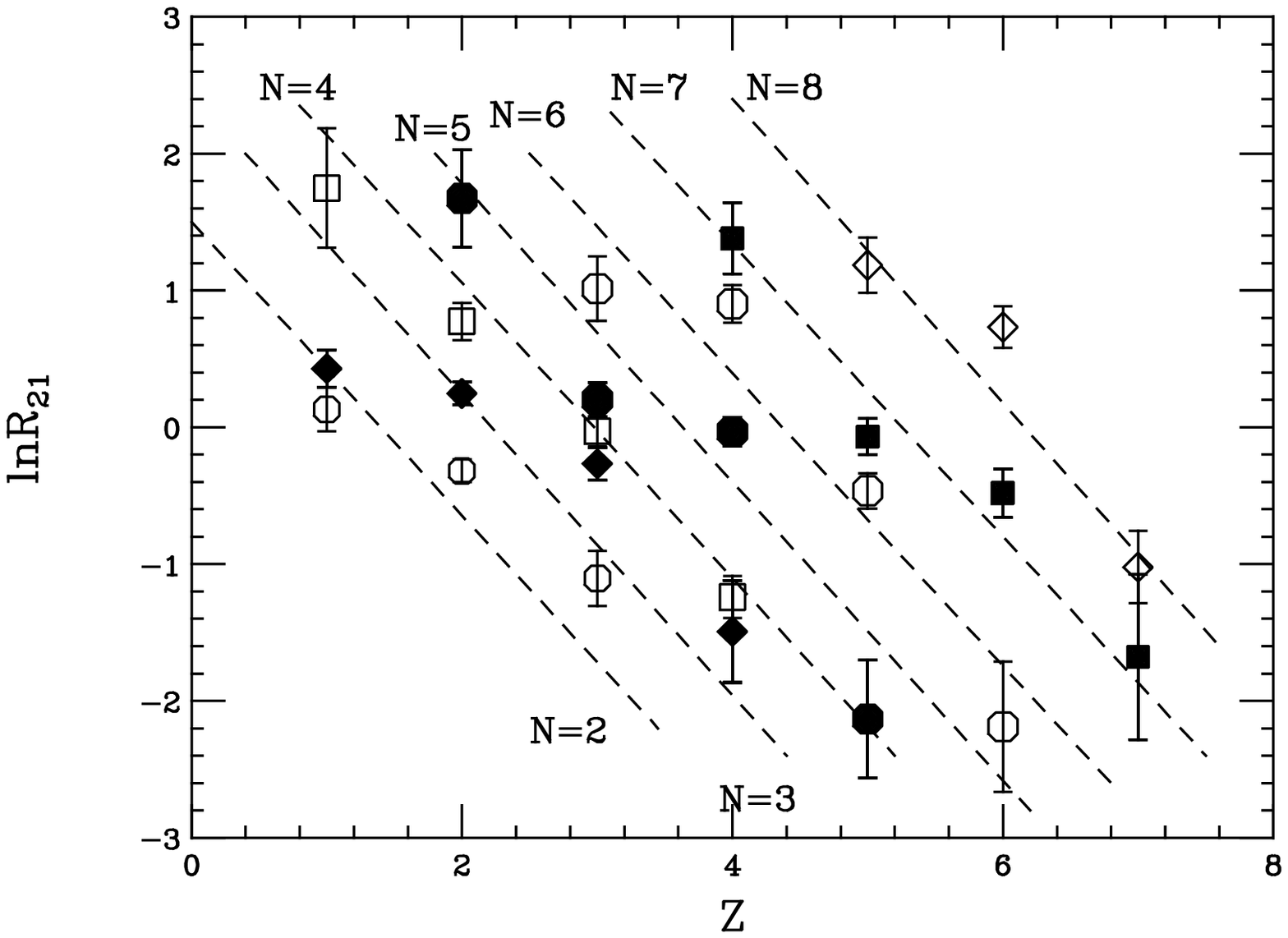}
%\vskip-1.2cm
\caption{Isoscaling in neck fragmentation: $lnR_{21}$ dependence on Z.}
\label{fig:isobeta}
\end{center}
\end{minipage}
%\vskip-0.6cm
\end{figure}

In Figs. \ref{fig:isoalpha}, (\ref{fig:isobeta})
is  shown the $N$ ($Z$) dependence of $lnR_{21}$,
for $Z=1$ to $Z=9$ light fragments
produced in the neck region, as obtained in the simulations
 with an $asystiff-EOS$ parametrization.
 $R_{21}$ is the yield ratio of the n-rich
$^{124}Sn+^{64}Ni$ vs. the n-poor $^{112}Sn+^{58}Ni$ system, 
see before Sect.\ref{isoscal}. 

We do  see the isoscaling signal, an 
exponential $N$- and $Z$-
dependence of the yield ratios with quite well defined $\alpha,\beta$
slopes.
Although we cannot use explicit equilibrium relations, like Eq.\ref{equil},
we still expect a symmetry energy dependence of the isoscaling
parameters. Since the fragment formation takes place in the neck region,
we  predict that its isospin content will dictate the values
of the isoscaling parameters.
Indeed, we may assume that for a
neutron poor system, closer to the symmetric case, the differences
between the various $asy-EOS$ on the isotopic and isotonic distributions
are reduced, and in a first approximation, identical. 
At variance, for the neutron rich
system, we have seen that passing from $asysoft$- to $superasystiff$-$EOS$
more neutron-rich
Neck-$IMF$s are formed. Therefore, the corresponding distributions have 
to be steeper. 
As reported in Table \ref{isotable}, a nice
increase (in the modulus) of the isoscaling parameters with the
increasing stiffness of the symmetry energy is observed.
\begin{table}[t]
\begin{center}
\begin{tabular}{|l|c|c|c|} \hline 
       & $asysoft$     & $asystiff$  & $superasystiff$   \\ \hline\hline
   $\alpha$ &   0.69   &     0.95     &   1.05       \\ \hline
   $\beta$  &  -0.67   &    -1.07     &  -1.18        \\ \hline
\end{tabular}
\end{center}
\vskip 0.2cm
\caption{\label{isotable} 
The isoscaling parameters $\alpha$ and $\beta$ in neck fragmentation
for three $asy-EOS$.}
\end{table}

\subsection{Charge Equilibration in Peripheral Collisions}

\subsubsection{Isospin diffusion in presence of inhomogenous density 
distributions}
%\addtocontents{toc}{\hspace{0.55cm}\thesubsubsection \hspace{0.12cm}
%Isospin diffusion in presence of inhomogenous density 
%distributions}
We turn now to discuss the isospin equilibration mechanism in even more 
peripheral collisions where only two primary fragments ($PLF-TLF$)
are observed in the exit channel. 
When  the ions in the entrance channel have
a different $N/Z$ ratio, we expect that isospin diffusion
will lead the system towards a more  uniform asymmetry
distribution. The degree of equilibration, correlated to the
interaction time, may provide insights on transport
properties of fermionic systems
\cite{UehlingPR43,HellundPR56}, in particular
on the diffusion coefficient of asymmetric nuclear matter,
\cite{AndersonPRB35,ShiPRC68}. 

Here we are focus on a charge asymmetric collision $^{124}Sn+^{112}Sn$,
at $50 AMeV$ bombarding energy, to which we refer as
the mixed system, $(M)$, where some data also exist, \cite{TsangPRL92}.
The simulations are performed for peripheral collisions
at impact parameters $b = 8,9,10 fm$. Binary events are selected.
We define the average interaction time, $t_c$, as the time elapsed
between the initial touching and the
moment when  the $PLF$ and  $TLF$
reseparate. From our simulations we obtain
$t_c \approx 120, 100, 80 fm/c$ respectively for the three 
impact parameters.
\begin{figure}
\centering
\includegraphics*[scale=0.45]{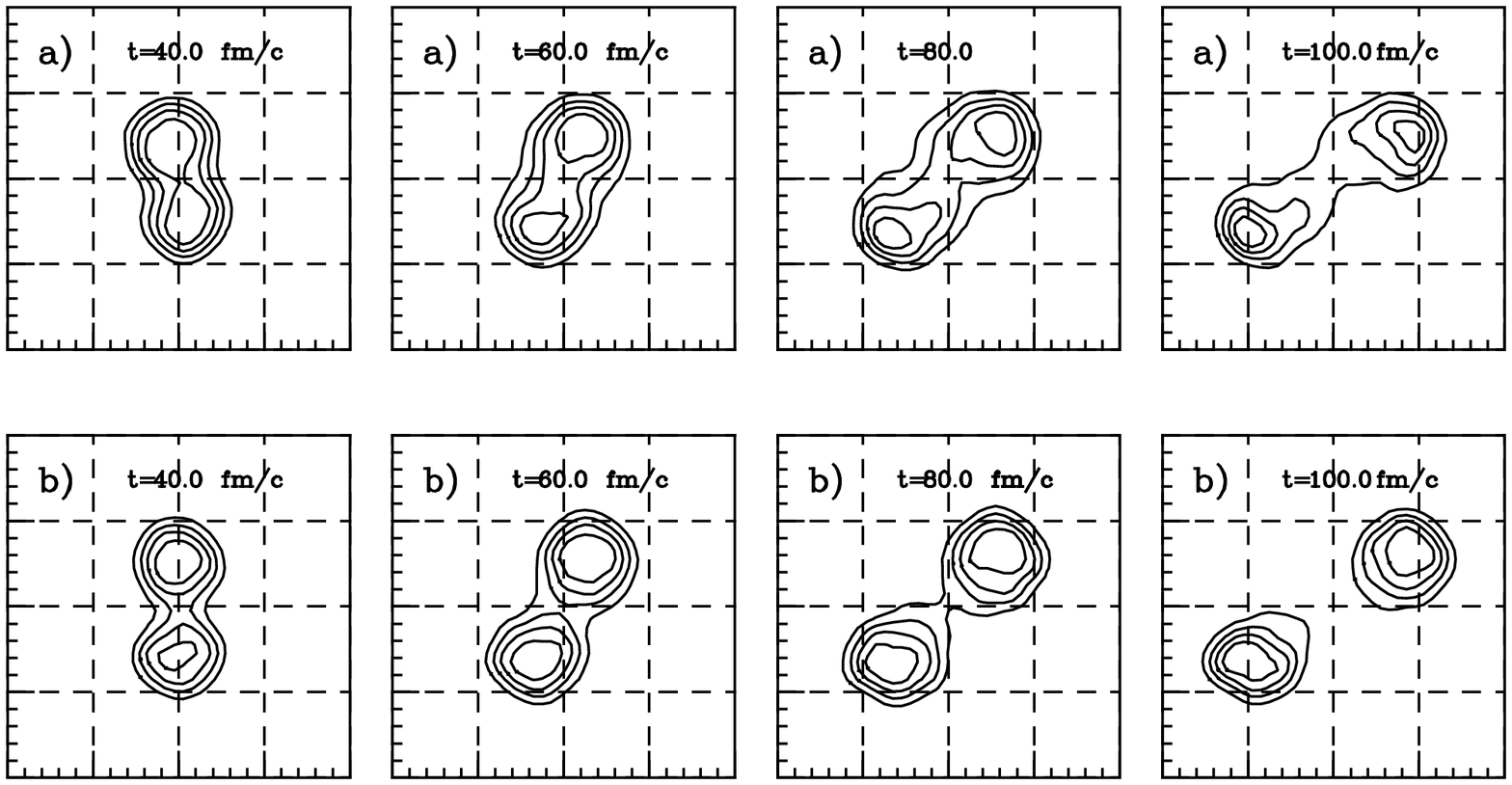}
\caption{$^{124}Sn+^{112}Sn$ collision at $50~AMeV$, $b=8fm$ (up) and 
$b=10fm$ (down): density contour plots.}
\label{difusden810}
\end{figure}
Typical density contour plots, at $b=8 fm$ and $b=10 fm$, 
are shown in Fig. \ref{difusden810}. In
the overlap region, after around $40 fm/c$, 
the formation of a lower density interface is evidenced.
The isospin migration takes place during the transient
contact between the two spectator regions with density
close to the normal one, separated  by a dilute
neck region. Thus we have concentration {\and} density gradients
ruling the isospin diffusion. In contrast, in deep-inelastic collisions
at lower energies, the isospin equilibration is driven
by the $N/Z$ difference between two interacting nuclei 
with an uniform density profile until the separation,
i.e. only the concentration gradient is active.

\begin{figure}
\centering
\includegraphics*[scale=0.50]{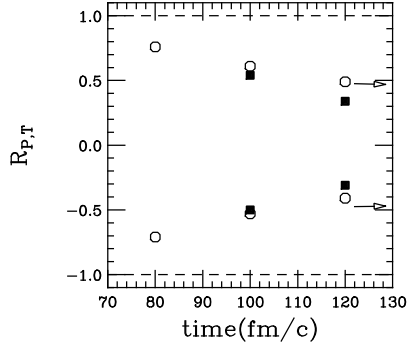}
\caption{$^{124}Sn+^{112}Sn$ $50~AMeV$ collision:
interaction time evolution of the projectile (upper) and target (lower)
isospin imbalance ratios. Squares: $Asysoft~ EOS$. Circles: $Superasystiff$.
The arrows correspond to the data of Ref.\cite{TsangPRL92}.}
\label{imbalance}
\end{figure}

In binary events 
the charge asymmetry of primary
projectile (target)- like fragments,
$PLF$ and $TLF$ provide the essential information
about the isospin equilibration rate.
At $b=8fm$, however, around $25 \%$ of the events
were ternary and as shown in the previous section
for even more central events this mechanism
becomes dominant. The $IMF$'s formation
in the overlap region will influence the
final isospin distribution,
rending more difficult the interpretation of the results.
Thus we select here only binary events.
We quantify the degree of equilibration through
the isospin imbalance ratio $R_i$ Eq.(\ref{imb}),  
\cite{TsangPRL92,RamiPRL84}.
The measured 
isospin dependent quantities are directly
the isospin content of the fragments at separation
 (in the $i=P,T$ rapidity regions),
as they result from the mixed reaction $124/112$,
from the reactions between neutron rich nuclei, ($124/124$), 
and between neutron poor nuclei, ($112/112$), respectively.
We report its dependence on 
$t_c$ for asysoft (squares)
and asysuperstiff (circles) $EOS$ in Fig. \ref{imbalance}.
A good degree of isospin equilibration corresponds to
a ``convergence'' to $0$ of both $R_{P}$ (upper curves) 
and $R_{T}$ (lower curves).
From our results we conclude that an asystiff-like
$EOS$ provides a better agreement to the experimental
observations, shown as arrows in Fig.\ref{imbalance}, 
 \cite{TsangPRL92,ShiPRC68}. 
In fact in the
$MSU$ experiment there is no particular selection on binary events.
We expect that the presence of events with  production of Neck-$IMFs$ 
will induce an apparent larger isospin equilibration since
more neutrons will migrate to the neck region (and viceversa for protons),
as discussed before.

A clear difference between the two equations of state
is evident especially for the longer interacting time, $b=8fm$ case. 
Smaller values
of the isospin imbalance ratios for asysoft $EOS$
point towards a faster equilibration rate.
In  Refs. \cite{TsangPRL92,ShiPRC68} a possible explanation was
proposed related to the observation that below normal
density the asysoft $EOS$ has a larger value of symmetry
energy. Therefore an enhanced isospin equilibration
has to be triggered if the diffusion takes place at lower density.
In fact the mechanism of charge equilibration
is more complicated at these energies due to
reaction dynamics (fast particle emissions,
density gradients, etc.), with interesting compensation effects,
 as shown in the following.

\subsubsection{Isospin diffusion and EOS dependence}
%\addtocontents{toc}{\hspace{0.55cm}\thesubsubsection \hspace{0.12cm}
%Isospin diffusion and EOS dependence}
Let us focus on the ``mixed'' $^{124}Sn+^{112}Sn$ case.
The isospin content of the two residues at separation
is determined by the interplay between the
nucleon emission from each ion 
and by the nucleon transfer through the neck:
\begin{eqnarray}
I_P = \frac{A_{0P}}{A_P}( I_{0P} - \frac{A_{gP}}{A_{0P}} I_{gP} -
\frac{A_{PT}}{A_{0P}} I_{PT} + \frac{A_{TP}}{A_{0P}} I_{TP})
\label{ipm} \\
I_{T} = \frac{A_{0T}}{A_T}( I_{0T} - 
\frac{A_{gT}}{A_{0T}}I_{gT} +
\frac{A_{PT}}{A_{0T}}I_{PT} -\frac{A_{TP}}{A_{0T}}I_{TP}) 
\label{itm}
\end{eqnarray}
Here $I_i, A_i$ represent asymmetries and masses for:
$i=P,T$, $PLF,TLF$ at separation; $i={0P},{0T}$, initial Projectile/Target;
$i={gP},{gT}$, fast emitted nucleons from Projectile/Target
(the label {\it $g$ stands for ``gas''}); $i={PT},{TP}$, nucleons transferred
from projectile (target) to target (projectile).

\begin{figure}
\centering
\includegraphics*[scale=0.50]{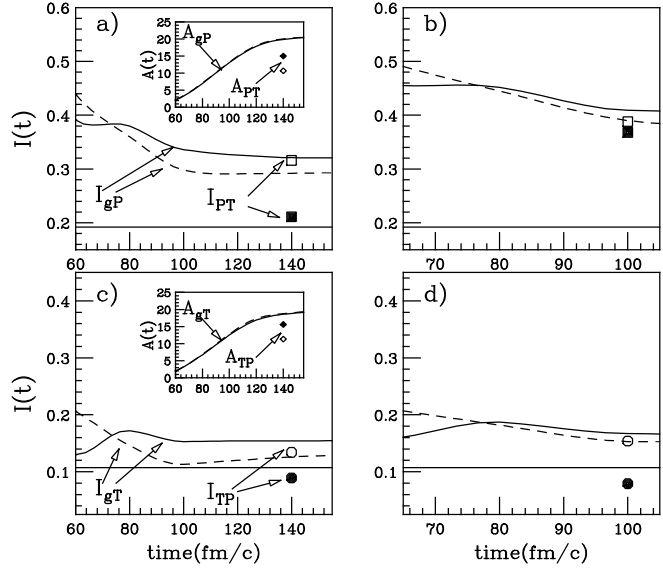}
\caption{$^{124}Sn+^{112}Sn$ $b=8fm$ (left) and 
$b=10fm$ (right) collision: time
evolution of isospin content and mass of projectile (target) 
pre-equilibrium emitted particles $I_{gP}$, $A_{gP}$ 
($I_{gT}$, $A_{gT}$).
The squares (circles) indicate the
asymmetry directly transferred 
from
projectile (target) to target (projectile) $I_{PT}$ ($I_{TP}$),
at separation time.
Asysoft: full symbols, solid lines. Superasystiff: empty symbols,
 dashed lines. Horizontal lines give the initial target and projectile 
asymmetries.}
\label{agasiso}
\end{figure}

In Fig. \ref{agasiso} we plot the time evolution of the
quantities $I_{gP}$, $I_{gT}$ and $A_{gP}$, $A_{gT}$
for asysoft and superasystiff $EOS$. We remark that $I_{gP}$ it
is much larger than $I_{0P} \simeq 0.19$. The same  is true for the
target but the difference is smaller. Consequently
the pre-equilibrium emission reduces the $N/Z$ difference
between the two nuclei thus competing with the transfer process.

When the two equations of state are compared a more neutron rich 
composition of pre-equilibrium is seen in the asysoft case 
since below normal density, from where most of the emitted nucleons
originate, the neutrons (protons) are
less (more) bound for this asy-$EOS$. 
The differences between the two asy-$EOS$ are 
reduced at larger impact parameters, as seen in the results
for $b=10fm$ in Fig. \ref{agasiso}.
The neutron rich nuclear skin seems to play a more important
role for such more peripheral collisions.   
For the projectile both preequilibrium emission and nucleon
transfer drive the system towards a more
symmetric configuration. The two processes tend
to compensate for the target. Therefore the projectile 
asymmetry has a more pronounced deviation
from the corresponding initial value in comparison to the target.  

From the Fig.\ref{agasiso} we realize that
asysoft interactions are very efficient in transferring isospin
to the ``gas'' while the isospin diffusion in the nuclear medium
appears more reduced. The opposite is happening with the stiff
symmetry term. The final result on isospin equilibration comes out
from a kind of compensation, with the asysoft case showing a
larger equilibration, see Fig.\ref{imbalance}.
The evidence that the isospin content of the free nucleon emission
is largely driving the isospin equilibration is extremely important.
We can then look for a consistent picture of several observables
ranging from isospin effects on fast nucleon emissions and collective
flows, see previous Sect.\ref{fastflows}, to the imbalance ratios of $N/Z$ or
other isospin dependent quantities. 

Before closing this discussion we would like to elaborate a little
more on the physics of isospin diffusion in the nuclear medium at 
the Fermi energies,
and in particular on the conclusion from the simulations of a much stronger
effectiveness of a stiff symmetry term.
From the Fig.\ref{agasiso} we see a clear dependence on the
asy-$EOS$ of the asymmetry transferred between the
two nuclei. In particular, for $b=8fm$ we observe that
\begin{eqnarray}
I_{PT}^{(supasystiff)} > I_{PT}^{(asysoft)} > I_{0P}=0.192 \nonumber \\
I_{TP}^{(supasystiff)} > I_{0T}=0.107 > I_{TP}^{(asysoft)} \label{inequal}
\end{eqnarray}
The origin of these inequalities lies
in the presence of density gradients due to the more dilute interface.
For the superasystiff $EOS$ the neutron migration
towards the  neck region is favored from both participants. This
explains simultaneously $I_{PT}^{(supasystiff)} > I_{0P}$ and
$I_{TP}^{(supasystiff)} > I_{0T}$.
For the asysoft $EOS$ this effect is  weakened because
the symmetry part of the mean-field 
for protons and neutrons does not change significantly
with density below saturation.

This argument can be made more explicit if we observe that
the proton and neutron migration is
dictated by the spatial gradients of
the corresponding chemical potentials $\mu_n(\rho_p,\rho_n,T)$
and $\mu_p(\rho_p,\rho_n,T)$ \cite{Balianbook}.
The currents of the two species can be expressed as follows:
\begin{eqnarray}
j_{n} = - Ct \nabla \mu_{n}(\rho_p,\rho_n,T) = - Ct [
\left({\partial \mu_n \over \partial \rho_n}\right)_{\rho_p,T} \nabla  \rho_n +
\left({\partial \mu_n \over \partial \rho_p}\right)_{\rho_n,T} \nabla  \rho_p ] \nonumber  \\
j_{p} = - Ct \nabla \mu_{p}(\rho_p,\rho_n,T) = - Ct [
\left({\partial \mu_p \over \partial \rho_n}\right)_{\rho_p,T} \nabla  \rho_n +
\left({\partial \mu_p \over \partial \rho_p}\right)_{\rho_n,T} \nabla  
\rho_p ], \nonumber
\end{eqnarray}
where $C$ is a constant.

In terms of Landau parameters, i.e.:  
\begin{equation}
N_q(T){\partial\mu_q \over \partial\rho_{q'}}
= \delta_{qq'} + F_0^{qq'},~~~~~~~q=n,p~~~~~~q'=n,p    \label{muF0}
\end{equation} 
we obtain:
\begin{eqnarray}
j_{n} &=& -Ct ([ \frac{1+I}{2} \frac{1+F_0^{nn}}{N_n} +  \frac{1-I}{2} \frac{F_0^{np}}{N_n} ] \nabla  \rho + 
 \frac{\rho}{2} [ \frac{1+F_0^{nn}}{N_n} - \frac{F_0^{np}}{N_n} ] \nabla I) \nonumber  \\ 
&=& -D_{1n} \nabla  \rho - D_{2n} \nabla I \\
j_{p} &=& - Ct ([ \frac{1+I}{2}\frac{F_0^{pn}}{N_p} +  \frac{1-I}{2} \frac{1+F_0^{pp}}{N_p} ] \nabla  \rho + 
 \frac{\rho}{2} [ \frac{F_0^{pn}}{N_p} - \frac{1+F_0^{pp}}{N_p} ] \nabla I) \nonumber \\ 
&=& - D_{1p} \nabla  \rho - D_{2p} \nabla I 
\end{eqnarray}
In the last expressions we have changed the variables $\rho_n,\rho_p$ to  $\rho=\rho_n + \rho_p$
and $I=(\rho_n - \rho_p)/ \rho$.

Hence it is clear that not only concentration gradients are important in
the dynamics of isospin migration and equilibration, but also
density gradients play an essential role.
Actually the final result comes out from a delicate balance between
the two effects. The stronger density dependence of the stiff symmetry
term below normal density enhances the density gradient
contribution, finally leading to the larger isospin diffusion of
Eqs.(\ref{inequal}). 
As already shown the influence of the symmetry term on fast nucleon emissions
can partially compensate the effect.

%\include{rep_bib}

%\end{document}

%% file: Chapter-6.tex
%\documentclass{elsart}
%\usepackage{epsfig}

%\usepackage{graphicx}

%\usepackage{amssymb}
%\tightenlines

%\begin{document}

\setcounter{figure}{0}
\setcounter{equation}{0}

\section{Effective interactions in the isovector channel: relativistic 
approach and the role of the $\delta$ meson} \label{qhd}

\markright{Chapter \arabic{section}: qhd}

The $QHD$ effective field model represents a very successful attempt to 
describe, in a fully relativistic picture, equilibrium and 
dynamical properties of nuclear systems at the hadronic 
level \cite{WaleckaAP83,SerotANP16,SerotIJMPE6}.
Consistent results have been obtained for the nuclear structure of finite 
nuclei \cite{SharmaAP231,RingPPNP37,TypelNPA656}, for the $NM$ 
Equation of State and liquid-gas 
phase transitions \cite{MuellerPRC52} and for the dynamics of 
nuclear collisions \cite{GiessenRPP56,KoJPG22}. Relativistic 
Random-Phase-Approximation ($RRPA$) 
theories have been developed to study the nuclear collective response 
\cite{DellaPRC44,HorowitzNPA531,MateraPRC49,VretenarNPA621,MaPRC55,MaNPA686}.

This report focusses on the dynamical response and static (equilibrium) 
properties of 
Asymmetric Nuclear Matter ($ANM$). We use a relativistic kinetic
 theory with the aim
of a transparent connection between the collective and reaction dynamics 
and the 
coupling to various channels of the nucleon-nucleon interaction. 
One of the main points of discussion is the relevance of the coupling
to a scalar isovector channel, the effective $\delta[a_0(980)]$ meson.
In fact the scalar-isovector coupling is not necessarily connected to
the exchange of a real $\delta$-meson, but such a channel appears automatically
in relativistic Hartree-Fock and Dirac-Brueckner-Hartree-Fock treatments
from exchange and correlation contributions \cite{JongPRC57,GrecoPRC64}.

The introduction of the isovector-scalar channel in covariant approaches
should play a key role in the effective interaction in asymmetric matter,
see ref.\cite{LiuboPRC65,GrecoPRC67}. This point has not received great 
attention before
for two main reasons:

i) The $\delta-$channel has not been considered {\it a priori}, just
on the basis of the weak contribution to the free Nucleon-Nucleon
interaction, \cite{NikolausPRC46,FurnstNPA627}. But in the spirit of the
$Effective~ Field~ Theory$  as a relativistic $Density~ Functional$ $Theory$,
(the $EFT/DFT$ framework \cite{FurnstCNPP2}), the relevance 
of this channel
could be completely different in nuclear matter, due to medium and
many-body effects, as noted before. In particular, we expect a large 
contribution
from exchange terms of the strongly coupled isoscalar channels,
\cite{GrecoPRC64,LiuboPRC65,GrecoPRC67}.

ii) The extension is not well supported by the existing data on exotic nuclei,
as remarked in the refs. 
\cite{HofmannPRC64,TypelNPA656,FurnstNPA671,BuervePRC65}. Clearly
these conclusions are mainly derived from the lack of
information on observables more sensitive to the density dependence
of the symmetry term. In particular finite nuclei studies cannot
easily disentangle the effects originating from different isovector mesons,
since the densities are mainly below saturation, see the discussion
in refs.\cite{TypelNPA656,HofmannPRC64,GaitanosNPA732}.

We note that recently , see the conclusions of refs.
\cite{FurnstNPA706,MadlandNPA741}, 
the $\delta$-channel has been reconsidered
as an interesting improvement of covariant approaches in the framework
of the $EFT/DFT$ philosophy. One of the main tasks of
our work is then to try to select the dynamical observables more sensitive
to it, \cite{LiuboPRC65,GrecoPRC67}. 

In this respect the results reported here on the collective response
and reaction dynamics can be useful in order to solve the
open problem of the determination of the scalar-isovector coupling.
Since contributions to this channel are mainly
coming from correlation effects \cite{GrecoPRC64}, the 
correct microscopic approach
should be to derive the relative coupling constant, in a $QHD$-Relativistic
Mean Field ($RMF$) framework, 
from Dirac-Brueckner-Hartree-Fock ($DBHF$) calculations. Several attempts 
have been 
recently performed, 
see \cite{JongPRC57,HofmannPRC64,SchillerEPJA11,MaPRC66},
 but the results 
are up to now not fully model independent.

An important outcome of our work is to show that the two effective couplings,
vector and scalar, in the isovector channel influence in a different way
the static (symmetry energy) and dynamic (collective response, reaction 
observables) properties
of asymmetric nuclear matter. This will open new possibilities for
a phenomenological determination of these fundamental quantities.

We will often derive transparent analytical results. In order to show
also some quantitative effects of the dynamical contribution of the
$\delta$-channel we have to fix in some way the corresponding coupling.
We have used a constant value (see Table \ref{qhdsets}) extracted 
from the $DBHF$ analysis
of refs.\cite{JongPRC57,HofmannPRC64}, where it actually appears not strongly
density dependent in a wide range of baryon densities.
Some results are also presented with Fock correlations
explicitly accounted for ($NLHF$ case, see following).

\subsection{QHD effective field theory}

We start from the {\it $QHD$} effective field picture of the hadronic 
phase of nuclear matter \cite{WaleckaAP83,SerotANP16,SerotIJMPE6}. 
To include the 
main dynamical 
degrees of freedom of the system we will consider the nucleons
coupled to the isoscalar scalar $\sigma$ and vector $\omega$ mesons
and to the isovector scalar $\delta$ and vector $\rho$ mesons.

The Lagrangian density for this model, including non--linear isoscalar/scalar 
$\sigma$-terms \cite{BogutaNPA505}, is given by:

\begin{eqnarray}
{ L} = {\bar {\psi}}[\gamma_\mu(i{\partial^\mu}-{g_\omega}{V}^\mu
- g_{\rho}{{\bf B}}^\mu \cdot  {\vec {\tau}} ) 
-(M-{g_\sigma}\phi-g_{\delta}{\vec {\tau}} \cdot {\vec {\delta}}~)]\psi +~~~~~~
\nonumber\\
  {1\over2}({\partial_\mu}\phi{\partial^\mu}\phi
- m_s^2 \phi^2) - {a \over 3} \phi^3 - {b \over 4} \phi^4
- {1 \over 4} W_{\mu\nu}
W^{\mu\nu} + {1 \over 2} m_v^2 {V}_\nu {{ V}^\nu}+ \nonumber\\
\frac{1}{2}(\partial_{\mu}{\vec {\delta}} \cdot \partial^{\mu}{\vec {\delta}}
-m_{\delta}^2{\vec {\delta}}^2)
- {1 \over 4} {\bf G}_{\mu\nu}
{\bf G}^{\mu\nu} + {1 \over 2} m_{\rho}^2 {{\bf B}}_\nu
{{{\bf B}}^\nu}~~~~~~~~~~~~~~~~~~~
\label{eq.1}    
\end{eqnarray}

where
$W^{\mu\nu}(x)={\partial^\mu}{{V}^\nu}(x)-
{\partial^\nu}{{V}^\mu}(x)~$ and
${\bf G}^{\mu\nu}(x)={\partial^\mu}{{{\bf B}}^\nu}(x)-
{\partial^\nu}{{{\bf B}}^\mu}(x)~.$

Here $\psi(x)$ is the nucleon
fermionic field, $\phi(x)$ and ${{V}^\nu}(x)$ represent neutral scalar
 and
vector boson fields, respectively. ${\vec {\delta}}(x)$ and 
${{{\bf B}}^\nu}(x)$
are the charged scalar and vector
fields and ${\vec {\tau}}$ denotes the isospin matrices .

From the Lagrangian, Eq.(\ref{eq.1}), with the Euler procedure a set of 
coupled equations of motion for the meson and nucleon fields can be 
derived. The basic 
approximation in nuclear matter applications consists in neglecting all the
terms containing derivatives of the meson fields with respect to the mass
contributions. Then the meson fields are simply connected to the operators
of the nucleon scalar and current densities by the following equations: 
\begin{equation}\label{Eq.4a}
{\widehat{\Phi}/f_\sigma} + A{\widehat{\Phi}^2}
+ B{\widehat{\Phi}^3}~=\bar\psi(x)\psi(x)~\equiv \widehat{\rho_S}
\end{equation}
\vskip -0.5cm
\begin{eqnarray}\label{Eq.4b}
{\widehat{ V}}^\mu(x)~\equiv g_{\omega} V^\mu~=~f_\omega\bar\psi(x)
{\gamma^\mu}\psi(x)~
\equiv f_\omega \widehat j_\mu~,\nonumber\\
{\widehat  {\bf  B}}^\mu(x)~\equiv g_{\rho} {\bf B}^\mu~=~f_{\rho}\bar\psi(x)
{\gamma^\mu} {\vec {\tau}}
\psi(x)~,\nonumber\\
{\widehat{\vec {\delta}}}(x)~\equiv g_{\delta}{\vec {\delta}}~=~f_{\delta}
\bar\psi(x){\vec {\tau}}\psi(x)
\end{eqnarray}
where $\widehat \Phi=g_\sigma\phi$, $f_\sigma = (g_\sigma/m_\sigma)^2$, 
$A = a/g_\sigma^3$, $B = b/g_\sigma^4$, $f_\omega = (g_\omega/m_\omega)^2$, 
$f_{\rho} = (g_{\rho}/m_{\rho})^2$, $f_{\delta} = (g_{\delta}/m_{\delta})^2$. 

For the nucleon fields we get a Dirac-like equation. Indeed after 
substituting Eqs.(\ref{Eq.4a},\ref{Eq.4b}) for the meson field operators, we 
obtain an equation which contains only nucleon field operators. The 
equations can be consistently solved in a Mean Field Approximation 
($RMF$), where most
applications have been performed, i.e. in a self-consistent Hartree scheme
 \cite{SerotANP16,SerotIJMPE6}.

The inclusion of Fock terms is conceptually important 
\cite{HorowitzNPA399,BouyssyPRC36}
since it automatically leads to contributions to various meson exchange 
channels, also
in absence of explicit coupling terms. 
A thorough study of the Fock contributions in a $QHD$ approach with non-linear
self-interacting terms has been recently performed \cite{GrecoPRC63}, in
particular for asymmetric matter \cite{GrecoPRC64}.
 
\subsubsection{Relativistic transport equations with Fock terms}
%\addtocontents{toc}{\hspace{0.55cm}\thesubsubsection \hspace{0.12cm}
%Relativistic transport equations with Fock terms}
We now discuss a kinetic approach consistent
with the previous approximation. We are concerned with a semiclassical 
description of nuclear dynamics, so that the nuclear medium is supposed
to be in states for which the nucleon scalar and current densities
are smooth functions of the space-time coordinates.
Within the $QHD$ model
we focus our analysis on a description of 
the many-body nuclear system in terms of one--body dynamics.
Correlation effects can be 
effectively included at the level of coupling constants, as noted in the 
discussion of the results.

We move to the quantum phase-space
introducing the Wigner transform of the one-body density matrix for the
fermion field \cite{DegrootRelKin,HakimNC6}.
The one--particle Wigner function is defined as:
$$[{\widehat F}(x,p)]_{\alpha\beta}=
{1\over(2\pi)^4}\int d^4Re^{-ip \cdot R}
\langle :\bar{\psi}_\beta(x+{R\over2})\psi_\alpha(x-{R\over2}):\rangle~,$$
where $\alpha$ and $\beta$ are double indices for spin and isospin. 
The brackets denote statistical averaging and the colons denote 
normal ordering.
The equation of motion is derived from
the Dirac field equation by using standard procedures
(see e.g.\cite{DegrootRelKin,HakimNC6}),
as:
\begin{eqnarray}\label{wig}
{i\over2}{\partial_\mu}[\gamma^\mu{\hat F}(x,p)]_
{\alpha\beta}+p_\mu[{\gamma^\mu}{\hat F}(x,p)]_{\alpha\beta}-
M{\hat F}_{\alpha\beta}(x,p) \nonumber\\
-{g_{\omega}\over(2\pi)^4}\int_R e^{-ip\cdot R}
<:\bar{\psi}_\beta(x_+){\gamma^\mu_{\alpha\gamma}}\psi_\gamma(x_-)
{V}_\mu(x_-):> \nonumber\\
+{g_{\sigma}\over(2\pi)^4}\int_R e^{-ip\cdot R}
<:\bar{\psi}_\beta(x_+)\psi_\alpha(x_-)\phi(x_-):> \nonumber\\
-{g_{\rho}\over(2\pi)^4}\int_R e^{-ip\cdot R}
<:\bar{\psi}_\beta(x_+){\gamma^\mu_{\alpha\gamma}}\psi_\gamma(x_-)
{\vec{\tau}} \cdot {\bf B}_\mu(x_-):> \nonumber\\
+{g_{\delta}\over(2\pi)^4}\int_R e^{-ip\cdot R}
<:\bar{\psi}_\beta(x_+)\psi_\alpha(x_-){\vec{\delta}}(x_-):> \nonumber\\
=0
\end{eqnarray}
with $x_+=x+{R\over2}$ and $x_-=x-{R\over2}$.
When we insert the Eqs.(\ref{Eq.4a},\ref{Eq.4b}) for the meson field operators
we clearly see the appearance of Fock contributions (at the lowest order in
density matrices).

The Wigner function is a matrix in spin and isospin space; in the case of
asymmetric $NM$ it is useful to decompose it into neutron and proton components.
Following the treatment of the Fock terms in non-linear $QHD$ introduced 
in Refs. \cite{GrecoPRC63,GrecoPRC64}, we obtain the 
 kinetic equation: 
\begin{eqnarray}\label{trans}
&&{i\over2}{\partial_\mu}{\gamma^\mu}{\hat F}^{(i)}(x,p)+\gamma^\mu
p^*_{\mu i}{\hat F}^{(i)}(x,p) - M^*_i{\hat F}^{(i)}(x,p)+\nonumber\\
&&{i\over2}\Delta
\left[\tilde f_\omega j_\mu(x)\gamma^\mu \pm \tilde f_\rho j_{3\mu}(x)
\gamma^{\mu}
- \tilde f_\sigma \rho_S(x) \mp \tilde f_\delta \rho_{S3}(x)\right]
{\hat F}^{(i)}(x,p)=0,
\end{eqnarray}
where $i=n,p$, and here and in the following upper and lower signs correspond
to protons and neutrons, respectively.
Here $\Delta={\partial_x} \cdot {\partial_p}$, with $\partial_x$ acting only 
on the first term of the products and  $\rho_{S3}=\rho_{Sp}-\rho_{Sn}$ and $j_{3\mu}(x)=j^p_{\mu}(x)
-j^n_{\mu}(x)$ 
are the isovector scalar density and the isovector baryon current, 
respectively. 
We have defined the kinetic momenta and effective masses as:
\begin{eqnarray}\label{psms}
&&p^*_{\mu i}=p_\mu-{\tilde f_\omega}j_\mu(x)\pm \tilde f_\rho{j}_{3\mu}(x)~,
   \nonumber\\
&&M^*_i=M- {\tilde f_\sigma}\rho_S(x) \pm \tilde f_\delta \rho_{S3}(x)~, 
\end{eqnarray}
with the effective coupling functions given by:
\begin{eqnarray}\label{cincoup}
&&{\tilde f}_\sigma={{\Phi} \over {\rho_S}}-{1\over 8}{{d\Phi(x)} \over 
{d\rho_S(x)}}
- {1\over {2\rho_S}}Tr {\widehat F}^{2}(x)
{{d^{2}\Phi(x)} \over {d\rho_S^{2}(x)}}
+{1\over 2}f_\omega+{3\over 2}f_{\rho}-{3\over 8}f_{\delta}~,\nonumber\\
&&{\tilde f}_\omega={1\over 8}{{d\Phi(x)} \over {d\rho_S(x)}}
+{5\over 4}f_\omega+{3\over 4}f_{\rho}+{3\over 8}f_{\delta}~,\nonumber\\
&&\tilde f_\delta=-{1\over 8}{{d\Phi(x)} \over {d\rho_S(x)}}
+{1\over 2}f_\omega-{1\over 2}f_{\rho}+{9\over 8}f_{\delta}~,\nonumber\\
&&\tilde f_\rho={1\over 8}{{d\Phi(x)} \over {d\rho_S(x)}}
+{1\over 4}f_\omega+{3\over 4}f_{\rho}-{1\over 8}f_{\delta}~,\nonumber\\
\end{eqnarray}
where $8~Tr {\widehat F}^{2}(x)=\rho_S^2+j_\mu j^\mu+\rho_{S3}^2+
j_{3\mu} j^{3\mu}$.
We recall that we are dealing with a 
transport equation so the 
currents and densities, in general, are varying functions 
of the space--time, at variance with the case of nuclear matter at 
equilibrium. Thus also the effective couplings are space, i.e. density,
 dependent. 

The expression of Eq.(\ref{psms}) for the effective mass 
embodies an isospin contribution from Fock terms even without 
a direct inclusion of 
the $\delta$ meson in the Lagrangian. 
As seen from Eqs.\ref{cincoup} 
the usual $RMF$ approximation (Hartree level) is recovered from the 
Hartree-Fock results, by changing the effective coupling functions 
$\tilde f_i (i=\sigma,\omega, \rho, \delta)$, Eqs.(\ref{cincoup}), to the 
the explicit coupling constants $f_i$.

\subsubsection{Equilibrium properties: the nuclear Equation of State}
%\addtocontents{toc}{\hspace{0.55cm}\thesubsubsection \hspace{0.12cm}
%Equilibrium properties: the nuclear Equation of State}
The energy density and pressure for symmetric and asymmetric 
nuclear matter and the $n,p$ effective masses 
are self-consistently calculated  
 in terms of the four boson coupling constants,
$f_i \equiv (\frac{g_i^2}{m_i^2})$, $i = \sigma, \omega, \rho, \delta$,
and the two parameters of the $\sigma$ self-interacting terms, 
$A \equiv \frac{a}{g_\sigma^3}$ and $B \equiv \frac{b}{g_\sigma^4}$,
\cite{LiuboPRC65,GrecoPRC67}. 
Here we will present results at the Hartree level. The extension to Fock 
contributions is easily performed following the scheme discussed before.

From the Lagrangian, Eq.(\ref{eq.1}), the energy-momentum tensor is derived 
in the mean field approximation as

\begin{eqnarray}\label{tens}
T_{\mu\nu}=i\bar{\psi}\gamma_{\mu}\partial_{\nu}\psi+[\frac{1}{2} 
 m_{\sigma}^2\phi^2+U(\phi)+\frac{1}{2}m_{\delta}^2\vec{\delta^2}
 \nonumber \\
-\frac{1}{2}m^2_{\omega}\omega_{\lambda} \omega^{\lambda}
-\frac{1}{2}m^2_{\rho}{\bf B}_{\lambda}{\bf B}^{\lambda}]g_{\mu\nu}.
\end{eqnarray}
with $U(\phi)$, the nonlinear potential of the $\sigma$ meson
$U(\phi)=\frac{1}{3}a\phi^{3}+\frac{1}{4}b\phi^{4}$.
By using the field equations for mesons, the equation of state for thermal
matter is derived, as in \cite{LiuboPRC65}, with energy density

\begin{eqnarray}\label{edens}
\epsilon=
\sum_{i=n,p}{2}\int \frac{{\rm d}^3k}{(2\pi)^3}E_{i}^*(k)
(n_i(k)+\bar{n_i}(k))
+\frac{1}{2}m_\sigma^2\phi^2 \nonumber \\
+U(\phi)
+\frac{g_{\omega}^2}{2m_\omega^2}\rho_B^2
+\frac{g_{\rho}^2}{2m_{\rho}^2}\rho_{B3}^2
+ \frac{g_{\delta}^2}{2m_{\delta}^2}\rho_{S3}^2,
\end{eqnarray}

and pressure

\begin{eqnarray}\label{press}
P =\sum_{i=n,p} \frac{2}{3}\int \frac{{\rm d}^3k}{(2\pi)^3}
\frac{k^2}{E_{i}^\star(k)}
(n_i(k)+\bar{n_i}(k)) -\frac{1}{2}m_\sigma^2\phi^2 \nonumber \\
-U(\phi) 
+\frac{g_{\omega}^2}{2m_\omega^2}\rho_B^2
+\frac{g_{\rho}^2}{2m_{\rho}^2}\rho_{B3}^2
-\frac{g_{\delta}^2}{2m_{\delta}^2}\rho_{S3}^2.
\end{eqnarray}

where
${E_i}^*=\sqrt{k^2+{{M_i}^*}^2}$ and the nucleon effective masses
are given in Eq.\ref{psms}. 
%\begin{equation}\label{eq.6}
%{M_i}^*=M_{N}-f_\sigma\rho_S \mp f_\delta\rho_{S3}~~~ 
%(-~proton, +~neutron).
%\end{equation}
The $n_i(k)$ and $\bar{n_i}(k)$ in Eqs.\ref{edens},\ref{press}
are the $p,n$ fermion and antifermion distribution functions:

\begin{eqnarray}\label{nnbar}
n_i(k)=\frac{1}{1+\exp\{({E_i}^*(k)-{\mu_i^*})/T \} }\,, \nonumber \\
\bar{n_i}(k)=\frac{1}{1+\exp\{({E_i}^*(k)+{\mu_i^*})/T \} }.
\end{eqnarray}

The effective chemical potentials $\mu_i^*$  are given in terms
of the vector meson mean fields 

\begin{eqnarray}\label{mu}
{\mu_i}=\mu_i^*  - f_\omega\rho_B \mp f_{\rho} \rho_{B3}~,
\end{eqnarray}

where $\mu_i$ are the thermodynamical chemical potentials 
$\mu_i=\partial\epsilon/\partial\rho_i$, which at zero temperature 
reduce to the Fermi energies $E_{Fi} \equiv \sqrt{k_{Fi}^2+{M_i^*}^2}$.
The baryon
densities $\rho_B$ and the scalar densities $\rho_S$  are given by 
($\gamma$ is the spin/isospin degeneracy)
\begin{eqnarray}\label{rhobs}
\rho_B=\gamma
\int\frac{{\rm d}^3k}{(2\pi)^3}(n(k)-\bar{n}(k))\,, \nonumber \\
\rho_S=\gamma
\int\frac{{\rm d}^3k}{(2\pi)^3}\frac{M^*}{E^*}(n(k)+\bar{n}(k)).
\end{eqnarray}
At the temperatures of interest here the antybaryon contributions
are actually negligible.

The isoscalar meson couplings are fixed from symmetric nuclear matter
properties at $T=0$: saturation density $\rho_0=0.15fm^{-3}$,
 binding energy $E/A = -16MeV$, nucleon effective mass $M^* = 0.7 M_N$
 ($M_N=939MeV$)  and incompressibility $K_V = 240 MeV$ 
at $\rho_0$. The fitted
$f_\sigma, f_\omega, A, B$ parameters are reported in Table \ref{qhdsets}. 
They are
quite standard for these minimal non-linear $RMF$ models, with
values compatible with microscopic $DBHF$ estimations 
\cite{JongPRC57,HofmannPRC64} in a wide range of densities.
In order to isolate the effects in the
isovector channel we will review results obtained using the same
isoscalar interaction and with the $\rho(NL\rho)-$ or the
$\rho+\delta(NL\rho\delta)-$ couplings in the isovector part
(refs.\cite{LiuboPRC65,GrecoPRC67}.
The symmetry energy at saturation is fixed to the Bethe-Weisz\"acker
fourth parameter $a_4=30.5~MeV$. $NLHF$ stands for 
the non-linear Hartree-Fock scheme described before.

\vskip 0.5cm
 \begin{table}
\begin{center}
\caption{Parameter sets of the $QHD$ models discussed here}
\begin{tabular}{ c c c c } \hline
$parameter$   &~ $NL\rho$  &~ $NL\rho\delta$    &~$NLHF$    \\ \hline
$f_\sigma~(fm^2)$  &~11.3    &~$same$           &~9.15       \\ \hline
$f_\omega~(fm^2)$  &~6.5    &~$same$            &~3.22       \\ \hline
$f_\rho~(fm^2)$    &~1.1     &~3.15             &~1.9        \\ \hline
$f_\delta~(fm^2)$  &~0.00     &~2.4            &~1.4         \\ \hline
$A~(fm^{-1})$     &~0.02    &~$same$            &~0.098     \\ \hline
$B$               &~-0.004  &~$same$            &~-0.021    \\ \hline
\end{tabular}
\label{qhdsets}
\end{center}
\vskip 0.2cm
\end{table}

\subsubsection{Symmetry energy}
%\addtocontents{toc}{\hspace{0.55cm}\thesubsubsection \hspace{0.12cm}
%Symmetry energy}
The symmetry energy in $ANM$ is defined from the expansion
of the energy per nucleon $E(\rho_B,I)$ in terms of the asymmetry
parameter $I$ defined as 
$I \equiv -\frac{\rho_{B3}}{\rho_B}=\frac{\rho_{Bn}-\rho_{Bp}}
{\rho_B}=\frac{N-Z}{A}$.
We have
\begin{equation}\label{eq.7}
E(\rho_B,I)~\equiv~\frac{\epsilon(\rho_B,I)}{\rho_B}~
=~E(\rho_B) + E_{sym}(\rho_B) I^2 \nonumber \\
+ O(I^4) +...
\end{equation}
and so in general
\begin{equation}\label{eq.8}
E_{sym}~\equiv~\frac{1}{2} \frac{\partial^2E(\rho_B,I)}
{\partial I^2} \vert_{I=0}~=~\frac{1}{2} \rho_B 
\frac{\partial^2\epsilon}{\partial\rho_{B3}^2}
 \vert_{\rho_{B3}=0}
\end{equation}

In the Hartree case an explicit expression for the symmetry energy is 
easily derived \cite{KubisPLB399,LiuboPRC65,GrecoPRC67}:
\begin{eqnarray}\label{eq.9}
E_{sym}(\rho_B) = \frac{1}{6} \frac{k_F^2}{E^*_F} + \frac{1}{2}
f_\rho \rho_B - \frac{1}{2} f_\delta 
\frac{{M^*}^2 \rho_B}{E_F^{* 2} \left[1+f_\delta A(k_F,M^*)\right]}
\nonumber \\
\equiv E_{sym}^{kin}+ E_{sym}^{pot}~, ~~~~~~~~~
\end{eqnarray}
where $k_F$ is the nucleon Fermi momentum corresponding to $\rho_B$, 
$E^*_F \equiv \sqrt{(k_F^2+{M^*}^2)}$ and $M^*$ is the effective
nucleon mass in symmetric $NM$, $M^*=M_N - g_\sigma \phi$.
$A(k_F,M^*)$ represents the integral
\begin{equation}\label{eq.10}
A(k_F,M^*)~\equiv~\frac{4}{(2\pi)^3} \int d^3k \frac{k^2}{ 
(k^2+{M^*}^2)^{3/2}}=3 \left(\frac{\rho_S}{M^*} - 
\frac{\rho_B}{E^*_F}\right)
\end{equation}

We remark that $A(k_F,M^*)$ is very small at low densities, 
and actually
can still be neglected up to a baryon density 
$\rho_B \simeq 3\rho_0$, \cite{LiuboPRC65,GrecoPRC67}.
Then in the density range of interest here we can use, to leading order,
a much simpler 
form of the symmetry energy, with more transparent  
$\delta$-meson effects:
\begin{equation}\label{eq.11}
E_{sym}(\rho_B) = \frac{1}{6} \frac{k_F^2}{E^*_F} + \frac{1}{2}
\left[f_\rho - f_\delta 
\left(\frac{M^*}{E^*_F}\right)^2\right] \rho_B
\end{equation}
We see that, when the $\delta$ is included, the empirical $a_4$ value actually
corresponds to the combination $[f_\rho - f_\delta (\frac{M}{E_F})^2]$
of the $(\rho,\delta)$ coupling constants. Therefore if
$f_\delta \not= 0$ we have to increase correspondingly the $\rho$-coupling 
(see Fig.1 of
ref.\cite{KubisPLB399}). 

In Table \ref{qhdsets} the $NL\rho$ set corresponds to $f_\delta=0$. In the 
$NL\rho\delta$ interaction
$f_\delta$ is chosen as $2.4fm^2$, roughly derived from the analysis 
of ref.\cite{HofmannPRC64}. As already noted before this choice is
not essential for our discussion: the aim of our work is 
to show the new dynamical effects of the $\delta$-meson
coupling and to select the corresponding most sensitive observables.

\begin{figure}[htb]
%\epsfysize=5.5cm
%\centerline{\epsfbox{lin2.ps}}
\begin{center}
\includegraphics*[scale=0.60]{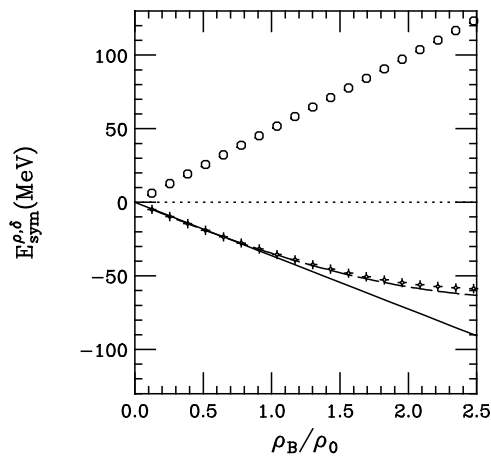}
\caption{
$\rho$- (open circles) and $\delta$- (crosses) contributions
to the potential symmetry energy, second and third terms of Eq.(\ref{eq.9}). 
The dashed line 
is the approximate $\delta$-contribution of Eq.(\ref{eq.11}).
The solid line is a linear extrapolation of the 
low density behaviour.}
\label{qhd1}
\end{center}
\end{figure}

In order to have the same $a_4$ at saturation we must increase the 
$\rho$-coupling constant
by a factor three, up to $f_\rho=3.15 fm^2$. Now the symmetry energy at 
saturation
density is actually built from the balance of scalar (attractive) and vector
(repulsive) contributions, with the scalar channel becoming weaker 
with increasing
baryon density. This is clearly shown in
Fig.\ref{qhd1}.This is indeed the isovector 
counterpart of the saturation
mechanism occurring in the isoscalar channel for symmetric nuclear matter. 
From this consideration we get a further  support to  the 
introduction
of the $\delta$-coupling in the symmetry energy evaluation.

\begin{figure}[htb]
%\epsfysize=5.5cm
%\centerline{\epsfbox{lin2.ps}}
\begin{center}
\includegraphics*[scale=0.50]{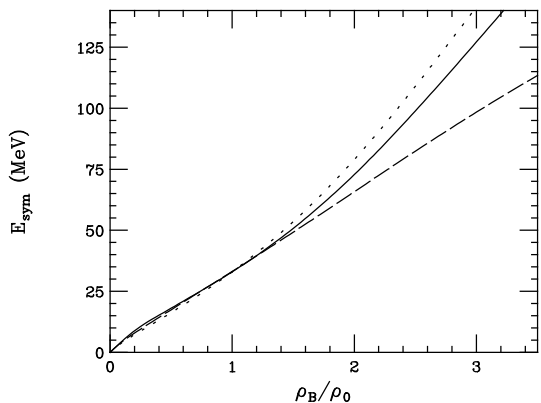}
\caption{Total (kinetic + potential) symmetry energy as a function of the
baryon density. Dashed line ($NL\rho$). Dotted line ($NL\rho\delta$).
Solid line $NLHF$.}
\label{qhd2}
\end{center}
\end{figure}

In  Fig.\ref{qhd2} we show the total symmetry energy for the 
different models.
At subnuclear densities, $\rho_B < \rho_0$, in both cases, 
$NL\rho$ and $NL\rho\delta$, from Eq.(\ref{eq.11}) we have an almost 
linear dependence 
of $E_{sym}$ on
the baryon density, since $M^* \simeq E_F$ as a  good approximation .
Around and above $\rho_0$ we see a steeper increase in the
$(\rho+\delta)$ case since $M^*/E_F$ is decreasing.

In conclusion when the $\delta$-channel is included the behaviour 
of the symmetry energy
is stiffer at high baryon density 
from the relativistic
mechanism discussed before. This is in fact due to a larger
contribution from the $\rho$ relative to the $\delta$ meson. We expect to
see these effects more clearly in the relativistic reaction dynamics at
intermediate energies, where higher densities are reached (see 
Sect.\ref{reldyn}.

\begin{figure}[htb]
%\epsfysize=5.5cm
%\centerline{\epsfbox{lin2.ps}}
\begin{center}
\includegraphics*[scale=0.60]{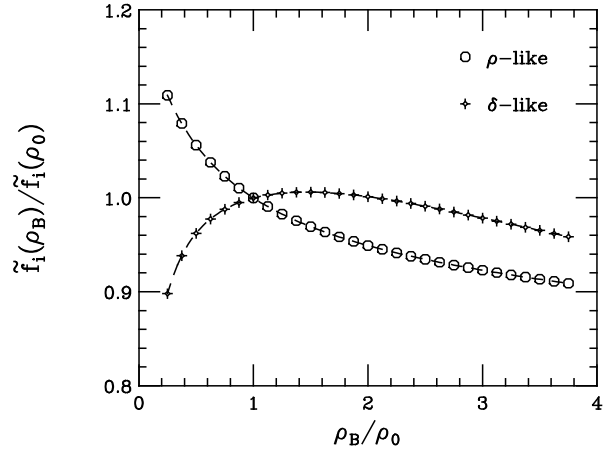}
\caption{Baryon density variation of the isovector 
effective coupling when the Fock terms are included, model $NLHF$.}
\label{qhd3}
\end{center}
\end{figure}

When the Fock terms are evaluated the new ``effective'' couplings 
Eqs.(\ref{cincoup})
naturally acquire a density dependence. This is shown in Fig.\ref{qhd3} 
for the isovector terms. The decrease of the $\rho$ 
coupling at high density accounts for the slight softening of the 
symmetry energy, Fig.\ref{qhd2}, in agreement with $DBHF$ expectations,
see \cite{HofmannPRC64,GaitanosNPA732}.
Details of the calculation can be found in Refs.\cite{GrecoPRC64,LiuboPRC65}.

\begin{figure}[htb]
%\epsfysize=5.5cm
%\centerline{\epsfbox{lin2.ps}}
\begin{center}
\includegraphics*[scale=0.65]{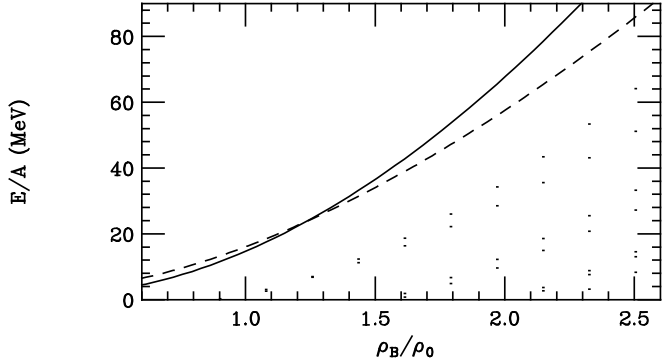}
\caption{
$EOS$ for pure neutron matter.
 Dashed line: $Nl\rho$. Solid line: $NL\rho\delta)$.}
\label{qhd4}
\end{center}
\end{figure}

In Fig.\ref{qhd4} we show the Equation of State (energy per nucleon) for pure
neutron matter ($I=1$) obtained with the two parameter Sets, 
$NL\rho$ and $NL\rho\delta$.
The values, in particular for the $NL\rho\delta$ case, are in good 
agreement with recent non-relativistic 
Quantum-Monte-Carlo
variational calculations with realistic $2-$ and $3-$body
forces \cite{FantoniPRL87}. The inclusion of the $\delta$-coupling
leads to a larger repulsion at baryon densities roughly above $1.5\rho_0$.
This would be of interest for the structure of neutron stars.
 It is also relevant for
the possibility of a transition to new forms of deconfined nuclear matter
 \cite{DitarX0210}.

\subsubsection{Symmetry Pressure and Symmetry Incompressibility}
%\addtocontents{toc}{\hspace{0.55cm}\thesubsubsection \hspace{0.12cm}
%Symmetry Pressure and Symmetry Incompressibility}

From the previous analysis we expect to see also 
interesting $\delta$-effects on
the slope ({\it symmetry pressure}) and curvature ({\it symmetry 
incompressibility})
of the symmetry energy around $\rho_0$. These quantities and their relevant 
physical meaning were already discussed in Sect.\ref{eos} in a non-relativistic
framework.

From Eq.(\ref{eq.11}) we get the potential contribution to the 
density variation
of $E_{sym}$ (after some algebra):
\begin{eqnarray}\label{eq.23}
\frac{\partial E_{sym}}{\partial\rho_B} \mid _{pot} =
 \frac{1}{2} [f_\rho - f_\delta (\frac{M^*}{E_F})^2] +
 \nonumber \\ 
~~~~~~~~~~~~~~~~~~~~~~~~~~~~~~f_\delta (\frac{M^* k_F}{E_F^2})^2
 [\frac{1}{3} - 
 \frac{\rho_B}{M^*}\frac{\partial M^*}{\partial\rho_B}].
\end{eqnarray}
Around normal density $\rho_0$ the first term is fixed by the
$a_4$ value (cfr. Eq.(\ref{eq.11})).
The second term (always positive since
$({\partial M^*}/{\partial\rho_B}) < 0$) gives
a net increase of the slope, due to the $\delta$-field introduction,
 as seen also in Fig.\ref{qhd2}. 
To be more quantitative, with our parametrization we get a potential
contribution to the slope $L$ of $45MeV$ from the first term
and a genuine $\delta$-contribution of about $20MeV$ from the 
second one. When we include also the kinetic part (from the
first term of Eq.(\ref{eq.11})) we have a total slope parameter going from
$L(\rho)=+84MeV$ to $L(\rho+\delta)=+103MeV$. 

We note again that the slope parameter, or equivalently the
{\it Symmetry Pressure} $P_{sym} \equiv \rho_0 L/3$, is of great
importance for structure properties, being linked to the
thickness of the neutron skin in n-rich (stable and/or unstable)
nuclei  \cite{Subneutron,BrownPRL85,HorowitzPRL86,FurnstNPA706} , 
and to the position
of the drip-line. Moreover the same parameter
gives an estimate of the shift of the saturation density
with asymmetry (at the lowest order in $I^2$),
easily obtained from a linear expansion around
the symmetric value $\rho_0(I=0)$ as shown in Sect.\ref{eos}, \cite{Erice98}. 

It is instructive to perform a similar analysis for the
curvature parameter $K_{sym}$. Now the potential
contribution is {\it exclusively} given by the $\delta$-meson,
with a definite positive sign, as we can see from the previous 
discussion.
It is a large effect on the total since the kinetic
part of the symmetry incompressibility is quite small \cite{Curvature}. 
Compared to
non-relativistic effective parametrizations, in Sect.\ref{eos},
when we add the $\delta$-meson,
we move from a linear to a roughly parabolic $\rho_B$-dependence of
the symmetry energy.

With our parameters we pass from a $K_{sym}(\rho) = +7 MeV$
(only kinetic) to a $K_{sym}(\rho+\delta) = +120 MeV$ \cite{Curvature}.
Thus this quantity appears extremely interesting to look at experimentally.
The problem is that the effect on the total incompressibility
of asymmetric matter, that likely could be easier to measure,
is not trivial. 
What really matters for the total incompressibility is the 
combination $(K_{sym}-6L)$, as shown in Sect.\ref{eos}, with the 
possibility of a compensation 
between the two terms. Just by chance this is actually what is
happening in
our calculations since for the above combination we get
$-497 MeV$ in the case of only $\rho$-coupling, and $-504 MeV$
when we add also the $\delta$-field.

\subsubsection{Finite Temperature Effects}
%\addtocontents{toc}{\hspace{0.55cm}\thesubsubsection \hspace{0.12cm}
%Finite Temperature Effects}
In the temperature range of interest in this paper, below the
critical temperature $T_c$ of the liquid-gas phase transition
of the order of $15-16$ MeV, 
temperature effects
on the symmetry properties are not expected to be large. Indeed
the contributions of antifermions, that could modify all the terms
with the scalar densities are still very reduced.
For both cases, $\rho$ and 
$(\rho+\delta)$, the temperature variation of symmetry energy is quite 
small. We have a reduction
mainly coming from the kinetic contribution due to the smoothing of the
$n/p$ Fermi distributions. We note that this result is in full agreement
with relativistic Brueckner-Hartree-Fock calculations \cite{HuberPRC57}.

The temperature effect on the nucleon mass splitting, given by a difference 
of $n/p$ scalar densities Eq.(\ref{psms}), is even smaller. In the following 
subsection we 
will study in detail the
phase diagram of heated asymmetric nuclear matter, focussing in
particular on the instability regions.

\subsection{Mechanical and chemical instabilities}

Heavy-ion collisions can provide the possibility of studying
equilibrated nuclear matter far away from normal conditions,
i.e. to sample new regions of the $NM$ phase diagram.
In particular the process of multifragmentation allows to probe
dilute nuclear matter at finite temperatures. In the symmetric case
we expect to see
a phase transition of first order of liquid-gas type, as suggested
from the very first equations of state built with effective interactions
\cite{SauerNPA264,BertschPLB126,JaqamanPRC29}.

As discussed extensively in Sect.\ref{rpa} in a non-relativistic framework, 
for asymmetric nuclear matter
a qualitatively new feature in the liquid-gas phase transition
is expected, the onset of a coupling to chemical instabilities 
(component separation). This will show
up in a novel nature of the unstable modes, the mixture of
density and charge fluctuations leading to an {\it Isospin Distillation}. 
Indeed, equilibrium thermodynamics 
as well as
non-equilibrium kinetics both  predict
that an asymmetric system will separate into more symmetric larger
fragments ("liquid" phase) and into neutron-rich light fragments 
("gas" phase).
Since the effect is driven by the isospin dependent part of the
nuclear equation of state, here we will look at the influence of
the $\delta$-coupling on this new liquid-gas phase transition.

As discussed earlier the instability condition
 of a two-component, $n/p$, thermodynamical system is given by

\begin{eqnarray}\label{instab}
\left(\frac{\partial P}{\partial \rho}\right)_{T,y}
\left(\frac{\partial \mu_{p}}{\partial y}\right)_{T,P} < 0,
\end{eqnarray}
where P is the pressure, $\mu_{p}$ is the proton chemical potential 
and $y$ the proton fraction $Z/A$, related to the asymmetry parameter
$I=1-2y$.
Eq.(\ref{instab}) is equivalent to constrain the free energy
to be a convex function in the space
of the $n,p$ density oscillations, $\delta\rho_n,\delta\rho_p$.
In charge symmetric matter
{\it isoscalar} (total density) $\delta\rho_n+\delta\rho_p$ and 
{\it isovector} (concentration)
 $\delta\rho_n-\delta\rho_p$ oscillations are not coupled and we
have two separate conditions for instability:
\begin{eqnarray}\label{mcinstab}
 \left(\frac{\partial P}{\partial \rho}\right)_{T,y} \leq 0,~~{\it mechanical},~ i.e.~ vs.~ density~ oscillations,~ and
 \nonumber \\
%\begin{eqnarray}\label{cinstab}
 \left(\frac{\partial \mu_{p}}{\partial y}\right)_{T,P} \leq 0,~~{\it chemical},~ i.e.~ vs.~ concentration~ oscillations.
\end{eqnarray}

In asymmetric matter the isoscalar and isovector modes are coupled
and the two separate inequalities do not anymore mantain a physical 
meaning, in the sense that they do not select the nature of the instability.
Inside the general condition Eq.(\ref{instab}) the corresponding
unstable modes are a mixing of density and concentration oscillations,
very sensitive to the charge dependent part of the nuclear interaction
in the various instability regions \cite{BaranPRL86}.

In dilute asymmetric $NM$ (n-rich) the normal unstable modes for all
realistic effective interactions are still {\it isoscalar-like},
i.e. in phase $n-p$ oscillations but with a larger proton
component. This leads to a more symmetric high density
(liquid) phase everywhere under the instability line defined by
Eq.(\ref{instab}) and consequentely to a more neutron-rich gas ({\it Isospin 
Distillation}). Such "chemical effect" is driven by the increasing 
symmetry repulsion going from low to roughly the saturation density
and so it appears rather
sensitive to the symmetry energy of the used effective interaction 
at subnuclear densities. 

In this section we study  the effect of the $\delta$-coupling on
the instability region given by Eq.(\ref{instab}) in dilute asymmetric nuclear
matter  and on the structure of the corresponding unstable modes.
We start from an identity valid for any binary thermodynamical system
\begin{eqnarray}\label{binid}
\left(\frac{\partial \mu_{p}}{\partial \rho_{p}}\right)_{T,\rho_{n}} 
\left(\frac{\partial \mu_{n}}{\partial \rho_{n}}\right)_{T,\rho_{p}}-
\left(\frac{\partial \mu_{p}}{\partial \rho_{n}}\right)_{T,\rho_{p}} 
\left(\frac{\partial \mu_{n}}{\partial \rho_{p}}\right)_{T,\rho_{n}} =
 \nonumber \\
\frac{1}{(1-y)\rho^2}
\left(\frac{\partial P}{\partial \rho}\right)_{T,y}
\left(\frac{\partial \mu_{p}}{\partial y}\right)_{T,P} 
\end{eqnarray}
where $\mu_{q}$,  $\rho_{q}$ ($q = n,p$) are 
respectively neutron/proton
 chemical potentials and densities. Thus from the
chemical potential on each isotherm, Eq.(\ref{mu}), we can easily 
compute the limits of the instability region in the $T,\rho_B$
plane for dilute asymmetric $NM$ in the two choices, without and with
the $\delta$-meson.

\begin{figure}
%\epsfysize=6.7cm
%\centerline{\epsfbox{liubo_fig7.ps}}
\begin{center}
\includegraphics*[scale=0.60]{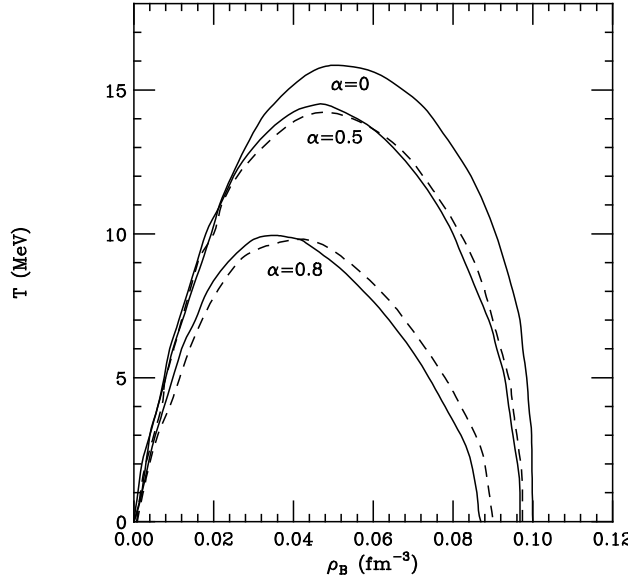}
\caption{
Limits of the instability region in the $T,\rho_B$ plane
for various asymmetries.
Dashed lines: $NL\rho)$. Solid lines: $NL\rho\delta)$ [$\alpha$ being the asymmetry parameter].}
\label{qhd6}
\end{center}
\end{figure}

The results are shown in Fig.\ref{qhd6}. 
The inclusion of a $\delta$-field (solid lines)
appears not to affect  
much the instability limits even at relatively
large asymmetry $I=0.8$ ($N \simeq 9Z$). We notice just a 
small reduction and a shift
to the left (lower densities) of the whole region: this can be related
to a slightly larger symmetry repulsion. 
We can understand the relatively small $\delta$-effect on the stability 
border by remembering that for low densities, well below $\rho_0$, 
the symmetry term has roughly the same linear behaviour in both $(\rho)$
and $(\rho+\delta)$ schemes, fixed by the $a_4$ parameter (see the
discussion after Eq.(\ref{eq.11})).  
Only for very large asymmetries
it appears
relatively easier in the $\delta$ case to be 
in the stable liquid 
phase.

\begin{figure}
%\epsfysize=6.7cm
%\centerline{\epsfbox{liubo_fig8.ps}}
\begin{center}
\includegraphics*[scale=0.55]{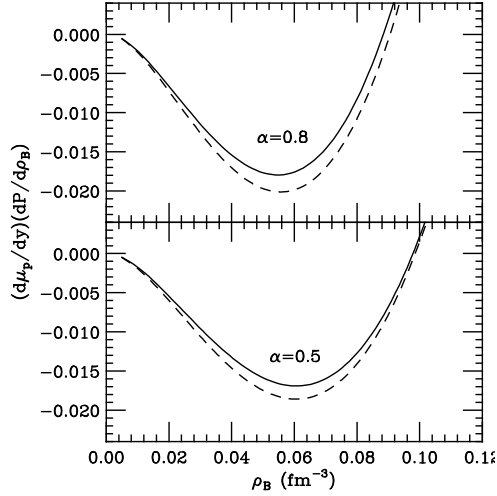}
\caption{
The quantity Eq.(\ref{instab}) inside the instability region
at $T=0$ and various asymmetries.
Dashed lines: $NL\rho$. Solid lines: $NL\rho\delta$.}
\label{qhd7}
\end{center}
\end{figure}

A larger difference can be seen  in the behavior of the
quantity Eq.(\ref{instab}) inside the instability region.
This is plotted in Fig.\ref{qhd7} for various asymmetries at zero temperature.
 The solid curves (with $\delta$-coupling) are systematically
above the dashed ones, possible signature of a weaker instability.
In fact this is not the case since, as shown in the following, the
$\delta$-meson is not affecting the low density unstable modes.
In order to better understand the origin of this effect we study
 the structure of the corresponding unstable modes. We follow the 
Landau dispersion relation approach to small amplitude oscillations
in Fermi liquids, as in Sect.\ref{rpa}. For a two component
($n,p$) matter the interaction is characterized by the Landau
parameters $F_0^{q,q'}, (q,q')=(n,p)$ defined as
\begin{equation}\label{eq.30}
N_q(T) \frac{\partial\mu_q}{\partial\rho_{q'}} \equiv \delta_{q,q'}
 + F_0^{q,q'}
\end{equation} 
where $N_q(T)$ represents the single particle level density
 at the Fermi energy. At zero temperature it has the simple form ($\hbar=c=1$)
$
N_q = \frac{k_{Fq} E^*_{Fq}}{\pi^2},~~~~~q = n, p.
$

In the symmetric case ($F_0^{nn}=F_0^{pp}, F_0^{np}=F_0^{pn}$), the 
Eqs.(\ref{mcinstab}) correspond to the two Pomeranchuk 
instability conditions
\begin{eqnarray}\label{eq.31}
F_0^s = F_0^{nn} + F_0^{np} < -1 ~~~~mechanical \nonumber \\
F_0^a = F_0^{nn} - F_0^{np} < -1 ~~~~chemical.
\end{eqnarray} 
From the dispersion relations $F_0^s$ will give the properties of the density
(isoscalar) modes while $F_0^a$ is related to the concentration 
(isovector) modes.
For asymmetric $NM$ we have corresponding generalized Landau
parameters $F_{0g}^s, F_{0g}^a$ which characterize the asymmetric collective
response. They
can be expressed as 
fixed combinations of the $F_0^{q,q'}$ for each baryon density,
asymmetry and temperature. This transformation
reduces Eq.(\ref{binid}) to a "diagonal" form  \cite{BaranPRL86}
\begin{eqnarray}\label{eq.32}
(1+F_{0g}^s)(1+F_{0g}^a) = 
\frac{4}{(1-y)\rho^2}\left(\frac{N_nN_p}{N_n+N_p}\right)^2
 \nonumber \\
\times\left(\frac{\partial P}{\partial \rho}\right)_{T,y}
\left(\frac{\partial \mu_{p}}{\partial y}\right)_{T,P} 
\end{eqnarray}

\begin{figure}
%\epsfysize=6.5cm
%\centerline{\epsfbox{liubo_fig10.ps}}
\begin{center}
\includegraphics*[scale=0.55]{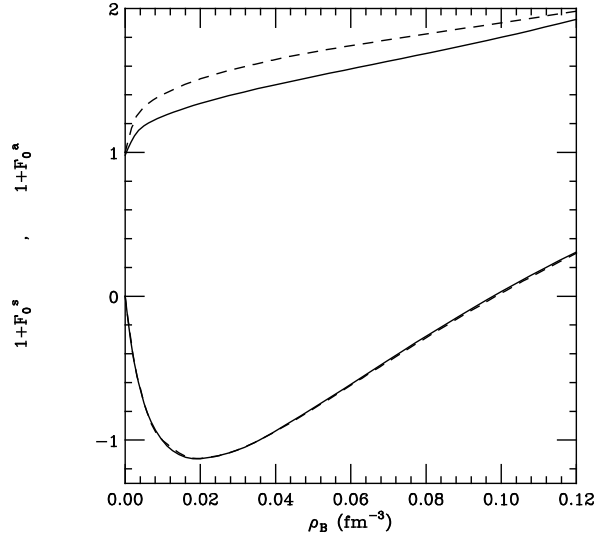}
\caption{
Behaviour of the generalized Landau parameters $F_{0g}^{a,s}$
inside the instability region at zero temperature and
asymmetry $I=0.5$ ($N=3Z$).
Dashed lines: $NL\rho$. Solid lines: $NL\rho\delta$.}
\vskip 0.5cm
\label{qhd8}
\end{center}
\end{figure}

As already discussed, in the unstable region of dilute asymmetric $NM$
we have {\it isoscalar-like} unstable modes with $1+F_{0g}^s<0$,
while the combination $1+F_{0g}^a$ will always stay positive.
In Fig.\ref{qhd8} we report a full calculation of these two quantities
in the unstable region at zero temperature, for asymmetry $I=0.5$,
 with and without
the $\delta$-coupling.
The $\delta$-meson almost does not affect at all the unstable mode,
given by the $F_{0g}^s$ parameter. Hence the limits of the instability
region, for $I=0.5$, are not changed 
(see the Figs.\ref{qhd6},\ref{qhd7}). We have a 
larger
effect on the $F_{0g}^a$ parameter which describes
"stable" {\it isovector-like} modes, that actually can propagate
as good zero-sound collective motions since $F_{0g}^a>0$.
This can be expected from the isovector nature of the $\delta$-meson. 
From here
we get the main differences seen in Fig.\ref{qhd7} for the  
Eq.(\ref{instab}) 
inside the
instability region, just the product of the two quantities plotted 
in Fig.\ref{qhd8} (from Eq.(\ref{eq.32})).

Another interesting aspect of the comparison between Fig.\ref{qhd7} and 
Fig.\ref{qhd8}
is the shift of the "maximum instability" density region. From the
thermodynamical condition reported in Fig.\ref{qhd7} it seems that the
largest instability (the most negative value) is around 
$\rho_B=0.06fm^{-3}$. In fact from Fig.\ref{qhd8} we see that the fastest
unstable mode, corresponding to the most negative Pomeranchuk
condition for $1+F_{0g}^s$, is actually present for more
dilute matter, around $\rho_B=0.02fm^{-3}$. This shows the
relevance of  the linear response analysis.

Finally the fact that the $\delta$-coupling mostly affects
the {\it stable,  isovector-like,} modes is of great interest for
possible effects on the isovector Giant Dipole Resonances
studied around normal density within the $RMF$ approach
in asymmetric systems. This will be shown in the next Section
on the relativistic linear response in asymmetric matter.

\subsection{Nucleon Effective Mass Splitting}

\begin{figure}
%\epsfysize=6.7cm
%\centerline{\epsfbox{liubo_fig6.ps}}
\begin{center}
\includegraphics*[scale=0.60]{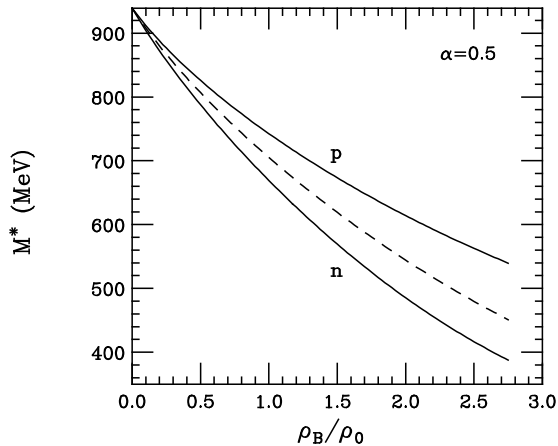}
\caption{
Neutron and proton effective masses vs. the baryon density 
for $I=0.5$ ($N=3Z$), $NL\rho\delta$ parameters. The dashed line corresponds to symmetric nuclear
matter.}
\label{qhd5}
\end{center}
\end{figure}

An important qualitatively new result of the $\delta$-meson coupling is
the $n/p$-effective mass splitting in asymmetric matter 
\cite{KubisPLB399,LiuboPRC65},
see Eq.(\ref{psms}). In Fig.\ref{qhd5} we report the baryon 
density dependence of the
$n/p$ effective masses for $I=0.5 ~(N=3Z)$ asymmetry, calculated 
with our $NL\rho\delta$ parameters, compared to the symmetric case \cite{Rhos}.
We see a splitting of the order of $15\%$ at normal density $\rho_0$,
 increasing with baryon density. Unfortunately, from the present nuclear 
data we have a very limited knowledge of this effect, due to the low
asymmetries available. However we can expect important effects
on transport properties ( fast particle emission, collective flows)
 of the dense and asymmetric $NM$ that will
be reached in Radioactive Beam collisions at intermediate energies.

The sign itself of the splitting would be very instructive, since it is a quite
controversial point (see the comments at the end of Sect.\ref{eos}). 
As we
can see from Eq.(\ref{psms}), in a relativistic approach in $n$-rich systems we have a neutron effective
mass always smaller than the proton one.  
We note again that a decreasing neutron 
effective mass in $n$-rich matter is a direct consequence of the relativistic mechanism
for the symmetry energy, i.e. the balance of scalar (attractive) 
and vector (repulsive)
contributions in the isovector channel.

\subsubsection{Dirac and Schr\"odinger Nucleon Effective Masses in 
Asymmetric Matter}
%\addtocontents{toc}{\hspace{0.55cm}\thesubsubsection \hspace{0.12cm}
%Dirac and Schr\"odinger Nucleon Effective Masses in 
%Asymmetric Matter}
The prediction of a definite $m^*_n<m^*_p$ effective mass splitting
in $RMF$ approaches, when a scalar $\delta-like$ meson is included,
is an important result that requires some further analysis, in 
particular relative to the controversial predictions of non-relativistic
approaches, see Sect.\ref{eos}.

Here we are actually discussing the {\it Dirac} effective masses 
$m^*_D(n,p)$,
i.e. the effective mass of a nucleon in the in-medium Dirac equation
with all the meson couplings. The relation to the {\it Schr\"odinger}
effective masses $m^*_S(n,p)$, i.e. the ``k-mass'' due to the momentum 
dependence
of the mean field in the non-relativistic in-medium Schr\"odinger
equation is not trivial, see 
\cite{BouyssyPRC36,SerotANP16,JaminonPRC22,JaminonPRC40}.
We will extend the argument of the refs.
\cite{JaminonPRC22,JaminonPRC40} to the case of
asymmetric matter.

We start from the simpler symmetric case without self-interacting terms.
The nucleon Dirac equation in the medium contains the scalar
self-energy $\Sigma_s = -f_\sigma \rho_S$ and the vector self-energy
(fourth component) $\Sigma_0 = f_\omega \rho_B$ and thus the
corresponding energy-momentum relation reads:
\begin{equation}\label{enmom}
(\epsilon + m - \Sigma_0)^2 = p^2 + (m + \Sigma_s)^2 = p^2 + {m_D^*}^2 
\end{equation}
i.e. a dispersion relation
\begin{equation}\label{disp}
\epsilon = - m + \Sigma_0 + \sqrt{p^2 + {m_D^*}^2} 
\end{equation}

From the total single particle energy $E = \epsilon + m$ expressed in the form
$E = \sqrt{k_{\infty}^2 + m^2}$, where $k_{\infty}$ is the relativistic
asymptotic momentum, using  Eq.(\ref{enmom}) we can get the relation
\begin{eqnarray}\label{ueff}
\frac{k_{\infty}^2}{2m} = \epsilon + \frac{\epsilon^2}{2m} =
 ~~~~~~~~~~~~~~~ ~~~~~~~~~~~~~~~ ~~~~~~~~~~~~~~~ \nonumber\\
  \frac{p^2}{2m} + \Sigma_s + \Sigma_0 + \frac{1}{2m}(\Sigma_s^2 - \Sigma_0^2)
 + \frac{\Sigma_0}{m}\epsilon \equiv \frac{p^2}{2m} + 
 U_{eff}(\rho_B,\rho_s,\epsilon)
\end{eqnarray}
i.e. a Schr\"odinger-type equation with a momentum dependent mean field
that with the dispersion relation Eq.(\ref{disp}) is written as
\begin{eqnarray}\label{unonrel}
U_{eff} =  \Sigma_s + \frac{1}{2m}(\Sigma_s^2 + \Sigma_0^2)
 + \frac{\Sigma_0}{m}\sqrt(p^2 + {m_D^*}^2) \nonumber\\
 \simeq \Sigma_s + \frac{\Sigma_0 m_D^*}{m} + 
 \frac{1}{2m}(\Sigma_s^2 + \Sigma_0^2) + \frac{p^2}{2m}\frac{\Sigma_0}{m_D^*}
\end{eqnarray}
The relation between Schr\"odinger
and Dirac nucleon effective masses is then
\begin{equation}\label{effmas}
m_{S}^* = \frac{m}{1 + \frac{\Sigma_0}{m_D^*}} = m_D^* \frac{m}{m + 
 \Sigma_s + \Sigma_0}
\end{equation}
Since at saturation the two self-energies are roughly compensating each other,
$\Sigma_s + \Sigma_0 \simeq -50MeV$ the two effective masses
are not much different, with the $S-mass$ slightly larger than the $D-mass$.

% $f_\sigma = (g_\sigma/m_\sigma)^2$, 
%$A = a/g_\sigma^3$, $B = b/g_\sigma^4$, $f_\omega = (g_\omega/m_\omega)^2$, 
%$f_{\rho} = (g_{\rho}/m_{\rho})^2$, $f_{\delta} = (g_{\delta}/m_{\delta})^2$. 
%$f_i \equiv (\frac{g_i^2}{m_i^2})$, $i = \sigma, \omega, \rho, \delta$,

%\begin{equation}\label{eq.6}
%{M_i}^\star=M_{N}-f_\sigma\rho_S \mp f_\delta\rho_{S3}~~~ 
%(-~proton, +~neutron).
%\end{equation}

%\begin{eqnarray}\label{mu}
%{\mu_i}=\mu_i^\star  - f_\omega\rho_B \mp f_{\rho} \rho_{B3}~~~
%(-~proton, +~neutron),
%$\end{eqnarray}

In the case of asymmetric matter, neutron-rich as always considered here,
we can have two cases:
\begin{itemize}
\item {\it Only $\rho~meson$ coupling}

Now the scalar part is not modified, we have the same scalar self energies
$\Sigma_s$ for neutrons and protons and so the same Dirac masses. The vector
self energies will show an isospin dependence with a new term
$\mp f_{\rho}\rho_{B3}$, repulsive for neutrons ($-$ sign, since we
use the definition $\rho_{(B,S)3} \equiv \rho_{(B,S)p}-\rho_{(B,S)n}$).
As a consequence we see a splitting at the level of the
Schr\"odinger masses since Eq.(\ref{effmas}) becomes 

\begin{equation}\label{effmasrho}
{m_{S}^*}(n,p) = m_D^* \frac{m}{m + 
 (\Sigma_s + \Sigma_0)_{sym} \mp f_{\rho}\rho_{B3}} 
\end{equation}

in the direction of $m_n^*<m_p^*$ (here and in the following upper signs are for neutrons). 

\item{\it $\rho+\delta$ coupling}

The above splitting is further enhanced by the direct effect of the
scalar isovector coupling, see Eq.(\ref{psms}). Then the
Schr\"odinger masses are

\begin{equation}\label{effmasrhodelta}
{m_{S}^*}(n,p) = 
({m_D^*}_{sym} \pm f_{\delta}\rho_{S3}) \frac{m}{m + 
 (\Sigma_s + \Sigma_0)_{sym} \mp (f_{\rho}\rho_{B3} - f_{\delta}\rho_{S3})}
\end{equation}

The new term in the denominator will further contribute to the $m*_n<m*_p$
splitting since we must have $f_{\rho}>f_{\delta}$ in order to get
a correct symmetry parameter $a_4$. The effect is larger at higher
baryon densities because of the decrease of the scalar $\rho_{S3}$, due
to the faster $\frac{m_{Dn}^*}{E_{Fn}^*}$ reduction of $\rho_{Sn}$.
\end{itemize}

Actually in a non-relativistic limit we can approximate in Eq.(\ref{ueff})
directly the energy $\epsilon$ with the asymptotic kinetic energy
leading to the much simpler relation:

\begin{equation}\label{nreleffmas}
m_{S}^* = \frac{m}{1 + \frac{\Sigma_0}{m}} \simeq {m - \Sigma_0}  
 = m_D^* - (\Sigma_s + \Sigma_0).
\end{equation}

In the case of asymmetric matter this leads to the more transparent
relations
\begin{itemize}
\item {\it Only $\rho~meson$ coupling}

\begin{equation}\label{nrelmasrho}
{m_{S}^*}(n,p) = m_D^* - (\Sigma_s + \Sigma_0)_{sym} \pm f_{\rho}\rho_{B3}
\end{equation}

\item{\it $\rho+\delta$ coupling}

\begin{equation}\label{nrelmasrhodelta}
{m_{S}^*}(n,p) = 
{m_D^*}_{sym} - (\Sigma_s + \Sigma_0)_{sym}
\pm (f_{\rho}\rho_{B3} - f_{\delta}\rho_{S3})
\end{equation}

\end{itemize}

In conclusion any relativistic effective field model will predict
the definite isospin splitting of the nucleon effective masses $m_n^*<m_p^*$
in the non-relativistic limit. We expect an increase of the difference
of the neutron/proton mean field at high momenta, with important
dynamical contributions that will enhance the transport effects of the 
symmetry energy.

\subsubsection{The Dirac-Lane Potential}
%\addtocontents{toc}{\hspace{0.55cm}\thesubsubsection \hspace{0.12cm}
%The Dirac-Lane Potential}
In the non-relativistic limit of the Eq.(\ref{ueff}) we can easily extract 
the neutron-proton mean optical potential using the isospin dependence
of the self-energies

\begin{eqnarray}\label{npself}
\Sigma_{0,q} = \Sigma_{0,sym} \mp f_{\rho}\rho_{B3}~,
\nonumber \\
\Sigma_{s,q} = \Sigma_{s,sym} \pm f_{\delta}\rho_{S3}~, 
\end{eqnarray}

With some algebra we get a compact form of the $Dirac-Lane~Potential$

\begin{eqnarray}\label{diraclane}
{U_{Dirac-Lane}} \equiv \frac{U_n-U_p}{2I} = \nonumber \\
\rho_0 \Big[ f_{\rho} (1- \frac{\Sigma_{0,sym}}{m}) - 
f_{\delta}{\frac{\rho_{S3}}{\rho_{B3}}}(1+ \frac{\Sigma_{s,sym}}{m})\Big]
 + f_{\rho} \frac{\rho_0}{m} \epsilon
\end{eqnarray}  

It is interesting to compare with the related discussion presented
in Sect.\ref{eos} for the non-relativistic effective forces, mainly of
Skyrme-like form.
First of all we predict a definite positive $E-slope$ given by 
the quantity $f_{\rho} \frac{\rho_0}{m}$, which is actually not large
within the simple $RMF$ picture described here. This is a obvious
consequence of the fact that the ``relativistic'' mass splitting is always
in the direction $m_n^*<m_p^*$.
The realistic magnitude of the effect could be larger if we take into account that
some explicit momentum dependence should be included in the scalar and
vector self energies, as discussed in ref.\cite{GiessenRPP56}.
An interesting point in this direction comes from the phenomenological
Dirac Optical Potential ($Madland-potential$) constructed in the refs.
\cite{KozackPRC39,KozackNPA509}, fitting simultaneously proton and
neutron (mostly total cross sections) data for collisions with a wide
range of nuclei at energies up to $100~MeV$. Recently this Dirac optical
potential has been proven to reproduce very well the new neutron 
scattering data on $^{208}Pb$ at $96~MeV$ \cite{KlugPRC6768} measured at
the Svendberg Laboratory in Uppsala.

\begin{figure}
%\epsfysize=6.5cm
%\centerline{\epsfbox{liubo_fig10.ps}}
\begin{center}
\includegraphics*[scale=0.55]{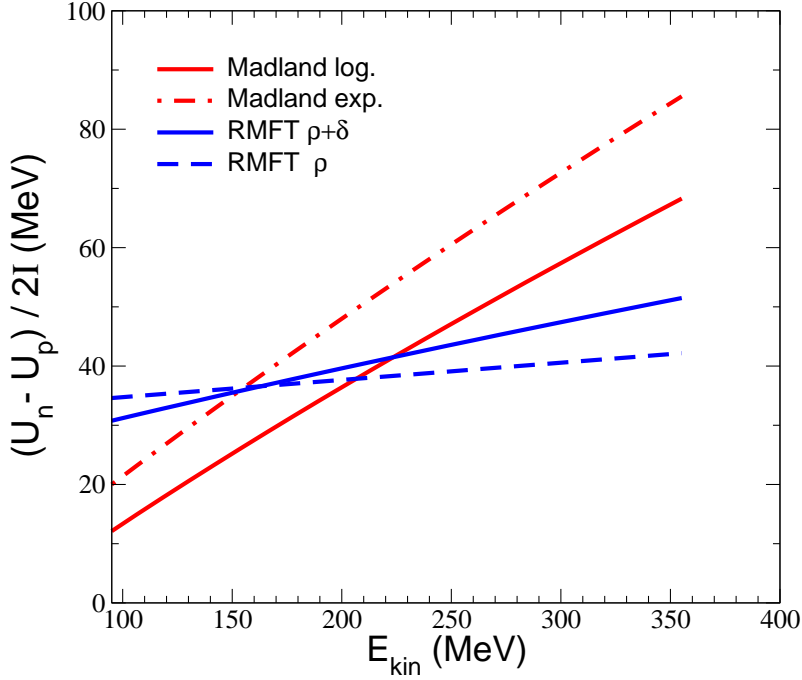}
\caption{
Energy dependence of the Dirac-Lane potential in the $RMF$ picture
(solid: $NL\rho$; dashed: $NL\rho\delta$) and in the phenomenologic
Dirac Optical Model of Madland et al. \cite{KozackPRC39,KozackNPA509},
see text.}
\label{dlane}
\end{center}
\end{figure}

The phenomenological $Madland-potential$ has different implicit 
momentum dependences 
($exp/log$) in the self-energies, \cite{KozackPRC39,KozackNPA509}.
In Fig.\ref{dlane} we show the corresponding energy dependences for the
$Dirac-Lane$ potentials, compared to our $NL\rho$ and $NL\rho\delta$
estimations. When we add the $\delta$-field we have a larger slope since
the $\rho-coupling$ should be increased.
The slopes of the phenomenological potentials are systematically
larger, interestingly similar to the ones of the $Skyrme-Lyon$ forces of the 
Fig.\ref{fig:elane} of Sect.\ref{eos}. We note that in the $Madland-potential$
the Coulomb interaction is included, i.e. an extra repulsive vector 
contribution for the protons.
The first, not energy dependent, term of Eq.(\ref{diraclane}) is directly
related to the symmetry energy at saturation, exactly like in the
non-relativistic case, see Eq.(\ref{lane}) of Sect.\ref{eos}.

The conclusion is that a good {\it systematic} measurement of the Lane 
potential in a wide
range of energies, and particularly around/above $100~MeV$,
would answer many fundamental questions in isospin physics.

%\include{rep_bib}

%\end{document}

%% file: Chapter-7.tex
%\documentclass{elsart}

%\usepackage{graphicx}

%\usepackage{amssymb}

%\begin{document}

\setcounter{figure}{0}
\setcounter{equation}{0}

\section{Collective modes of asymmetric nuclear matter in 
the relativistic approach}\label{relin}

\markright{Chapter \arabic{section}: relin}

\subsection{Linear Response Equations}

We discuss here collective oscillations 
that propagate in cold nuclear matter under the influence of the 
mean field dynamics.
These studies can be considered a relativistic extension of the method
introduced by Landau to study liquid-$^3He$ 
\cite{LandauSP89,AbrikosovRPP22,BaymBook78}    and
recently applied to investigate stable and unstable modes in nuclear matter
\cite{PethickAP183,ColonnaPLB428,BaranNPA632}.
 The starting point is
the relativistic kinetic transport equation of Sect.\ref{qhd}, 
 Eq.(\ref{trans}).
 We look for solutions corresponding to small 
oscillations of the Wigner function ${\hat F}(x,p)$ around the 
equilibrium value. 
Therefore we put 
\begin{equation}
{\hat F}(x,p)={\hat H}(p)+{\hat G}(x,p)
\end{equation}
where $\widehat H(p)$ is the distribution function at equilibrium 
and 
$\widehat G(x,p)$ represents its fluctuations, ref.\cite{GrecoPRC67}.
 
In the linear approximation, i.e. neglecting terms of second order in 
$\widehat G(x,p)$, 
the equations for the Wigner functions become 
\begin{eqnarray}\label{elin1}
&&{i\over2}{\partial_\mu}\gamma^\mu{\hat G}_{(i)}(x,p)+
\bigl({\Pi_\mu} \pm \tilde f_\rho\,b_{\mu}\bigr)\gamma^\mu{\hat G}_{(i)}(x,p)
-M^*_i\,{\hat G}_{(i)}(x,p)=\nonumber\\
&&(1-{i\over2}\Delta)\bigl(\hat{F}(x) \mp \hat{F}_{3}(x)\bigr)
{\hat H}_{(i)}(p)~, 
\end{eqnarray}
for neutrons and protons ($i=n,p$, upper/lower signs),
 where $M^*_i=M- \tilde f_\sigma \rho_S \pm 
\tilde f_\delta\,\rho_{S3}$. 
The quantities ${\hat {F}}(x)$ and ${\hat {F}}_{3}(x)$  
are the isoscalar and the isovector components of the self--consistent field: 
\begin{eqnarray}
\hat{F}(x)=-8{\tilde f_\sigma}G(x)
+8{\tilde f_\omega}\gamma_\mu{G^\mu}(x) 
-8{{\partial {\tilde f_\sigma}} \over {\partial \rho_S}} \rho_S G(x)
-8{{\partial {\tilde f_\sigma}} \over {\partial j_\mu}} \rho_S  
G^\mu(x)\nonumber\\ 
-8{{\partial {\tilde f_\sigma}} \over {\partial \rho_{S3}}} \rho_S G_3(x)
-8{{\partial {\tilde f_\sigma}} \over {\partial j_{3\mu}}} \rho_S G_3^\mu(x) 
+8{{{\partial {\tilde f_\omega}} \over {\partial \rho_S}}} 
\gamma_\mu j^\mu G(x)
~,
\end{eqnarray}

\begin{equation}
\hat{F}_{3}(x)=-8 \tilde f_\delta G_{3}(x)
+8 \tilde f_\rho\gamma_\mu G^\mu_{3}(x)
-8{{\partial {\tilde f_\delta}} \over {\partial \rho_S}}~\rho_{S3}~G(x)
+8{{\partial {\tilde f_\rho}} \over {\partial \rho_S}} \gamma_\mu j_3^\mu 
G(x)~.
\end{equation}

Following the notation of Sect.\ref{qhd} the coupling functions $\tilde f_i$
are in general density dependent due to the Fock contributions.
The Hartree approximation is recovered by vanishing all the 
derivative terms in 
the quantities $\hat{F}(x)$ and $\hat{F}_{3}(x)$ (except 
${{\partial {\tilde f_S}} \over {\partial \rho_S}}$, since still 
$\tilde f_\sigma=\Phi(\rho_S)/\rho_S$). 

We obtain equations for the collective oscillations by  
multiplying Eqs. (\ref{elin1}) by $\gamma_{\lambda}$. After
performing the traces,  we equate to zero both the real and 
imaginary parts of 
the result \cite{DellaPRC44,MateraPRC49}. Furthermore, by  
Fourier transforming and integrating over four--momentum, we get the 
set of equations  for the scalar and vector 
fluctuation of each species, \cite{GrecoPRC67}.
 We note that the formalism developed here in Hartree-Fock approximation
 includes a
 linear response theory valid also for any other 
approach to $QHD$ beyond $RMF$. Namely when we consider in general  
a dependence on the baryon density of all meson-nucleon couplings,
like in the Density Dependent Hadronic ($DDH$) model
\cite{FuchsPRC52,FuchsNPA589,TypelNPA656,TypelPRC67},
inspired by the Dirac-Brueckner-Hartree-Fock ($DBHF$) approach to account for
many-body correlations.

The set of equations developed in the Hartree-Fock approximation
includes the ones corresponding to the usual Hartree approximation 
($RMF$). As already mentioned, it
is easily obtained by considering the coupling $\tilde f_i$ to each channel
equal to the coupling constant of the corresponding meson.
The result is of appreciable plainer structure due to 
the constant value of all couplings, except $\tilde f_\sigma$ 
\cite{phinote}.
Since the physics results become more transparent, in the following
 we will stay in the
 Hartree scheme, keeping well in mind that the Fock 
contributions can be easily
included. In fact they amount to having some extra contributions in the 
various 
interaction channels
without qualitative modifications of the physical response.

The normal collective modes are plane waves, characterized by the wave vector
($k^\mu=(k^0,0,0,\mid \bf{k}\mid))$. They are determined by solving a 
set of homogeneous linear equations. The solutions 
correspond only to longitudinal waves
and do not depend on
$k^0$ and $\vert {\bf k} \vert$ separately, but only on the ratio
$$v_s=\frac{k^0}{\vert {\bf k} \vert}~.$$
The zero-sound velocities are given by those values of $v_s$ for which the
relevant determinant of the set  of equations
 vanishes, i.e. by the dispersion relations. Correspondingly the 
neutron/proton structure of the eigenvectors (normal modes) can 
be derived. Again it should be noted
that in asymmetric nuclear matter isoscalar and isovector components are
mixed in the normal modes. Here this can be argued by the fact that in each  
linearized equation (\ref{elin1}) both proton/neutron densities  
and currents are appearing, \cite{GrecoPRC67}.
However we also recall that one can still identify  
isovector-like excitations as the modes where neutrons and protons move
out of phase, while isoscalar-like modes are characterized by neutrons
and protons moving in phase \cite{ColonnaPLB428,BaranPRL86}.   

\subsection{The Role of Scalar/Vector Fields in the Dynamical Response}
Before showing numerical results for the dynamical response of 
asymmetric nuclear matter
in various baryon density regions and using the different effective 
interactions,
we would like to analyse in more detail the structure of the 
relativistic linear response 
theory in order to clearly pin down the role of each meson coupling.

\subsubsection{Isovector Response}
%\addtocontents{toc}{\hspace{0.55cm}\thesubsubsection \hspace{0.12cm}
%Isovector Response}
One may perhaps expect that once the asymmetry parameter $a_4$ is fixed, the
velocity of sound is also fixed \cite{MatsuiNP365}. 
On the other hand our results will clearly show a different
dynamical response, e.g. with or without the $\delta$-meson channel, 
for interactions
which give $exactly~the~same~a_4~parameter$. 
Essentially this is due to the fact that the response depends on the slope
of the symmetry energy for isovector oscillations around the saturation point.
 In order to get a complete quantitative understanding of this 
effect we will first consider
the case of symmetric nuclear matter in the Hartree scheme, where 
the dispersion
relations assume a particularly transparent analytical form.

In this case it is also possible to decouple the collective modes 
into {\it pure} isoscalar and isovector oscillations, 
 \cite{ColonnaPLB428,BaranPRL86} as also seen in  
Sect.3. 
 After a straightforward
rearrangement  we have a dispersion relation  
for the isovector modes.
Now in order to find the zero--sound velocity one has to evaluate the
determinant of a $2 \times 2$ matrix (and not a $4\times 4$, as in the 
asymmetric  $NM$). After a straightforward
rearrangement  we have a dispersion relation \cite{GrecoPRC67} 
for the isovector modes
\begin{equation}\label{det1}
1+N_F\left[f_\rho(1-v_s^2)- f_\delta\frac{M^{*2}}{E^{*2}_F} 
\left({1-f_\delta A(k_F,M^*)}-f_\rho\, \frac{\rho_S}{M^*}\,v_s^2\right)
\right]\varphi(s)=0~~.
\end{equation}
Here $N_F=\frac{2k_F E^*_F}{\pi^2}$ is the density of states at the
Fermi surface and 
$s\equiv v_s/v_F$. $\varphi(s)$ is the usual Lindhard function of the
Landau Fermi Liquid response theory:
$$\varphi(s)=1-{s\over 2}ln\left|{s+1\over s-1}\right|+{i\over 2}
\,\pi s\,\theta(1-s)~$$
The quantity $A(k_F,M^*)$ is the same integral discussed in 
the Sect.\ref{qhd}, Eq.(\ref{eq.10}).

In fact the structure of the dispersion relation is the same for the 
isoscalar excitations,
of course one has to replace the isovector fluctuations
($\delta \rho_{B3},~\delta \rho_{S3}$)
 with the isoscalar
ones ($\delta \rho_B,~\delta \rho_{S}$), and the coupling constants
of isovector mesons with those of the isoscalar mesons.

At this point we can make the following approximation
$$v_s^2 \simeq  v_F^2 = \frac{k^2_F}{E^{*2}_F}\,,$$
to evaluate the expression inside the square brackets in Eq.(\ref{det1}).
Looking at numerical results shown later (Fig.\ref{lin5}) this is a good 
approximation to within $3\%$ for all effective interactions considered.
As in Sect.6 we also neglect the integral $A(k_F,M^*)$.  
Then Eq.(\ref{det1}) assumes the transparent form:
\begin{equation}\label{det2}
1+ \frac{6\,E^*_F}{k^2_F}\left[E_{sym}^{pot}-\frac{f_\rho}{2}
\frac{k^2_F}{E^{*2}_F}\left(1-f_\delta\,\frac{M^*}{E^{*2}_F}\,\rho_S\right)
\rho_B\right]\varphi(s)=0\,.
\end{equation}
The potential part of the symmetry energy explicitly appears in
the dispersion relations, but {\it together with an important correction term} 
which exhibits a different $f_\rho,f_\delta$ structure with respect to 
that of $E_{sym}^{pot}$,
Eq.(.....) of Sect.6. 
{\it We can easily have interactions with the same
$a_4$ value at normal density but with very different isovector response}.
E.g. when including the $\delta$ channel we have to increase the
$f_\rho$ coupling in order to have the same $a_4$, as discussed in
Sect.\ref{qhd}, but now the ``restoring force'' 
(coefficient of the 
Lindhard function in the Eq.(\ref{det2})) will be  reduced.

A similar effect has been pointed out in
a detailed non-relativistic $Skyrme-RPA$ study of the Giant Dipole
Resonance in heavy nuclei ($^{208}Pb$) using effective interactions
with various isovector terms \cite{ReinhardNPA649}. 
A separate sensitivity
of the average resonance frequencies on the symmetry energy $a_4$
and on its slope has been found. In a covariant scheme we can see from
Eq.(\ref{det2}) that such behaviour can be achieved only by using two 
isovector fields, at the lowest order.
This result shows more generally that a dynamical observable can be
more sensitive to the microscopic structure of the isovector interaction
than static properties.
For instance in a careful study of the neutron distributions,
 \cite{FurnstNPA706},
 it is clearly shown that these ``equilibrium'' observables are
almost equally correlated to value, slope and curvature of the symmetry
term.

\subsubsection{Isoscalar Response}
%\addtocontents{toc}{\hspace{0.55cm}\thesubsubsection \hspace{0.12cm}
%Isoscalar Response}
We already noted that for symmetric $NM$ there is a close 
analogy between the isoscalar and the isovector response in 
the $RMF$ approach. 
In the isoscalar
degree of freedom the compressibility will play the same role of the symmetry
energy in the dispersion relation equations. 
Also in this case we will have an important correction term 
coming from the interplay of the scalar and vector channel.

The Eq.(\ref{det1}) now becomes \cite{phinote}:
\begin{equation}\label{dcomp}
1+N_F\left[f_\omega(1-v_s^2)- f_\sigma\frac{M^{*2}}{E^{*2}_F} 
\left({1-f_\sigma A(k_F,M^*)}-f_\omega\, \frac{\rho_S}{M^*}\,v_s^2\right)
\right]\varphi(s)=0
\end{equation}
that can be reduced with the same approximations to the isoscalar 
equivalent of the Eq.(\ref{det2}):
\begin{equation}\label{dcomp2}
1+ \frac{E^*_F}{3\,k_F^2}\left[K_{NM}^{pot}-9\,f_{\omega}
\frac{k^2_F}{E^{*2}_F}\left(1-f_\sigma\,\frac{M^*}{E^{*2}_F}\,\rho_S\right)
\rho_B\right]\varphi(s)=0\,.
\end{equation}
Here the $K_{NM}^{pot}$ is the potential part of the nuclear matter 
compressibility
that in the Hartree scheme has the simple structure 
 \cite{phinote}(see  Eq.(16)
of ref. \cite{MatsuiNP365})
\begin{equation}\label{eq.21}
K_{NM}(\rho_B) = \frac{3\,k_F^2}{E^*_F} + 9
\left[f_\omega - f_\sigma 
\left(\frac{M^*}{E^*_F}\right)^2\right] \rho_B 
\equiv K^{kin}_{NM}+K^{pot}_{NM}\,.
\end{equation}

With this analogy, the previous discussion 
can be transferred to isoscalar 
oscillations with the role of $E_{sym}$ now ``played'' by the compressibility.
In fact in the isoscalar sector one always takes into account both the scalar
and vector fields in any $RMF$ model. However, the coupling 
constant $f_\omega$ can assume
very different values depending on the chosen value for effective masses
$M^*_0$. 
This is easy to understand since in the $RMF$ limit the saturation binding
energy has the simple form 
$$E/A(\rho_0)=E^*_F+f_\omega\rho_0-M_N$$
where $M_N$ is the bare nucleon mass. Thus in order to
have the same saturation values
of $\rho_0$, $E/A(\rho_0)$, when we decrease $M^*_0$ we have to 
increase $f_\omega$.
We then come to the natural conclusion that two $EOS$ with 
different effective masses,
 even if the compressibilities are the same, are expected to have 
different dynamical behaviour. This is a very general feature present also
in non-relativistic approaches.

From studies in $RMF$  on monopole
resonances in finite nuclei  it seems that a 
higher value of compressibility is required with respect to non-relativistic
calculations. Many authors state that this certainly 
demands for a clarification \cite{MaPRC55,MaNPA686,NiksicPRC66,VretenarPRC68}. 
Even if the monopole resonance
is not directly connected to the isoscalar collective mode in nuclear matter,
our discussion nicely suggests to look at the interplay between effective mass
and compressibility.
For example we can estimate by means of the dispersion relation 
Eq.(\ref{dcomp2}) that we can have  
a shift between the compressibility and the ``effective compressibility'' of
the order of $\sim 100\, MeV$.
%among different parametrization with the same $K_{NM}$. 
Therefore
an effective interaction e.g.  with $K \sim 300\, AMeV$ can reproduce 
the same frequencies
of another one with $K \sim 200\, AMeV$ (and a slightly larger $M^*_0$).

\subsubsection{Landau Parameters}
%\addtocontents{toc}{\hspace{0.55cm}\thesubsubsection \hspace{0.12cm}
%Landau Parameters} 
Next we  discuss the relativistic equations for 
collective modes
in terms of the Landau parameters. Interesting features will appear from the 
comparison to the analogous  non-relativistic case of Sect.\ref{rpa}. 
We will focus first on the isovector response, but, as  
 shown before,
 the structure of the results will be absolutely similar in 
the isoscalar channel.

The general non--relativistic expression for dispersion relation of 
isovector modes
can be found in ref.\cite{AbrikosovRPP22}:
\begin{equation}\label{nonr1}
1+\left[F^a_0+\frac{F_1^a}{1+1/3\,F^a_1}\,s^2 \right]\varphi(s)=0
\end{equation}
where $F^a_0$ is the ``isovector'' combination of the 
Landau $F_0$ parameters for 
neutrons and protons $F_0^a=F^{nn}_0-F^{np}_0$, already introduced in 
Sect.\ref{eos}.
$F^a_1$ are the equivalent quantities for the momentum dependent part of 
the mean field.
In the relativistic approach, for symmetric nuclear matter, we have:
\begin{eqnarray}\label{f1a}
&&F_0^a= F_\rho-F_\delta\,\frac{M^{*2}}{E^{*2}_F}\frac{1}
{1+f_\delta\,A(k_F,M^*)}\nonumber\\
&&F_1^{a}=-F_\rho\frac{v^2_F}{1+ \frac{1}{3}F_\rho\,v^2_F}\,,
\end{eqnarray}
where $F_i=N_F f_i\,(i=\rho,\delta)$ with $N_F=2N_{n,p}$. Note that the $F_1^a$
contribution results only from the vector coupling. 
By using the expression of $E_{sym}^{pot}$ 
(Eq.(\ref{eq.11}) of Sect.\ref{qhd}), 
 we can write 
Eq.(\ref{nonr1})
in the same form of Eqs.(\ref{det1}, \ref{det2}). The result is a similar 
expression but without
the term in $f_\delta$ inside the brackets 
in Eq.(\ref{det2}). As said before this is not the leading term, however, 
around saturation
density it amounts to about $10\%$ of the total correction.

 Moreover turning to the analogy with 
isoscalar channel, the corresponding term is now the coupling of 
the $\sigma$ field, which is  much larger.
This purely relativistic contribution could be up to a $20\%$. 
We underline this point because generally the linear
response in $RMF$ is discussed calculating the Landau parameters and then 
using these values directly in
the non-relativistic expression for collective modes 
\cite{MatsuiNP365,CaillonNPA696}.

From the analysis in terms of the Landau parameters, 
we can describe
the effect of the scalar-vector coupling competition  
previously discussed in the
following way. The symmetry energy fixes the $F^a_0$, in fact:
\begin{equation}\label{esim}
E_{sym}=\frac{k_F^2}{6\,E^*_F}(1+F_0^a),
\end{equation}   
but in the dynamical response also the $F^a_1$ enters, linked to the momentum
dependence of the mean field, mostly given by the vector meson coupling.
The results are completely analogous in the isoscalar channel, with the
compressibility given by
\begin{equation}
K_{NM}=\frac{3\,k_F^2}{E^*_F}(1+F_0^s)
\end{equation} 
with the ``isoscalar'' combination $F^s_0=F^{nn}_0+F^{np}_0$. 
The relativistic forms of
the isoscalar Landau parameters are exactly the same as in Eq.(\ref{f1a}), just
substituting the $\delta,\rho$ coupling constants for the $\sigma,\omega$ ones
\cite{phinote}.

\subsection{Isovector Collective Modes in Asymmetric Nuclear Matter}
We now turn to Asymmetric Nuclear Matter.
We discuss first results for the isovector collective oscillations,
which are driven by the symmetry energy terms of the nuclear $EOS$.
The aim is mainly to investigate the effect of the scalar-isovector channel.
This is normally not included in studying the isovector modes, or 
generally the properties of asymmetric matter in a relativistic approach,
while it should be naturally present on the basis of the analysis 
presented in Sect.\ref{qhd}.
Moreover we emphasize again that a Hartree-Fock scheme leads 
to the presence
of a scalar-isovector channel in any case, {\it even without the 
inclusion of the $\delta$-meson field} \cite{GrecoPRC63,GrecoPRC64}.

We will first discuss results obtained in the Non-Linear Hartree scheme 
($NLH$) including
either both the isovector ($\rho+\delta$) mesons ($NL\rho\delta$ 
parametrization) or only 
the $\rho$ meson, ($NL\rho$).
Even though the Hartree approximation has a simpler structure, it contains 
all the physical
effects we want to point out. Finally from the complete 
Hartree-Fock ($NLHF$) calculations we will confirm the dynamical contribution
of the scalar isovector channel.
For $NLH$ calculations  we use the parametrizations of Table \ref{qhdsets}, 
 Set($NL\rho$)
 and Set($NL\rho\delta$), of Sect.\ref{qhd}. 
In the Hartree-Fock case the coupling 
constant $f_\delta$ is adjusted to the value 
$\tilde f_\delta(\rho_0)=2.0 fm^2$ of the $NLHF$ model,
Eqs.(\ref{cincoup}) of Sect.\ref{qhd}.

\begin{figure}[htb]
\begin{center}
%\epsfysize=7.0cm
%\centerline{\epsfbox{relin1.ps}}
\includegraphics[scale=0.75]{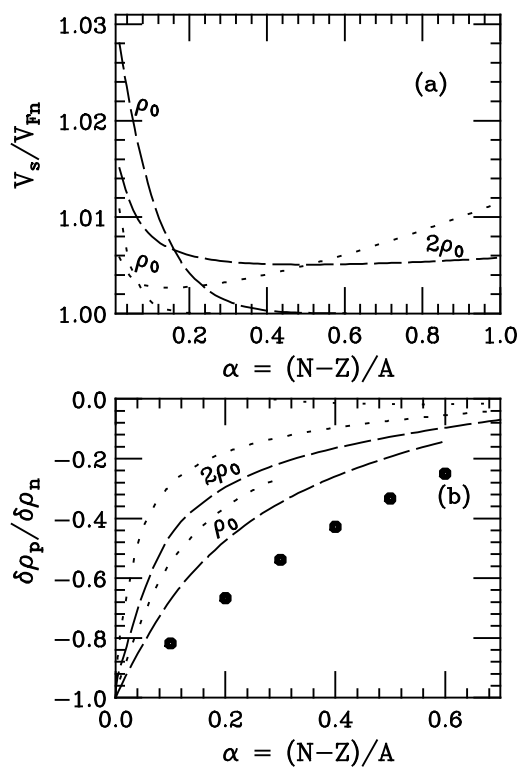}
\caption{Isovector-like modes: (a) Ratio of zero sound velocities 
to the neutron
Fermi velocity $V_{Fn}$ as a function of the asymmetry parameter $I$ for
two values of baryon density. Long dashed line: $NL\rho$. 
Dotted line: $NL\rho\delta$. (b) Corresponding ratios of proton 
and neutron amplitudes. All lines
are labelled with the baryon density, $\rho_0=0.16 fm^{-3}$. The full circles
in panel (b) represent the trivial behaviour of $-(\rho_p/\rho_n)$
vs. asymmetry parameter.}
\label{lin3}
\end{center}
\end{figure}

In Fig.\ref{lin3}a we show the sound velocities in the Hartree approximation
as a function of the asymmetry parameter $I$ for different baryon
densities. We actually plot the sound velocities in units of the neutron Fermi 
velocities. This is physically convenient: when the ratio is approaching
$1.00$ we can expect that this ``zero'' sound will not propagate due
to the strong coupling to the ``chaotic'' single particle motions 
(``Landau damping''). This quantity then will also directly give a measure
of the ``robustness'' of the collective mode.
Dotted lines refer
to calculations including ($\rho+\delta$) mesons, long-dashed lines
correspond to the case with only the $\rho$ meson.  Calculations are
performed for $\rho_B = \rho_0$ and $\rho_B = 2~\rho_0$. 
We note that the results of the two models differ already
at zero asymmetry, $I=0$. 
At normal density ($\rho_0$ curves),
in spite of the 
fact that the symmetry energy coefficient, $a_4=E_{sym}(\rho_0)$, 
is exactly the same in the two cases, significant differences are observed in
the response of the system. 
From Fig.\ref{lin3}(a) we expect a reduction of the 
frequency for the bulk isovector dipole mode in stable nuclei when the 
scalar isovector channel ($\delta-like$) is present. Moreover we note that, in
the $NL\rho$ case, the excitation of isovector modes persists up to higher
asymmetries at saturation density.
These are non-trivial features, related to the different way scalar and vector
fields are entering in the dynamical response of the nuclear system. Such
behaviours are therefore present in both collective responses, isoscalar and
isovector, as seen in the previous subsection.

Differences are observed even at 
$\rho_B = 2\rho_0$, where however also the symmetry energy is 
different. At higher densities a larger $E_{sym}$ is obtained in the case 
including the $\delta$ meson and this leads to 
a compensation of the 
effect observed at normal nuclear density. In particular, at higher
asymmetries $I$ the collective excitation becomes more robust
for $ NL\rho\delta$. 
Differences are observed also in the "chemical" structure of the mode, 
represented by the ratio $\delta \rho_p/\delta \rho_n$, which is plotted in 
Fig.\ref{lin3}(b).
The ratio of the out of phase $n/p$ oscillations is not following the ratio
of the $n/p$ densities for a given asymmetry, as shown by the full circles 
in the
figure. We systematically see a larger amplitude of the neutron oscillations.
The effect is more pronounced when the $\delta$ (scalar-isovector) channel
is present (dotted lines).

\subsubsection{Disappearance of the Isovector Modes}
\begin{figure}[htb]
\begin{center}
%\epsfysize=7.0cm
%\centerline{\epsfbox{relin2.ps}}
\includegraphics[scale=0.75]{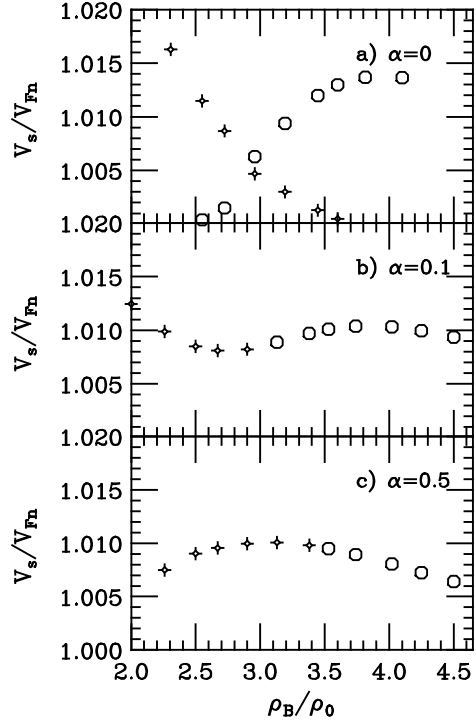}
\caption{Sound phase velocities of the propagating collective 
mode vs. the baryon density ($NL\rho$ case). Crosses: isovector-like. 
Open circles:
isoscalar-like. (a): symmetric matter. (b): asymmetric matter, $I=0.1$.
(c): asymmetric matter, $I=0.5$.}
\label{lin5}
\end{center}
\end{figure}

\begin{figure}[htb]
\begin{center}
%\epsfysize=4.7cm
%\centerline{\epsfbox{relin3.ps}}
\includegraphics[scale=0.75]{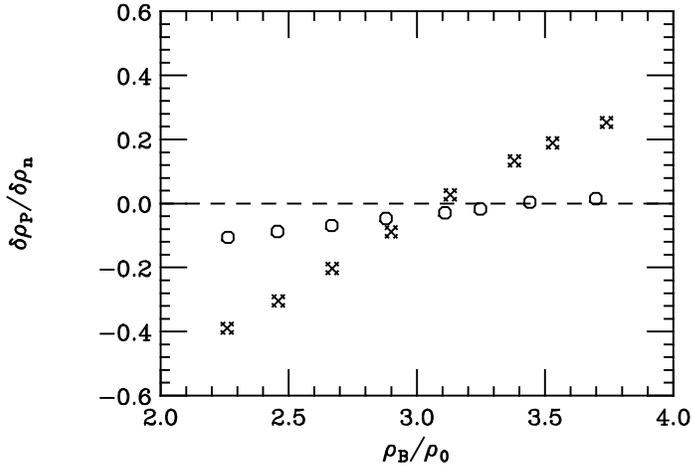}
\caption{Ratio of protons and neutron amplitudes in the propagating mode, for
different asymmetries, as a function of the baryon density around 
the $\rho_B^{cross}$ ($NL\rho$ case). Crosses: $I=0.1$, Fig.\ref{lin5}b. 
Open circles: 
$I=0.5$, Fig.\ref{lin5}c.}
\label{lin6}
\end{center}
\end{figure}

%\addtocontents{toc}{\hspace{0.55cm}\thesubsubsection \hspace{0.12cm}
%Disappearance of the Isovector Modes}
For asymmetric matter it is found that for all interactions with
increasing baryon density the isovector modes disappear: we call the density,
 where this occurs,
$\rho_B^{cross}$. E.g. from Fig.\ref{lin3}(b) we see that 
the ratio $\delta \rho_p/\delta \rho_n$ tends very quickly to zero with 
increasing
baryon density, almost for all asymmetries. Around this transition 
density we expect to have an almost $pure~ neutron~wave$ propagation of 
the sound.
In Figs.\ref{lin5} and \ref{lin6}
we show the results of the $NL\rho$ case, but the effect is clearly 
present in all the models.

In Fig.\ref{lin5} we give the zero-sound velocities as a function of density
for a definite asymmetry.
For symmetric matter we have a real crossing of the two sound velocities, 
isoscalar and isovector, as shown in Fig.\ref{lin5}(a). Above $\rho_B^{cross}$
the isoscalar mode is the most robust.

For asymmetric matter we observe a transition in the structure of 
the propagating normal mode, from isovector-like to isoscalar-like, 
Fig.\ref{lin5}(b,c).
Similar effects have been seen in a non-relativistic picture
\cite{ColonnaPLB428}.
This mechanism is analysed in  Fig.\ref{lin6} for the ratio of proton to
neutron amplitudes (including the sign).
In fact we see that the proton component of the propagating sound
is quite small in a relatively wide region around the ``transition'' 
baryon density, a feature becoming more relevant with increasing asymmetry, 
see the open 
circle line. This is quite interesting since it could open the possibility
of observing some experimental signatures of the $neutron~wave$ effect.
 
For a given asymmetry $I$ the value of $\rho_B^{cross}$ is different
for the models considered, as can be argued by the behaviour of 
$\delta \rho_p/\delta \rho_n$ at $2\rho_0$ in 
Fig.\ref{lin3}(b).
E.g. for $I=0.1\,~NL\rho\delta$ has the lower value 
($\rho_B^{cross}\simeq 2.4 \rho_0 $), while $NL\rho$ has the higher one 
($\rho_B^{cross}\simeq 3.0 \rho_0$).
This is again related to the reduction of the isovector 
restoring force when the
scalar-isovector channel ($\delta$-like) is present, as discussed before.
 
\subsection{Isoscalar Collective Modes in Asymmetric Nuclear Matter}

\subsubsection{Exotic high baryon density modes}
%\addtocontents{toc}{\hspace{0.55cm}\thesubsubsection \hspace{0.12cm}
%Exotic high baryon density modes}
From the previous analysis we have seen the isoscalar-like excitations 
to become
dominant at high baryon density, above the $\rho_B^{cross}$ introduced before.

\begin{figure}[htb]
\begin{center}
%\epsfysize=7.0cm
%\centerline{\epsfbox{relin4.ps}}
\includegraphics[scale=0.75]{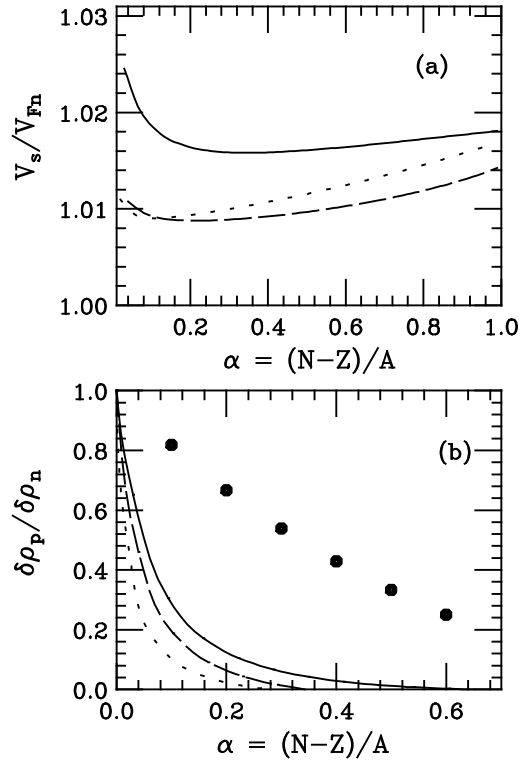}
\caption{Sound velocity (a) and chemical composition (b) of 
isoscalar-like modes at $\rho_B=3.5\rho_0$.
Solid line: $NLHF$. Long dashed line: $NL\rho$. Dotted line: $NL\rho\delta$. 
 The full circles in panel (b)
represent the behaviour of $\rho_p/\rho_n$ vs. $I$.}
\label{lin7}
\end{center}
\end{figure}

Some results are shown in Fig.\ref{lin7}.
It should be noticed that 
the frequency of the isoscalar-like modes is essentially related to the 
compressibility of the system at the considered density. 
In Fig.\ref{lin7}(a) we display the sound velocity obtained in Hartree and 
Hartree-Fock calculations at $\rho_B = 3.5~\rho_0$, as a function 
of the asymmetry $I$.   
The differences observed among calculations performed within the Hartree or
Hartree-Fock scheme for symmetric matter are due to a different behaviour of   
the associated equation of state at high density.  
At $I=0$ the two Hartree models have exactly the
same isoscalar mean fields, but for asymmetric nuclear matter
the different behaviour of the symmetry energy
leads to a different compressibility. The case 
$NL\rho\delta$, which has the stiffer $E_{sym}$ 
(resulting in a greater incompressibility for $I\,>\,0$) with respect to
$ NL\rho$, shows 
also a larger increase of $v_s/v_{Fn}$ with density.

Differences are also observed in the chemical composition
of the mode (Fig.\ref{lin7}(b)). The black circles show the behaviour 
of $\rho_p/\rho_n$ vs. $\alpha$. Note the $almost~pure$ $neutron~wave$ 
structure 
of the propagating
sound, since the oscillations of
protons appear strongly damped ($\delta\rho_p/\delta\rho_n\ll \rho_p/\rho_n$),
see the previous comments. 

Before closing this discussion we have to remark that the isoscalar-like modes
at high baryon density are vanishing if the nuclear $EOS$ becomes softer. This
is indeed the results of two recent models, 
 \cite{CaillonNPA696,TypelLMU02}, where
the nuclear compressibility is decreasing at high baryon density because of a 
reduction of the isoscalar  vector channel contribution. In ref.
\cite{CaillonNPA696}
this is due to
self-interacting high order terms for the $\omega$ meson, while in 
ref.\cite{TypelLMU02} to a decreasing
density-dependent $f_\omega$ coupling. 

Finally we note that all causality violation problems (superluminal sound
velocities) observed in the non relativistic results at high 
baryon density, see
 ref.\cite{MatsuiNP365} and Fig.3c in ref.\cite{ColonnaPLB428}, are 
completely absent 
in the relativistic approach, as seen in the high density trends 
in Fig.\ref{lin5}.

\subsubsection{Isospin Distillation in Dilute Matter}
\begin{figure}[htb]
\begin{center}
%\epsfysize=7.0cm
%\centerline{\epsfbox{relin5.ps}}
\includegraphics[scale=0.75]{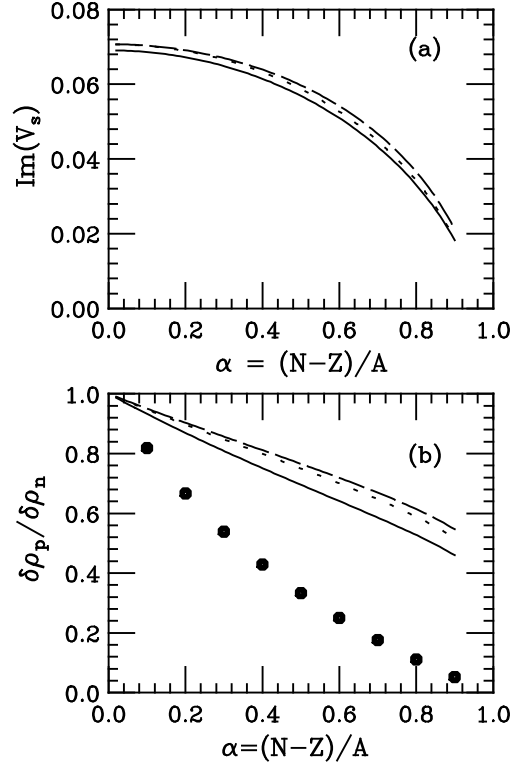}
\caption{Isoscalar--like unstable modes at $\rho_B=0.4\,\rho_0$ : 
Imaginary sound velocity (a), in $c$-units, and ratio of proton and neutron 
amplitudes (b)  as a function of the
asymmetry. Solid line: $NLHF$. Dotted line: 
 $NL\rho\delta$. Long dashed line: $NL\rho$. The full circles in panel (b)
represent the behaviour of $\rho_p/\rho_n$ vs. the asymmetry parameter.} 
\label{lin8}
\end{center}
\end{figure}

%\addtocontents{toc}{\hspace{0.55cm}\thesubsubsection \hspace{0.12cm}
%Isospin Distillation in Dilute Matter}
As a good check of the relativistic approach we have also investigated 
the response of the system in the region of spinodal
instability associated with the liquid-gas phase transition,
 which occurs at low densities. In this region
an isoscalar unstable mode can be found, with 
imaginary sound velocity, that gives rise to an exponential growth of the
fluctuations. The latter can represent a dynamical mechanism for
the multi-fragmentation
process observed in heavy-ion collisions, see the discussion in 
Sects.\ref{rpa} and \ref{fermi}. 
We find this kind of solution. 
In Fig.\ref{lin8} 
we show the imaginary sound velocity and the ratio 
$\delta\rho_p/\delta\rho_n$ as a function of
the initial asymmetry for this collective mode. For all the 
interactions this ratio is different from  
the corresponding $\rho_p/\rho_n$ of the initial asymmetry. This is 
exactly the chemical effect associated with the new instabilities in
dilute asymmetric matter \cite{MuellerPRC52,BaranPRL86}.
In particular  
it is seen that, when isoscalar-like modes become unstable,
the ratio $\delta\rho_p/\delta\rho_n$ 
becomes $larger$ than the ratio  $\rho_p/\rho_n$ (at variance with 
the stable modes at high densities, see Fig.\ref{lin7}). Hence proton
oscillations are relatively larger than neutron oscillations leading
to a more symmetric liquid phase and to a more neutron rich gas phase, 
during the disassembly of the system. 
This is the {\it Isospin Distillation} effect in fragmentation,
 as already studied in detail in the previous Sects.\ref{rpa} and 
 \ref{fermi} in
a non-relativistic frame.
 
In conclusion we note that in dilute asymmetric $NM$ we 
can distinguish two regions 
of instability, mechanical (cluster formation) and chemical (component
separation). There is however no discontinuity in the structure
of the unstable modes which are developing. For all realistic
effective nuclear interactions (relativistic and not) the nature of the 
unstable normal modes at low densities is {\it always isoscalar-like},
i.e. with neutrons and protons oscillating in phase, although
with the neutron distillation effect discussed before, 
refs. \cite{BaranPRL86,DitNPA722}.

\subsection{General comments from the ANM collective response}
We have shown that from a detailed study of the collective response of
asymmetric nuclear systems it would be possible to obtain information on
the $Lorentz$ structure of the in-medium interaction in the isovector channel.
In fact we have singled out some qualitative new effects of 
the $\delta$-meson-like 
channel on the dynamical response of $ANM$. Essentially, our investigation 
indicates
that even if the symmetry energy is fixed, the dynamical 
response is affected by
its internal $Lorentz$ structure, i.e. the presence or not of 
an isovector-scalar field.
This is implemented by the explicit introduction of an effective $\delta$-meson
 and/or by the Fock term contributions. Both mechanisms are absent in 
the present relativistic $RPA$
calculations for finite nuclei. In the spirit of the $EFT/DFT$ approach
\cite{FurnstCNPP2} it would be interesting to see the effect of 
an isovector scalar field extension,
at the lowest order, on the existing covariant $RPA$ results.
 In general we see a close analogy 
in the structure of the linear response equations for isoscalar/isovector
 modes:
\begin{itemize}
\item Same form of the dispersion relations.
\item Parallel role of $E_{sym}^{pot}$ and $K^{pot}_{NM}$ in the determination
of the restoring force.
\item Parallel structure of the corrections due to the scalar-vector meson
competition.
\end{itemize} 

This appears to be a beautiful ``mirror'' structure of the 
relativistic approach that seems to nicely  support the introduction of a 
$\delta$-meson-like coupling 
in the isovector channel, at least from a formal point of view. 
We like to remind that the same ``mirror'' structure
of the relativistic picture is remarked in Sect.\ref{qhd} for
equilibrium properties, saturation binding and symmetry energy, the 
$a_1$ and $a_4$ parameters of the Bethe-Weisz\"acker mass formula.

%\include{rep_bib}

%\end{document}

%% file: Chapter-8.tex
%\documentclass{elsart}

%\usepackage{graphicx}

%\usepackage{amssymb}

%\begin{document}

\setcounter{figure}{0}
\setcounter{equation}{0}
\section{Relativistic Heavy Ion Collisions: the covariant structure of
the symmetry term}\label{reldyn}

\markright{Chapter \arabic{section}: reldyn}

Intermediate energy Heavy Ion Collisions, $HIC$, open the unique
possibility to explore the Equation of State ($EOS$) of
nuclear matter far from saturation, in particular the
density dependence of the symmetry energy $E_{sym}(\rho_B)$ . 
Within a relativistic
transport model it is possible to see that the
isovector-scalar $\delta$-meson, which affects the high density
behavior of the symmetry term, influences the
isospin dynamics. The effect is greatly enhanced by a relativistic
mechanism related to the covariant nature of the fields
contributing to the isovector channel. 

An increasing $E_{sym}(\rho_B)$ leads to a more proton-rich
neutron star whereas a decreasing one would make it more pure in
neutron content.
As a consequence the chemical composition and 
cooling mechanism of protoneutron stars 
\cite{LattimerPRL66,SumiyashiAPJ422}, mass-radius
correlations  \cite{PrakashPRL61,EngivikPRL73}, critical densities for kaon
condensation in dense stellar matter \cite{LeePR275,KubisAPPB30}
as well as the possibility of a mixed quark-hadron phase \cite{KutscheraPRC62}
in neutrons stars will all be rather different.
It has recently been argued by means of simple thermodynamics
considerations
that even the onset of a quark-deconfined phase at high baryon density
could present a sensitivity  to the
behaviour of $E_{sym}(\rho_B)$ even for not very large asymmetries
 \cite{DitarX0210}.

In the previous Sections we have seen how the search for 
$E_{sym}(\rho_B)$ around saturation
density has driven a lot of theoretical and experimental efforts.
We have seen how $HIC$s at Fermi energies
can give the possibility to extract some information on the symmetry
term
of the nuclear Equation of State ($EOS$) in region below and/or sligthly
above
the normal density.
Here we will focus our attention on Relativistic $HIC$s 
trying to select reaction observables
particularly sensitive to the symmetry energy at higher
density, where furthermore we cannot have complementary investigations
from nuclear structure like in the case of the
low density behaviour. We stress again that $HIC$s provide the unique
way to create asymmetric matter at high density in terrestrial
laboratories. Moreover effective interactions for high momentum
nucleons can be probed.
Calculations within transport approaches
 show that $HIC$s around $1AGeV$ allow to
reach a transient state of matter with more than twice the normal 
baryon density. Moreover, although the data are mostly of inclusive type
(and the colliding nuclei not very neutron rich), quite clearly a dependence
of some observables on charge asymmetry is emerging.

Collective flow, particle production and isospin equilibration results for 
reactions induced by stable and radioactive beams
are discussed.
The elliptic flows of nucleons and light isobars appear to be 
quite sensitive to the microscopic structure of the symmetry term, 
in particular for particles
with large transverse momenta, since they represent an earlier
emission from a compressed source.
Thus future, more exclusive, experiments with relativistic radioactive
beams should be able to set stringent
constraints on the density dependence of the symmetry energy far
from ground state nuclear matter.

Moreover we show that in fact a {\it relativistic} description of the nuclear
mean field can account for an enhancement of isospin effects during the
dynamics of heavy-ion collisions. 
The isospin dependence of collective flows and pion production has been 
already discussed
in a non-relativistic framework \cite{BaoPRL85,ScalonePLB461,UmaPRC57,BaoPRC67}
  using very different
$EOS$ with opposite behaviours of the symmetry term at high densities,
increasing repulsion ({\it asy-stiff}) vs. increasing attraction
 ({\it asy-soft}). The main new result shown here, in a fully 
relativistic scheme, is the importance at higher energies 
of the microscopic covariant structure 
of the effective interaction in the isovector channel: {\it effective forces 
with very similar symmetry terms can give rise to very different isospin 
effects in relativistic heavy ion
collisions}.

In recent years some efforts have been devoted to the
effects of the scalar-isovector channel in finite nuclei, 
 \cite{TypelNPA656,FurnstNPA671,BuervePRC65}.
Such investigations have not
shown a clear evidence for the $\delta$-field and this can
be understood considering that in finite nuclei one can test the
interaction properties mainly
below the normal density, where the effect of the $\delta-$channel on 
symmetry energy
and on the effective masses is indeed small \cite{LiuboPRC65,GrecoPRC67}
 and  eventually
could be absorbed into non linear
terms of the $\rho$ field. 
In Sect.\ref{relin} we have also seen that the dynamical collective response
is not much affected, in particular the spinodal instabilities. We can expect
just some weak effects on the isovector-like dipole modes.

Here we show that heavy-ion collisions around $1AGeV$ with
radioactive
beams can provide instead a unique opportunity to spot the presence of
the scalar isovector channel. In fact, due to the large counterstreaming
nuclear currents one may even exploit the
different Lorentz nature of a scalar and a vector field.
\begin{figure}
\begin{center}
%\epsfysize=8.5cm
%\centerline{\epsfbox{esym-nl2rbuu.ps}}
\includegraphics[scale=0.75]{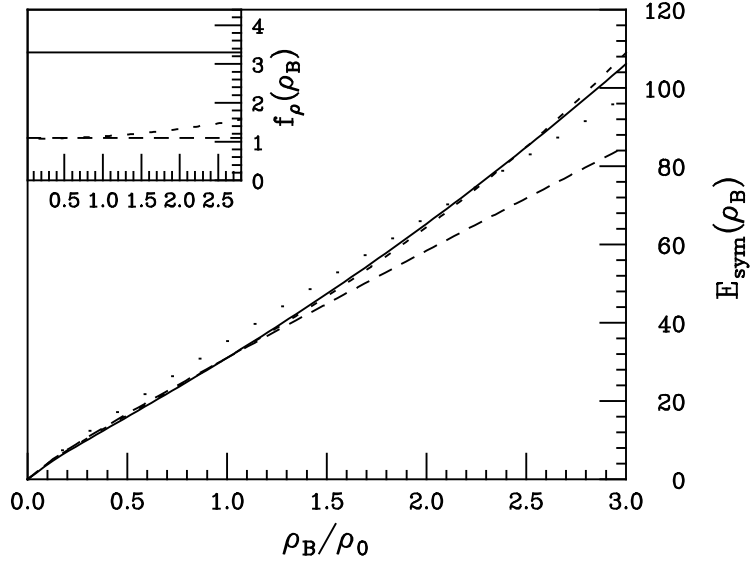}
\caption{Total (kinetic+potential) symmetry energy (in MeV) 
as a function of the
baryon
density.  Solid: $NL\rho\delta$. Dashed: $NL\rho$. Short Dashed:
$NL-D\rho$. In the insert
 the density behaviour of the $\rho$ coupling, $f_\rho$ (in $fm^2$), 
for the three
models is shown.}
\label{reldyn1}
\end{center}
\end{figure}

When both $\rho$-like and $\delta$-like channels are considered
$E_{sym}(\rho_B)$ can be written as (Sect.\ref{qhd}):
\begin{equation}\label{eq.2}
E_{sym}(\rho_B) = \frac{1}{6} \frac{k_F^2}{E^*_F} + \frac{1}{2}
\left[f_\rho - f_\delta
\left(\frac{M^*}{E^*_F}\right)^2\right] \rho_B\equiv
E_{sym}^{kin}+ E_{sym}^{pot}
\end{equation}
with $E^*=\sqrt{k_F^2+M^{*2}}$, $M^*$ the
effective Dirac mass and $f_{\rho,\delta}=(g_{\rho,\delta}/m_{\rho,\delta})^2$
are the coupling constants of the isovector channels.
We see that, when the $\delta$-meson is included, the observed
$a_4=E_{sym}(\rho_0)$
value actually assigns  the combination
$[f_\rho - f_\delta (\frac{M^*}{E^*_F})^2]$
of the $(\rho,\delta)$ coupling constants, \cite{LiuboPRC65,GrecoPRC67}.
In Fig.\ref{reldyn1} we show the density dependence of the symmetry energy
for three different $Non-Linear-RMF$ models: 
two of these, 
one including only the
$\rho$ field ($NL\rho$), the other with $\rho$ and $\delta$
fields ($NL\rho\delta$), were already introduced in 
 Sect.\ref{qhd} (see Table \ref{qhdsets}). 
The last, ($NLD\rho$), includes only a
$\rho$ field but with a covariant density dependence of $f_\rho$,
\cite{TypelNPA656,FuchsPRC52}. This is tuned to give
at high density the same $E_{sym}(\rho_B)$ of the the ($NL\rho\delta$)
case.
As shown in the following, the latter model is useful for disentagling 
in the reaction dynamics the 
effects due to a
difference in $E_{sym}(\rho_B)$
from those directly linked to the strength of the $\rho$ vector field 
\cite{GrecoPLB562}.
Thus these models parametrize the isovector mean field either by only
the vector field with $f_{\rho}=1.1 fm^2$, or with a balance
between a vector field with $f_{\rho}=3.3 fm^2$ and a scalar one with
$f_{\delta}=2.4 fm^2$, or finally by
a normal density coupling $f_{\rho}(\rho_0)=1.1 fm^2$, at saturation, 
but with
an increasing density dependence as shown in Fig.\ref{reldyn1} (insert).
We stress again that in $NL\rho\delta$ the symmetry energy results
from a balance between a scalar attraction, ($\delta-like$),
and a vector repulsion, ($\rho-like$), where the $\rho$-coupling is now 
roughly three times larger 
than in the $NL\rho$ case.

The strength of  $f_\delta$ is fixed relatively well
by $DBHF$ \cite{HofmannPRC64,DalenarX0407}  
and $DHF$ \cite{GrecoPRC64} calculations.
Therefore the effects described in the following are not artificially
enhanced, but based on a reliable estimate available at the
moment. In any case the aim of our work is to present some
qualitative new features expected in the reaction dynamics,
in particular for collective flows, particle production and 
isospin stopping power,
which result from the introduction
of a scalar effective field in the isovector channel. 

We finally note that the use of a $RMF$ approach with fixed couplings 
in high
density regions generally requires some caution due to the density
dependence of the virtual meson couplings because of 
correlations beyond the mean field scheme. In particular a softer $EOS$
for the isoscalar part is favoured, as suggested from experimental analyses
 \cite{DanielNPA673}, from $DBHF$ predictions and from reaction simulations 
for collisions
of charge symmetric ions at intermediate energies \cite{GaitanosEPJA12}.
In the calculations presented here the $(\sigma,\omega)$ coupling constants
have been modified accordingly, see details in ref. \cite{GaitanosNPA732}.

\subsection{Relativistic transport simulations}

For the theoretical description of heavy ion collisions we solve
the covariant transport equation of the Boltzmann type 
 \cite{KoPRL59,GiessenRPP56}  within the 
Relativistic Landau
Vlasov ($RLV$) method \cite{FuchsNPA589} (for the Vlasov part) 
and applying
a Monte-Carlo procedure for the collision term. $RLV$ is a test particle
method using covariant Gaussians in phase space for the test particles.
The collision term includes elastic and inelastic processes involving
the production/absorption of the $\Delta(1232 MeV)$ and $N^{*}(1440
MeV)$ resonances as well as their decays into one- and two-pion channels.
Details about the cross sections for all the possible channels can
be found in ref.\cite{HuberNPA573}.
An explicit isospin-dependent Pauli blocking term for the fermions is employed.

For the following discussion it is useful to recall  how a
relativistic Vlasov equation can be obtained from Wigner Function
dynamics of Sect.\ref{qhd}.
The neutron/proton Wigner functions are expanded in terms of
components with definite transformation properties. Consistently with 
the effective fields
included in the minimal model one can limit the expansion to 
scalar and vector parts:
$$
{\hat F}^{(i)}(x,p) = F_S^{(i)}(x,p) + \gamma_\mu F^{{(i)}\mu}(x,p),~~~i=n,p.
$$
From the kinetic equation (\ref{trans}) of Sect.\ref{qhd} 
we obtain after some algebra
a relation between the vector and scalar components
$$
F^{{(i)}\mu}~\equiv~ p_i^{{*}\mu} {{F_S^{(i)}} \over {M_i^*}}
$$
and thus a transport equation of Vlasov type 
for the scalar part $f_i(x,p^{* \mu})
 \equiv F_S^i/M_i^*$:
\begin{equation}\label{vlarel}
\{p_{\mu i}^{*}\partial^\mu + [p_{\nu i}^{*} F_i^{\mu\nu} + 
 M_i^*(\partial^\mu M_i^*)] {\partial_\mu}^{p^*} \} f_i(x,p^{{*}\mu})~=~0
\end{equation}
with the field tensors
$$
F_i^{\mu\nu} \equiv \partial^\mu p_i^{{*}\nu} + \partial^\nu p_i^{{*}\mu}. 
$$
The trajectories of test particles obey to the following equation
of motion:
\begin{eqnarray}\label{eqmot}
&&\frac{d}{d\tau}x_i^\mu=\frac{p^*_i(\tau)}{M^*_i(x)}~,\nonumber\\
&&\frac{d}{d\tau}p^{*\mu}_i=\frac{p^*_{i\nu}(\tau)}{M^*_i(x)}
{F}_i^{\mu\nu}\left(x_i(\tau)\right)+\partial^\mu M^*_i(x)~.
\end{eqnarray}

In order to have an idea of the dynamical effects of the
covariant structure of the interactive fields, we
write down, with some approximations, the ``force'' acting on a particle. 
Since we are interested in isospin contributions we will take into account 
only the isovector part of the interaction \cite{GrecoPLB562}:
$$
\frac{{d\vec p}^{\,*}_i}{d\tau}=\pm f_{\rho} \frac{p_{i\nu}}{M^*_i}
\left[\vec\nabla J_{3}^{\nu}-\partial^\nu\vec{J_3} \right]
\mp f_\delta \nabla \rho_{S3}, ~~~~(p/n) %[(+,-)p, (-,+)n]
$$
\begin{equation}\label{force}
\approx 
\pm f_{\rho} \frac{E^*_i}{M^*_i}
\vec\nabla \rho_{3}
\mp f_\delta \vec\nabla \rho_{S3}
\end{equation}
In the second line we have neglected the contribution coming from 
the gradient 
of the
 current in the transverse direction, i.e. terms like
 $(\nabla_l J_{3m})$ ($l \not\equiv m=x,y,z$) and the derivative of the 
current with respect to time.
This form will be used below to interpret collective flow. 

\subsection{Collective flows}

Collective flows in heavy ion collisions give important information
on the dynamic response of dense, hot and asymmetric nuclear matter
 \cite{StoeckerPR137,DasguptaPT46,DanielSCI298,RHIC-v2}. 
In particular the proton-neutron
differential flow $F^{pn}(y)$ \cite{BaoPRL82}
has been found to be a very useful probe of the isovector part of the $EOS$
since it appears rather insensitive to the isoscalar potential
and to the in medium nuclear cross section. 
The definition of the $F^{pn}(y)$ is
\begin{equation}\label{eq.3}
F^{pn}(y)\equiv \frac{1}{N(y)} \sum_{i=1}^{N(y)} p_{x_{i}} \tau_i
\end{equation}
where $N(y)$ is the total number of free nucleons at the
rapidity $y$, $p_{x_{i}}$ is the transverse momentum of
particle $i$ in the reaction plane, and $\tau_i$ is +1 and -1
for protons and neutrons, respectively.

\begin{figure}[ht]
\begin{center}
%\epsfysize=6.5cm
%\centerline{\epsfbox{sn132_15b6-40.ps}}
\includegraphics[scale=0.55]{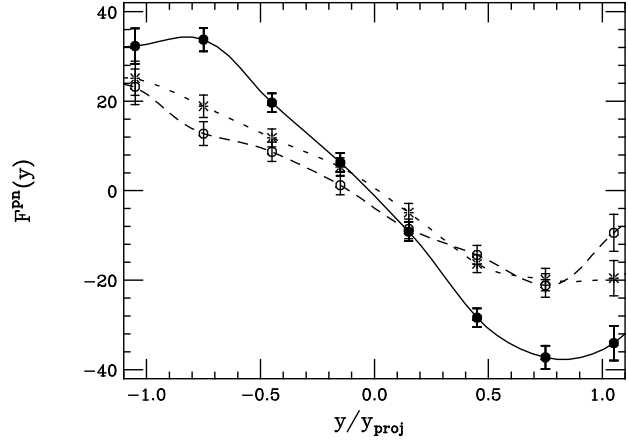}
\caption{ $^{132}Sn+^{132}Sn$
reaction at $1.5~AGeV~(b=6fm)$: Proton-neutron differential transverse  
collective flow (in $MeV/c$), vs. rapidity, for the three different models 
for the isovector mean fields.
Full circles and solid line: $NL\rho\delta$.
Open circles and dashed line: $NL\rho$.
Stars and short dashed line : $NLD\rho$.
}
\label{reldyn2}
\end{center}
\end{figure}

A typical result for the very asymmetric $^{132}Sn+^{132}Sn$ reaction 
at $1.5AGeV$ 
 (semicentral collisions) is
shown in Fig.\ref{reldyn2}. The error bars represent the statistical 
fluctuations due
to the Monte-Carlo nature of the simulations.
We notice that the differential flow in case of the
$NL\rho\delta$ presents a stiffer
behaviour relative to the $NL\rho$ model,
as expected from the more repulsive 
symmetry energy $E_{sym}(\rho_B)$ at high baryon densities, in 
Fig.\ref{reldyn1}.
On the other hand it may be 
surprising that a relatively small
difference at $2\rho_0$ results in a such different collective flows.

The calculation can be repeated using the
$NLD\rho$ interaction, i.e. with only a $\rho$ contribution {\it but}
tuned to reproduce the same $EOS$ of the $NL\rho\delta$ case.
The results, short-dashed curve of Fig.\ref{reldyn2}, are very similar to the
ones of the $NL\rho$ interaction. 
Therefore we understand the large flow effect as mainly due to 
the different strengths of the vector-isovector field 
in the $NL\rho\delta$
 and the $NL\rho,NLD\rho$ models. 
In fact if a source is moving the
vector field is enhanced (essentially by the local 
 $\gamma \equiv \frac{E^*}{M^*}$ Lorentz factor)
relative to the scalar one.

In the second line of Eq.(\ref{force}) we have written the
isovector contribution to the effective force. 
We are actually interested in the difference between the forces acting on 
a neutron and on a proton.
Oversimplifying the $HIC$ dynamics we consider locally neutrons and  protons
with the same $\gamma$  factor (i.e. with
the same velocity). Then Eq.(\ref{force}) can be expressed approximately
in the following
transparent form ($\rho_{S3} \simeq \frac{M^*}{E^*}\rho_{3}$):
\begin{equation}\label{pndiff}
\frac{{d\vec p}^{\,*}_p}{d\tau}-\frac{{d\vec p}^{\,*}_n}{d\tau}\simeq 2\left
[\gamma f_{\rho}
- \frac{f_{\delta}}{\gamma}\right]\vec\nabla \rho_{3}~=~
\left[ \frac{4}{\rho_B}
E_{sym}^{pot} + 2(\gamma-1)f_{\rho}\right]\vec\nabla \rho_{3}
\end{equation}
where $\gamma \equiv \frac{E^*}{M^*}$ is the local Lorentz 
factor of the collective motion 
and Eq.(\ref{eq.2}) has been used.

Keeping in mind that
$NL\rho\delta$
has a three times larger $\rho$-coupling it is clear that dynamically
the vector-isovector
mean field acting during the $HIC$ is much larger than the one of the
$NL\rho,NLD\rho$ cases.
Then the isospin effect is mostly caused by 
the different
Lorentz structure of the
``interaction'' which results in a dynamical breaking of the balance
between
the $\rho$ vector and $\delta$ scalar fields, present in nuclear matter
at equilibrium,\cite{GrecoPLB562}.  This effect is analogous to 
the interplay between
the isoscalar vector- and scalar-fields which is seen in the
magnitude and energy dependence of the real part of the optical
potential,\cite{GaitanosEPJA12}.
To characterize the effect on differential collective flows we have
calculated
the slope $dF^{pn}(y)/d(y/y_{proj})$ at mid-rapidity.
Its value is $46.7MeV/c$ for $NL\rho\delta$
and $23.4MeV/c$ for $NL\rho$, i.e. a factor two difference.
Calculations performed at lower beam energies show
that below $500AMeV$ there is no essential difference in the differential
flow predictions among the models
discussed here. The effect from the strength of
$\rho$ field  starts to become important around $1AGeV$, as expected from 
a relativistic mechanism.

\begin{figure}[htb]
\begin{center}
%\epsfysize=6.5cm
%\centerline{\epsfbox{sn132_15v2pnb6.ps}}
\includegraphics[scale=0.55]{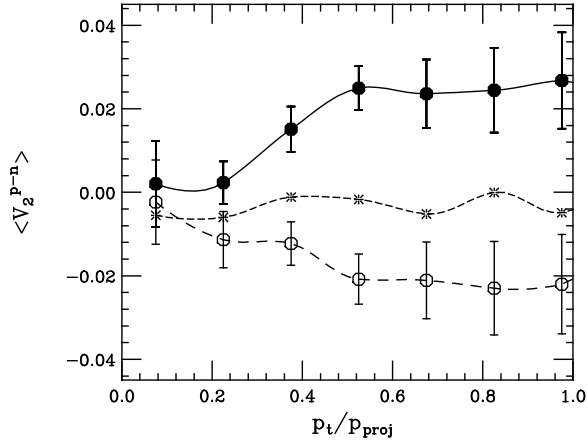}
\caption{Difference between neutron and proton elliptic flow
as a function of the transverse momentum in the
$^{132}Sn+^{132}Sn$ reaction at 1.5 AGeV (b=6fm) in the
rapidity range $-0.3 \leq y/y_{proj} \leq 0.3$.
Full circles and solid line: $NL\rho\delta$.
Open circles and dashed line: $NL\rho$.
Stars and short dashed line: $NLD\rho$.
The statistical error bars of the $NLD\rho$ curve are similar to the 
other cases and not shown.}
\label{reldyn3}
\end{center}
\end{figure}
Another interesting observable is the elliptic flow $v_{2}(y,p_t)$, the second 
coefficient in a Fourier series expansion of the
azimuthal emission distribution, see Sect.\ref{fastflows},
$$v_2=<\frac{p^2_x-p^2_y}{p^2_t}>$$
where $p_t=\sqrt{p^2_x+p^2_y}$ is the transverse momentum
 \cite{OlliPRD46,DanielNPA673}.
A negative value of $v_2$ corresponds to the predominant emission of
matter perpendicular to the reaction plane (the so-called $squeeze-out$ flow). 
The $p_t$-dependence of
$v_2$,
 which has been recently investigated by various groups
 \cite{DanielNPA673,DanielSCI298,LarionovPRC62,GaitanosEPJA12} is
very sensitive to the high density behavior of the $EOS$ since highly
energetic
particles ($p_t \ge 0.5$) originate from the initial compressed 
phase of the collision,\cite{GaitanosEPJA12}.

In Fig.\ref{reldyn3} we present the $p_t$ dependence of
the proton-neutron difference of the elliptic flow in the same
very exotic  $^{132}Sn+^{132}Sn$ reaction at $1.5AGeV$ 
 (semicentral collisions) for mid-rapidity emissions. 
The larger error bars correspond to a reduced
statistics when a selection on different $p_t$ bins is introduced,
increasing for larger $p_t$ values due to the 
smaller number of contributions.
From Fig.\ref{reldyn3} we see that 
with the $(\rho+\delta)$
dynamics the high-$p_t$ neutrons show a much larger $squeeze-out$.
This is fully consistent with an early emission (more spectator shadowing)
due to the larger repulsive $\rho$-field. We understand this large 
effect since the relativistic enhancement discussed above is
relevant expecially at the first stage of the collision.
The $v_2$ observable, which is a good {\it chronometer} of the reaction
dynamics, appears to be particularly sensitive to the Lorentz structure
of the effective interaction.

\begin{figure}[htb]
\begin{center}
\includegraphics[scale=0.75]{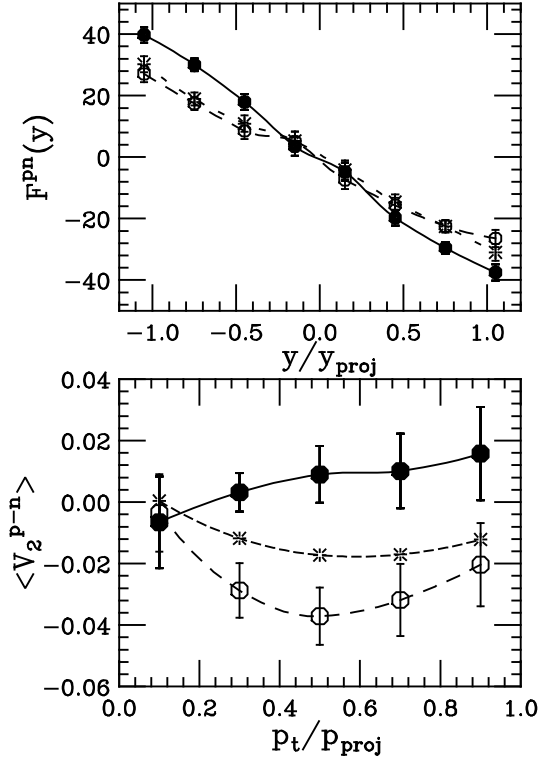}
\caption{$^{132}Sn+^{124}Sn$
reaction at $1.5~AGeV~(b=6fm)$ from the three different models for the
isovector mean fields.
Top: as in Fig.\ref{reldyn2}. Bottom: as in Fig.\ref{reldyn3}.
Full circles and solid line: $NL\rho\delta$.
Open circles and dashed line: $NL\rho$.
Stars and short dashed line: $NL-D\rho$.
Error bars: see the text and the previous caption.
}
\label{reldyn4}
\end{center}
\end{figure}

The same set of simulations has been repeated in ref.\cite{GrecoPLB562}
 for the more realistic
$^{132}Sn+^{124}Sn$
reaction at $1.5~AGeV~(b=6fm)$, that likely could be studied with 
the new planned radioactive beam facilities at intermediate energies.
The results are shown in Fig.\ref{reldyn4}. The effect of the 
different structure of the 
isovector channel is still quite clear, of course with a reduction
due to the smaller isospin density in the interaction region.
Particularly evident is again the splitting in the high $p_t$
region of the elliptic flow.

\subsection{$\pi^-/\pi^+$ Ratios}

\begin{figure}
\begin{center}
\includegraphics*[scale=0.50]{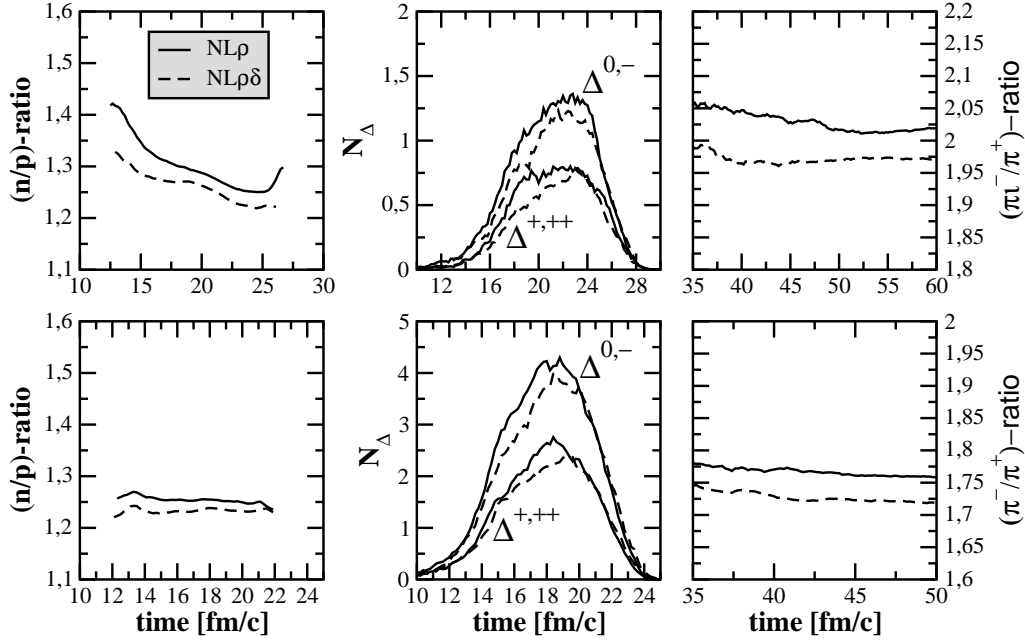}
%\vskip-1.3cm
\caption{ Central $Au+Au$ collisions at $0.6AGeV$ (upper) and $1.0AGeV$
(bottom). Time evolution of $n/p$ ratio and $\Delta$ resonance production
in high density regions ($\rho/\rho_0 \ge 2.0$) (first two columns) and
of the total $\pi^-/\pi^+$ ratio (right).
Solid lines: $NL\rho$. Dashed lines: $NL\rho\delta$.
}
\label{reldyn5}
\end{center}
%\vskip-1.5cm
\end{figure}

Observable effects originating from the high density symmetry energy 
are related to differences in neutron and proton 
densities. Thus, we consider in the following the ratio of neutrons to protons 
as a function of time and space. One expects isospin effects on the different 
isospin channels of pions, since they are produced in $nn$, $pp$ 
and $np$ collisions 
via the decay of $\Delta$ and $N^{*}$ resonances. 
Using the same relativistic transport code the
$\pi^-$ vs. $\pi^+$ production for central $Au+Au$ collisions at different 
energies can be evaluated, see \cite{GaitanosNPA732}. Results are shown 
in Fig.\ref{reldyn5}.
For the $NL\rho\delta$ model the neutrons are emitted much earlier
than protons from the high density phase due to a more repulsive mean field
together with a lower $n$-effective mass. This 
effect is responsible for the reduction of the $n/p$-ratio in the
residual system and, particularly, it 
influences the particle production,\cite{UmaPRC57,BaoPRC67}.
 
The four isospin states of the $\Delta$-resonance are produced in 
different scattering channels, when the threshold energy is available, e.g. 
$nn \rightarrow p\Delta^{-},n\Delta^{0},~pp \rightarrow p\Delta^{+},n\Delta
^{++},\cdots$. Thus
$\Delta^{0,-}$ ($\Delta^{+,++}$) resonances are mainly formed in energetic 
$nn$- ($pp$-) collisions. Therefore, the $n/p$-ratio
 is indirectly 
related to that of the particle production. 

The $n/p$ effective mass
splitting in asymmetric matter can also directly affect the resonance 
production 
because of threshold effects. 
We note 
that resonances will also have isospin dependent in-medium effective masses
that can be related to the nucleon effective masses through
the isospin coupling coefficients in the process 
$\Delta \leftrightarrow \pi{N}$, \cite{BaoNPA708,GaitanosNPA732}.
In terms of the self-energies $\Sigma$ we have:
\begin{eqnarray}
& & \Sigma_i (\Delta^-) = \Sigma_i(n) \nonumber \\
& & \Sigma_i (\Delta^0) = \frac{2}{3}\Sigma_i(n) + \frac{1}{3}\Sigma_i(p)
\nonumber \\
& & \Sigma_i (\Delta^+) = \frac{1}{3}\Sigma_i(n) + \frac{2}{3}\Sigma_i(p)
 \nonumber \\
& & \Sigma_i (\Delta^{++}) = \Sigma_i(p)~~~~~and~~~ \nonumber \\
& & \Sigma_i ({N^*}^{(+,0)}) = \Sigma_i(p,n). 
\label{weights}
\quad
\end{eqnarray}
where $i=scalar~and~vector$.

Thus the effective 
masses of the four isospin states of the
$\Delta$-resonance will be different when a $\delta$-meson is included
in the calculation. We have seen    
from
general considerations (see Sect.\ref{qhd}), that, in $n$-rich systems, a 
isovector scalar meson field
is leading to a neutron effective mass smaller than that for protons, 
in particular 
at high baryon densities. Consequently in the process
$nn \rightarrow p\Delta^{-}$ less energy will be available for the
 $\Delta^-$ production in the $NL\rho\delta$ case.
We like to note that this mechanism is not strictly linked to the 
density behavior
of the symmetry energy. E.g., in density dependent coupling models we
can have at high densities a lower $E_{sym}(\rho)$ because the
$\rho$-meson coupling is decreasing but still a large $m^*_p-m^*_n$
splitting if the $\delta$-meson coupling stays constant or it is slightly
increasing, as e.g. in the $DBHF$ calculations of ref.\cite{GaitanosNPA732}.

In the $\Delta$-resonance multiplicities  
(middle 
figures) one thus sees a decrease of the $\Delta^{0,-}$ isospin 
states due to the 
effect of the isovector-scalar $\delta$-meson. The pions 
 are mainly produced 
from resonance decays, e.g. as
$\Delta^{-} \rightarrow n \pi^{-},~\Delta^{0} \rightarrow p\pi^{-},
~\Delta^{+} \rightarrow n\pi^{+}, \cdots$. Therefore the decrease of 
$\Delta^{0,-}$ resonances reduces 
the production of negative encharged pions which consequently decreases 
the $\pi^{-}/\pi^{+}$-ratio 
(right part in Fig.~\ref{reldyn5}). This ratio then appears to be sensitive 
to the isospin term of equation of state at high densities. 
One should note that 
both models used here exhibit a similar asy-stiff behavior at 
high densities (see Fig.\ref{reldyn1}). We deduce that the
$n/p$ effective mass splitting mechanism described above
is very important in inducing isospin effects on the particle production
 \cite{masses}.

\begin{figure}[t]
\begin{center}
\includegraphics*[scale=0.50]{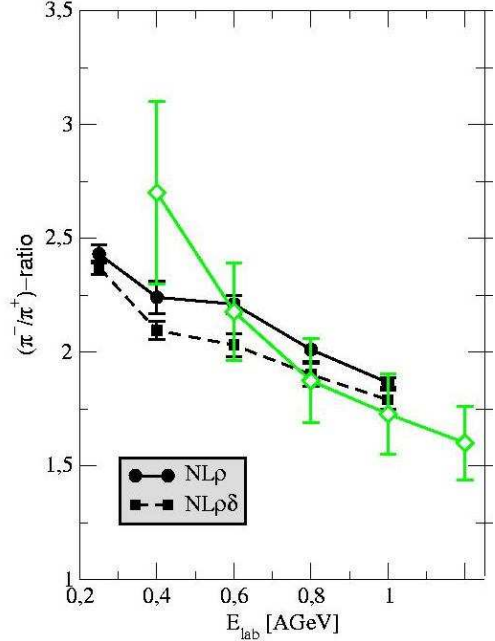}
\caption{Energy dependence of the ($\pi^{-}/\pi^{+}$)-ratio in central
$Au+Au$ collisions. Results of $NL\rho$ (solid) and
$NL\rho\delta$ (dashed) parametrizations. The grey bars correspond to some very
preliminary data from the $FOPI-GSI$ collaboration \cite{FOPI}.
}
\label{reldyn6}
\end{center}
\end{figure}

A comparison of total  $\pi^{+} / \pi^{-}$ ratio with preliminary data,
at different energies, is 
presented in Fig.\ref{reldyn6}, in central $Au+Au$ collisions.
We note that the $\delta$ influence (difference between
solid and dashed curves) is slightly decreasing with the beam energy. This is 
an indication of the $n/p$ effective mass splitting mechanism
in $\Delta^-$ production, expected
to be more important at the threshold.  

Generally isospin effects on pion production are decreasing
with incident energy, as seen in Fig.\ref{reldyn6}, 
due to secondary collisions,
i.e. pion absorption and $\Delta$-rescattering,
which are essentially not isospin-dependent.  
Thus the isospin dependence
of pion production will be on average moderated.
This trend was already found in earlier studies 
using a non-relativistic Quantum-Molecular Dynamics ($QMD$) 
approach \cite{UmaPRC57}.

For the same systems new very preliminary data on pion production 
have recently been reported from the $FOPI$-collaboration 
at $GSI$ \cite{FOPI}, shown as grey bars in the 
Fig.\ref{reldyn6}.  
Generally our simulations provide a qualitative 
good description of the 
pion ratio with respect to beam energy, except perhaps at the lower energies. 
The inclusion of a
$\delta$ meson in the iso-vector part of the equation of state 
improves the comparison, at least at higher energies.
In any case it appears that $\pi^-/\pi^+$-ratios might be a
probe of the high density symmetry energy and the role of the
virtual $\delta$-meson. It would be helpful to look at more exclusive data
on pion production in asymmetric systems,\cite{GaitanosNPA732}.

The effect of the $\delta$-field as discussed above is enhanced in the 
momentum spectra, i.e. the rapidity and transverse momentum distributions,
due to the interplay of the isospin effects 
with the Coulomb interaction which affects differently
the $\pi^+$ and $\pi^-$ distributions. Thus a clearer and more observable
effect of the $\delta$-field is predicted in the flow observables of
the $(\pi^{-}/\pi^{+})$-ratios, as a function of energy.

\subsection{Isospin Transparency}

Isospin equilibration is seen in the collisions of systems with different
asymmetries. It has been extensively investigated by the $FOPI$ collaboration
as a signature of trasparency \cite{RamiPRL84,HongPRC66}.
Here we investigate it also with respect to the sensitivity to the 
$iso-EOS$.  
The idea is to study 
colliding systems with the same mass number but different $N/Z$ ratio 
In particular, 
a combination of ${}^{96}_{44}Ru,~N/Z=1.18$ and ${}^{96}_{40}Zr,~N/Z=1.4$ 
has been 
used as projectile/target in experiments at intermediate energies of $0.4$ and 
$1.528~AGeV$ \cite{RamiPRL84,HongPRC66}. 
The degree of stopping or transparency has been determined from 
the rapidity dependence of the {\it imbalance ratio} for the 
mixed reactions 
$Ru(Zr)+Zr(Ru)$: $R(y^{(0)})=N^{RuZr}(y^{(0)})/N^{ZrRu}(y^{(0)})$, 
where $N^{i}(y^{(0)})$ is the particle yield  
at a given rapidity for $Ru+Zr$ and $Zr+Ru$ with $i=RuZr,~ZrRu$. The 
observable $R$ is measured for different particle species, 
like protons, neutrons, light fragments such as $t$ and ${}^{3}He$ and 
produced particles such as pions ($\pi^{0,\pm}$), etc. It characterizes 
different stopping scenarios.

In the proton case, moving from
target to $cm$ rapidity, $R(p)$ rises 
(positive slope) 
for partial transparency, 
falls (negative slope) for full rebound scenarios and it is flat 
when full stopping and total isospin mixing is achieved in the collision.
An opposite behavior will appear for neutrons. 
Indeed we recall that in the full transparency limit $R$ should 
have the value of 
$R(p)=Z^{Zr}/Z^{Ru}=40/44=0.91$ and $R(n)=N^{Zr}/N^{Ru}=56/52=1.077$ for 
protons and 
neutrons at target rapidity, respectively. 
Therefore, $R(p)$ can be regarded as a sensitive 
observable with respect to isospin diffusion, i.e. to properties of the 
symmetry term. 

For the investigation of isospin equilibration  we have 
analyzed transport results obtained in the same way as carried out in the 
$FOPI$ experiment, refs.\cite{Bormio04,GaitanosPLB595}.
In particular, central events are selected through the observable $ERAT$, 
the ratio of the mean transverse to the mean longitudinal 
kinetic energy. For this a phase space coalescence algorithm 
for fragment 
production in the final state is used, and the observable 
$ERAT$ is calculated for each fragment. 
A detailed 
description of this analysis can be found in ref.\cite{GaitanosEPJA12}, 
where it was shown 
that the  charged particle multiplicity 
and $ERAT$ distributions fit well the 
experimental data, as an important check of the phenomenological 
phase space coalescence model for fragment production. 
Thus the same centrality cuts
as in the $FOPI$ experiments \cite{HongPRC66} can be used.

\begin{figure}
\begin{center}
\includegraphics[scale=0.55]{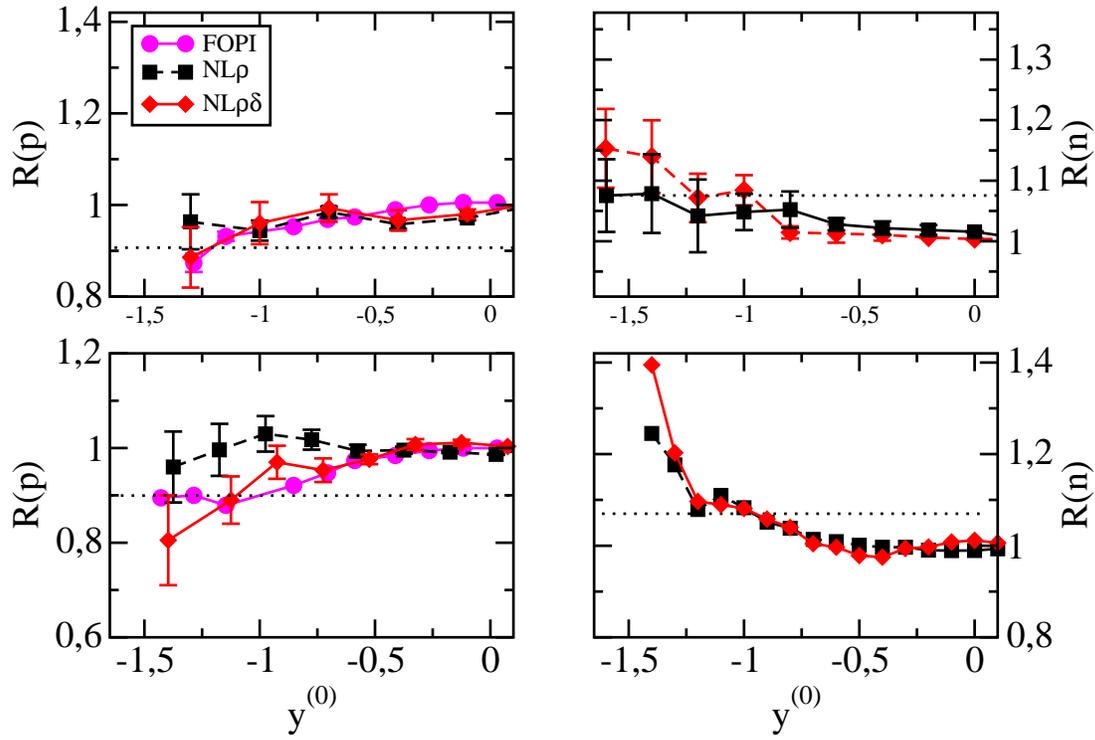}
\caption{\label{reldyn7} 
The imbalance ratio $R(y^{(0)})=\frac{N^{RuZr}(y^{(0)})}{N^{ZrRu}(y^{(0)})}$ 
as 
function of the normalized rapidity $y^{(0)}=y/y_{proj}$ of free protons 
(left panels) and free neutrons (right panels) for the $NL\rho$ 
and $NL\rho\delta$ models, for central ($b\leq 2~fm$) 
$Ru(Zr)+Zr(Ru)$-collisions 
at $0.4$ (top) and $1.528~AGeV$ (bottom) beam energies. 
The experimental data are taken from the $FOPI$ collaboration 
\cite{HongPRC66}. The ratio in the initial system is shown by the horizontal
dotted line. 
}
\end{center}
\end{figure}
\begin{figure}
\begin{center}
\includegraphics[scale=0.30]{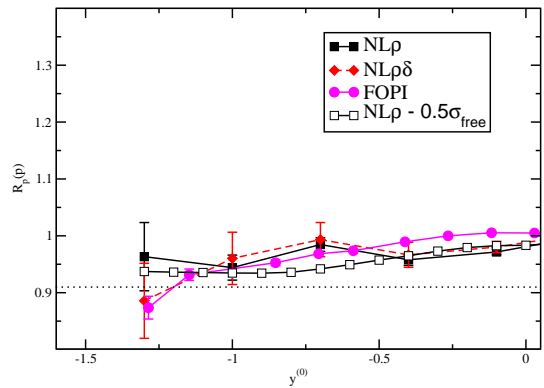}
\caption{\label{reldyn8} 
Proton imbalance ratio as 
function of the normalized rapidity for central ($b\leq 2~fm$) 
$Ru(Zr)+Zr(Ru)$-collisions 
at $0.4~AGeV$ as in the top-left part of Fig.\ref{reldyn7}.
A curve is added (empty squares) corresponding to a $NL\rho$ calculation
with reduced nucleon-nucleon cross sections (half the free values,
 $\sigma_{NN}(E) = \frac{1}{2} \sigma_{free}$).  
}
\end{center}
\end{figure}
In Fig. \ref{reldyn7}
 we show the rapidity dependence of 
the imbalance ratio
for free protons ($R(p)$) and 
free neutrons ($R(n)$) at the two energies
$0.4$ and $1.528~AGeV$.
The imbalance ratio approaches 
unity at mid-rapidity for all particle types due to  
symmetry 
in the mid-rapidity region, as expected in central collisions. 

Going from target- to mid-rapidity the ratio nicely rises for protons,
 and decreases for neutrons, a good signature of isospin transparency. 
The effect is more evident for the $NL\rho\delta$ interaction.
 The observed difference between 
the two models can be understood  
since within the $NL\rho\delta$ picture 
neutrons experience
a more repulsive iso-vector mean field, particularly at high densities, than 
 protons, and consequently much less nucleon stopping in the 
colliding system.

However the influence of the iso-$EOS$ on the imbalance ratio of protons 
and neutrons is 
not very large. At low intermediate energies ($0.4~AGeV$) 
one deals with moderate compressions of $\rho_{B} < 2 \cdot \rho_{sat}$ 
where the differences in the iso-vector $EOS$ arising from 
the $\delta$ meson are small. 
At higher 
incident energies ($1.528~AGeV$, Fig.\ref{reldyn7}-bottom)  
where a larger effect is expected,  
 we actually see a slightly higher 
isospin effect on the imbalance ratios, at least for protons.  
In fact with increasing beam energy the 
opening of inelastic channels via the 
production/decay of $\Delta$ resonances through pions,
as discussed in the last section,  
also contribute to the final result. 
This interpretation is 
confirmed by other studies \cite{BaoPRC67}. 

Reduced in-medium Nucleon-Nucleon ($NN$) cross sections, in particular
$\sigma_{np}$, will also increase the isospin transparency.
This possibility can be investigated considering a factor
of two reduction, $\sigma = \frac{1}{2} \sigma_{free}$, of the free 
 $NN$-cross section values used before, \cite{GaitanosPLB595}. We note 
that such a reduction
represents rather an upper limit of in-medium effects as compared to
recent microscopic Dirac-Brueckner estimations \cite{FuchsPRC64}.
In Fig.\ref{reldyn8} we show the results for the proton 
imbalance ratios at $0.4~AGeV$ with $0.5\sigma_{free}$ in the $NL\rho$
case. We see an overall slightly increased transparency but not enough
to reproduce the trend of the experimental values in the target rapidity 
region. On the other hand the reduction of the $NN$ cross sections,
and in particular of $\sigma_{(np)}$, leads too large a
transparency in the proton rapidity distributions for central collisions
of the charge symmetric $Ru~+~Ru$ case; as is seen in our
simulations and already remarked in previous $IQMD$ calculations,
\cite{HongPRC66}. Thus can exclude further reductions
of the $NN$ cross sections.

Since the calculations are performed with the same $EOS$ for the
symmetric nuclear matter, the same compressibility and momentum dependence,
the observed transparency appears to be uniquely
related to $isovector-EOS$ effects, i.e. to the isospin dependence
of the nucleon self-energies at high baryon densities. 
\begin{figure}[t]
\begin{center}
\includegraphics[scale=0.30]{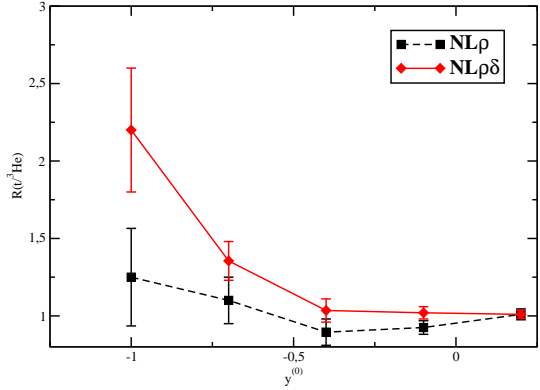}
\caption{Imbalance ratio of $t$ to ${}^{3}He$ for the 
same collision as in Fig.~\ref{reldyn7} for $0.4~AGeV$ beam energy. 
}
\label{reldyn9}
\end{center}
\end{figure}
\begin{figure}[t]
\begin{center}
\includegraphics[scale=0.30]{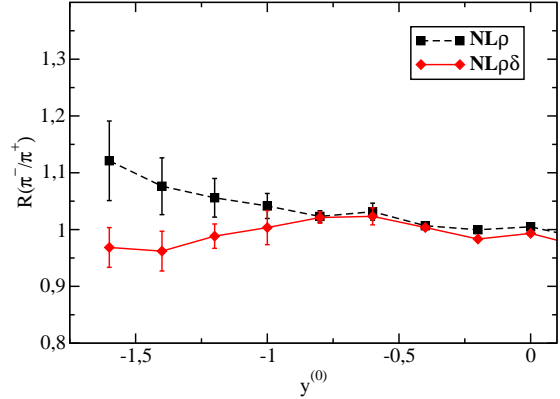}
\caption{ Imbalace ratio for the ratio of negative 
to positive charged pions for the 
same collision as in Fig.~\ref{reldyn7} for $0.4~AGeV$ beam energy. 
}
\label{reldyn10}
\end{center}
\end{figure}
The fact that protons {\it and} neutrons
exhibit an {\it opposite behavior} for the imbalance 
ratios at target 
rapidity suggests that the detection of the imbalance observable 
$R(t/{}^{3}He)$, i.e. the double ratio of the 
$t/{}^{3}He$ yield should reveal a larger sensitivity. 
The correct and practical 
method how to properly describe light fragment formation is still 
controversial. We report here on results obtained in ref.
\cite{GaitanosPLB595} with 
simplest algorithm, 
namely a phase space coalescence model \cite{GaitanosEPJA12}. 

Fig. \ref{reldyn9} shows the imbalance ratio $R(t/{}^{3}He)$ 
for central 
collisions at $0.4~AGeV$ incident energy. The isospin effect 
from the 
inclusion of the iso-vector, scalar $\delta$ meson in the 
$NL\rho\delta$ model 
is found to be very large near target rapidities. We note that the 
effect indeed can be 
hardly seen from the separate imbalance ratios for protons and neutrons 
at the same 
rapidity (see Fig. \ref{reldyn7} top panels) apart from the difficulties 
of neutron detections. 
It would be therefore of great interest to experimentally measure 
directly this quantity.

Finally, another sensitive observable should be, from the
discussion in the last Section, 
the imbalance 
ratio of charged pions 
$R(\pi^{-}/\pi^{+})$ (Fig. \ref{reldyn10}). 
At variance with the previous results for neutrons and light isobars, 
this ratio is reduced at target rapidity with the $NL\rho\delta$ model.
This effect is consistent with our understanding of the $\pi^{+}/\pi^{-}$-
ratios.  
Pions are produced from the decay of $\Delta$ resonances formed
during the high density phase, see Fig.\ref{reldyn5}. 
The $\pi^{-}$ abundance is then linked to the neutron-excess of the 
high density matter,  
as discussed in the previous Section. We recall that 
the contribution 
of the $\delta$ 
meson leads to a more repulsive field for neutrons at supra-normal densities 
and consequently to less neutron collisions and finally to  
a smaller $\pi^{-}/\pi^{+}$ ratio. 

We have analyzed the isospin 
transparency in relativistic collisions.
We have observed that this observable is sensitive to the microscopic Lorentz
structure of the symmetry term. Effective interactions with symmetry energies
which are 
not much different at 2-3 times normal density $\rho_{sat}$ 
predict large differences in the $isospin-transparencies$, depending on
the relative contribution of the various charged vector and scalar fields.
Intermediate energy heavy-ion collisions
with radioactive beams can give information on the symmetry energy 
at high baryon density and on its detailed microscopic structure. 
We have shown that such experiments provide
a unique tool to investigate the strength of the $\delta-like$ field.
The sensitivity is enhanced relative to the static property $E_{sym}(\rho_B)$ 
because of the covariant nature of the fields
involved in collision dynamics.

%\include{rep_bib}
%\end{document}

%% file: Chapter-9.tex
%\documentclass{elsart}
%\usepackage{epsfig}

%\usepackage{graphicx}

%\usepackage{amssymb}
%\tightenlines

% nuovi comandi
% 2 su 2pigreco al cubo
%\newcommand{\norm}{\frac{2}{(2\pi)^3}}
% parentesi quadre
%\newcommand{\qd}[1]{\left[ #1 \right]}
% parentesi tonde
%\newcommand{\td}[1]{\left( #1 \right)}

%\newcommand{\itg}[1]{\norm \int d^3k f_{#1}(k)}
% integrale di fn
%\newcommand{\ienne}[1]{\itg{n} #1}
% integrale di fp
%\newcommand{\ipi}[1]{\itg{p} #1}
% integrale di fn g
%\newcommand{\ien}{\ienne{g(k,\Lambda)}}
% integrale di fp g
%\newcommand{\iz}{\ipi{g(k,\Lambda)}}

%\setlength{\unitlength}{1cm}

% integrale di ftau g
%\newcommand{\itau}{\itg{\tau}g(k,\Lambda)}
% integrale di ftau g
%\newcommand{\itaup}{\itg{\tau ^\prime} g(k,\Lambda)}

% rozero
%\newcommand{\rz}{\rho_{_0}}
% rho su rozero
%\newcommand{\ra}{\td{ \frac{\rho}{\rho_{_0}} }}
% densita' di energia per A e B
%\newcommand{\ene}[1]{\qd{
% \td{\frac{1}{2}x_{#1}}\rho^2
%-\td{\frac{1}{2}+x_{#1}} \td{\rho_n^2+\rho_p^2} }}
% 0.5 + x0(x3)
%\newcommand{\umd}[1]{ \td{ \frac{1}{2}+x_{#1} } }

%\newcommand{\inew}[1]{\mathcal{I}_#1 }

%\begin{document}

\setcounter{figure}{0}
\setcounter{equation}{0}
\section{Conclusion and outlook}\label{out}

\markright{Chapter \arabic{section}: out}

Nuclear reactions with neutron-rich (or radioactive) nuclei have opened the
possibility to learn about the behaviour of the nuclear interaction and, in
particular, of the symmetry energy in a wide spectrum of conditions of
density and temperature.  This study appears extremely important: indeed
a deeper understanding of the behaviour of neutrons and protons in a
charge asymmetric nuclear medium is essential to test and to extend 
our present knowledge of the nuclear interaction and is of highest 
importance for the modelling of astrophysical processes, like supernova 
explosion or neutron stars.

In the context of nuclear reactions, one has to identify the most sensitive
observables that may provide information on the behaviour of the symmetry 
energy in several conditions of density and temperature. 
This is the main line followed along this Report. 

In Section \ref{eos} we discuss 
the symmetry energy dependence around normal density
and how this can affect important properties of neutron-rich nuclei, such as
compressibility (and monopole frequency), saturation density and neutron 
skin. 

The behaviour of asymmetric matter at low density has been investigated in
Section \ref{rpa}. We discuss in 
particular the phase diagram of asymmetric matter 
and the relevant features of instabilities. 
This subject appears important in connection to the possibility to observe a
liquid-gas phase transition in violent heavy ion reactions, where, after the
initial collisional shock, low density regions can be easily reached during
the expansion phase. New important features, such as the isospin distillation
effect, are predicted and some experimental evidences have already 
appeared along this direction, though a more careful analysis is still needed, 
to disantangle among other possible contributions to the distillation, 
such as pre-equilibrium effects, and to really prove the mechanism driving 
the fragmentation process. 

In Section \ref{fastflows} we have focused on features of the early stage of
the reaction dynamics between neutron-rich nuclei. 
Pre-equilibrium emission and collective flows appear particularly sensitive 
also to the the momentum-dependent part of the interaction. 
In asymmetric matter a splitting of neutron and proton effective
masses is observed. The sign of the splitting is quite controversial, since
the behaviour of neutron and proton optical potential at large energy 
has not been experimentally measured yet. 
Hence it appears very important to try to extract information on this 
fundamental question from nuclear collisions, where one can use 
probes, such as pre-equilibrium particles, particularly sensitive to
the high density phase, where also high momenta are reached.   
 
In Section \ref{fermi}, we have explored several 
fragmentation mechanisms, occurring
at the Fermi energies, in the framework of a stochastic mean-field approach.  
We discuss the features of multifragmentation in neutron-rich systems, 
and in particular the isotopic content of fragments. This can be connected
to the behaviour (the slope) of the symmetry energy at low density. 
For semi-peripheral reactions an interesting neutron enrichment of the
overlap (``neck'') region appears, due to the neutron migration from 
higher (spectator regions) to lower (neck) density regions. Also this effect
is  nicely connected to the slope of the symmetry energy. 
A careful comparison with experimental data would give important indications
on the fragmentation mechanism and on the behaviour of the symmetry energy. 

In sections 6-7 we have discussed static properties and 
dynamical mechanisms in the context of relativistic approaches
including isovector, both vector and scalar, channels 
(the $\rho$ and $\delta$ mesons). 
The contribution of these two channels to static or dynamical properties
has a different weight, leading to interesting effects on the frequency
of the stable isovector modes, that is not simply related to the value of the
symmetry energy (see Sections \ref{qhd}, \ref{relin}). 
In the low density region, instabilities and the distillation mechanims are 
observed, in close parallelism with the results of the non-relativistic
treatment discussed above. 

Finally in Section \ref{reldyn} we discuss isospin 
effects in relativistic heavy ion 
collisions. 
The observable consequences of the inclusion of the  $\delta$ meson are
enhanced by the Lorentz structure of the effective nuclear interaction in 
the isovector channel. 
In particular, effects %of the inclusion of the $\delta$ meson
on particle production, such as pions or kaons, 
light particle collective flows and
isospin diffusion are investigated. 
Once again, from the comparison with experimental data,   
one could get important information on the structure of the isovector 
interaction.

\subsection{Outlook: The Eleven Observables}
In conclusion, 
the study of isospin effects on static and dynamical nuclear properties
appears as a very rich and stimulating field.
Several probes can be used to get an insight on the behaviour of the
symmetry energy in different conditions of density and temperature.
A joint effort, from the experimental and theoretical side, should allow
to extract relevant information on fundamental properties of the nuclear
interaction. 
We like to suggest a selection
of {\it Eleven Observables}, from low to relativistic energies, that 
we expect particularly sensitive
to the microscopic structure of the {\it in medium }interaction
in the isovector channel, i.e. to the symmetry energy and its
``fine structure'':

\noindent
{\it 1. Competition of Reaction Mechanisms}. Interplay of low-energy
 dissipative mechanisms, e.g. fusion (incomplete) vs.
 deep-inelastic vs. neck fragmentation: a stiff symmetry term leads to
 a  more repulsive dynamics.(Sect.\ref{fermi})

\noindent
{\it 2. Energetic particle production}. N/Z of fast nucleon emission: 
symmetry repulsion of the
 neutron/proton mean field in various density regions. Moreover at the Fermi
energies we expect to see also effects from the $n/p$ splitting of
the effective masses. Even the spectra and yields of hard photons
produced via $(n,p)$ bremmstrahlung should be sensitive to the
density and momentum dependence of the symmetry fields.(Sect.\ref{fastflows})

\noindent
{\it 3. Neutron/Proton correlation functions}. Time-space structure
 of the fast particle emission and its relation to the baryon density
 of the source. Again combined effects of density and momentum
dependence of the symmetry term are expected (Sect.\ref{fastflows}). 

\noindent
{\it 4. E-slope of the Lane Potential}. A systematic study of the
energy dependence of the $(n/p)$ optical potentials on asymmetric nuclei
will shed lights on the effective mass splitting, at least around
normal density.(Sects.\ref{eos} and \ref{qhd})

\noindent
{\it 5. Isospin Distillation (Fractionation)}. Isospin content
of the Intermediate Mass Fragments in central collisions. Test of 
the symmetry term in dilute matter and connection to 
the possibility to observe a liquid-gas phase transition.(Sect.\ref{fermi})

\noindent
{\it 6. Properties of Neck-Fragments}. Mid-rapidity $IMF$ produced
in semicentral collisions: correlations between $N/Z$, alignement and
size. Isospin effects on the reaction dynamics and ``Isospin
 Migration''.(Sect.\ref{fermi}) 

\noindent
{\it 7. Isospin Diffusion}. Measure of charge equilibration
in the ``spectator'' region in semicentral collisions. Test of
the interplay between concentration and density gradients in
the isospin dynamics.(Sect.\ref{fermi})

\noindent
{\it 8. Neutron-Proton Collective Flows}. Together with
 light isobar flows. Check of symmetry transport effects.
Test of the momentum dependence (relativistic structure) of the
interaction in the isovector channel.
 Measurements also for different $p_t$ selections. 
 (Sects.\ref{fastflows} and \ref{reldyn})

\noindent
{\it 9. Isospin Transparency}. Measure of isospin properties in
Projectile/Target rapidity regions in central collisions
of ``mirror'' ions at intermediate energies. Similar effects, but due to
the nucleon-nucleon cross section are expected to be smaller
(Sect.\ref{reldyn}).

\noindent
{\it 10. $\pi^-/\pi^+$ Yields}. Since $\pi^-$ are mostly produced
in $nn$ collisions we can expect a reduction for highly repulsive
symmetry terms at high baryon density. Importance of a $p_t$ selection.
Similar studies for mesons with smaller rescattering effects
would be of great interest. (Sect.\ref{reldyn})

\noindent
{\it 11. Deconfinement Precursors}. Signals of a mixed phase formation
 (quark-bubbles) in high baryon density regions reached with asymmetric 
 $HIC$ at intermediate energies, versus the properties of the 
interaction considered (Sect.\ref{reldyn}).

We stress again the richness of the phenomenology and nice opportunities of
getting several cross-checks from completely different experiments.

For the points $3,~5,~6,~7,~8,~9,~10$
from the transport simulations discussed here we presently get
some indications of {\it asy-stiff} behaviors, i.e. increasing
repulsive density dependence of the symmetry term, but not more
fundamental details. Moreover all the available data are obtained
with stable beams, i.e. within low asymmetries.

\vskip 1.0cm
{\bf Acnowledgements}
\vskip 0.5cm
This report is deeply related to ideas and results partially reached in
very pleasant and fruitful collaborations with very nice people:
L.W.Chen, G.Fabbri, 
G.Ferini, Th.Gaitanos, C.M.Ko, A.B.Larionov, B.A.Li, R.Lionti, 
B.Liu, S.Maccarone, F.Matera, M.Zielinska-Pfabe', J.Rizzo, L.Scalone 
and H.H.Wolter. We have learnt a lot 
from all of them in physics as well as in human relationships.

We are particularly grateful to H.H.Wolter for many stimulating suggestions
during the preparation of the report and for a very accurate reading of the
manuscript.

Finally we warmly thank our experimental colleagues for the richness of
mutual interaction and for the exciting possibility of discussing new
results even before publication. We like to mention in particular the
Medea/Multics and Chimera Collaboration at the LNS-Catania and the
FOPI Collaboration at the GSI-Darmstadt. 
%very special thanks are due
%to P.Sapienza, E.De Filippo, A.Pagano, J.Wilczynski and W.Reisdorf.

%\end{document}

%% file: Bibliography.tex
%\documentclass{elsart}
%\usepackage{epsfig}

%\usepackage{graphicx}

%\usepackage{amssymb}

%\begin{document}

%\end{document}

%% file: report_1104.bbl
\begin{thebibliography}{00}
\bibitem{bred} The reduced impact parameter is defined as 
$b_{red} \equiv b/b_{max}$ where $b_{max}$ is the sum of the 
 two nuclear radii.
\bibitem{Curvature} We note that in the non-relativistic case the
kinetic contribution to $K_{sym}$ is actually negative, since it has
the form $-\pi^2\rho_0/(Mk_F)$ which is of the order $\sim -30 MeV$. 
In the relativistic 
case it becomes slightly positive due to the density dependence of the
nucleon effective mass.
\bibitem{Rhos} The solid lines in Fig.4 are not fully symmetric with
respect to the $\alpha=0$ dashed line. This is due to the fact that,
for fixed baryon density, the total scalar density which gives
the $\sigma$ field contribution (first part of the Eq.(6)) is
also slightly $N/Z$ dependent.
\bibitem{phinote} We recall that, when the $\phi$ field has self-interaction 
terms, instead
of $f_\sigma$ one has to use $\frac{d\phi}{d\rho_S}$.
\bibitem{masses}
In our framework this is a characteristic relativistic effect that should
be somewhat included in the corresponding non-relativistic models. 
The importance of this contribution was not realized so far in 
previous non-relativistic studies \cite{BaoPRC67}, where no isospin effects
were included in the momentum dependent part of the effective forces. 
In this case the differences
in the used symmetry 
energy parametrizations at supra-normal densities were very 
pronounced and therefore they
resulted in very large 
effects with respect to earlier neutron emission and particle production.
\end{thebibliography}
